# Applied Conformal Field Theory

Paul Ginsparg[†]

Lyman Laboratory of Physics
Harvard University
Cambridge, MA 02138

# Applied Conformal Field Theory

Les Houches Lectures, P. Ginsparg, 1988

**Contents**

These lectures consisted of an elementary introduction to conformal field theory, with some applications to statistical mechanical systems, and fewer to string theory. They were given to a mixed audience of experts and beginners (more precisely an audience roughly 35% of which was alleged to have had no prior exposure to conformal field theory, and a roughly equal percentage alleged to be currently working in the field), and geared in real time to the appropriate level. The division into sections corresponds to the separate (1.5 hour) lectures, except that 7 and 8 together stretched to three lectures, and I have taken the liberty of expanding some rushed comments at the end of 9.

It was not my intent to be particularly creative in my presentation of the material, but I did try to complement some of the various introductory treatments that already exist. Since these lectures were given at the beginning of the school, they were intended to be more or less self-contained and generally accessible. I tried in all cases to emphasize the simplest applications, but not to duplicate excessively the many review articles that already exist on the subject.

More extensive applications to statistical mechanical models may be found in J. Cardy's lectures in this volume, given concurrently, and many string theory applications of conformal field theory were covered in D. Friedan's lectures, which followed. The standard reference for the material of the first three sections is [1]. Some of the review articles that have influenced the presentation of the early sections are listed in [2]. A more extensive (physicist-oriented) review of affine Kac-Moody algebras, discussed here in section 9, may be found in [3]. Throughout I have tried to include references to more recent papers in which the interested reader may find further references to original work. Omitted references to relevant work are meant to indicate my prejudices rather than my ignorance in the subject.

I am grateful to the organizers and students at the school for insisting on the appropriate level of pedagogy and for their informative questions, and to P. di Francesco and especially M. Goulian for most of the answers. I thank numerous participants at the conformal field theory workshop at the Aspen Center for Physics (Aug., 1988) for comments on the manuscript, and thank S. Giddings, G. Moore, R. Plesser, and J. Shapiro for actually reading it. Finally I acknowledge the students at Harvard who patiently sat through a dry run of this material (and somewhat more) during the spring of 1988. This work was supported in part by NSF contract PHY-82-15249, by DOE grant FG-84ER40171, and by the A. P. Sloan foundation.

# 1. Conformal theories in $d$ dimensions

Conformally invariant quantum field theories describe the critical behavior of systems at second order phase transitions. The canonical example is the Ising model in two dimensions, with spins $\sigma_i = \pm 1$ on sites of a square lattice. The partition function $Z = \sum_{\{\sigma\}} \exp(-E/T)$ is defined in terms of the energy $E = -\epsilon \sum_{\langle ij \rangle} \sigma_i \sigma_j$, where the summation $\langle ij \rangle$ is over nearest neighbor sites on the lattice. This model has a high temperature disordered phase (with the expectation value $\langle \sigma \rangle = 0$) and a low temperature ordered phase (with $\langle \sigma \rangle \neq 0$). The two phases are related by a duality of the model, and there is a $2^{\text{nd}}$ order phase transition at the self-dual point. At the phase transition, typical configurations have fluctuations on all length scales, so the field theory describing the model at its critical point should be expected to be invariant at least under changes of scale. In fact, critical theories are more generally invariant under the full conformal group, to be introduced momentarily. In three or more dimensions, conformal invariance does not turn out to give much more information than ordinary scale invariance. But in two dimensions, the conformal algebra becomes infinite dimensional, leading to significant restrictions on two dimensional conformally invariant theories, and perhaps ultimately giving a classification of possible critical phenomena in two dimensions.

Two dimensional conformal field theories also provide the dynamical variable in string theory. In that context conformal invariance turns out to give constraints on the allowed spacetime (i.e. critical) dimension and the possible internal degrees of freedom. A classification of two dimensional conformal field theories would thus provide useful information on the classical solution space of string theory, and might lead to more propitious quantization schemes.

## 1.1. Conformal group in $d$ dimensions

We begin here with an introduction to the conformal group in $d$-dimensions. The aim is to exhibit the constraints imposed by conformal invariance in the most general context. In section 2 we shall then restrict to the case of two dimensional Euclidean space, which will be the focus of discussion for the remainder.

We consider the space $\mathbf{R}^d$ with flat metric $g_{\mu\nu} = \eta_{\mu\nu}$ of signature $(p, q)$ and line element $ds^2 = g_{\mu\nu} dx^\mu dx^\nu$. Under a change of coordinates, $x \to x'$, we have $g_{\mu\nu} \to g'_{\mu\nu}(x') = \frac{\partial x^\alpha}{\partial x'^\mu} \frac{\partial x^\beta}{\partial x'^\nu} g_{\alpha\beta}(x)$. By definition, the conformal group is the subgroup of coordinate transformations that leaves the metric invariant up to a scale change,

$$g_{\mu\nu}(x) \to g'_{\mu\nu}(x') = \Omega(x) g_{\mu\nu}(x) \ . \qquad (1.1)$$

These are consequently the coordinate transformations that preserve the angle $v \cdot w / (v^2 w^2)^{1/2}$ between two vectors $v, w$ (where $v \cdot w = g_{\mu\nu} v^\mu w^\nu$). We note that the Poincaré group, the semidirect product of translations and Lorentz transformations of flat space, is always a subgroup of the conformal group since it leaves the metric invariant ($g'_{\mu\nu} = g_{\mu\nu}$).

The infinitesimal generators of the conformal group can be determined by considering the infinitesimal coordinate transformation $x^\mu \to x^\mu + \epsilon^\mu$, under which

$$ds^2 \to ds^2 + (\partial_\mu \epsilon_\nu + \partial_\nu \epsilon_\mu) dx^\mu dx^\nu \ .$$

To satisfy (1.1) we must require that $\partial_\mu \epsilon_\nu + \partial_\nu \epsilon_\mu$ be proportional to $\eta_{\mu\nu}$,

$$\partial_\mu \epsilon_\nu + \partial_\nu \epsilon_\mu = \frac{2}{d}(\partial \cdot \epsilon)\eta_{\mu\nu} \ , \qquad (1.2)$$

where the constant of proportionality is fixed by tracing both sides with $\eta^{\mu\nu}$. Comparing with (1.1) we find $\Omega(x) = 1 + (2/d)(\partial \cdot \epsilon)$. It also follows from (1.2) that

$$\bigl(\eta_{\mu\nu}\Box + (d-2)\partial_\mu\partial_\nu\bigr)\partial \cdot \epsilon = 0 \ . \qquad (1.3)$$

For $d > 2$, (1.2) and (1.3) require that the third derivatives of $\epsilon$ must vanish, so that $\epsilon$ is at most quadratic in $x$. For $\epsilon$ zeroth order in $x$, we have

a) $\epsilon^\mu = a^\mu$, i.e. ordinary translations independent of $x$.

There are two cases for which $\epsilon$ is linear in $x$:

b) $\epsilon^\mu = \omega^\mu{}_\nu x^\nu$ ($\omega$ antisymmetric) are rotations,

and

c) $\epsilon^\mu = \lambda x^\mu$ are scale transformations.

Finally, when $\epsilon$ is quadratic in $x$ we have

d) $\epsilon^\mu = b^\mu x^2 - 2x^\mu b \cdot x$, the so-called special conformal transformations. (these last may also be expressed as $x'^\mu/x'^2 = x^\mu/x^2 + b^\mu$, i.e. as an inversion plus translation). Locally, we can confirm that the algebra generated by $a^\mu \partial_\mu$, $\omega^\mu{}_\nu \epsilon^\nu \partial_\mu$, $\lambda x \cdot \partial$, and $b^\mu(x^2 \partial_\mu - 2x^\mu x \cdot \partial)$ (a total of $p + q + \frac{1}{2}(p+q)(p+q-1) + 1 + (p+q) = \frac{1}{2}(p+q+1)(p+q+2)$ generators) is isomorphic to $SO(p+1, q+1)$ (Indeed the conformal group admits a nice realization acting on $\mathbf{R}^{p,q}$, stereographically projected to $S^{p,q}$, and embedded in the light-cone of $\mathbf{R}^{p+1,q+1}$.).

Integrating to finite conformal transformations, we find first of all, as expected, the Poincaré group

$$\begin{aligned} x \to x' &= x + a \\ x \to x' &= \Lambda x \quad (\Lambda^\mu{}_\nu \in SO(p,q)) \end{aligned} \qquad (\Omega = 1) \ . \qquad (1.4a)$$

Adjoined to it, we have the dilatations

$$x \to x' = \lambda x \qquad (\Omega = \lambda^{-2}) \ , \qquad (1.4b)$$

and also the special conformal transformations

$$x \to x' = \frac{x + bx^2}{1 + 2b \cdot x + b^2 x^2} \qquad \bigl(\Omega(x) = (1 + 2b \cdot x + b^2 x^2)^2\bigr) \ . \qquad (1.4c)$$

Note that under (1.4c) we have $x'^2 = x^2/(1 + 2b \cdot x + b^2 x^2)$, so that points on the surface $1 = 1 + 2b \cdot x + b^2 x^2$ have their distance to the origin preserved, whereas points on the exterior of this surface are sent to the interior and vice-versa. (Under the finite transformation (1.4c) we also continue to have $x'^\mu/x'^2 = x^\mu/x^2 + b^\mu$.)

*1.2. Conformal algebra in 2 dimensions*

For $d = 2$ and $g_{\mu\nu} = \delta_{\mu\nu}$, (1.2) becomes the Cauchy-Riemann equation

$$\partial_1 \epsilon_1 = \partial_2 \epsilon_2 \ , \qquad \partial_1 \epsilon_2 = -\partial_2 \epsilon_1 \ .$$

It is then natural to write $\epsilon(z) = \epsilon^1 + i\epsilon^2$ and $\overline{\epsilon}(\overline{z}) = \epsilon^1 - i\epsilon^2$, in the complex coordinates $z, \overline{z} = x^1 \pm ix^2$. Two dimensional conformal transformations thus coincide with the analytic coordinate transformations

$$z \to f(z) \ , \qquad \overline{z} \to \overline{f}(\overline{z}) \ , \qquad (1.5)$$

the local algebra of which is infinite dimensional. In complex coordinates we write

$$ds^2 = dz\, d\overline{z} \to \left|\frac{\partial f}{\partial z}\right|^2 dz\, d\overline{z} \ , \qquad (1.6)$$

and have $\Omega = |\partial f/\partial z|^2$.

To calculate the commutation relations of the generators of the conformal algebra, i.e. infinitesimal transformations of the form (1.5), we take for basis

$$z \to z' = z + \epsilon_n(z) \qquad \overline{z} \to \overline{z}' = \overline{z} + \overline{\epsilon}_n(\overline{z}) \qquad (n \in \mathbf{Z}) \, ,$$

where

$$\epsilon_n(z) = -z^{n+1} \qquad \overline{\epsilon}_n(\overline{z}) = -\overline{z}^{m+1} \, .$$

The corresponding infinitesimal generators are

$$\ell_n = -z^{n+1} \partial_z \qquad \overline{\ell}_n = -\overline{z}^{n+1} \partial_{\overline{z}} \qquad (n \in \mathbf{Z}) \, . \qquad (1.7)$$

The $\ell$'s and $\overline{\ell}$'s are easily verified to satisfy the algebras

$$[\ell_m, \ell_n] = (m-n)\ell_{m+n} \qquad [\overline{\ell}_m, \overline{\ell}_n] = (m-n)\overline{\ell}_{m+n} \, , \qquad (1.8)$$

and $[\ell_m, \overline{\ell}_n] = 0$. In the quantum case, the algebras (1.8) will be corrected to include an extra term proportional to a central charge. Since the $\ell_n$'s commute with the $\overline{\ell}_m$'s, the local conformal algebra is the direct sum $\mathcal{A} \oplus \overline{\mathcal{A}}$ of two isomorphic subalgebras with the commutation relations (1.8).

Since two independent algebras naturally arise, it is frequently useful to regard $z$ and $\overline{z}$ as independent coordinates. (More formally, we would say that since the action of the conformal group in two dimensions factorizes into independent actions on $z$ and $\overline{z}$, Green functions of a 2d conformal field theory may be continued to a larger domain in which $z$ and $\overline{z}$ are treated as independent variables.) In terms of the original coordinates $(x^1, x^2) \in \mathbf{R}^2$, this amounts to taking instead $(x^1, x^2) \in \mathbf{C}^2$, and then the transformation to $z, \overline{z}$ coordinates is just a change of variables. In $\mathbf{C}^2$, the surface defined by $\overline{z} = z^*$ is the 'real' surface on which we recover $(x, y) \in \mathbf{R}^2$. This procedure allows the algebra $\mathcal{A} \oplus \overline{\mathcal{A}}$ to act naturally on $\mathbf{C}^2$, and the 'physical' condition $\overline{z} = z^*$ is left to be imposed at our convenience. The real surface specified by this condition is preserved by the subalgebra of $\mathcal{A} \oplus \overline{\mathcal{A}}$ generated by $\ell_n + \overline{\ell}_n$ and $i(\ell_n - \overline{\ell}_n)$. In the sections that follow, we shall frequently use the independence of the algebras $\mathcal{A}$ and $\overline{\mathcal{A}}$ to justify ignoring anti-holomorphic dependence for simplicity, then reconstruct it afterwards by adding terms with bars where appropriate.

We have been careful thus far to call the algebra (1.8) the local conformal algebra. The reason is that the generators are not all well-defined globally on the Riemann sphere $S^2 = \mathbf{C} \cup \infty$. Holomorphic conformal transformations are generated by vector fields

$$v(z) = -\sum_n a_n \ell_n = \sum_n a_n z^{n+1} \partial_z \, .$$

Non-singularity of $v(z)$ as $z \to 0$ allows $a_n \neq 0$ only for $n \geq -1$. To investigate the behavior of $v(z)$ as $z \to \infty$, we perform the transformation $z = -1/w$,

$$v(z) = \sum_n a_n \left(-\frac{1}{w}\right)^{n+1} \left(\frac{dz}{dw}\right)^{-1} \partial_w = \sum_n a_n \left(-\frac{1}{w}\right)^{n-1} \partial_w \, .$$

Non-singularity as $w \to 0$ allows $a_n \neq 0$ only for $n \leq 1$. We see that only the conformal transformations generated by $a_n \ell_n$ for $n = 0, \pm 1$ are globally defined. The same considerations apply to anti-holomorphic transformations.

In two dimensions the global conformal group is defined to be the group of conformal transformations that are well-defined and invertible on the Riemann sphere. It is thus generated by the globally defined infinitesimal generators $\{\ell_{-1}, \ell_0, \ell_1\} \cup \{\overline{\ell}_{-1}, \overline{\ell}_0, \overline{\ell}_1\}$. From (1.7) and (1.4) we identify $\ell_{-1}$ and $\overline{\ell}_{-1}$ as generators of translations, $\ell_0 + \overline{\ell}_0$ and $i(\ell_0 - \overline{\ell}_0)$ respectively as generators of dilatations and rotations (i.e. generators of translations of $r$ and $\theta$ in $z = re^{i\theta}$), and $\ell_1, \overline{\ell}_1$ as generators of special conformal transformations. The finite form of these transformations is

$$z \to \frac{az+b}{cz+d} \qquad \overline{z} \to \frac{\overline{a}\,\overline{z}+\overline{b}}{\overline{c}\,\overline{z}+\overline{d}} \, , \qquad (1.9)$$

where $a, b, c, d \in \mathbf{C}$ and $ad - bc = 1$). This is the group $SL(2, \mathbf{C})/\mathbf{Z}_2 \approx SO(3,1)$, also known as the group of projective conformal transformations. (The quotient by $\mathbf{Z}_2$ is due to the fact that (1.9) is unaffected by taking all of $a, b, c, d$ to minus themselves.) In $SL(2, \mathbf{C})$ language, the transformations (1.4) are given by

$$\text{translations}: \begin{pmatrix} 1 & B \\ 0 & 1 \end{pmatrix} \qquad \text{rotations}: \begin{pmatrix} e^{i\theta/2} & 0 \\ 0 & e^{-i\theta/2} \end{pmatrix}$$

$$\text{dilatations}: \begin{pmatrix} \lambda & 0 \\ 0 & \lambda^{-1} \end{pmatrix} \qquad \text{special conformal}: \begin{pmatrix} 1 & 0 \\ C & 1 \end{pmatrix} \, ,$$

where $B = a^1 + ia^2$ and $C = b^1 - ib^2$.

The distinction encountered here between global and local conformal groups is unique to two dimensions (in higher dimensions there exists only a global conformal group). Strictly speaking the only true conformal group in two dimensions is the projective (global) conformal group, since the remaining conformal transformations of (1.5) do not have global inverses on $\mathbf{C} \cup \infty$. This is the reason the word algebra rather than the word group appears in the title of this subsection.

The global conformal algebra generated by $\{\ell_{-1}, \ell_0, \ell_1\} \cup \{\overline{\ell}_{-1}, \overline{\ell}_0, \overline{\ell}_1\}$ is also useful for characterizing properties of physical states. Suppose we work in a basis of eigenstates of the two operators $\ell_0$ and $\overline{\ell}_0$, and denote their eigenvalues by $h$ and $\overline{h}$ respectively. Here $h$ and $\overline{h}$ are meant to indicate independent (real) quantities, not complex conjugates of one another. $h$ and $\overline{h}$ are known as the conformal weights of the state. Since $\ell_0 + \overline{\ell}_0$ and $i(\ell_0 - \overline{\ell}_0)$ generates dilatations and rotations respectively, the scaling dimension $\Delta$ and the spin $s$ of the state are given by $\Delta = h + \overline{h}$ and $s = h - \overline{h}$. In later sections, we shall generalize these ideas to the full quantum realization of the algebra (1.8).

### 1.3. Constraints of conformal invariance in d dimensions

We shall now return to the case of an arbitrary number of dimensions $d = p + q$ and consider the constraints imposed by conformal invariance on the $N$-point functions of a quantum theory. In what follows we shall prefer to employ the jacobian,

$$\left|\frac{\partial x'}{\partial x}\right| = \frac{1}{\sqrt{\det g'_{\mu\nu}}} = \Omega^{-d/2} , \qquad (1.10)$$

to describe conformal transformations, rather than directly the scale factor $\Omega$ of (1.1). For dilatations (1.4b) and special conformal transformations (1.4c), this jacobian is given respectively by

$$\left|\frac{\partial x'}{\partial x}\right| = \lambda^d \quad \text{and} \quad \left|\frac{\partial x'}{\partial x}\right| = \frac{1}{(1 + 2b \cdot x + b^2 x^2)^d} . \qquad (1.11)$$

We define a theory with conformal invariance to satisfy some straightforward properties:

1) There is a set of fields $\{A_i\}$, where the index $i$ specifies the different fields. This set of fields in general is infinite and contains in particular the derivatives of all the fields $A_i(x)$.

2) There is a subset of fields $\{\phi_j\} \subset \{A_i\}$, called "quasi-primary", that under global conformal transformations, $x \to x'$ (i.e. elements of $O(p+1, q+1)$), transform according to

$$\phi_j(x) \to \left|\frac{\partial x'}{\partial x}\right|^{\Delta_j/d} \phi_j(x') , \qquad (1.12)$$

where $\Delta_j$ is the dimension of $\phi_j$ (the $1/d$ compensates the exponent of $d$ in (1.10)). The theory is then covariant under the transformation (1.12), in the sense that the correlation functions satisfy

$$\begin{aligned}&\langle \phi_1(x_1) \ldots \phi_\nu(x_n) \rangle \\ &= \left|\frac{\partial x'}{\partial x}\right|^{\Delta_1/d}_{x=x_1} \cdots \left|\frac{\partial x'}{\partial x}\right|^{\Delta_n/d}_{x=x_n} \langle \phi_1(x'_1) \ldots \phi_n(x'_n) \rangle .\end{aligned} \qquad (1.13)$$

3) The rest of the $\{A_i\}$'s can be expressed as linear combinations of the quasi-primary fields and their derivatives.

4) There is a vacuum $|0\rangle$ invariant under the global conformal group.

The covariance property (1.13) under the conformal group imposes severe restrictions on 2- and 3-point functions of quasi-primary fields. To identify independent invariants on which $N$-point functions might depend, we construct some invariants of the conformal group in $d$ dimensions. Ordinary translation invariance tells us that an $N$-point function depends not on $N$ independent coordinates $x_i$, but rather only on the differences $x_i - x_j$ ($d(N-1)$ independent quantities). If we consider for simplicity spinless objects, then rotational invariance furthermore tells us that for $d$ large enough, there is only dependence on the $N(N-1)/2$ distances $r_{ij} \equiv |x_i - x_j|$. (As we shall see, for a given $N$-point function in low enough dimension, there will automatically be linear relations among coordinates that reduce the number of independent quantities.) Next, imposing scale invariance (1.4b) allows dependence only on the ratios $r_{ij}/r_{kl}$. Finally, since under the special conformal transformation (1.4c), we have

$$|x_1' - x_2'|^2 = \frac{|x_1 - x_2|^2}{(1 + 2b \cdot x_1 + b^2 x_1^2)(1 + 2b \cdot x_2 + b^2 x_2^2)} \;, \qquad (1.14)$$

only so-called cross-ratios of the form

$$\frac{r_{ij}\, r_{kl}}{r_{ik}\, r_{jl}} \qquad (1.15)$$

are invariant under the full conformal group. The number of independent cross-ratios of the form (1.15), formed from $N$ coordinates, is $N(N-3)/2$ [4]. (To see this, use translational and rotational invariance to describe the $N$ coordinates as $N-1$ points in a particular $N-1$ dimensional subspace, thus characterized by $(N-1)^2$ independent quantities. Then use rotational, scale, and special conformal transformations of the $N-1$ dimensional conformal group, a total of $(N-1)(N-2)/2 + 1 + (N-1)$ parameters, to reduce the number of independent quantities to $N(N-3)/2$.)

According to (1.13), the 2-point function of two quasi-primary fields $\phi_{1,2}$ in a conformal field theory must satisfy

$$\langle \phi_1(x_1)\, \phi_2(x_2) \rangle = \left| \frac{\partial x'}{\partial x} \right|_{x=x_1}^{\Delta_1/d} \left| \frac{\partial x'}{\partial x} \right|_{x=x_2}^{\Delta_2/d} \langle \phi_1(x_1')\, \phi_2(x_2') \rangle \;. \qquad (1.16)$$

Invariance under translations and rotations (1.4a) (for which the jacobian is unity) forces the left hand side to depend only on $r_{12} \equiv |x_1 - x_2|$. Invariance under the dilatations $x \to \lambda x$ then implies that

$$\langle \phi_1(x_1)\, \phi_2(x_2) \rangle = \frac{C_{12}}{r_{12}^{\Delta_1 + \Delta_2}} \;,$$

where $C_{12}$ is a constant determined by the normalization of the fields. Finally, using the special conformal transformation (1.14) for $r_{12}$ and (1.11) for its jacobian, we find that (1.16) requires that $\Delta_1 = \Delta_2$ if $c_{12} \neq 0$, and hence

$$\langle \phi_1(x_1)\, \phi_2(x_2) \rangle = \begin{cases} \dfrac{c_{12}}{r_{12}^{2\Delta}} & \Delta_1 = \Delta_2 = \Delta \\ 0 & \Delta_1 \neq \Delta_2 \;. \end{cases} \qquad (1.17)$$

The 3-point function is similarly restricted. Invariance under translations, rotations, and dilatations requires

$$\langle \phi_1(x_1)\, \phi_2(x_2)\, \phi_3(x_3) \rangle = \sum_{a,b,c} \frac{C_{abc}}{r_{12}^a\, r_{23}^b\, r_{13}^c} \;,$$

where the summation (in principle this could be an integration over a continuous range) over $a, b, c$ is restricted such that $a + b + c = \Delta_1 + \Delta_2 + \Delta_3$. Then covariance under the special conformal transformations (1.4c) in the form (1.14) requires $a = \Delta_1 + \Delta_2 - \Delta_3$, $b = \Delta_2 + \Delta_3 - \Delta_1$, and $c = \Delta_3 + \Delta_1 - \Delta_2$. Thus the 3-point function depends only on a single constant $C_{123}$,

$$\langle \phi_1(x_1)\, \phi_2(x_2)\, \phi_3(x_3) \rangle = \frac{C_{123}}{r_{12}^{\Delta_1 + \Delta_2 - \Delta_3}\, r_{23}^{\Delta_2 + \Delta_3 - \Delta_1}\, r_{13}^{\Delta_3 + \Delta_1 - \Delta_2}} \;. \qquad (1.18)$$

It might seem at this point that conformal invariant theories are rather trivial since the Green functions thus far considered are entirely determined up to some constants. The $N$-point functions for $N \geq 4$, however, are not so fully determined since they begin to have in general a dependence on the cross-ratios (1.15). The 4-point function, for example, may take the more general form

$$G^{(4)}(x_1, x_2, x_3, x_4) = F\left( \frac{r_{12}\, r_{34}}{r_{13}\, r_{24}}, \frac{r_{12}\, r_{34}}{r_{23}\, r_{41}} \right) \prod_{i<j} r_{ij}^{-(\Delta_i + \Delta_j) + \Delta/3} \;, \qquad (1.19)$$

where $F$ is an arbitrary function of the $4(4-3)/2 = 2$ independent cross-ratios, and $\Delta = \sum_{i=1}^{4} \Delta_i$. $N$-point functions in general are thus functions of the $N(N-3)/2$ independent cross-ratios and global conformal invariance alone cannot give any further information about these functions. In two dimensions, however, the local conformal group provides additional constraints that we shall study in the next section.

## 2. Conformal theories in 2 dimensions

### 2.1. Correlation functions of primary fields

We now apply the general formalism of section 1 to the special case of two dimensions, as introduced in subsection 1.2. Recall from (1.6) that the line element $ds^2 = dz\, d\overline{z}$ transforms under $z \to f(z)$ as

$$ds^2 \to \left( \frac{\partial f}{\partial z} \right) \left( \frac{\partial \overline{f}}{\partial \overline{z}} \right) ds^2 \;.$$

We shall generalize this transformation law to the form

$$\Phi(z,\overline{z}) \to \left(\frac{\partial f}{\partial z}\right)^h \left(\frac{\partial \overline{f}}{\partial \overline{z}}\right)^{\overline{h}} \Phi(f(z),\overline{f}(\overline{z})) , \qquad (2.1)$$

where $h$ and $\overline{h}$ are real-valued (and $\overline{h}$ again does not indicate the complex conjugate of $h$). (2.1) is equivalent to the statement that $\Phi(z,\overline{z})dz^h d\overline{z}^{\overline{h}}$ is invariant. It is similar in form to the tensor transformation property

$$A_{\mu\ldots\nu}(x) \to \frac{\partial x'^\alpha}{\partial x^\mu} \cdots \frac{\partial x'^\beta}{\partial x^\nu} A_{\alpha\ldots\beta}(x') ,$$

under $x \to x'$. In two dimensional complex coordinates, a tensor $\Phi_{zzz\ldots\overline{z}\overline{z}}(z,\overline{z})$, with $m$ lower $z$ indices and $n$ lower $\overline{z}$ indices, would transform as (2.1) with $h=m$, $\overline{h}=n$.

The transformation property (2.1) defines what is known as a primary field $\Phi$ of conformal weight $(h,\overline{h})$. Not all fields in conformal field theory will turn out to have this transformation property — the rest of the fields are known as secondary fields. A primary field is automatically quasi-primary, i.e. satisfies (1.12) under global conformal transformations. (A secondary field, on the other hand, may or may not be quasi-primary. Quasi-primary fields are sometimes also termed $SL(2,\mathbf{C})$ primaries.). Infinitesimally, under $z \to z + \epsilon(z)$, $\overline{z} \to \overline{z} + \overline{\epsilon}(\overline{z})$, we have from (2.1)

$$\delta_{\epsilon,\overline{\epsilon}} \Phi(z,\overline{z}) = \left((h\partial\epsilon + \epsilon\partial) + (\overline{h}\,\overline{\partial}\,\overline{\epsilon} + \overline{\epsilon}\,\overline{\partial})\right)\Phi(z,\overline{z}) , \qquad (2.2)$$

where $\overline{\partial} \equiv \partial_{\overline{z}}$.

Now the 2-point function $G^{(2)}(z_i,\overline{z}_i) = \langle \Phi_1(z_1,\overline{z}_1)\Phi_2(z_2,\overline{z}_2) \rangle$ is supposed to satisfy the infinitesimal form of (1.13),

$$\delta_{\epsilon,\overline{\epsilon}} G^{(2)}(z_i,\overline{z}_i) = \langle \delta_{\epsilon,\overline{\epsilon}} \Phi_1, \Phi_2 \rangle + \langle \Phi_1, \delta_{\epsilon,\overline{\epsilon}} \Phi_2 \rangle = 0 ,$$

giving the partial differential equation

$$\Big((\epsilon(z_1)\partial_{z_1} + h_1\partial\epsilon(z_1)) + (\epsilon(z_2)\partial_{z_2} + h_2\partial\epsilon(z_2)) \\ + (\overline{\epsilon}(\overline{z}_1)\partial_{\overline{z}_1} + h_1\partial\overline{\epsilon}(\overline{z}_1) + \overline{\epsilon}(\overline{z}_2)\partial_{\overline{z}_2} + h_2\partial\overline{\epsilon}(\overline{z}_2))\Big)G^{(2)}(z_i,\overline{z}_i) = 0 . \qquad (2.3)$$

Then paralleling the arguments that led to (1.17), we use $\epsilon(z)=1$ and $\overline{\epsilon}(\overline{z})=1$ to show that $G^{(2)}$ depends only on $z_{12} = z_1 - z_2$, $\overline{z}_{12} = \overline{z}_1 - \overline{z}_2$; then use $\epsilon(z)=z$ and $\overline{\epsilon}(\overline{z})=\overline{z}$ to require $G^{(2)} = C_{12}/(z_{12}^{h_1+h_2}\overline{z}_{12}^{\overline{h}_1+\overline{h}_2})$; and finally $\epsilon(z)=z^2$ and $\overline{\epsilon}(\overline{z})=\overline{z}^2$ to require $h_1=h_2=h$, $\overline{h}_1=\overline{h}_2=\overline{h}$. The result is that the 2-point function is constrained to take the form

$$G^{(2)}(z_i,\overline{z}_i) = \frac{C_{12}}{z_{12}^{2h}\,\overline{z}_{12}^{2\overline{h}}} . \qquad (2.4)$$

To make contact with (1.17), we consider bosonic fields with spin $s = h - \overline{h} = 0$. In terms of the scaling weight $\Delta = h + \overline{h}$, we see that (2.4) is equivalent to

$$G^{(2)}(z_i,\overline{z}_i) = \frac{C_{12}}{|z_{12}|^{2\Delta}} .$$

The 3-point function $G^{(3)} = \langle \Phi_1 \Phi_2 \Phi_3 \rangle$ is similarly determined, by arguments parallel to those leading to (1.18), to take the form

$$G^{(3)}(z_i,\overline{z}_i) = C_{123} \frac{1}{z_{12}^{h_1+h_2-h_3} z_{23}^{h_2+h_3-h_1} z_{13}^{h_3+h_1-h_2}} \\ \cdot \frac{1}{\overline{z}_{12}^{\overline{h}_1+\overline{h}_2-\overline{h}_3} \overline{z}_{23}^{\overline{h}_2+\overline{h}_3-\overline{h}_1} \overline{z}_{13}^{\overline{h}_3+\overline{h}_1-\overline{h}_2}} , \qquad (2.5)$$

where $z_{ij} = z_i - z_j$. As in (1.18), the 3-point function depends only on a single constant. This is because three points $z_1, z_2, z_3$ can always be mapped by a conformal transformation to three reference points, say $\infty, 1, 0$, where we have $\lim_{z_1 \to \infty} z_1^{2h_1} \overline{z}_1^{2\overline{h}_1} G^{(3)} = C_{123}$. The coordinate dependence for general $z_1, z_2, z_3$ can be reconstructed by conformal invariance. For all fields taken to be spinless, so that $s_i = h_i - \overline{h}_i = 0$, (2.5) correctly reduces to (1.18) with $\Delta_i = h_i + \overline{h}_i$ and $r_{ij} = |z_{ij}|$.

As in (1.19), the 4-point function, on the other hand, is not so fully determined just by conformal invariance. Global conformal invariance allows it to take the form

$$G^{(4)}(z_i,\overline{z}_i) = f(x,\overline{x}) \prod_{i<j} z_{ij}^{-(h_i+h_j)+h/3} \prod_{i<j} \overline{z}_{ij}^{-(\overline{h}_i+\overline{h}_j)+\overline{h}/3} , \qquad (2.6)$$

where $h = \sum_{i=1}^{4} h_i$, $\overline{h} = \sum_{i=1}^{4} \overline{h}_i$. In (2.6) the cross-ratio $x$ is defined as $x = z_{12}z_{34}/z_{13}z_{24}$. (We note that this cross-ratio is annihilated by the differential operator $\sum_{i=1}^{4} \epsilon(z_i)\partial_{z_i}$ so the analog of (2.3) leaves the function $f$ undetermined.) In two dimensions, the two cross-ratios of (1.19) are linearly related (because 4 points constrained to be coplanar must satisfy an additional linear relation). The six possible cross ratios of the form (1.15), constructed from four $z_i$'s, are given by

$$x = \frac{z_{12}z_{34}}{z_{13}z_{24}}, \qquad 1 - x = \frac{z_{14}z_{23}}{z_{13}z_{24}}, \qquad \frac{x}{1-x} = \frac{z_{12}z_{34}}{z_{14}z_{23}},$$

and their inverses. With respect to the argument that fixed the form of the 3-point function (2.5), we can understand the residual $x$ dependence of (2.6) by recalling that global conformal transformations only allow us to fix three coordinates, so the best we can do is to take say $z_1, z_2, z_3, z_4 = \infty, 1, x, 0$.

In (2.4)–(2.6), the $h_i$'s and $\overline{h}_i$'s are in principle arbitrary. Later on we shall see how they may be constrained by unitarity. We shall also formulate differential equations which, together with monodromy conditions, allow one in principle to determine all the unknown functions (generalizing the $f$ of (2.6)) for arbitrary $N$-point functions in a given theory.

## 2.2. Radial quantization and conserved charges

To probe more carefully the consequences of conformal invariance in a two dimensional quantum field theory, we enter into some of the details of the quantization procedure. We begin with flat Euclidean "space" and "time" coordinates $\sigma^1$ and $\sigma^0$. In Minkowski space, the standard light-cone coordinates would be $\sigma^0 \pm \sigma^1$. In Euclidean space the analogs are instead the complex coordinates $\zeta, \overline{\zeta} = \sigma^0 \pm i\sigma^1$. The two dimensional Minkowski space notions of left- and right-moving massless fields become Euclidean fields that have purely holomorphic or anti-holomorphic dependence on the coordinates. For this reason we shall occasionally call the holomorphic and anti-holomorphic fields left- and right-movers respectively. To eliminate any infrared divergences, we compactify the space coordinate, $\sigma^1 \equiv \sigma^1 + 2\pi$. This defines a cylinder in the $\sigma^1, \sigma^0$ coordinates.

Next we consider the conformal map $\zeta \to z = \exp\zeta = \exp(\sigma^0 + i\sigma^1)$ that maps the cylinder to the complex plane coordinatized by $z$ (see fig. 1.) Then infinite past and future on the cylinder, $\sigma^0 = \mp\infty$, are mapped to the points $z = 0, \infty$ on the plane. Equal time surfaces, $\sigma^0$=const, become circles of constant radius on the $z$-plane, and time reversal, $\sigma^0 \to -\sigma^0$, becomes $z \to 1/z^*$. To build up a quantum theory of conformal fields on the $z$-plane, we will need to realize the operators that implement conformal mappings of the plane. For example dilatations, $z \to e^a z$, on the cylinder are just the time translations $\sigma^0 \to \sigma^0 + a$. So the dilatation generator on the conformal plane can be regarded as the Hamiltonian for the system, and the Hilbert space is built up on surfaces of constant radius. This procedure for defining a quantum theory on the plane is known as radial quantization[5]. It is particularly useful for two dimensional conformal field theory in the Euclidean regime since it facilitates use of the full power of contour integrals and complex analysis to analyze short distance expansions, conserved charges, etc. Our intuition for manipulations in this scheme will frequently come from referring things back to the cylinder.

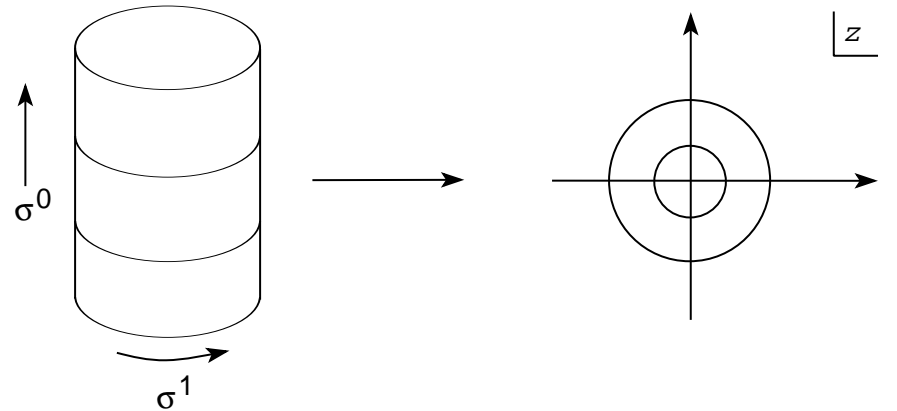

Fig. 1. Map of the cylinder to the plane

Symmetry generators in general can be constructed via the Noether prescription. A $d+1$ dimensional quantum theory with an exact symmetry has an associated conserved current $j^\mu$, satisfying $\partial_\mu j^\mu = 0$. The conserved charge $Q = \int d^d x\, j_0(x)$, constructed by integrating over a fixed-time slice, generates, according to $\delta_\epsilon A = \epsilon[Q, A]$, the infinitesimal symmetry variation in any field $A$. In particular, local coordinate transformations are generated by charges constructed from the stress-energy tensor $T_{\mu\nu}$, in general a symmetric divergence-free tensor. In conformally invariant theories, $T_{\mu\nu}$ is also traceless. This follows from requiring the conservation $0 = \partial \cdot j = T^\mu{}_\mu$ of the dilatation current $j_\mu = T_{\mu\nu} x^\nu$ (associated to the ordinary scale transformations $x^\mu \to x^\mu + \lambda x^\mu$). The current associated to other infinitesimal conformal transformations is $j_\mu = T_{\mu\nu} \epsilon^\nu$, where $\epsilon^\mu$ satisfies (1.2). This current as well has an automatically vanishing divergence, $\partial \cdot j = \frac{1}{2} T^\mu{}_\mu (\partial \cdot \epsilon) = 0$, due to the traceless condition on $T_{\mu\nu}$.

To implement the conserved charges on the conformal $z$-plane, we introduce the necessary complex tensor analysis. The flat Euclidean plane ($g_{\mu\nu} = \delta_{\mu\nu}$) in complex coordinates $z = x + iy$ has line element $ds^2 = g_{\mu\nu}\, dx^\mu dx^\nu = dx^2 + dy^2 = dz\, d\bar z$. The components of the metric referred to complex coordinate frames are thus $g_{zz} = g_{\bar z \bar z} = 0$ and $g_{z\bar z} = g_{\bar z z} = \frac{1}{2}$, and the components of the stress-energy tensor referred to these frames are $T_{zz} = \frac{1}{4}(T_{00} - 2iT_{10} - T_{11})$, $T_{\bar z \bar z} = \frac{1}{4}(T_{00} + 2iT_{10} - T_{11})$, and $T_{z\bar z} = T_{\bar z z} = \frac{1}{4}(T_{00} + T_{11}) = \frac{1}{4} T^\mu{}_\mu$. The conservation law $g^{\alpha\mu} \partial_\alpha T_{\mu\nu} = 0$ gives two relations, $\partial_{\bar z} T_{zz} + \partial_z T_{\bar z z} = 0$ and $\partial_z T_{\bar z \bar z} + \partial_{\bar z} T_{z \bar z} = 0$. Using the traceless condition $T_{z\bar z} = T_{\bar z z} = 0$, these imply

$$\partial_{\bar z} T_{zz} = 0 \quad \text{and} \quad \partial_z T_{\bar z \bar z} = 0 \ .$$

The two non-vanishing components of the stress-energy tensor

$$T(z) \equiv T_{zz}(z) \quad \text{and} \quad \overline{T}(\bar z) = T_{\bar z \bar z}(\bar z)$$

thus have only holomorphic and anti-holomorphic dependences. We shall find numerous properties of conformal theories on the $z$-plane to factorize similarly into independent left and right pieces.

It is natural to expect $T$ and $\overline{T}$, the remnants of the stress-energy tensor in complex coordinates, to generate local conformal transformations on the $z$-plane. In radial quantization, the integral of the component of the current orthogonal to an "equal-time" (constant radius) surface becomes $\int j_0(x)\, dx \to \int j_r(\theta)\, d\theta$. Thus we should take

$$Q = \frac{1}{2\pi i} \oint \Big( dz\, T(z) \epsilon(z) + d\bar z\, \overline{T}(\bar z) \bar\epsilon(\bar z) \Big) \tag{2.7}$$

as the conserved charge. The line integral is performed over some circle of fixed radius and our sign conventions are such that both the $dz$ and the $d\bar z$ integrations are taken in the counter-clockwise sense. Note that (2.7) is a formal expression that cannot be evaluated until we specify what other fields lie inside the contour.

The variation of any field is given by the "equal-time" commutator with the charge (2.7),

$$\delta_{\epsilon,\bar\epsilon} \Phi(w,\bar w) = \frac{1}{2\pi i} \oint \Big[dz\, T(z)\, \epsilon(z)\,, \Phi(w,\bar w)\Big] + \Big[d\bar z\, \overline{T}(\bar z)\, \bar\epsilon(\bar z)\,, \Phi(w,\bar w)\Big] \ . \tag{2.8}$$

Now products of operators $A(z) B(w)$ in Euclidean space radial quantization are only defined for $|z| > |w|$. (In general, recall that to continue any Minkowski space Green function

$$\langle A_1(x_1, t_1) \ldots A_n(x_n, t_n) \rangle$$

to Euclidean space, we let $A(x,t) \to e^{H\tau} A(x,0) e^{-H\tau}$, where $t = i\tau$. In a theory with energy bounded from below, the Euclidean space Green function

$$\langle A_1(x_1, 0) e^{-H(\tau_1 - \tau_2)} A_2(x_2, 0) \ldots e^{-H(\tau_{n-1} - \tau_n)} A(x_n, 0) \rangle$$

is guaranteed to converge only for operators that are time-ordered, i.e. for which $\tau_j > \tau_{j+1}$. The analytic continuation of time-ordered Euclidean Green functions then gives the desired solution to the Minkowski space equations of motion on the cylinder. In a Euclidean space functional integral formulation, Green functions

$$\langle \phi_1 \ldots \phi_n \rangle = \int_\varphi \exp(-S[\varphi]) \varphi_1 \ldots \varphi_n \Big/ \int_\varphi \exp(-S[\varphi])$$

are computed in terms of dummy integration variables $\varphi$, which automatically calculate the time-ordered (convergent) result.) Thus we define the radial ordering operation $R$ as

$$R\big(A(z)B(w)\big) = \begin{cases} A(z)B(w) & |z| > |w| \\ B(w)A(z) & |z| < |w| \end{cases} \qquad (2.9)$$

(or with a minus sign for fermionic operators). This allows us to define the meaning of the commutators in (2.8). The equal-time commutator of a local operator $A$ with the spatial integral of an operator $B$ will become the contour integral of the radially ordered product, $\big[\int dx\, B, A\big]_{\text{E.T.}} \to \oint dz\, R\big(B(z)A(w)\big)$.

In fig. 2 we have represented the contour integrations that we need to perform in order to evaluate the commutator in (2.8). We see that the difference combines into a single integration about a contour drawn tightly around the point $w$. (The reader might derive further insight into the map (fig. 1) from the cylinder to the plane by pulling back fig. 2 to the cylinder and seeing what it looks like in terms of equal time $\sigma^0$ contours.) We may thus rewrite (2.8) in the form

$$\begin{aligned}\delta_{\epsilon,\overline{\epsilon}}\Phi(w,\overline{w}) &= \frac{1}{2\pi i}\left(\oint_{|z|>|w|} - \oint_{|z|<|w|}\right)\Big(dz\,\epsilon(z)R\big(T(z)\Phi(w,\overline{w})\big) \\ &\qquad\qquad + d\overline{z}\,\overline{\epsilon}(\overline{z})R\big(\overline{T}(\overline{z})\Phi(w,\overline{w})\big)\Big) \\ &= \frac{1}{2\pi i}\oint \Big(dz\,\epsilon(z)R\big(T(z)\Phi(w,\overline{w})\big) + d\overline{z}\,\overline{\epsilon}(\overline{z})R\big(T(\overline{z})\Phi(w,\overline{w})\big)\Big) \\ &= h\,\partial\epsilon(w)\Phi(w,\overline{w}) + \epsilon(w)\partial\Phi(w,\overline{w}) \\ &\qquad + \overline{h}\,\overline{\partial}\overline{\epsilon}(\overline{w})\,\Phi(w,\overline{w}) + \overline{\epsilon}(\overline{w})\,\overline{\partial}\Phi(w,\overline{w})\ ,\end{aligned}$$

where in the last line we have substituted the desired result, i.e. the result of the transformation (2.1) in the case of infinitesimal $f(z) = z + \epsilon(z)$. In order that the charge (2.7) induce the correct infinitesimal conformal transformations, we infer that the short distance singularities of $T$ and $\overline{T}$ with $\Phi$ should be

$$R\big(T(z)\Phi(w,\bar w)\big) = \frac{h}{(z-w)^2}\Phi(w,\overline{w}) + \frac{1}{z-w}\partial_w\Phi(w,\overline{w}) + \ldots$$

$$R\big(\overline{T}(\overline{z})\Phi(w,\overline{w})\big) = \frac{\overline{h}}{(\overline{z}-\overline{w})^2}\Phi(w,\overline{w}) + \frac{1}{\overline{z}-\overline{w}}\partial_{\overline{w}}\Phi(w,\overline{w}) + \ldots\ .$$

These short distance properties can be taken to define the quantum stress-energy tensor. They are naturally realized by standard canonical definitions of the stress-energy tensor in two dimensions (since they ordinarily result in generators of coordinate transformations). In a moment, we shall confirm how all of this works in some specific examples.

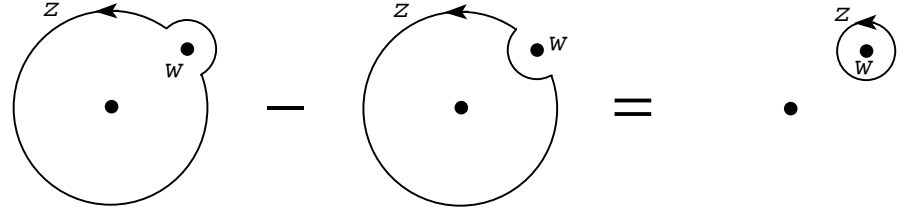

Fig. 2. Evaluation of "equal-time" commutator on the conformal plane.

We see that the transformation law (2.1) for primary fields leads to a short distance operator product expansion for the holomorphic and anti-holomorphic stress-energy tensors, $T$ and $\overline{T}$, with a primary field. From now on we shall drop the $R$ symbol and consider the operator product expansion itself as a shorthand for radially ordered products. The operator product expansion that defines the notion of a primary field is abbreviated as

$$T(z)\Phi(w,\bar w) = \frac{h}{(z-w)^2}\Phi(w,\overline{w}) + \frac{1}{z-w}\partial_w\Phi(w,\overline{w}) + \ldots$$

$$\overline{T}(\overline{z})\Phi(w,\overline{w}) = \frac{\overline{h}}{(\overline{z}-\overline{w})^2}\Phi(w,\overline{w}) + \frac{1}{\overline{z}-\overline{w}}\partial_{\overline{w}}\Phi(w,\overline{w}) + \ldots\ , \qquad (2.10)$$

and encodes the conformal transformation properties of $\Phi$. In the next section, we shall see how operator product expansions are also equivalent to canonical commutators of the modes of the fields.

We pause at this point to recall some of the standard lore concerning operator product expansions[6]. In general, the singularities that occur when operators approach one another are encoded in operator product expansions of the form

$$A(x)B(y) \sim \sum_i C_i(x-y) O_i(y) \ , \qquad (2.11)$$

where the $O_i$'s are a complete set of local operators and the $C_i$'s are (singular) numerical coefficients. Ordinarily (2.11) is an asymptotic expansion, but in a conformal theory it has been argued to converge (since $e^{-\ell/|z-w|}$ type terms that would be expected if the series did not converge require a dimensional parameter $\ell$, absent in a conformal field theory). For operators of fixed scaling dimension $d$ in (2.11), we can determine the coordinate dependence of the $C_i$'s by dimensional analysis to be $C_i \sim 1/|x-y|^{d_A+d_B-d_{O_i}}$.

In two dimensional conformal field theories, we can always take a basis of operators $\phi_i$ with fixed conformal weight. If we normalize their 2-point functions (2.4) as

$$\langle \phi_i(z,\overline{z}) \, \phi_j(w,\overline{w}) \rangle = \delta_{ij} \, \frac{1}{(z-w)^{2h_i}} \frac{1}{(\overline{z}-\overline{w})^{2\overline{h}_i}} \ , \qquad (2.12)$$

then the operator product coefficients $C_{ijk}$ defined by

$$\phi_i(z,\overline{z}) \, \phi_j(w,\overline{w}) \sim \sum_k C_{ijk} \, (z-w)^{h_k-h_i-h_j} \, (\overline{z}-\overline{w})^{\overline{h}_k-\overline{h}_i-\overline{h}_j} \, \phi_k(w,\overline{w}) \quad (2.13)$$

are symmetric in $i,j,k$. By taking the limit as any two of the $z_i$'s in the 3-point function $\langle \phi_i \phi_j \phi_k \rangle$ approach one another, and using (2.12), it is easy to show that the $C_{ijk}$'s of (2.13) coincide precisely with the numerical factors in the 3-point functions (2.5).

*2.3. Free boson, the example*

We shall now illustrate the formalism developed thus far in the case of a single massless free boson, also known as the gaussian model. We use the string theory normalization for the action,

$$S = \int \mathcal{L} = \frac{1}{2\pi} \int \partial X \, \overline{\partial} X \ , \qquad (2.14)$$

so that $X(z,\overline{z})$ has propagator $\langle X(z,\overline{z}) X(w,\overline{w}) \rangle = -\frac{1}{2} \log |z-w|$. (This is calculated using $z = \frac{1}{2}(\sigma^1 + i\sigma^0)$, and integration measure $2i dz \wedge d\overline{z} = d\sigma^1 \wedge d\sigma^0$

in (2.14), although ultimately only the normalization of the propagator itself is important in what follows.) The standard statistical mechanical convention (see e.g. section 4.2 of Cardy's lectures) uses instead a factor of $g/4\pi$ in front in the action (2.14). For solutions of the equations of motion, we find that $X(z,\overline{z}) = \frac{1}{2}\bigl(x(z) + \overline{x}(\overline{z})\bigr)$ splits into two pieces with only holomorphic and anti-holomorphic dependence respectively. (These are the massless left-movers and right-movers. To avoid any ambiguity we could write $x_L(z)$ and $x_R(\overline{z})$, but the meaning is usually clear from context.) These pieces have propagators

$$\langle x(z) x(w) \rangle = -\log(z-w), \qquad \langle \overline{x}(\overline{z}) \, \overline{x}(\overline{w}) \rangle = -\log(\overline{z}-\overline{w}) \ . \qquad (2.15)$$

Note that the field $x(z)$ is not itself a conformal field, but its derivative, $\partial x(z)$, has leading short distance expansion

$$\partial x(z) \, \partial x(w) = -\frac{1}{(z-w)^2} + \dots \ , \qquad (2.16)$$

inferred by taking two derivatives of (2.15). We see from the scaling properties of the right hand side of (2.16) that $\partial x(z)$ has a chance to be a (1,0) conformal field.

Concentrating for the moment on the holomorphic dependence of the theory, we define the stress-energy tensor $T(w)$ via the normal-ordering prescription

$$\begin{aligned} T(w) &= -\frac{1}{2} : \partial x(z) \partial x(w) : \\ &\equiv -\frac{1}{2} \lim_{z \to w} \left[ \partial x(z) \partial x(w) + \frac{1}{(z-w)^2} \right] \ . \end{aligned} \qquad (2.17)$$

Using the Wick rules and Taylor expanding, we can compute the singular part of

$$\begin{aligned} T(z) \, \partial x(w) &= -\frac{1}{2} : \partial x(z) \partial x(z) : \, \partial x(w) \\ &= -\frac{1}{2} \partial x(z) \langle \partial x(z) \partial x(w) \rangle \cdot 2 + \dots \\ &= \partial x(z) \frac{1}{(z-w)^2} + \dots \\ &= \bigl( \partial x(w) + (z-w) \partial^2 x(w) \bigr) \frac{1}{(z-w)^2} + \dots \ , \end{aligned}$$

in the limit $z \to w$. We find

$$T(z)\partial x(w) \sim \frac{\partial x(w)}{(z-w)^2} + \frac{1}{z-w}\partial^2 x(w) + \ldots ,$$

in accord with (2.10) for a (1,0) primary field. Moreover substituting in (2.8), we see that

$$\left[\oint \frac{dz}{2\pi i} T(z)\epsilon(z) , \partial x(w)\right] = \oint \frac{dz}{2\pi i}\epsilon(z)\left(\frac{\partial x(w)}{(z-w)^2} + \frac{\partial^2 x(w)}{z-w} + \ldots\right)$$
$$= \partial\epsilon(w)\partial x(w) + \epsilon(w)\partial^2 x(w) .$$

This is all as expected since under $z \to z + \epsilon$, we have $x(z) \to x(z+\epsilon) = x(z) + \epsilon \partial x(z)$, and consequently $\partial x(z) \to \partial x(z) + \partial \epsilon \partial x(z) + \epsilon \partial^2 x(z)$. The above result is just the statement that $\partial x$ transforms as in (2.1) as a tensor of mass dimension $h = 1$.

As another illustration of (2.10), we consider the operator $:\exp i\alpha x(w):$. The normal ordering symbol is meant to remind us not to contract the $x(w)$'s in the expansion of the exponent with one another. (This prescription is equivalent to a multiplicative wave function renormalization, and for convenience we will frequently drop the normal ordering symbol in the following). Taking the operator product expansion with $T(z)$ as $z \to w$, we find the leading singular behavior

$$-\frac{1}{2}\big(\partial x(z)\big)^2 \, e^{i\alpha x(w)} = -\frac{1}{2}\big(\langle\partial x(z) i\alpha x(w)\rangle\big)^2 e^{i\alpha x(w)}$$
$$-\frac{1}{2} 2\, \partial x(z)\langle\partial x(z) i\alpha x(w)\rangle\, e^{i\alpha x(w)}$$
$$= \frac{\alpha^2/2}{(z-w)^2} e^{i\alpha x(w)} + \frac{i\alpha \partial x(z)}{z-w} e^{i\alpha x(w)} \qquad (2.18)$$
$$= \frac{\alpha^2/2}{(z-w)^2} e^{i\alpha x(w)} + \frac{1}{z-w} \partial e^{i\alpha x(w)} .$$

$\exp(i\alpha x)$ is thus a primary field of conformal dimension $h = \alpha^2/2$.

This result could also be inferred from the 2-point function

$$\left\langle e^{i\alpha x(z)} e^{-i\alpha x(w)}\right\rangle = e^{\alpha^2 \langle x(z) x(w)\rangle} = \frac{1}{(z-w)^{\alpha^2}} , \qquad (2.19)$$

where the first equality is a general property of free field theory, and the second equality follows from the specifically two dimensional logarithmic behavior (2.15) (recall that the propagator in $d > 2$ spacetime dimensions goes instead as $\int d^d p \exp(ipx)/p^2 \sim 1/x^{d-2}$). We see that the logarithmic divergence of the scalar propagator leads to operators with continuously variable anomalous dimensions in two dimensions, even in free field theory.

Identical considerations apply equally to anti-holomorphic operators, such as $\overline{\partial}\overline{x}(\overline{z})$ and $\exp(i\alpha \overline{x}(\overline{z}))$. Their operator products with $\overline{T}(\overline{z}) = \frac{1}{2}:\overline{\partial}\overline{x}(\overline{z})\overline{\partial}\overline{x}(\overline{z}):$ shows them to have conformal dimensions $(0, 1)$ and $(0, \alpha^2/2)$. More generally if we took an action $S = \frac{1}{2\pi}\int \partial X^\mu \overline{\partial} X^\mu$ with a vector of fields $X^\mu(z, \overline{z}) = \frac{1}{2}(x^\mu(z) + \overline{x}^\mu(\overline{z}))$, then $\langle x^\mu(z) x^\nu(w)\rangle = -\delta^{\mu\nu}\log(z-w)$ and $\exp\big(\pm i\alpha^\mu x^\mu(z)\big)$ for example has conformal dimension $(\alpha \cdot \alpha/2, 0)$.

Before closing this introduction to massless scalars in two dimensions, we should dispel an occasional unwarranted confusion concerning the result of [7], which states that the Goldstone phenomenon does not occur in two dimensions. In the present context this does not mean that there is anything particularly peculiar about massless scalar fields, only that they are not Goldstone bosons. Although it appears that (2.14) has a translation symmetry $X \to X + a$ that can be spontaneously broken, this symmetry is an illusion at the quantum level. That is because the field $X$ is itself ill-defined due to the incurable infrared logarithmic divergence of its propagator. $\partial_\mu X$ is of course well defined but is not sensitive to the putative symmetry breaking. Exponentials of $X$ as in (2.19) can also be defined by appropriate extraction of wave function normalization, but their non-vanishing correlation functions all have simple power law falloff, and again show no signal of symmetry breakdown. This is all consistent with the result of [7].

*2.4. Conformal Ward identities*

We complete our discussion of conformal formalities by writing down the conformal Ward identities satisfied by correlations functions of primary fields $\phi_i$. Ward identities are generally identities satisfied by correlation functions as a reflection of symmetries possessed by a theory. They are easily derived in the functional integral formulation of correlation functions for example by requiring
that they be independent of a change of dummy integration variables. The
Ward identities for conformal symmetry can thus be derived by considering the
behavior of $n$-point functions under a conformal transformation. This should
be considered to take place in some localized region containing all the operators
in question, and can then be related to a surface integral about the boundary
of the region.

For the two dimensional conformal theories of interest here, we shall instead
implement this procedure in the operator form of the correlation functions. By
global conformal invariance, these correlation functions satisfy (compare with
(1.13))

$$\langle \phi_1(z_1,\overline{z}_1)\ldots\phi_n(z_n,\overline{z}_n)\rangle$$
$$= \prod_j \left(\partial f(z_j)\right)^{h_j}\left(\overline{\partial}\,\overline{f}(\overline{z}_j)\right)^{\overline{h}_j}\langle \phi_1(w_1,\overline{w}_1)\ldots\phi_n(w_n,\overline{w}_n)\rangle\ , \quad (2.20)$$

with $w = f(z)$ and $\overline{w} = \overline{f}(\overline{z})$ of the form (1.9). To gain additional information
from the *local* conformal algebra, we consider an assemblage of operators at
points $w_i$ as in fig. 3, and perform a conformal transformation in the interior
of the region bounded by the $z$ contour by line integrating $\epsilon(z)T(z)$ around
it. By analyticity, the contour can be deformed to a sum over small contours
encircling each of the points $w_i$, as depicted in the figure. The result of the
contour integration is thus

$$\left\langle \oint \frac{dz}{2\pi i}\,\epsilon(z)T(z)\,\phi_1(w_1,\overline{w}_1)\ldots\phi_n(w_n,\overline{w}_n)\right\rangle$$
$$= \sum_{j=1}^n \left\langle \phi_1(w_1,\overline{w}_1)\ldots\left(\oint \frac{dz}{2\pi i}\,\epsilon(z)T(z)\phi_j(w_j,\overline{w}_j)\right)\ldots\phi_n(w_n,\overline{w}_n)\right\rangle \quad (2.21)$$
$$= \sum_{j=1}^n \left\langle \phi(w_1,\overline{w}_1)\ldots\delta_\epsilon \phi_j(w_j,\overline{w}_j)\ldots\phi_n(w_n,\overline{w}_n)\right\rangle\ .$$

In the last line we have used the infinitesimal transformation property

$$\delta_\epsilon \phi(w,\overline{w}) = \oint \frac{dz}{2\pi i}\,\epsilon(z)T(z)\phi(w,\overline{w}) = \big(\epsilon(w)\partial + h\partial\epsilon(w)\big)\phi(w,\overline{w})\ ,$$

encoded in the operator product expansion (2.10).

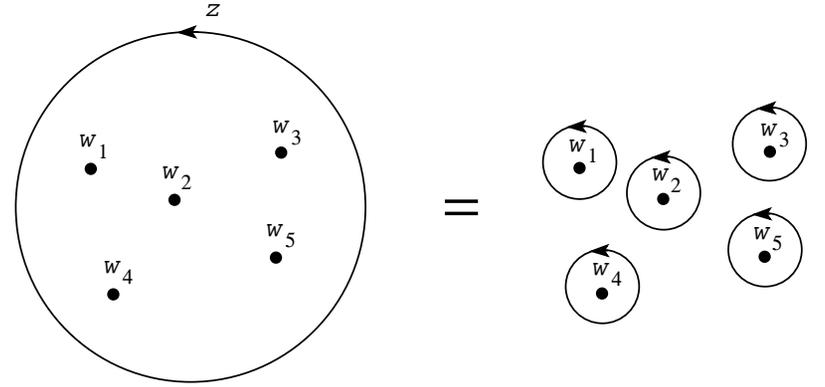

Fig. 3. Another deformed contour

Since (2.21) is true for arbitrary $\epsilon(z)$ and $\oint d\overline{z}\,T(z) = 0$, we can write an
unintegrated form of the conformal Ward identities,

$$\langle T(z)\phi_1(w_1,\overline{w}_1)\ldots\phi_n(w_n,\overline{w}_n)\rangle$$
$$= \sum_{j=1}^n \left(\frac{h_j}{(z-w_j)^2} + \frac{1}{z-w_j}\frac{\partial}{\partial w_j}\right)\langle \phi_1(w_1,\overline{w}_1)\ldots\phi_n(w_n,\overline{w}_n)\rangle\ . \quad (2.22)$$

This states that the correlation functions are meromorphic functions of $z$ with
singularities at the positions of inserted operators. The residues at these singularities are simply determined by the conformal properties of the operators.
Later on we shall use (2.22) to show that 4-point correlation functions involving
so-called degenerate fields satisfy hypergeometric differential equations.

## 3. The central charge and the Virasoro algebra

*3.1. The central charge*

Not all fields satisfy the simple transformation property (2.1) under conformal transformations. Derivatives of fields, for example, in general have more
complicated transformation properties. A secondary field is any field that has
higher than the double pole singularity (2.10) in its operator product expansion with $T$ or $\overline{T}$. In general, the fields in a conformal field theory can be grouped into families $[\phi_n]$ each of which contains a single primary field $\phi_n$ and an infinite set of secondary fields (including its derivative), called its descendants. These comprise the irreducible representations of the conformal group, and the primary field can be regarded as the highest weight of the representation. The set of all fields in a conformal theory $\{A_i\} = \sum_n [\phi_n]$ may be composed of either a finite or infinite number of conformal families.

An example of a field that does not obey (2.1) or (2.10) is the stress-energy tensor. By performing two conformal transformations in succession, we can determine its operator product with itself to take the form

$$T(z)T(w) = \frac{c/2}{(z-w)^4} + \frac{2}{(z-w)^2} T(w) + \frac{1}{z-w} \partial T(w) \,. \tag{3.1}$$

The $(z-w)^{-4}$ term on the right hand side, with coefficient $c$ a constant, is allowed by analyticity, Bose symmetry, and scale invariance. Its coefficient cannot be determined by the requirement that $T$ generate conformal transformations, since that only involves the commutator of $T$ with other operators. Apart from this term, (3.1) is just the statement that $T(z)$ is a conformal field of weight (2,0). The constant $c$ is known as the central charge and its value in general will depend on the particular theory under consideration. Since $\langle T(z)T(0)\rangle = (c/2)/z^4$, we expect at least that $c \geq 0$ in a theory with a positive semi-definite Hilbert space.

Identical considerations apply to $\overline{T}$, so that

$$\overline{T}(\overline{z})\,\overline{T}(\overline{w}) = \frac{\overline{c}/2}{(\overline{z}-\overline{w})^4} + \frac{2}{(\overline{z}-\overline{w})^2}\overline{T}(\overline{w}) + \frac{1}{\overline{z}-\overline{w}}\overline{\partial}\,\overline{T}(\overline{w}) \,, \tag{3.2}$$

where $\overline{c}$ is in principle an independent constant. (Later on we shall see that modular invariance constrains $c - \overline{c} = 0 \bmod 24$.) A theory with a Lorentz-invariant, conserved 2-point function $\langle T_{\mu\nu}(p)\, T_{\alpha\beta}(-p)\rangle$ requires $c = \overline{c}$. This is equivalent to requiring cancellation of local gravitational anomalies[8], allowing the system to be consistently coupled to two dimensional gravity. In heterotic string theory, for example, this is achieved by adding ghosts to the system so that $c = \overline{c} = 0$.

In general, the infinitesimal transformation law for $T(z)$ induced by (3.1) is

$$\delta_\epsilon T(z) = \epsilon(z)\,\partial T(z) + 2\partial\epsilon(z)\,T(z) + \frac{c}{12}\partial^3 \epsilon(z) \,.$$

It can be integrated to give

$$T(z) \to (\partial f)^2\, T\big(f(z)\big) + \frac{c}{12}\, S(f,z) \tag{3.3}$$

under $z \to f(z)$, where the quantity

$$S(f,z) = \frac{\partial_z f\, \partial_z^3 f - \frac{3}{2}(\partial_z^2 f)^2}{(\partial_z f)^2}$$

is known as the Schwartzian derivative. It is the unique weight two object that vanishes when restricted to the global $SL(2,\mathbf{R})$ subgroup of the two dimensional conformal group. (It also satisfies the composition law $S(w,z) = (\partial_z f)^2 S(w,f) + S(f,z)$.) The stress-energy tensor is thus an example of a field that is quasi-primary, i.e. $SL(2,\mathbf{C})$ primary, but not (Virasoro) primary.

We can readily calculate (3.1) for the free boson stress-energy tensor (2.17), $T(z) = -\frac{1}{2}{:}\partial x(z)\partial x(z){:}$. The result is

$$\begin{aligned}
&T(z)T(w) \\
&= \left(-\frac{1}{2}\right)^2 \left\{ 2\big(\langle\partial x(z)\partial x(w)\rangle\big)^2 + 4\partial x(z)\partial x(w)\,\langle\partial x(z)\partial x(w)\rangle + \ldots \right\} \\
&= \frac{1/2}{(z-w)^4} + \frac{2}{(z-w)^2}\left(-\frac{1}{2}\big(\partial x(w)\big)^2\right) + \frac{1}{z-w}\partial\left(-\frac{1}{2}\big(\partial x(w)\big)^2\right) \,,
\end{aligned}$$

and thus the leading term in (3.1) is normalized so that a single free boson has $c = 1$.

A variation on (2.17) is to take instead

$$T(w) = -\frac{1}{2}{:}\partial x(z)\partial x(w){:} + i\sqrt{2}\alpha_0\,\partial^2 x(z) \,. \tag{3.4}$$

The extra term is a total derivative of a well-defined field and does not affect the status of $T(z)$ as a generator of conformal transformations. Using (2.16) and proceeding as above, we can show that the $T(z)$ of (3.4) satisfies (3.1) with central charge

$$c = 1 - 24\alpha_0^2 \ .$$

We see that the effect of the extra term in (3.4) is to shift $c < 1$ for $\alpha_0$ real. Since the stress-energy tensor in (3.4) has an imaginary part, the theory it defines is not unitary for arbitrary $\alpha_0$. For particular values of $\alpha_0$, it turns out to contain a consistent unitary subspace. (In section 4, we will discuss the role played by unitarity in field theory and statistical mechanical models and also implicitly identify the relevant values of $\alpha_0$.)

The modification of $T(z)$ in (3.4) is interpreted as the presence of a 'background charge' $-2\alpha_0$ at infinity. This is created by the operator $:\exp\bigl(-i2\sqrt{2}\alpha_0 x(\infty)\bigr):$, so we take as out-state

$$\langle(-2\alpha_0)| = \frac{\langle 0|V_{-2\alpha_0}(\infty)}{\langle 0|V_{-2\alpha_0}(\infty)\,V_{2\alpha_0}(0)|0\rangle} \ ,$$

where $V_\beta(z) \equiv\, :\exp\bigl(i\sqrt{2}\beta x(z)\bigr):$. Thus the only non-vanishing correlation functions of strings of operators $V_{\beta_j}(z)$ are those with $\sum_j \beta_j = 2\alpha_0$. $n$-point correlation functions may be derived by sending a $V_{-2\alpha_0}(z)$ to infinity in an $n+1$-point function. For example, the result (2.19) for the 2-point function is modified to

$$\langle V_\beta(z)\, V_{2\alpha_0-\beta}(w)\rangle = \frac{1}{(z-w)^{2\beta(\beta-2\alpha_0)}} \ .$$

The operators in this 2-point function are regarded as adjoints of one another in the presence of the background charge, and each thus has conformal weight $h = \beta(\beta - 2\alpha_0)$. We arrive at the same result (rather than simply $h = \beta^2$) by calculating the conformal weight of the operator $V_\beta(z)$ as in (2.18), only using the modified definition (3.4) of $T(z)$. This formalism was anticipated in ancient times[9] and has more recently been used to great effect[10] to calculate correlation functions of the $c < 1$ theories to be discussed in the next section. These and other applications are described in more detail in Zuber's lectures.

### 3.2. The free fermion

Another free system that will play a major role later on here is that of a free massless fermion. With both chiralities, we write the action

$$S = \frac{1}{8\pi}\int \bigl(\psi\overline{\partial}\psi + \overline{\psi}\partial\overline{\psi}\bigr) \ . \qquad (3.5)$$

The equations of motion determine that $\psi(z)$ and $\overline{\psi}(\overline{z})$ are respectively the left- and right-moving "chiralities". (Recall that in 2 Euclidean dimensions the Dirac operator can be represented as

$$\not\partial = \sigma_x\partial_x + \sigma_y\partial_y = \begin{pmatrix} & \partial_x - i\partial_y \\ \partial_x + i\partial_y & \end{pmatrix} \sim \begin{pmatrix} & \partial \\ \overline{\partial} & \end{pmatrix} \ ,$$

so that the operators $\partial, \overline{\partial}$ are picked out by the chirality projectors $\frac{1}{2}(1 \pm \sigma_z)$.) The normalization of (3.5) is chosen so that the leading short distance singularities are

$$\psi(z)\psi(w) = -\frac{1}{z-w} \ , \qquad \overline{\psi}(\overline{z})\overline{\psi}(\overline{w}) = -\frac{1}{\overline{z}-\overline{w}} \ .$$

This system has holomorphic and anti-holomorphic stress-energy tensors

$$T(z) = \frac{1}{2}:\psi(z)\partial\psi(z): \ , \qquad \overline{T}(\overline{z}) = \frac{1}{2}:\overline{\psi}(\overline{z})\,\overline{\partial}\,\overline{\psi}(\overline{z}):$$

that satisfy (3.1) with $c = \overline{c} = \frac{1}{2}$. From the $T(z)\psi(w)$ and $\overline{T}(\overline{z})\overline{\psi}(\overline{w})$ operator products we verify that $\psi$ and $\overline{\psi}$ are primary fields of conformal weight $(\frac{1}{2}, 0)$ and $(0, \frac{1}{2})$.

### 3.3. Mode expansions and the Virasoro algebra

It is convenient to define a Laurent expansion of the stress-energy tensor,

$$T(z) = \sum_{n\in\mathbf{Z}} z^{-n-2}L_n \ , \qquad \overline{T}(\overline{z}) = \sum_{n\in\mathbf{Z}} \overline{z}^{-n-2}\overline{L}_n \ , \qquad (3.6)$$

in terms of modes $L_n$ (which are themselves operators). The exponent $-n-2$ in (3.6) is chosen so that for the scale change $z \to z/\lambda$, under which $T(z) \to$ $\lambda^2 T(z/\lambda)$, we have $L_{-n} \to \lambda^n L_{-n}$. $L_{-n}$ and $\overline{L}_{-n}$ thus have scaling dimension $n$. (3.6) is formally inverted by the relations

$$L_n = \oint \frac{dz}{2\pi i} z^{n+1} T(z) , \qquad \overline{L}_n = \oint \frac{d\overline{z}}{2\pi i} \overline{z}^{n+1} \overline{T}(\overline{z}) . \tag{3.7}$$

To compute the algebra of commutators satisfied by the modes $L_n$ and $\overline{L}_n$, we employ a procedure for making contact between local operator products and commutators of operator modes that will repeatedly prove useful. The commutator of two contour integrations $[\oint dz, \oint dw]$ is evaluated by first fixing $w$ and deforming the difference between the two $z$ integrations into a single $z$ contour drawn tightly around the point $w$, as in fig. 2. In evaluating the $z$ contour integration, we may perform operator product expansions to identify the leading behavior as $z$ approaches $w$. The $w$ integration is then performed without further subtlety. For the modes of the stress-energy tensor, this procedure gives

$$\begin{aligned}
[L_n, L_m] &= \left( \oint \frac{dz}{2\pi i} \oint \frac{dw}{2\pi i} - \oint \frac{dw}{2\pi i} \oint \frac{dz}{2\pi i} \right) z^{n+1} T(z) w^{m+1} T(w) \\
&= \oint \frac{dz}{2\pi i} \oint \frac{dw}{2\pi i} z^{n+1} w^{m+1} \left( \frac{c/2}{(z-w)^4} + \frac{2T(w)}{(z-w)^2} + \frac{\partial T(w)}{z-w} + \ldots \right) \\
&= \oint \frac{dw}{2\pi i} \Big( \frac{c}{12}(n+1)n(n-1) w^{n-2} w^{m+1} \\
&\qquad + 2(n+1) w^n w^{m+1} T(w) + w^{n+1} w^{m+1} \partial T(w) \Big) .
\end{aligned}$$

(where the residue of the first term results from $\frac{1}{3!} \partial_z^3 z^{n+1}|_{z=w} = \frac{1}{6}(n+1)n(n-1) w^{n-2}$). Integrating the last term by parts and combining with the second term gives $(n-m) w^{n+m+1} T(w)$, so performing the $w$ integration gives

$$[L_n, L_m] = (n-m) L_{n+m} + \frac{c}{12}(n^3 - n) \delta_{n+m,0} . \tag{3.8a}$$

The identical calculation for $\overline{T}$ results in

$$[\overline{L}_n, \overline{L}_m] = (n-m) \overline{L}_{n+m} + \frac{\overline{c}}{12}(n^3 - n) \delta_{n+m,0} . \tag{3.8b}$$

Since $T(z)$ and $\overline{T}(\overline{z})$ have no power law singularities in their operator product, on the other hand, we have the commutation

$$[L_n, \overline{L}_m] = 0 . \tag{3.8c}$$

In (3.8a–c) we find two copies of an infinite dimensional algebra, called the Virasoro algebra, originally discovered in the context of string theory [11]. Every conformally invariant quantum field theory determines a representation of this algebra with some value of $c$ and $\overline{c}$. For $c = \overline{c} = 0$, (3.8a,b) reduces to the classical algebra (1.8). The form of the algebra may be altered a bit by shifting the $L_n$'s by constants. In (3.8a) this freedom is exhausted by the requirement that the subalgebra $L_{-1}, L_0, L_1$ satisfy

$$[L_{\mp 1}, L_0] = \mp L_{\mp 1} \quad [L_1, L_{-1}] = 2L_0 ,$$

with no anomaly term. The global conformal group $SL(2, \mathbf{C})$ generated by $L_{-1,0,1}$ and $\overline{L}_{-1,0,1}$ thus remains an exact symmetry group despite the central charge in (3.8).

### 3.4. In- and out-states

To analyze further the properties of the modes, it is useful to introduce the notion of adjoint,

$$\left[ A(z, \overline{z}) \right]^\dagger = A\left( \frac{1}{\overline{z}}, \frac{1}{z} \right) \frac{1}{\overline{z}^{2h}} \frac{1}{z^{2\overline{h}}} , \tag{3.9}$$

(on the real surface $\overline{z} = z^*$), for Euclidean-space fields that correspond to real (Hermitian) fields in Minkowski space. Although (3.9) might look strange, it is ultimately justified by considering the continuation back to the Minkowski space cylinder, as described in section 2.2. The missing factors of $i$ in Euclidean-space time evolution, $A(x, \tau) = e^{H\tau} A(x, 0) e^{-H\tau}$, must be compensated in the definition of the adjoint by an explicit Euclidean-space time reversal, $\tau \to -\tau$. As discussed earlier, this is implemented on the plane by $z \to 1/z^*$. The additional $z, \overline{z}$ dependent factors on the right hand side of (3.9) are required to give the adjoint the proper tensorial properties under the conformal group.

We derive further intuition by considering in- and out-states in conformal field theory. In Euclidean field theory we ordinarily associate states with operators via the identification

$$|A_{\rm in}\rangle = \lim_{\sigma^0 \to -\infty} A(\sigma^0, \sigma^1)|0\rangle = \lim_{\sigma^0 \to -\infty} e^{H\sigma^0} A(\sigma^1)|0\rangle .$$

Since time $\sigma^0 \to -\infty$ on the cylinder corresponds to the origin of the $z$-plane, it is natural to define in-states as

$$|A_{\text{in}}\rangle \equiv \lim_{z,\overline{z}\to 0} A(z,\overline{z})|0\rangle \ .$$

To define $\langle A_{\text{out}}|$ we need to construct the analogous object for $z \to \infty$. Conformal invariance, however, allows us relate a parametrization of a neighborhood about the point at $\infty$ on the Riemann sphere to that of a neighborhood about the origin via the map $z = 1/w$. If we call $\widetilde{A}(w,\overline{w})$ the operator in the coordinates for which $w \to 0$ corresponds to the point at $\infty$, then the natural definition is

$$\langle A_{\text{out}}| \equiv \lim_{w,\overline{w}\to 0} \langle 0|\widetilde{A}(w,\overline{w}) \ . \quad (3.10a)$$

Now we need to relate $\widetilde{A}(w,\overline{w})$ to $A(z,\overline{z})$. Recall that for primary fields we have under $w \to f(w)$

$$\widetilde{A}(w,\overline{w}) = A\bigl(f(w),\overline{f}(\overline{w})\bigr) \bigl(\partial f(w)\bigr)^h \bigl(\overline{\partial}\,\overline{f}(\overline{w})\bigr)^{\overline{h}} \ ,$$

so that in particular under $f(w) = 1/w$ we have

$$\widetilde{A}(w,\overline{w}) = A\!\left(\frac{1}{w},\frac{1}{\overline{w}}\right)(-w^{-2})^h (-\overline{w}^{-2})^{\overline{h}} \ .$$

The definition (3.9) of adjoint then gives the natural relation between $\langle A_{\text{out}}|$ and $|A_{\text{in}}\rangle$ (up to a spin dependent phase ignored here for convenience),

$$\begin{aligned}\langle A_{\text{out}}| &= \lim_{w,\overline{w}\to 0} \langle 0|\widetilde{A}(w,\overline{w}) && \text{definition} \\ &= \lim_{z,\overline{z}\to 0} \langle 0| A\!\left(\frac{1}{z},\frac{1}{\overline{z}}\right)\frac{1}{z^{2h}}\frac{1}{\overline{z}^{2\overline{h}}} && \text{conformal invariance} \\ &= \lim_{z,\overline{z}\to 0} \langle 0| \bigl[A(\overline{z},z)\bigr]^\dagger && \text{adjoint} \\ &= \left[\lim_{z,\overline{z}\to 0} A(\overline{z},z)|0\rangle\right]^\dagger \\ &= |A_{\text{in}}\rangle^\dagger \ .\end{aligned} \quad (3.11)$$

Occasionally we shall be sloppy and write the out-state in the form $\langle A_{\text{out}}| \equiv \lim_{z,\overline{z}\to\infty} \langle 0|A(z,\overline{z})$ — this should be recognized as shorthand for

$$\langle A_{\text{out}}| \equiv \lim_{z,\overline{z}\to\infty} \langle 0|A(z,\overline{z})\, z^{2h}\overline{z}^{2\overline{h}} \ , \quad (3.10b)$$

as follows from the definition (3.10a) and the second line of (3.11). (Eqns. (3.10a,b) are actually correct for any quasi-primary field, since we only make use of the $SL(2,\mathbf{C})$ transformation $w \to 1/w$ to define the out-state. For general secondary fields, on the other hand, the slightly more complicated expression may be found for example in [12].)

(We point out that in defining our in- and out-states by means of fields of well-defined scaling dimension, we are proceeding somewhat differently than in ordinary perturbative field theory calculations. The procedure here defines asymptotic states that are eigenstates of the exact Hamiltonian of the system, rather than eigenstates of some fictitious asymptotically non-interacting Hamiltonian. Our ability to do this in conformal field theories in two dimensions stems from their providing non-trivial examples of solvable quantum field theories. If we could implement such a prescription in non-trivial 3+1 dimensional field theories, we of course would. We also point out that the correspondence between operators and states in field theory is not ordinarily one-to-one — in massive field theories, for example, more than one operator typically creates the same state as $\sigma^0 \to -\infty$. In conformal field theory, the number of fields and states with any fixed conformal weight is ordinarily finite so by orthogonalization we can associate a unique field with each state.)

Note that for the stress-energy tensor, equality of

$$T^\dagger(z) = \sum \frac{L_m^\dagger}{\overline{z}^{m+2}} \quad \text{and} \quad T\!\left(\frac{1}{\overline{z}}\right)\frac{1}{\overline{z}^4} = \sum \frac{L_m}{\overline{z}^{-m-2}}\frac{1}{\overline{z}^4}$$

results in

$$L_m^\dagger = L_{-m} \ . \quad (3.12)$$

(3.12) should be regarded as the condition that $T(z)$ is hermitian. Hermiticity of $\overline{T}(\overline{z})$ equivalently results in $\overline{L}_m^\dagger = \overline{L}_{-m}$.

Other important conditions on the $L_n$'s can be derived by requiring the regularity of

$$T(z)|0\rangle = \sum_{m \in \mathbf{Z}} L_m \, z^{-m-2}|0\rangle$$

at $z=0$. Evidently only terms with $m \leq -2$ are allowed, so we learn that

$$L_m|0\rangle = 0\,, \qquad m \geq -1\,. \tag{3.13a}$$

From (3.11) we have also that $\langle 0|L_m^\dagger = 0$, $m \geq -1$. $L_{0,\pm 1}|0\rangle = 0$ is the statement that the vacuum is $SL(2,\mathbf{R})$ invariant, and we see that this follows directly just from the requirement that $z=0$ be a regular point (the rest of the vanishing $L_m|0\rangle = 0$, $m \geq 1$, come along for free). From (3.12) we find $L_m^\dagger|0\rangle = 0$, $m \leq 1$, and thus from (3.11) that

$$\langle 0|L_m = 0\,, \qquad m \leq 1\,. \tag{3.13b}$$

The states $L_{-n}|0\rangle$ for $n \geq 2$, on the other hand, are in principle non-trivial Hilbert space states that transform as part of some representation of the Virasoro algebra.

The only generators in common between (3.13a,b), annihilating both $\langle 0|$ and $|0\rangle$, are $L_{\pm 1, 0}$. It is easy to show, using the commutation relations (3.8a), that this is the only finite subalgebra of the Virasoro algebra for which this is possible. Identical results apply as well for the $\overline{L}_n$'s, and we shall call the vacuum state $|0\rangle$, annihilated by both $L_{\pm 1, 0}$ and $\overline{L}_{\pm 1, 0}$, the $SL(2, \mathbf{C})$ invariant vacuum. (Strictly speaking we could denote this as the tensor product $|0\rangle \otimes |0\rangle$ of two $SL(2,\mathbf{R})$ invariant vacuums, but any ambiguity in the symbol $|0\rangle$ is ordinarily resolved by context.)

The conditions (3.13) together with the commutation rules (3.8a) can be used to verify that

$$\langle T(z)\, T(w)\rangle = \langle 0| \sum_{n \in \mathbf{Z}} L_n\, z^{-n-2} \sum_{m \in \mathbf{Z}} L_m\, w^{-m-2} |0\rangle = \frac{c/2}{(z-w)^4}\,, \tag{3.14}$$

giving an easy way to calculate $c$ in some theories. Similarly, we can compute all higher point correlation functions of the form

$$\begin{aligned}\langle T(w_1)\cdots T(w_n)\overline{T}(\overline{z}_1)\cdots \overline{T}(\overline{z}_m)\rangle \\ = \langle T(z_1)\cdots T(z_n)\rangle \langle \overline{T}(\overline{z}_1)\cdots \overline{T}(\overline{z}_m)\rangle\,,\end{aligned} \tag{3.15}$$

by substituting the mode expansions (3.6) and commuting the $L_n$'s with $n$ positive (negative) to the right (left). We can also see the condition $c > 0$ to result from the algebra (3.8a), and the relations (3.13a) and (3.12):

$$\frac{c}{2} = \langle 0|[L_2, L_{-2}]|0\rangle = \langle 0|L_2\, L_2^\dagger|0\rangle \geq 0\,,$$

since the norm satisfies $\|L_2^\dagger|0\rangle\|^2 \geq 0$ in a positive Hilbert space.

*3.5. Highest weight states*

Let us now consider the state

$$|h\rangle = \phi(0)|0\rangle \tag{3.16}$$

created by a holomorphic field $\phi(z)$ of weight $h$. From the operator product expansion (2.10) between the stress-energy $T$ and a primary field $\phi$ we find

$$[L_n, \phi(w)] = \oint \frac{dz}{2\pi i}\, z^{n+1} T(z)\phi(w) = h(n+1)w^n \phi(w) + w^{n+1}\partial\phi(w)\,, \tag{3.17}$$

so that $[L_n, \phi(0)] = 0$, $n > 0$. The state $|h\rangle$ thus satisfies

$$L_0|h\rangle = h|h\rangle \qquad L_n|h\rangle = 0,\ n > 0\,. \tag{3.18a}$$

More generally, an in-state $|h, \overline{h}\rangle$ created by a primary field $\phi(z,\overline{z})$ of conformal weight $(h, \overline{h})$ will also satisfy (3.18a) with the replacements $L \to \overline{L}$, $h \to \overline{h}$. Since $L_0 \pm \overline{L}_0$ are the generators of dilatations and rotations, we identify $h \pm \overline{h}$ as the scaling dimension and Euclidean spin of the state.

Any state satisfying (3.18a) is known as a highest weight state. States of the form $L_{-n_1}\cdots L_{-n_k}|h\rangle$ ($n_i > 0$) are known as descendant states. The out-state $\langle h|$, defined as in (3.10), evidently satisfies

$$\langle h|L_0 = h\langle h| \qquad \langle h|L_n = 0,\ n < 0\,. \tag{3.18b}$$

The states $\langle h|L_{n_1}\cdots L_{n_k}$ $(n_i > 0)$ are the descendants of the out-state. Using (3.12), (3.18), and (3.8a), we evaluate

$$\begin{aligned}\langle h|L_{-n}^{\dagger} L_{-n}|h\rangle &= \langle h|[L_n, L_{-n}]|h\rangle \\ &= 2n\langle h|L_0|h\rangle + \frac{c}{12}(n^3 - n)\langle h|h\rangle \\ &= \left(2nh + \frac{c}{12}(n^3 - n)\right)\langle h|h\rangle \ . \end{aligned} \quad (3.19)$$

Again, this quantity must be positive if the Hilbert space has a positive norm. For $n$ large this tells us that we must have $c > 0$, and for $n = 1$ this requires that $h \geq 0$. In the latter case we also see that $h = 0$ only if $L_{-1}|h\rangle = 0$, i.e. only if $|h\rangle$ is identically the $SL(2, \mathbf{R})$ invariant vacuum $|0\rangle$.

We can also show for $c = 0$ that the Virasoro algebra has no interesting unitary representations. From (3.19), we see that all states $L_{-n}|0\rangle$ would have zero norm and hence should be set equal to zero. Moreover for arbitrary $h$ if we consider[13] the matrix of inner products in the 2×2 basis $L_{-2n}|h\rangle$, $L_{-n}^2|h\rangle$, we find a determinant equal to $4n^3h^2(4h - 5n)$. For $h \neq 0$ this quantity is always negative for large enough $n$. Thus for $c = 0$ the only unitary representation of the Virasoro algebra is completely trivial: it has $h = 0$ and all the $L_n = 0$.

It follows from (3.17) that a field $\phi$ with conformal weight $(h, 0)$ is purely holomorphic. We first note from (3.17) adapted to the anti-holomorphic case that $[\overline{L}_{-1}, \phi] = \overline{\partial}\phi$, then argue as in (3.19) to show that the norm of the state $\overline{L}_{-1}\phi|0\rangle = 0$, and hence that $\overline{\partial}\phi = 0$. To see what (3.16) means in terms of modes, we generalize the mode expansions (3.6) to arbitrary holomorphic primary fields $\phi(z)$ of weight $(h, 0)$,

$$\phi(z) = \sum_{n \in \mathbf{Z}-h} \phi_n z^{-n-h} \ ,$$

again chosen so that $\phi_{-n}$ has scaling weight $n$. The modes satisfy

$$\phi_n = \oint \frac{dz}{2\pi i} z^{h+n-1} \phi(z) \ .$$

Regularity of $\phi(z)|0\rangle$ at $z = 0$ requires $\phi_n|0\rangle = 0$ for $n \geq -h + 1$, generalizing the case $h = 2$ in (3.13a). From (3.16) we see that the state $|h\rangle$ is created by the mode $\phi_{-h}$: $|h\rangle = \phi_{-h}|0\rangle$. To check that the states $\phi_n|0\rangle$ have the correct $L_0$ values, we use (3.17) to calculate the commutator

$$\begin{aligned}[L_n, \phi_m] &= \oint \frac{dw}{2\pi i} w^{h+m-1}\bigl(h(n+1)w^n\phi(w) + w^{n+1}\partial\phi(w)\bigr) \\ &= \oint \frac{dw}{2\pi i} w^{h+m+n-1}\bigl(h(n+1) - (h+m+n)\bigr)\phi(w) \\ &= \bigl(n(h-1) - m\bigr)\phi_{m+n} \ . \end{aligned} \quad (3.20)$$

So $[L_0, \phi_m] = -m\phi_m$, consistent for example with $L_0|h\rangle = L_0\phi_{-h}|0\rangle = h|h\rangle$.

Before turning to a detailed consideration of descendant fields, we show how the formalism of this subsection may be used to derive the generalization of (2.3) to $n$-point functions. We first use the $SL(2, \mathbf{C})$ invariance of the vacuum, $U|0\rangle = |0\rangle$ for $U \in SL(2, \mathbf{C})$, to derive (1.13) (or rather (2.20)) in the form

$$\langle 0|U^{-1}\phi_1 U \cdots U^{-1}\phi_n U|0\rangle = \langle 0|\phi_1 \cdots \phi_n|0\rangle \ , \quad (3.21)$$

where the $\phi_i$'s are quasi-primary fields (i.e. satisfy

$$U^{-1}\phi(z, \overline{z})U = \bigl(\partial f(z)\bigr)^h \bigl(\overline{\partial}\overline{f}(\overline{z})\bigr)^{\overline{h}} \phi\bigl(f(z), \overline{f}(\overline{z})\bigr) \ ,$$

for $f$ of the form (1.9)). Infinitesimally, (3.21) takes the obvious form

$$0 = \langle 0|[L_k, \phi_1(z_1)]\ldots\phi_n(z_n)|0\rangle + \cdots + \langle 0|\phi_1(z_1)\ldots[L_k, \phi_n(z_1)]|0\rangle \ ,$$

for $k = 0, \pm 1$. Using (3.17) we write this equivalently as

$$\begin{aligned}&\sum_{i=1}^{n} \partial_i \langle 0|\phi_1(z_1)\ldots\phi_n(z_n)|0\rangle = 0 \\ &\sum_{i=1}^{n} (z_i\partial_i + h_i)\langle 0|\phi_1(z_1)\ldots\phi_n(z_n)|0\rangle = 0 \\ &\sum_{i=1}^{n} (z_i^2\partial_i + 2z_ih_i)\langle 0|\phi_1(z_1)\ldots\phi_n(z_n)|0\rangle = 0 \ , \end{aligned} \quad (3.22)$$

implying respectively invariance under translations, dilatations, and special conformal transformations. We also point out that (3.21) applies as well to the correlation functions (3.15) even though $T$ is not a primary field. Recall that the Schwartzian derivative $S(f, z)$ of (3.3) vanishes for the global transformations (1.9), implying that $T$ is quasi-primary, and that suffices to show that its correlation functions transform covariantly under $SL(2, \mathbf{C})$.

*3.6. Descendant fields*

As mentioned at the beginning of this section, representations of the Virasoro algebra start with a single primary field. Remaining fields in the representation are given by successive operator products with the stress-energy tensor. Together all these fields comprise a representation $[\phi_n]$. (In terms of modes, the descendant fields are obtained by commuting $L_{-n}$'s with primary fields.) Acting on the vacuum, the descendant fields create descendant states. We shall see that the conformal ward identities give differential equations that determine the correlation functions of descendant fields in terms of those of primaries. The utility of organizing a two dimensional conformal field theory in terms of conformal families, i.e. irreducible representations of the Virasoro algebra, is that the theory may then be completely specified by the Green functions of the primary fields.

We extract the descendant fields $\widehat{L}_{-n}\phi$, $n > 0$, from the less singular parts of the operator product expansion of $T(z)$ with a primary field,

$$T(z)\phi(w,\overline{w}) \equiv \sum_{n \geq 0}(z-w)^{n-2}\widehat{L}_{-n}\phi(w,\overline{w})$$
$$= \frac{1}{(z-w)^2}\widehat{L}_0\phi + \frac{1}{z-w}\widehat{L}_{-1}\phi + \widehat{L}_{-2}\phi + (z-w)\widehat{L}_{-3}\phi + \ldots \quad (3.23)$$

The fields
$$\widehat{L}_{-n}\phi(w,\overline{w}) = \oint \frac{dz}{2\pi i}\frac{1}{(z-w)^{n-1}}\,T(z)\phi(w,\overline{w}) \quad (3.24)$$

are sometimes also denoted as $\phi^{(-n)}$ (and in the presence of larger algebraic structures are called Virasoro descendants to avoid ambiguity). The conformal weight of the descendant field $\widehat{L}_{-n}\phi$ is $(h+n, \overline{h})$. Note from (2.10) that the first two descendant fields are given by $\phi^{(0)} = \widehat{L}_0\phi = h\phi$ and $\phi^{(-1)} = \widehat{L}_{-1}\phi = \partial\phi$.

A simple example of a descendant field is
$$\bigl(\widehat{L}_{-2}1\bigr)(w) = \oint \frac{dz}{2\pi i}\frac{1}{z-w}T(z)1 = T(w) \ .$$
Thus $1^{(-2)}(w) = \bigl(\widehat{L}_{-2}1\bigr)(w) = T(w)$, and we see that the stress-energy tensor is always a level 2 descendant of the identity operator. This explains why the operator product (3.1) of the stress-energy tensor with itself does not take the canonical form (2.10) of that for a primary field.

For $n > 0$, primary fields satisfy $\widehat{L}_n\phi = 0$. The first few descendant fields, ordered according to their conformal weight, are

$$\begin{array}{ccl}
\underline{\text{level}} & \underline{\text{dimension}} & \underline{\text{field}} \\
0 & h & \phi \\
1 & h+1 & \widehat{L}_{-1}\phi \\
2 & h+2 & \widehat{L}_{-2}\phi,\ \widehat{L}_{-1}^2\phi \\
3 & h+3 & \widehat{L}_{-3}\phi,\ \widehat{L}_{-1}\widehat{L}_{-2}\phi,\ \widehat{L}_{-1}^3\phi \\
& \cdots & \\
N & h+N & P(N) \text{ fields} ,
\end{array} \quad (3.25)$$

where the number at level $N$ is given by $P(N)$, the number of partitions of $N$ into positive integer parts. $P(N)$ is given in terms of the generating function

$$\frac{1}{\prod_{n=1}^{\infty}(1-q^n)} = \sum_{N=0}^{\infty} P(N)\, q^N \ , \quad (3.26)$$

where $P(0) \equiv 1$. The fields in (3.25) arise from repeated short distance expansion of the primary field $\phi$ with $T(z)$, and constitute the conformal family $[\phi]$ based on $\phi$. Since $\widehat{L}_{-1}\psi = \partial\psi$ for any field $\psi$, $[\phi]$ naturally contains in particular all derivatives of each of its fields.

All the correlation functions of the secondary fields are given by differential operators acting on those of primary fields. For example if we let $z \to w_n$ in (2.22), expand in powers of $z - w_n$, and use the definition (3.23) of secondary fields, we find

$$\begin{aligned}
&\bigl\langle \phi_1(w_1,\overline{w}_1)\ldots\phi_{n-1}(w_{n-1},\overline{w}_{n-1})\bigl(\widehat{L}_{-k}\phi\bigr)(z,\overline{z})\bigr\rangle \\
&\quad = \mathcal{L}_{-k}\bigl\langle \phi_1(w_1,\overline{w}_1)\ldots\phi_{n-1}(w_{n-1},\overline{w}_{n-1})\,\phi(z,\overline{z})\bigr\rangle \ ,
\end{aligned} \quad (3.27a)$$

where the differential operator (for $k \geq 2$) is defined by

$$\mathcal{L}_{-k} = -\sum_{j=1}^{n-1} \left( \frac{(1-k)h_j}{(w_j - z)^k} + \frac{1}{(w_j - z)^{k-1}} \frac{\partial}{\partial w_j} \right) . \qquad (3.27b)$$

The $\mathcal{L}$'s provide a differential realization of (3.8a) with $c = 0$. With $z = \overline{z} = 0$, we see from (3.24) and (3.7) that $\widehat{L}_{-k}\phi(0) \to L_{-k}\phi(0)$. Thus (3.27) can also be derived at $z = 0$ by using (3.17) to commute $L_{-k}$ to the left, and then using the highest weight property (3.13b) of the out vacuum. (Although (2.22) was derived for $|z|$ greater than all the $|w_i|$'s, it is easy to show either by contour integral methods or by substituting the mode expansion for $T$ and commuting $L$'s that it remains true for any ordering of the arguments). By the same methods, the generalization of (3.27) to correlation functions involving one arbitrary secondary field is

$$\langle 0|\phi_1(w_1,\overline{w}_1)\ldots \phi_{n-1}(w_{n-1},\overline{w}_{n-1}) \widehat{L}_{-k_1}\ldots \widehat{L}_{-k_\ell} \phi(z,\overline{z})|0\rangle$$
$$= \mathcal{L}_{-k_1}\ldots \mathcal{L}_{-k_\ell} \langle 0|\phi_1(w_1,\overline{w}_1)\ldots \phi_{n-1}(w_{n-1},\overline{w}_{n-1}) \phi(z,\overline{z})|0\rangle . \qquad (3.28)$$

In principle one can write down expressions for correlation functions of arbitrary secondary fields in terms of those for primaries, but there is no convenient closed form expression in the most general case. A particular case of interest is the 2-point function. If we take orthogonal primary fields as in (2.12), then it follows directly from (2.22) that the 2-point functions of descendants of different primary fields must vanish.

A problem related to calculating correlation functions of secondary fields is to write the operator product coefficients (2.13) for descendants in terms of those for primaries. Let us consider (2.13) with $\phi_i$ and $\phi_j$ primary fields, and group together all the secondary fields belonging to the conformal family $[\phi_p]$ in the summation to write

$$\phi_i(z,\overline{z})\phi_j(w,\overline{w}) =$$
$$\sum_{p\{k\overline{k}\}} C_{ijp}^{\{k\overline{k}\}} z^{(h_p - h_i - h_j + \Sigma_\ell k_\ell)} \overline{z}^{(\overline{h}_p - \overline{h}_i - \overline{h}_j + \Sigma_\ell \overline{k}_\ell)} \phi_p^{\{k\overline{k}\}}(w,\overline{w}) . \qquad (3.29)$$

Here we have labeled the descendants

$$\widehat{L}_{-k_1}\cdots \widehat{L}_{-k_n} \widehat{\overline{L}}_{-\overline{k}_1}\cdots \widehat{\overline{L}}_{-\overline{k}_m} \phi_p$$

of a primary field $\phi_p$ by $\phi_p^{\{k\overline{k}\}}$, and we assume the normalization (2.12). The operator product coefficients in this normalization are symmetric and from (2.5) coincide with the numerical factor in the 3-point function

$$\langle \phi_i | \phi_j(z,\overline{z}) | \phi_p \rangle = \langle \phi_i(\infty) \phi_j(z,\overline{z}) \phi_p(0) \rangle = C_{ijp}\, z^{h_i - h_j - h_p}\, \overline{z}^{\overline{h}_i - \overline{h}_j - \overline{h}_p} ,$$

where these fields are either primary or secondary. Using (3.28) in the case of the 3-point function for fields as in (3.29) (or by performing a conformal transformation on both sides of (3.29) and comparing terms), one can show[1] that

$$C_{ijp}^{\{k\overline{k}\}} = \mathbf{C}_{ijp}\, \beta_{ij}^{p\{k\}}\, \overline{\beta}_{ij}^{p\{\overline{k}\}} , \qquad (3.30)$$

where the $\mathbf{C}_{ijp}$'s are the operator product coefficients for primary fields, and $\beta_{ij}^{p\{k\}}$ ($\overline{\beta}_{ij}^{p\{\overline{k}\}}$) is a function of the four parameters $h_i$, $h_j$, $h_p$, and $c$ ($\overline{h}_i, \overline{h}_j, \overline{h}_p$, and $\overline{c}$) determined entirely by conformal invariance (and can in principle be computed mechanically). Moreover the 3-point function for any three descendant fields can be determined from that of their associated primaries (although as noted after (3.28), the explicit form of the relation is awkward to write down in all generality). The primary $\mathbf{C}_{ijp}$'s thus determine the allowed non-vanishing 3-point functions for any members of the families $[\phi_i]$, $[\phi_j]$, and $[\phi_p]$.

We see that the complete information to specify a two dimensional conformal field theory is provided by the conformal weights $(h_i, \overline{h}_i)$ of the Virasoro highest weight states, and the operator product coefficients $\mathbf{C}_{ijk}$ between the primary fields that create them. Everything else follows from the values of these parameters, which themselves cannot be determined solely on the basis of conformal symmetry.

### 3.7. Duality and the bootstrap

To determine the $\mathbf{C}_{ijk}$'s and $h$'s, we need to apply some dynamical principle to obtain additional information. Up to now, we have considered only the local constraints imposed by the infinite conformal algebra. Associativity of the operator algebra (2.13), on the other hand, imposes global constraints on correlation functions. To see how this works, we consider evaluating the 4-point function

$$\langle \phi_i(z_1,\overline{z}_1)\phi_j(z_2,\overline{z}_2)\phi_\ell(z_3\overline{z}_3)\phi_m(z_4,\overline{z}_4)\rangle \tag{3.31}$$

in two ways. First we take $z_1 \to z_2$, $z_3 \to z_4$, and find the schematic result depicted in the left hand side of fig. 4, where the sum over $p$ is over both primary and secondary fields. (3.31) can alternatively be evaluated by taking $z_1 \to z_3$, $z_2 \to z_4$, and we have represented this result diagrammatically in the right hand side of fig. 4. Associativity of the operator algebra implies that these two methods of calculating the 4-point function should give the same result. Their equality is a necessary consistency requirement, known as crossing symmetry of the 4-point function.

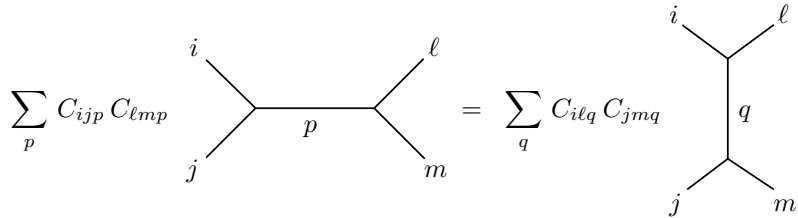

Fig. 4. Crossing symmetry

In fig. 4, we thus have an infinite number of equations that the $C_{ijk}$'s must satisfy. The sum over all the descendant states can be performed in principle, and the relations in fig. 4 become algebraic equations for the $C_{ijk}$'s. These very strong constraints were originally suggested to give a means of characterizing all conformally invariant systems in $d$ dimensions (the procedure of solving the relations of fig. 4 to find conformal field theories is known as 'the conformal bootstrap'). This program however proved too difficult to implement in practice. In two dimensions the problem becomes somewhat more tractable, since one need only consider the primary fields, vastly reducing the number of independent quantities in the problem. There remains however the possibility of encountering an unmanageable number of primary fields, and as well one must still evaluate the objects represented diagrammatically in fig. 4. In [1], it was shown that there are certain special $c, h$ values where things simplify dramatically (such values were also noted in [14]), as we shall discuss momentarily.

First we need to convert fig. 4 to an analytic expression. We can write the contribution to the 4-point function from 'intermediate states' belonging only to the conformal family $[\phi_p]$ as $\mathcal{F}_{ij}^{\ell m}(p|x)\overline{\mathcal{F}}_{ij}^{\ell m}(p|\overline{x})$. This amplitude is represented in fig. 5, and we are for simplicity taking $z_1, z_2, z_3, z_4 = 0, x, 1, \infty$ in the 4-point function (3.31). The amplitude projected onto a single conformal family takes a factorized form because the sums over descendants in the holomorphic and anti-holomorphic families $[\phi_p]$ and $[\overline{\phi}_p]$ (generated by $T$ and $\overline{T}$) are independent. The $\mathcal{F}_{ij}^{\ell m}(p|x)$ depend on the parameters $h_i$, $h_j$, $h_\ell$, $h_m$, $h_p$, and $c$, and are known as "conformal blocks" since any correlation function can be built from them.

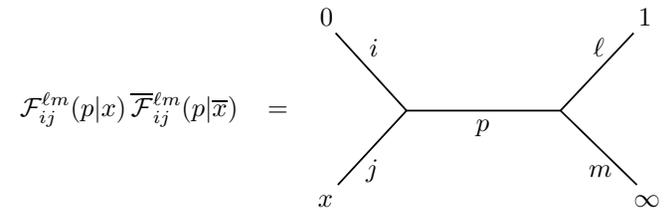

Fig. 5. Single channel amplitude

In terms of the conformal blocks, we can write an analytic form of the diagrammatic equations fig. 4 as

$$\sum_p C_{ijp}C_{\ell mp}\mathcal{F}_{ij}^{\ell m}(p|x)\overline{\mathcal{F}}_{ij}^{\ell m}(p|\overline{x})$$
$$= \sum_q C_{i\ell q}C_{jmq}\mathcal{F}_{i\ell}^{jm}(q|1-x)\overline{\mathcal{F}}_{i\ell}^{jm}(q|1-\overline{x}) \ . \tag{3.32}$$

If we know the conformal blocks $\mathcal{F}$, then (3.32) yields a system of equations that determine the $C_{ijk}$'s and $h,\overline{h}$'s. This has not been carried out in general but at the special values of $c,h$ mentioned earlier, the $\mathcal{F}$'s can be determined as solutions of linear differential equations (that result from the presence of so-called null states). In section 5, we shall see some examples of how this works.

The particular values of $c$ for which things simplify, as mentioned above, take the form
$$c = 1 - \frac{6(m'-m)^2}{mm'} \ ,$$
where $m$ and $m'$ are two coprime positive integers. In [1], these models were called 'minimal models', and it was shown that they possessed a closed operator algebra with only a finite number of primary fields. For these models the bootstrap equation (3.32) can be solved completely, and everything about these conformal field theories can be determined in principle. These models thus realize an old hope[15] that the most singular part of the operator product expansion should define a closed, finite-dimensional algebra of primary fields in a theory. We shall see in the next section that imposing as well the criterion of unitary selects an even smaller subset of these models (with $m' = m + 1$), known as the unitary discrete series. In section 9, we shall see how the fusion rules for their closed operator algebras can be calculated.

The relation represented in fig. 4 is also known as 'duality of the 4-point function' (not to be confused with various other forms of duality that appear in these notes). This notion of duality generalizes to the $n$-point correlation functions
$$\langle \phi_1(z_1,\overline{z}_1) \ldots \phi_n(z_n,\overline{z}_n) \rangle$$
of sensible conformal field theories on arbitrary genus Riemann surfaces. The requirement of duality states that any such correlation function should 1) be a single-valued real analytic function of the $z_i$'s and the moduli of the Riemann surface, and 2) be independent of the basis of conformal blocks used to compute it. Requirement 2) generalizes (3.32) and insures that the correlation function is not sensitive to the particular decomposition of the Riemann surface into thrice-punctured spheres (and also that it be independent of the order of the $\phi_i$'s). Pictorially this generalizes fig. 4 to $n$-point functions, and is discussed further in the contribution of Dijkgraaf to these proceedings.

## 4. Kac determinant and unitarity

### 4.1. The Hilbert space of states

We now return to consider more carefully the Hilbert space of states of a conformal field theory. For the time being it will be sufficient to consider only the holomorphic half of the theory. We recall that a highest weight state $|h\rangle = \phi(0)|0\rangle$, satisfying $L_0|h\rangle = h|h\rangle$, is created by acting with a primary field $\phi$ of conformal weight $h$ on the $SL(2,\mathbf{R})$ invariant vacuum $|0\rangle$, which satisfies $L_n|0\rangle = 0$, $n \geq -1$. We have seen from (3.19) that a positive Hilbert space requires $h \geq 0$. Descendant states are created by acting on $|h\rangle$ with a string of $L_{-n_i}$'s, $n_i > 0$. These states can also be regarded to result from the action of a descendant field acting on the vacuum, e.g.

$$L_{-n}|h\rangle = L_{-n}\big(\phi(0)|0\rangle\big) = \big(\widehat{L}_{-n}\phi\big)(0)|0\rangle = \phi^{(-n)}(0)|0\rangle \ .$$

We wish to verify that every sensible representation of the Virasoro algebra is characterized by such a highest weight state. Generally we are interested in scaling operators, i.e. operators of fixed conformal weight, whose associated states diagonalize the action of $L_0$. Thus we focus on eigenstates $|\psi\rangle$ of $L_0$, say with $L_0|\psi\rangle = h_\psi|\psi\rangle$. Then since $[L_0, L_n] = -nL_n$, we have $L_0 L_n|\psi\rangle = (h_\psi - n)L_n|\psi\rangle$ and $L_n$ lowers the eigenvalue of $L_0$ for $n > 0$. But dilatation in $z$ on the plane, generated by $L_0 + \overline{L}_0$, corresponds to translation in $\sigma^0$ on the cylinder, generated by the energy $H$. $L_0 + \overline{L}_0$ should thus be bounded below in any sensible quantum field theory. Since $L_0$ and $\overline{L}_0$ reside in independent holomorphic and anti-holomorphic algebras, they must be separately bounded from below. By acting with $L_n$'s, we must therefore ultimately reach a state annihilated by $L_n$, $n > 0$ (and similarly by $\overline{L}_n$). This state is the highest weight, or primary, state, that we have been calling $|h\rangle$. We see that we can regard the $L_n$'s, $n > 0$, as an infinite number of harmonic oscillator annihilation operators and the $L_n^\dagger = L_{-n}$'s as creation operators. The representation theory of the Virasoro algebra thus resembles that of $SU(2)$, with $L_0$ playing the role of $J^3$ and the $L_{\pm n}$'s playing the roles of an infinite number of $J^\mp$'s.

We also wish to show that every state in a positive Hilbert space can be expressed as a linear combination of primary and descendant states. Suppose not, i.e. suppose that there exists a state $|\lambda\rangle$ that is not a descendant of a highest weight state. Then in a positive metric theory, we can decompose $|\lambda\rangle = |\delta\rangle + |\psi\rangle$, where $|\psi\rangle$ is orthogonal to all descendants $|\delta\rangle$. If $|\psi\rangle$ has $L_0$ eigenvalue $h_\psi$, let $K = [h_\psi]$ (the greatest integer part). Now consider some order $K$ combination of the $L_{n_i}$'s (such that $\sum n_i = K$ for any term), symbolically denoted $L_K$. Then $|h\rangle = L_K|\psi\rangle$ is a highest weight state with $h = h_\psi - K$ (it must be annihilated by all the $L_n$'s, $n > 0$, since otherwise they would create a state with $h < 0$). But we also have $\langle h|h\rangle = \langle\psi|L_K^\dagger|h\rangle = 0$, since $\langle\psi|$ is orthogonal to all descendants. It follows that $|h\rangle = 0$. We next consider the state $L_{(K-1)}|\psi\rangle = |h+1\rangle$, where $L_{(K-1)}$ is order $(K-1)$ in the $L_n$'s. The same argument as above shows that $|h+1\rangle$ too must be highest weight but have zero norm, and consequently must vanish. By induction we find that $|\psi\rangle$ itself is a highest weight state, concluding the argument.

With this characterization of the Hilbert space of states in hand, we turn to a more detailed consideration of the state representations of the Virasoro algebra. (Via the correspondence between states and fields, we could equally proceed in terms of the fields (3.25), but framing the discussion in terms of states turns out to be slightly more convenient for our purposes.) Starting from a highest weight state $|h\rangle$, we build the set of states

$$
\begin{array}{ccl}
\text{level} & \text{dimension} & \text{state} \\
0 & h & |h\rangle \\
1 & h+1 & L_{-1}|h\rangle \\
2 & h+2 & L_{-2}|h\rangle,\ L_{-1}^2|h\rangle \\
3 & h+3 & L_{-3}|h\rangle,\ L_{-1}L_{-2}|h\rangle,\ L_{-1}^3|h\rangle \\
& \cdots & \\
N & h+N & P(N) \text{ states} ,
\end{array}
\quad (4.1)
$$

known as a Verma module. We are not guaranteed however that all the above states are linearly independent. That depends on the structure of the Virasoro algebra (3.8a) for given values of $h$ and $c$. A linear combination of states that vanishes is known as a null state, and the representation of the Virasoro algebra with highest weight $|h\rangle$ is constructed from the above Verma module by removing all null states (and their descendants).

(It is useful at this point to contrast the situation in two dimensions with that of higher dimensions, where the conformal algebra is finite dimensional. The finite dimensional analog in two dimensions is the closed $SL(2,\mathbf{C})$ subalgebra generated by $L_{0,\pm 1}, \overline{L}_{0,\pm 1}$. Its irreducible representations are much smaller than those of the full infinite dimensional Virasoro algebra. In general an irreducible representation of the Virasoro algebra contains an infinite number of $SL(2,\mathbf{C})$ representations, whose behavior is thereby tied together. It is this additional structure that enables a more extensive analysis of conformal theories in two dimensions.)

Let us now consider the consequences of a linear combination of states that vanishes. At level 1, the only possibility is that

$$L_{-1}|h\rangle = 0 ,$$

but this just implies that $h = 0$, i.e. $|h\rangle = |0\rangle$. At level 2, on the other hand, it may happen that

$$L_{-2}|h\rangle + aL_{-1}^2|h\rangle = 0$$

for some value of $a$. By applying $L_1$ to the above equation, we derive a consistency condition,

$$[L_1, L_{-2}]|h\rangle + a[L_1, L_{-1}^2]|h\rangle = 3L_{-1}|h\rangle + a(L_{-1}2L_0 + 2L_0 L_{-1})|h\rangle$$
$$= \bigl(3 + 2a(2h+1)\bigr)L_{-1}|h\rangle = 0 ,$$

which requires that $a = -3/2(2h+1)$. By applying $L_2$, we find that

$$[L_2, L_{-2}]|h\rangle + a[L_2, L_{-1}^2]|h\rangle = \left(4L_0 + \frac{c}{12}6\right)|h\rangle + 3aL_1 L_{-1}|h\rangle$$
$$= (4h + c/2 + 6ah)|h\rangle = 0 ,$$

so that the central charge must satisfy $c = 2(-6ah - 4h) = 2h(5-8h)/(2h+1)$. We conclude that a highest weight state $|h\rangle$ of the Virasoro algebra at this value of $c$ satisfies

$$\left(L_{-2} - \frac{3}{2(2h+1)}L_{-1}^2\right)|h\rangle = 0 \ . \tag{4.2}$$

Such a state $|h\rangle$, with a null descendant at level 2, is also called degenerate at level 2.

For a degenerate primary field, the analogous statement is

$$\left(\widehat{L}_{-2} - \frac{3}{2(2h+1)}\widehat{L}_{-1}^2\right)\phi = 0 \ .$$

By (3.27), correlation functions of such a field are annihilated by the differential operator $\mathcal{L}_{-2} - \frac{3}{2(2h+1)}\mathcal{L}_{-1}^2$. To express this differential equation in a form that will prove useful later, we write $\widehat{L}_{-2}\phi = -a\widehat{L}_{-1}^2\phi = -a\frac{\partial^2}{\partial z^2}\phi$ for a field $\phi$ degenerate at level 2. From the definition (3.23), as $z \to w$,

$$\widehat{L}_{-2}\phi(w,\overline{w}) = T(z)\phi(w,\overline{w}) - \frac{h\phi(w,\overline{w})}{(z-w)^2} - \frac{\partial \phi(w,\overline{w})}{z-w} - \cdots \ ,$$

together with (2.22) in the limit $z \to w_1$, we derive

$$\begin{aligned}
&-a\frac{\partial^2}{\partial w_1^2}\langle \phi_1(w_1,\overline{w}_1)\ldots\phi_n(w_n,\overline{w}_n)\rangle \\
&= \left\langle \left(T(z)\phi_1(w_1,\overline{w}_1) - \frac{h\phi_1(w_1,\overline{w}_1)}{(z-w_1)^2} - \frac{\partial \phi_1(w_1,\overline{w}_1)}{z-w_1}\right) \right. \\
&\qquad \left. \cdot \phi_2(w_2,\overline{w}_2)\ldots\phi_n(w_n,\overline{w}_n) \right\rangle_{z\to w_1} \\
&= \sum_{j\neq 1}\left(\frac{h_j}{(w_1-w_j)^2} + \frac{1}{w_1-w_j}\frac{\partial}{\partial w_j}\right)\langle \phi_1(w_1,\overline{w}_1)\ldots\phi_n(w_n,\overline{w}_n)\rangle \ .
\end{aligned} \tag{4.3}$$

This is a second order differential equation for any $n$-point function involving a primary field $\phi_1$ with a null descendant at level 2. In the case of 4-point functions, the solutions to (4.3) are expressible in terms of standard hypergeometric functions. In section 5, we shall show how monodromy conditions can be used to select particular solutions that are physically relevant.

### 4.2. Kac determinant

At any given level, the quantity to calculate to determine more generally whether there are any non-trivial linear relations among the states is the matrix of inner products at that level. A zero eigenvector of this matrix gives a linear combination with zero norm, which must vanish in a positive definite Hilbert space. At level 2, for example, we work in the 2×2 basis $L_{-2}|h\rangle$, $L_{-1}^2|h\rangle$, and calculate

$$\begin{pmatrix} \langle h|L_2 L_{-2}|h\rangle & \langle h|L_1^2 L_{-2}|h\rangle \\ \langle h|L_2 L_{-1}^2|h\rangle & \langle h|L_1^2 L_{-1}^2|h\rangle \end{pmatrix} = \begin{pmatrix} 4h + c/2 & 6h \\ 6h & 4h(1+2h) \end{pmatrix} \ . \tag{4.4a}$$

We can write the determinant of this matrix as

$$2(16h^3 - 10h^2 + 2h^2c + hc) = 32\big(h - h_{1,1}(c)\big)\big(h - h_{1,2}(c)\big)\big(h - h_{2,1}(c)\big) \ , \tag{4.4b}$$

where $h_{1,1}(c) = 0$ and $h_{1,2}, h_{2,1} = \frac{1}{16}(5-c) \mp \frac{1}{16}\sqrt{(1-c)(25-c)}$. The $h = 0$ root is actually due to the null state at level 1, $L_{-1}|0\rangle = 0$, which implies also the vanishing $L_{-1}(L_{-1}|0\rangle) = 0$. This is a general feature: if a null state $|h+n\rangle = 0$ occurs at level $n$, then at level $N$ there are $P(N-n)$ null states $L_{-n_1}\cdots L_{-n_k}|h+n\rangle = 0$ (with $\sum_i n_i = N-n$). Thus a null state for some value of $h$ that first appears at level $n$ implies that the determinant at level $N$ will have a $\big[P(N-n)\big]^{\text{th}}$ order zero for that value of $h$ (and the first term in the product (4.4b) can be reexpressed as $\big(h - h_{1,1}(c)\big)^{P(1)}$ to reflect its origin).

At level $N$, the Hilbert space consists of all states of the form

$$\sum_{\{n_i\}} a_{n_1\cdots n_k} L_{-n_1}\cdots L_{-n_k}|h\rangle \ ,$$

where $\sum_i n_i = N$. We can pick $P(N)$ basis states as in (4.1), and the level $N$ analog of (4.4a,b) is to take the determinant of the $P(N){\times}P(N)$ matrix $M_N(c,h)$ of inner products of the form

$$\langle h|L_{m_\ell}\cdots L_{m_1} L_{-n_1}\cdots L_{-n_k}|h\rangle$$

(where $\sum_{i=1}^\ell m_i = \sum_{j=1}^k n_j = N$). If det $M_N(c,h)$ vanishes, then there exists a linear combination of states with zero norm for that $c, h$. If negative, then the determinant has an odd number of negative eigenvalues (i.e. at least one). The representation of the Virasoro algebra at those values of $c$ and $h$ includes states of negative norm, and is consequently not unitary.

The formula generalizing (4.4b),

$$\det M_N(c,h) = \alpha_N \prod_{pq \leq N} (h - h_{p,q}(c))^{P(N-pq)} , \qquad (4.5a)$$

is due to Kac and was proven in [16]. The product in (4.5a) is over all positive integers $p, q$ whose product is less than or equal to $N$, and $\alpha_N$ is a constant independent of $c$ and $h$. The $h_{p,q}(c)$'s are most easily expressed by reparametrizing $c$ in terms of the (in general complex) quantity

$$m = -\frac{1}{2} \pm \frac{1}{2}\sqrt{\frac{25-c}{1-c}} .$$

Then the $h_{p,q}$'s of (4.5) are given by

$$h_{p,q}(m) = \frac{[(m+1)p - mq]^2 - 1}{4m(m+1)} . \qquad (4.5b)$$

(For $c < 1$ we conventionally choose the branch $m \in (0, \infty)$ — in any event the determinant (4.5a) is independent of the choice of branch since it can be compensated by the interchange $p \leftrightarrow q$ in (4.5b).) We easily verify that (4.5) reduces to (4.4b) for $N = 2$. We also note that $c$ is given in terms of $m$ by $c = 1 - 6/m(m+1)$. Finally we point out that the values of the $h_{p,q}$'s in (4.5b) possess the symmetry $p \to m - p$, $q \to m + 1 - q$.

Although (4.5) can be proven by relatively straightforward methods, we shall not undertake to reproduce a complete proof since only the result itself will be needed in what follows. Here we briefly indicate how the proof goes[16][17]. To begin with one writes down an explicit set of states parametrized by integers $p, q$, shows that they are null, and calculates their eigenvalue $h$. Since $\det M_N(c, h)$ is a polynomial in $h$, it can be determined up to a constant by its zeros in $h$ and their multiplicities. Making use of the observation after (4.4b) that a zero of $\det M_n$ leads to a multiplicity $P(N - n)$ zero of $\det M_N$, the explicit enumeration of states shows that $\det M_N$ has at least all the zeros appearing on the right hand side of (4.5a). To show that this is indeed the full polynomial, i.e. that there are no other zeroes, it suffices to show that the order of the r.h.s. of (4.5a) coincides with the order $\nu_N$ of $\det M_N(c, h)$ as a polynomial in $h$. This latter order can be determined by noting that the highest power of $h$ in $\det M_N(c, h)$ comes from the product of the diagonal elements of the matrix $M_N(c, h)$ (these elements result in the maximum number of $L_0$'s generated by commuting $L_k$'s through an identical set of $L_{-k}$'s). The diagonal element for a state $L_{-n_1} \cdots L_{-n_k}|h\rangle$ gives a contribution proportional to $h^k$. The order of $\det M_N(c, h)$ is thus given by

$$\nu_N = \sum_{\{n_1 + \ldots + n_k = N\}} k = \sum_{pq \leq N} P(N - pq) ,$$

where the summation on the left is over all $\{n_i > 0\}$ with $\sum_{i=1}^{k} n_i = N$, and the right hand side follows from a standard number theoretic identity. We see that the order of the polynomial on the right hand side of (4.5a) coincides with that of $\det M_N(h, c)$, showing that the states explicitly exhibited in [16],[17] exhaust all the zeros and hence determine the determinant up to a constant.

*4.3. Sketch of non-unitarity proof*

Now we are ready to investigate the values of $c$ and $h$ for which the Virasoro algebra has unitary representations[18]. In field theory, unitarity is the statement of conservation of probability and is fundamental. In statistical mechanical systems, it does not necessarily play as central a role. There unitarity is expressed as the property of reflection positivity, and consequently the existence of a hermitian transfer matrix. Statistical mechanical systems that can be described near a second order phase transition by an effective field theory of a local order parameter, however, are always expected to be described by a unitary theory. Higher derivative interactions which might spoil unitarity of a Lagrangian theory are generically irrelevant operators, and do not survive to the long distance effective theory. For the remainder here, we will thus restrict attention to unitary theories. (That is not to say, however, that unitary theories necessarily exhaust *all* cases of interest. The $Q \to 0$ limit of the $Q$-state Potts model, for example, useful in studying percolation, is not described by a local order parameter and is not a unitary theory. The Yang-Lee edge singularity also appears in a non-unitary theory, in this case due to the presence of an imaginary field.)

The analysis of unitary representations of the Virasoro algebra proceeds from a study of the Kac determinant (4.5). As mentioned in the previous subsection, if the determinant is negative at any given level it means that there are negative norm states at that level and the representation is not unitary. If the determinant is greater than or equal to zero, further investigation can determine whether or not the representation at that level is unitary.

In the region $c > 1$, $h \geq 0$, it is easy to see that there are no zeroes of the Kac determinant (4.5) at any level. For $1 < c < 25$, $m$ is not real, and the $h_{p,q}$'s of (4.5b) either have an imaginary part or (for $p = q$) are negative. For $c \geq 25$ we can choose the branch $-1 < m < 0$ and find that all the $h_{p,q}$'s are negative. Now we can show that the non-vanishing of $\det M_N$ in this region implies that all the eigenvalues of $M_N$ are positive. This is because for $h$ large, the matrix becomes dominated by its diagonal elements (as shown at the end of the previous subsection, these are highest order in $h$). Since these matrix elements are all positive, the matrix has all positive eigenvalues for large $h$. But since the determinant never vanishes for $c > 1$, $h \geq 0$, all of the eigenvalues must stay positive in the entire region.

On the boundary $c = 1$, the determinant vanishes at the points $h = n^2/4$ but does not become negative. Thus the Kac determinant (4.5) poses no obstacle in principle to having unitary representations of the Virasoro algebra for any $c \geq 1$, $h \geq 0$.

Only the region $0 < c < 1$, $h > 0$ is delicate to treat, although all steps in the argument are elementary. First we draw the vanishing curves $h = h_{p,q}(c)$ in the $h, c$ plane (see fig. 6), by reexpressing (4.5b) in the form

$$h_{p,q}(c) = \frac{1-c}{96}\left[\left((p+q) \pm (p-q)\sqrt{\frac{25-c}{1-c}}\right)^2 - 4\right] . \quad (4.5b')$$

(In this form it is evident that the convention for which branch in $m$ is chosen is compensated by the interchange $p \leftrightarrow q$). The behavior near $c = 1$ is determined by taking $c = 1 - 6\epsilon$ which gives, to leading order in $\epsilon$,

$$h_{p,q}(c = 1 - 6\epsilon) = \frac{1}{4}(p-q)^2 + \frac{1}{4}(p^2-q^2)\sqrt{\epsilon} \quad (p \neq q)$$
$$h_{p,p}(c = 1 - 6\epsilon) = \frac{1}{4}(p^2-1)\epsilon .$$

By analyzing the curves (4.5b'), it is easy to show that one may connect any point in the region $0 < c < 1$, $h > 0$ to the $c > 1$ region by a path that crosses a single vanishing curve of the Kac determinant at some level. The vanishing is due to a single eigenvalue crossing through zero, so the determinant reverses sign passing through the vanishing curve and there must be a negative norm state at that level. This excludes unitary representations of the Virasoro algebra at all points in this region, except those on the vanishing curves themselves where the determinant vanishes. A more careful analysis[18] of the determinant shows that there is an additional negative norm state everywhere on the vanishing curves except at certain points where they intersect, as indicated in fig. 6.

This discrete set of points, where unitary representations of the Virasoro algebra are not excluded, occur at values of the central charge

$$c = 1 - \frac{6}{m(m+1)} \qquad m = 3, 4, \ldots \qquad (4.6a)$$

($m = 2$ is the trivial theory $c = 0$). To each such value of $c$ there are $m(m-1)/2$ allowed values of $h$ given by

$$h_{p,q}(m) = \frac{[(m+1)p - mq]^2 - 1}{4m(m+1)} \qquad (4.6b)$$

where $p, q$ are integers satisfying $1 \leq p \leq m-1$, $1 \leq q \leq p$.

Thus we see that the *necessary* conditions for unitary highest weight representations of the Virasoro algebra are $(c \geq 1, h \geq 0)$ or $(4.6a,b)$. That the latter of these two conditions is also sufficient, i.e. that there indeed exist unitary representations of the Virasoro algebras for these discrete values of $c, h$, was shown in [19] via a coset space construction (to be discussed in section 9). The overall status of conformal field theories with $c \geq 1$ is not as yet well understood, and much effort is currently being expended to develop more powerful techniques to investigate them.

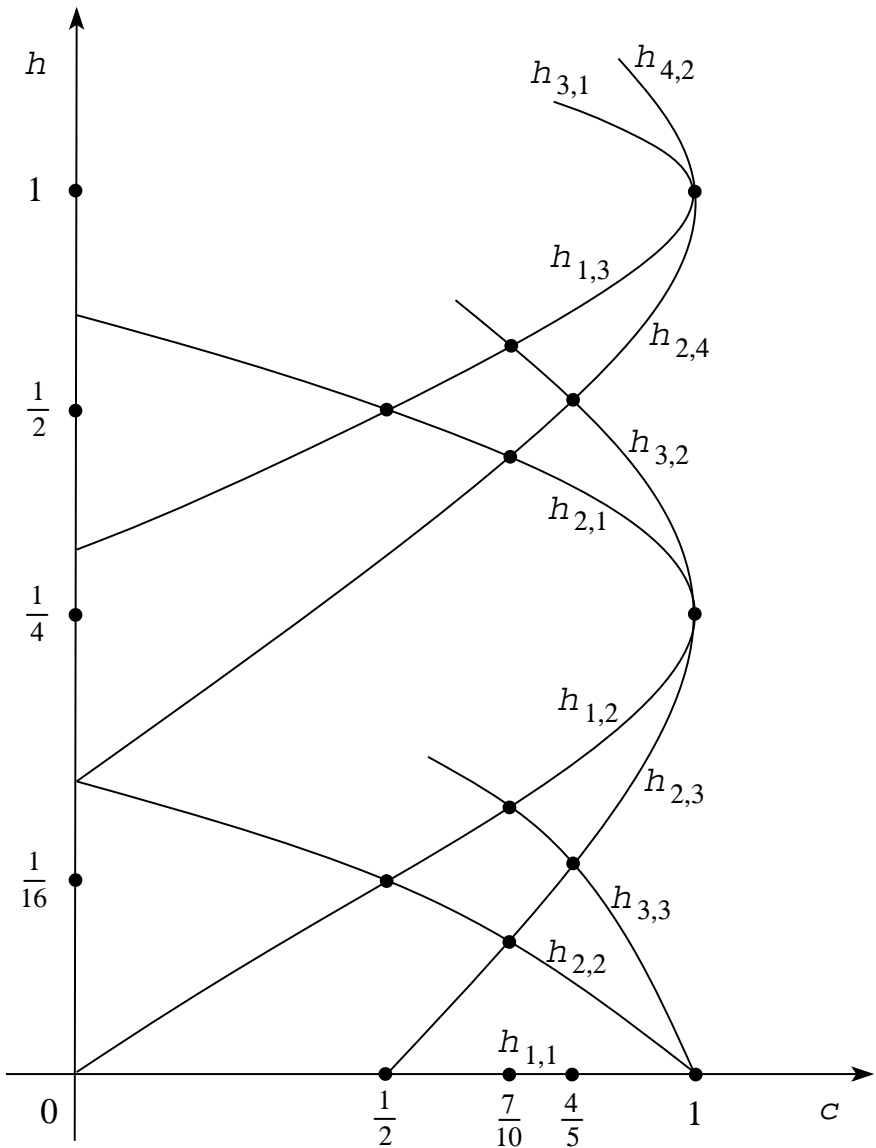

Fig. 6. First few vanishing curves $h = h_{p,q}(c)$ in the $h, c$ plane.

### 4.4. Critical statistical mechanical models

We pause here to emphasize the import of $(4.6a, b)$. The representation theory of the Virasoro algebra in principle allows us to describe the possible scaling dimensions of fields of two dimensional conformal field theories, and thereby the possible critical indices of two dimensional systems at their second order phase transitions. In the case of unitary systems with $c \leq 1$, this has turned out to give a complete classification of possible two dimensional critical behavior. We shall later see how to identify the particular representations of the Virasoro algebra which occur in the description of a given two dimensional system at its critical point. (In Cardy's lectures (section 3.2), we have already seen how to calculate the central charge of the $Q$-state Potts model.)

While the $c < 1$ discrete series distinguishes a set of representations of the Virasoro algebra, it is not obvious that these should be realized by readily constructed statistical mechanical model at their critical points. The first few members of the series $(4.6a)$ with $m = 3, 4, 5, 6$, i.e. central charge $c = \frac{1}{2}, \frac{7}{10}, \frac{4}{5}, \frac{6}{7}$, were associated in [18] respectively with the critical points of the Ising model, tricritical Ising model, 3-state Potts model, and tricritical 3-state Potts model, by comparing the allowed conformal weights $(4.6b)$ with known scaling dimensions of operators in these models. The first of these, $m = 3$, we will treat in great detail in the next section. In general, there may exist more than one model at a given discrete value of $c < 1$, corresponding to different consistent subsets of the full unitarity-allowed operator content $(4.6b)$.

By coincidence, at roughly the same time as the unitarity analysis, the authors of [20] had constructed a new series of exactly solvable models of RSOS (restricted solid-on-solid) type. The critical points of these models models were quickly identified[21] to provide particular realizations of all members of the discrete series $(4.6a)$. The RSOS models of [20] are defined in terms of height variables $\ell_i$ that live at the sites of a square lattice. The heights are subject to the restriction $\ell_i = 1, \ldots, m$, and nearest neighbor heights are also constrained to satisfy $\ell_i = \ell_j \pm 1$. $m$ is here an integer parameter that characterizes different models. The Boltzmann weights for the models are given in terms of four-height interactions around each plaquette of the lattice (known as 'IRF' interactions for 'interactions round a face'). These weights are defined so that each model has a second order phase transition at a self-dual point. The continuum limit theory of the RSOS model with heights restricted to take values from 1 to $m$ turns out to give a realization of the Virasoro algebra with central charge $c = 1 - 6/m(m+1)$. (The nearest neighbor constraint in the case $m = 3$, for example, causes the lattice to decompose to an even sublattice on which $\ell_i = 2$ for all sites, and an odd sublattice on which $\ell_i = 1, 3$. The even sublattice decouples, and the remaining 2-state model on the odd sublattice is the Ising model.) Other models of RSOS type were later constructed[22] and have critical points also described by unitary representations of the Virasoro algebra with $c < 1$, but have a different operator content than the models of [20]. For example, the model of [20] with $m = 5$ ($c = 4/5$) is in the universality class of the tetracritical Ising model, whereas a model of [22] with the same value of $c$ is in the universality class of the 3-state Potts model (these two may be associated respectively to the Dynkin diagrams of $A_5$ and $D_4$). We shall return to say a bit more about these models in section 9.

*4.5. Conformal grids and null descendants*

To prepare for our discussion of the operator content in later sections, we need a convenient way of organizing the allowed highest weights $h_{p,q}$ of (4.6b). As noted, the $h_{p,q}$ are invariant under $p \to m - p$, $q \to m + 1 - q$. Thus if we extend the range of $q$ to $1 \leq q \leq m$, we will have a total of $m(m-1)$ values of $h_{p,q}$ with each appearing exactly twice. It is frequently convenient to arrange this extended range in an $(m-1) \times m$ "conformal grid" with columns labeled by $p$ and rows by $q$. For the cases $m = 3$ (Ising model, $c = 1/2$), $m = 4$ (tricritical Ising model, $c = 7/10$), and $m = 5$ (3-state Potts model, $c = 4/5$), we find the conformal weights tabulated in fig. 7. Note that the symmetry in $p$ and $q$ mentioned above means that the diagram is left invariant by a rotation by $\pi$ about its center. The singly-counted set of operators with $q \leq p$ are those below the $q = p$ diagonal in fig. 7. Another way of eliminating the double counting is to restrict to operators with $p + q$ even — this selects operators in a checkerboard pattern starting from the identity operator at lower left.

$\uparrow$
$q$

| $\frac{1}{2}$ | 0 |
|---|---|
| $\frac{1}{16}$ | $\frac{1}{16}$ |
| 0 | $\frac{1}{2}$ |

$p \rightarrow$

| $\frac{3}{2}$ | $\frac{7}{16}$ | 0 |
|---|---|---|
| $\frac{3}{5}$ | $\frac{3}{80}$ | $\frac{1}{10}$ |
| $\frac{1}{10}$ | $\frac{3}{80}$ | $\frac{3}{5}$ |
| 0 | $\frac{7}{16}$ | $\frac{3}{2}$ |

| 3 | $\frac{7}{5}$ | $\frac{2}{5}$ | 0 |
|---|---|---|---|
| $\frac{13}{18}$ | $\frac{21}{40}$ | $\frac{1}{40}$ | $\frac{1}{8}$ |
| $\frac{2}{3}$ | $\frac{1}{15}$ | $\frac{1}{15}$ | $\frac{2}{3}$ |
| $\frac{1}{8}$ | $\frac{1}{40}$ | $\frac{21}{40}$ | $\frac{13}{8}$ |
| 0 | $\frac{2}{5}$ | $\frac{7}{5}$ | 3 |

Fig. 7. Conformal grids for the cases $m = 3, 4, 5$ ($c = \frac{1}{2}, \frac{7}{10}, \frac{4}{5}$).

In general we have seen from the Kac determinant formula that the primary state with $L_0$ eigenvalue $h_{p,q}$ has a null descendant at level $pq$. For the three allowed values $h_{1,1} = 0$, $h_{2,1} = \frac{1}{2}$, and $h_{1,2} = \frac{1}{16}$ at $m = 3$, the associated null states at levels one and two were determined to be

$$L_{-1}|0\rangle = 0 \qquad (4.7a)$$

and (from (4.2))

$$\left(L_{-2} - \frac{3}{2(2h_{2,1}+1)}L_{-1}^2\right)|\tfrac{1}{2}\rangle = \left(L_{-2} - \frac{3}{4}L_{-1}^2\right)|\tfrac{1}{2}\rangle = 0$$
$$\left(L_{-2} - \frac{3}{2(2h_{1,2}+1)}L_{-1}^2\right)|\tfrac{1}{16}\rangle = \left(L_{-2} - \frac{4}{3}L_{-1}^2\right)|\tfrac{1}{16}\rangle = 0 \ . \qquad (4.7b)$$

For higher values of $m$, null states begin to occur at higher levels $pq$. For $m = 4$, for example, the state $|h_{3,1}\rangle = |\tfrac{3}{2}\rangle$ has a null descendant at level three, and is thus annihilated by a linear combination of $L_{-3}$, $L_{-2}L_{-1}$, and $L_{-1}^3$, as easily determined by applying the commutation rules of the Virasoro generators with $c = 7/10$.

## 5. Identification of $m = 3$ with the critical Ising model

The unitary representation theory of the Virasoro algebra plays the same role in studying two dimensional critical phenomena as representation theory of finite and Lie groups plays in other branches of physics. Once the relevant symmetry group of a physical system has been identified, the analysis of its spectrum and interactions is frequently reduced to a straightforward exercise in group representation theory and branching rules. For a given critical statistical mechanical model, the 2-point correlation functions allow an identification of the scaling weights of the operators in the theory and in many cases that is sufficient to identify the relevant representation of the Virasoro algebra. We have already mentioned that the discrete unitary series with $c < 1$, for example, provides a set of possibilities for 2d critical behavior that can be matched up with that of known statistical mechanical systems.

We shall now make explicit the identification of the first member of the discrete unitary series, i.e. the case $m = 3$ with $c = 1/2$, with the Ising model at its critical point. Up to now we have concentrated on the analytic dependence $T(z)$ of the stress-energy tensor. The physical systems we shall consider here also have a non-trivial $\overline{T}(\overline{z})$ with central charge $\overline{c} = c$. The primary fields in our theory are thus described by the two scaling weights $h$ and $\overline{h}$ (the eigenvalues of the associated highest weight state under $L_0$ and $\overline{L}_0$). The simplest possibility is to consider the left-right symmetric fields $\Phi_{p,q}(z,\overline{z}) = \phi_{p,q}(z)\overline{\phi}_{p,q}(\overline{z})$ with conformal weights $(h,\overline{h})$

$$\Phi_{1,1}: \ (0,0) \qquad \Phi_{2,1}: \ (\tfrac{1}{2},\tfrac{1}{2}) \qquad \Phi_{1,2}: \ (\tfrac{1}{16},\tfrac{1}{16}) \qquad (5.1)$$

(we shall later infer that this is the only possibility allowed by modular invariance for the theory on a torus).

*5.1. Critical exponents*

The $(0,0)$ field above is present in every theory and is identified as the identity operator. To compare the remaining fields in (5.1) with those present in the conventional description of the Ising model on a lattice, we need to make a brief digression into some of the standard lore of critical phenomena. (For a review of the material needed here, see [23].) Suppose we have a system with an order parameter $\sigma$ (such as the spin ($\sigma = \pm 1$) in the Ising model). Suppose further that the system has a 2$^{\mathrm{nd}}$ order transition separating a high temperature (disordered) phase with $\langle \sigma \rangle = 0$ from a low temperature (ordered) phase with $\langle \sigma \rangle \neq 0$. In the high temperature phase the 2-point function of the order parameter will fall off exponentially $\langle \sigma_n \sigma_0 \rangle \sim \exp(-|n|/\xi)$, where the correlation length $\xi$ depends on the temperature (we see $\xi^{-1}$ can be regarded as a mass for the theory). At the critical point the correlation length diverges (theory becomes massless) and the 2-point function instead falls off as a power law

$$\langle \sigma_n \sigma_0 \rangle \sim \frac{1}{|n|^{d-2+\eta}},$$

where $d$ is the dimension of the system and this expression defines the critical exponent $\eta$. Another exponent, $\nu$, can be defined in terms of the 4-point function at criticality

$$\langle \varepsilon_n \varepsilon_0 \rangle \sim \langle \sigma_n \sigma_{n+1} \sigma_0 \sigma_1 \rangle \sim \frac{1}{|n|^{2(d-1/\nu)}} \qquad (5.2)$$

(more precisely $\varepsilon_n$ should be defined by averaging over all nearest neighbor sites to $n$, but for our purposes here any one nearest neighbor, which we denote $n+1$, suffices).

The critical exponents calculated for the two dimensional Ising model are $\eta = 1/4$, $\nu = 1$. Therefore the 2-point function behaves as

$$\langle \sigma_n \sigma_0 \rangle \sim \frac{1}{|n|^{1/4}} \sim \frac{1}{r^{2\Delta_\sigma}},$$

where the $r$ dependence is appropriate for the 2-point function of a conformal field of scaling dimension $\Delta_\sigma = h_\sigma + \overline{h}_\sigma$ and spin $s_\sigma = h_\sigma - \overline{h}_\sigma = 0$. We see that $\Delta_\sigma = 2h_\sigma = 2\overline{h}_\sigma = 1/8$ and hence the $(\tfrac{1}{16},\tfrac{1}{16})$ field in (5.1) should be identified with the spin $\sigma$ of the Ising model. The energy operator, on the other hand, satisfies

$$\langle \varepsilon_n \varepsilon_0 \rangle \sim \frac{1}{|n|^{2\Delta_\varepsilon}}.$$

Its scaling weight, then, can be identified from (5.2) with $\nu = 1$ as $d - 1/\nu = 1 = \Delta_\varepsilon = h_\varepsilon + \overline{h}_\varepsilon$. Thus the $(\tfrac{1}{2},\tfrac{1}{2})$ field in (5.1) should be identified with the energy operator of the Ising model. This completes the identification of the primary fields in the Ising model, which turns out to have a total of only three conformal families.

(Although we have chosen to introduce the exponents $\eta$ and $\nu$ in terms of critical correlation functions, we mention that many exponents are also defined in terms of off-critical correlation functions. Different definitions of the same exponent are related by the scaling hypothesis. The critical exponent $\nu$, for example, is defined alternatively in terms of the divergence of the correlation length close to criticality,
$$\xi \sim t^{-\nu} ,$$
where $t = (T - T_c)/T_c$ parametrizes the deviation of temperature from the critical temperature $T_c$. Another common critical exponent is defined similarly in terms of the divergence of the specific heat,
$$C \sim t^{-\alpha} ,$$
near the critical point.

Now according to the scaling hypothesis, the divergence of all thermodynamic quantities at the critical point is due to their dependence on the correlation length $\xi$. Dimensional analysis thus allows us to find relations between critical exponents. For example the free energy density has dimension $(\text{length})^{-d}$ in $d$-dimensions so we find
$$f \sim \xi^{-d} \sim t^{\nu d} .$$
The specific heat, on the other hand, is given by
$$C \sim \frac{\partial^2 f}{\partial t^2} \sim t^{\nu d - 2} ,$$
so the scaling hypothesis implies the relation $\alpha = 2 - \nu d$. Finally the energy density itself satisfies
$$\varepsilon \sim \frac{\partial f}{\partial t} \sim t^{\nu d - 1} \sim \xi^{-(\nu d - 1)/\nu} , \qquad (5.3)$$
and comparing with (5.2) we see that the scaling hypothesis implies coincidence of the two definitions of $\nu$.

To make the relationship more precise, we consider the continuum limit of the correlation function
$$\langle \varepsilon(r)\varepsilon(0) \rangle = \frac{1}{r^p} g(r/\xi)$$
close to criticality. Then the specific heat satisfies
$$C \sim \frac{\partial^2 f}{\partial t^2} \sim \int d^d r \, \langle \varepsilon(r)\varepsilon(0) \rangle \sim \xi^{d-p} \sim t^{-\nu(d-p)} \sim t^{-\alpha} ,$$
so that $p = d - \alpha/\nu = 2(d - 1/\nu)$. At the critical point, $\xi \to \infty$, and $\langle \varepsilon(r)\,\varepsilon(0) \rangle = g(0)/r^p = g(0)/r^{2(d-1/\nu)}$, in accord with the definition (5.2).

We note from (5.3) that in two dimensions the scaling weight of a spinless energy operator is $h_\varepsilon = \overline{h}_\varepsilon = (1 - \alpha)/(2 - \alpha)$. For other magnetization type operators, one can define exponents $\beta$ by $m \sim t^\beta$, and proceeding as above we find
$$m \sim t^\beta \sim \xi^{-\beta/\nu} \sim \xi^{-d\beta/(2-\alpha)} .$$
For spinless magnetization type operators in two dimensions, we thus have $h_m = \overline{h}_m = \beta/(2 - \alpha)$. The reader might benefit from repeating the argument of the preceding paragraph to see how the exponent $\beta$ may be alternatively defined via a 2-point function at the critical point.)

In (3.5), we introduced another $c = \overline{c} = \frac{1}{2}$ system consisting of free fermions $\psi(z)$ and $\overline{\psi}(\overline{z})$. In [24], it is shown that the Ising model can generally be written as a theory of a free lattice fermion. At the critical point the fermion becomes massless and renormalizes onto a massless continuum fermion. The free fermion system (3.5) thus turns out to be equivalent to the critical Ising model field theory. From the standpoint of the free fermion description of the Ising critical point, we see that the energy operator corresponds to the $(\frac{1}{2}, \frac{1}{2})$ field $\psi(z)\overline{\psi}(\overline{z})$. Moving away from criticality by adding a perturbation proportional to the energy operator thus corresponds to adding a mass term $\delta m\,\psi(z)\overline{\psi}(\overline{z})$. The emergence of the $(\frac{1}{16}, \frac{1}{16})$ field $\sigma$ in the fermionic language, on the other hand, is not as immediately obvious. In section 6 we shall see why a field of that weight should naturally occur. In section 7 we shall further exploit the free fermion representation of the Ising model to investigate its spectrum.

As described in Cardy's lectures, the Ising model also possesses a disorder operator $\mu$, dual to the spin $\sigma$. Since the critical point occurs at the self-dual point of the model, at the critical point the field $\mu(z, \overline{z})$ will have the same conformal weights and operator algebra as the spin field $\sigma(z, \overline{z})$. Thus the full operator content of the Ising model includes two $(\frac{1}{16}, \frac{1}{16})$ fields, although the two are not mutually local (and neither is local with respect to the fermions $\psi$, $\overline{\psi}$). Both $\sigma$ and $\mu$ are each individually local, on the other hand, with respect to the energy operator $\varepsilon$.

5.2. *Critical correlation functions of the Ising model*

Since, as noted after (3.30), the non-vanishing operator products for any members of conformal families are determined by those of the primaries, it is possible to write "fusion rules" $[\phi_i][\phi_j] = \sum_k [\phi_k]$ for conformal families. They determine which conformal families $[\phi_k]$ may have their members occurring in the operator product between any members of conformal families $[\phi_i]$ and $[\phi_j]$. In the case of the Ising model, we write the three conformal families associated to the primary fields of (5.1) as $1$, $[\epsilon]$, and $[\sigma]$. The fusion rules allowed by the spin reversal ($\sigma \to -\sigma$) and duality ($\varepsilon \to -\varepsilon$) symmetries of the critical Ising model are

$$[\sigma][\sigma] = 1 + [\varepsilon]$$
$$[\sigma][\varepsilon] = [\sigma] \qquad (5.4)$$
$$[\varepsilon][\varepsilon] = 1 \ .$$

We shall shortly confirm that 4-point correlation functions in the critical Ising model are consistent with the non-vanishing operator products represented by (5.4).

In the conformal field theory description of the critical point, both the energy and spin (order/disorder) operators of (5.1) have null descendants at level 2. That means that any correlation function of these operators will satisfy a second order differential equation. Specifically from (4.7*b*) we see that correlation functions involving either $\mu$ or $\sigma$ will be annihilated by the differential operator $(\mathcal{L}_{-2} - \frac{4}{3}\mathcal{L}_{-1}^2)$. From (4.3), we find furthermore that any correlation function of $\sigma$'s and $\mu$'s,

$$G^{(2M,2N)} = \big\langle \sigma(z_1, \overline{z}_1) \cdots \sigma(z_{2M}, \overline{z}_{2M})$$
$$\mu(z_{2M+1}, \overline{z}_{2M+1}) \cdots \mu(z_{2M+2N}, \overline{z}_{2M+2N}) \big\rangle \ ,$$

will satisfy the differential equations ($i = 1, \ldots, 2M + 2N$)

$$\left[ \frac{4}{3} \frac{\partial^2}{\partial z_i^2} - \sum_{j \ne i}^{2M+2N} \left( \frac{1/16}{(z_i - z_j)^2} + \frac{1}{z_i - z_j} \frac{\partial}{\partial z_j} \right) \right] G^{(2M,2N)} = 0 \ , \qquad (5.5)$$

and similarly for $z_i \to \overline{z}_i$.

Here we shall illustrate (following Appendix E of [1]) how these differential equations can be used to determine the 4-point function $G^{(4)}$ of four $\sigma$'s at the critical point of the Ising model. The constraints of global conformal invariance discussed in section 2 first of all require that

$$G^{(4)} = \big\langle \sigma(z_1, \overline{z}_1)\sigma(z_2, \overline{z}_2)\sigma(z_3, \overline{z}_3)\sigma(z_4, \overline{z}_4) \big\rangle$$
$$= \left( \frac{z_{13} z_{24}}{z_{12} z_{23} z_{34} z_{41}} \right)^{1/8} \overline{\left( \frac{z_{13} z_{24}}{z_{12} z_{23} z_{34} z_{41}} \right)}^{1/8} F(x, \overline{x}) \qquad (5.6)$$

where $x = z_{12} z_{34} / z_{13} z_{24}$ is the conformally invariant cross-ratio and $z_{ij} = z_i - z_j$. (To facilitate comparison with the conventional Ising model result I have absorbed some additional $x$ dependence in the prefactor to $F$ in (5.6) with respect to the canonical form of 4-point functions given in (2.6). The result is also frequently cited in terms of the prefactor in (5.6) written in the equivalent form $\left| z_{13} \, z_{24} \, x(1-x) \right|^{-1/4}$.)

(5.5) then yields the second order ordinary differential equation

$$\left( x(1-x) \frac{\partial^2}{\partial x^2} + (\tfrac{1}{2} - x) \frac{\partial}{\partial x} + \frac{1}{16} \right) F(x, \overline{x}) = 0 \qquad (5.7)$$

satisfied by $F$ (and a similar equation with $x \to \overline{x}$). (5.7) has regular singular points at $x = 0, 1, \infty$ and the exponents at these singular points can be obtained by standard asymptotic analysis. The two independent solutions are expressible as hypergeometric functions which in the case at hand reduce to the elementary functions $f_{1,2}(x) = (1 \pm \sqrt{1-x})^{1/2}$. Taking also into account the $\overline{z}$ dependence, $G^{(4)}$ takes the form

$$G^{(4)} = \left|\frac{z_{13}z_{24}}{z_{12}z_{23}z_{34}z_{41}}\right|^{1/4} \sum_{i,j=1}^{2} a_{ij}\, f_i(x) f_j(\overline{x}) \,. \tag{5.8}$$

But when $\overline{x}$ is the complex conjugate of $x$, single-valuedness of $G^{(4)}$ allows only the linear combination $a(|f_1(x)|^2 + |f_2(x)|^2)$. The resulting expression agrees with that derived directly in the critical Ising model[25].

Now that we have determined the 4-point function, it is possible to identify the coefficient $C_{\sigma\sigma\varepsilon}$ in the operator product expansion

$$\sigma(z_1,\overline{z}_1)\sigma(z_2,\overline{z}_2) \sim \frac{1}{z_{12}^{1/8}\overline{z}_{12}^{1/8}} + C_{\sigma\sigma\varepsilon}\, z_{12}^{3/8}\overline{z}_{12}^{3/8}\,\varepsilon(z_2,\overline{z}_2) + \ldots\,, \tag{5.9}$$

where the first term fixes the normalization conventions for the $\sigma$'s. (5.9) implies that (5.6) must behave in the $x \to 0$ limit as

$$G^{(4)} \sim \frac{1}{|z_{12}|^{1/4}} \frac{1}{|z_{34}|^{1/4}} + C_{\sigma\sigma\varepsilon}^2 \frac{|z_{12}|^{3/4}|z_{34}|^{3/4}}{|z_{24}|^2} + \ldots\,. \tag{5.10}$$

Comparison of the first term above with the leading small $x$ behavior of (5.8) determines that $a = a_{11} = a_{22} = \frac{1}{2}$, i.e.

$$G^{(4)} = \frac{1}{2}\left|\frac{z_{13}z_{24}}{z_{12}z_{23}z_{34}z_{41}}\right|^{1/4} \left(\left|1+\sqrt{1-x}\right| + \left|1-\sqrt{1-x}\right|\right)\,. \tag{5.11}$$

Comparing the next leading terms of (5.10) and (5.11) as $x \to 0$ we find $C_{\sigma\sigma\varepsilon} = \frac{1}{2}$. The non-vanishing operator product coefficients considered thus far are consistent with the fusion rules (5.4).

Similar methods may be used to obtain the other 4-point functions. Instead of (5.6), we can calculate

$$\begin{aligned}G^{(2,2)} &= \langle \sigma(z_1,\overline{z}_1)\mu(z_2,\overline{z}_2)\sigma(z_3,\overline{z}_3)\mu(z_4,\overline{z}_4)\rangle \\ &= \left|\frac{z_{13}z_{24}}{z_{12}z_{23}z_{34}z_{41}}\right|^{1/4} F(x,\overline{x})\,.\end{aligned} \tag{5.12}$$

$G^{(2,2)}$ satisfies the same differential equation (5.7), only now we require the solution to be double-valued as $z_1$ is taken around $z_2$ ($x$ taken around 0). This allows another solution with $a_{21} = -a_{12}$, $a_{11} = a_{22} = 0$. In the limit $x \to \infty$ ($z_1 \to z_3$, $z_2 \to z_4$), we have $G^{(2,2)} \sim \langle(\sigma(z_1,\overline{z}_1)\sigma(z_3,\overline{z}_3)\rangle\langle(\mu(z_2,\overline{z}_2)\mu(z_4,\overline{z}_4)\rangle = |z_{13}z_{24}|^{-1/4}$, the same leading behavior as in (5.10). This determines $a_{21} = -a_{12} = \frac{i}{2}$, i.e.

$$\begin{aligned}G^{(2,2)} = \frac{i}{2}\left|\frac{z_{13}z_{24}}{z_{12}z_{23}z_{34}z_{41}}\right|^{1/4} &\left[\left(1-\sqrt{1-x}\right)^{1/2}\left(1+\sqrt{1-\overline{x}}\right)^{1/2} \right.\\ &\left. - \left(1+\sqrt{1-x}\right)^{1/2}\left(1-\sqrt{1-\overline{x}}\right)^{1/2}\right]\,.\end{aligned} \tag{5.13}$$

In the next section we will use the non-leading terms in (5.13) to determine some of the operator product coefficients involving $\sigma$ and $\mu$.

In principle one can use the $(p,q) \to (m-p, m+1-q)$ symmetry of (4.5b) to generate both an order $pq$ and an order $(m-p)(m+1-q)$ differential equation for correlation functions involving a $\phi_{p,q}$ operator. In some cases[26], combining the two equations allows one to derive a lower order differential equation for correlation functions involving the field in question. For the ($m=3$) Ising model, for example, this procedure gives both second and third order differential equations for correlation functions involving the operator $\varepsilon = \Phi_{2,1}$. These can be combined to give readily solved first-order partial differential equations for the 4-point functions $\langle\varepsilon\varepsilon\varepsilon\varepsilon\rangle$ and $\langle\varepsilon\varepsilon\sigma\sigma\rangle$.

### 5.3. Fusion rules for $c < 1$ models

Although rather cumbersome in general, the above differential equation method in principle gives the correlation functions of any set of degenerate operators and can be used to determine the operator product coefficients $C_{ijk}$ (for the 3-state Potts model this has been carried out in [27]). A different method, based on the background charge ideas described after (3.4), gives instead integral representations for the correlation functions which have been studied extensively in [10]. Again the results for the 4-point functions can be used to infer the $C_{ijk}$'s.

Applied directly to the 3-point functions, the above differential equation method does not determine the $C_{ijk}$'s, but does give useful selection rules that determine which are allowed to be non-vanishing. For example, the 3-point function $\langle \phi_{2,1}(z_1) \phi_{p,q}(z_2) \phi_{p',q'}(z_3) \rangle$ is annihilated by the second order differential operator $\mathcal{L}_{-2} - \frac{3}{2(2h_{2,1}+1)} \mathcal{L}_{-1}^2$. If we substitute the operator product expansion for $\phi_{2,1}(z_1)$ and $\phi_{p,q}(z_2)$ into this differential equation and consider the most singular term as $z_1 \to z_2$, the characteristic equation gives a quadratic relation between $h_{p,q}$ and $h_{p',q'}$ which is satisfied only for $p' = p \pm 1$ and $q' = q$. For 3-point functions involving $\phi_{1,2}$, we find similar the selection rule $p' = p$ and $q' = q \pm 1$.

By considering multiple insertions of $\phi_{1,2}$ and $\phi_{2,1}$ and using associativity of the operator product expansion, it is possible to derive the general selection rules for non-vanishing $\langle \phi_{p_1,q_1} \phi_{p_2,q_2} \phi_{p_3,q_3} \rangle$. If we choose the $\phi_{p,q}$'s of fig. 7 with $p = 1, \ldots, m-1$, $q = 1, \ldots, m$, and $p+q$ even, these selection rules may be expressed as

$$\phi_{p_1,q_1} \times \phi_{p_2,q_2} = \sum_{\substack{p_3 = |p_1-p_2|+1}}^{\substack{\min(p_1+p_2-1,\\ 2m-1-(p_1+p_2))}} \sum_{\substack{q_3 = |q_1-q_2|+1}}^{\substack{\min(q_1+q_2-1,\\ 2m+1-(q_1+q_2))}} \phi_{p_3,q_3} . \qquad (5.14)$$

The selection rules take a more intuitive form reexpressed in terms of 'spins' $p_i = 2j_i + 1$, $q_i = 2j'_i + 1$. They then resemble $SU(2)$ branching rules, i.e. allowed $j_3$ are those that appear in the decomposition of $j_1 \times j_2$ considered as representations of $SU(2)$ (and cyclic permutations). The same conditions must be satisfied by the $j'$'s. These conditions allow, among other things, non-vanishing $C_{ijk}$'s only for all $p$'s odd (all vector-like) or two even, one odd (two spinor-like, one vector-like). The selection rules are not quite those of $SU(2)$ because of the upper restriction involving $m$ on the summations. In fact they are the selection rules instead for what is known as affine $SU(2)$ (at levels $k = m-2$ and $m-1$ respectively for $p$ and $q$). We will derive the selection rules (5.14) from this point of view when we discuss affine algebras and the coset construction of these models in section 9.

We have deliberately written (5.14) in a notation slightly different from (5.4). (5.14) involves only the holomorphic parts of the fields and determines a commutative associative algebra. In general we write such fusion rules as[28]

$$\phi_i \times \phi_j = \sum_k N_{ij}{}^k \phi_k , \qquad (5.15)$$

where the $\phi_i$'s denote a set of primary fields. In the event that the chiral algebra is larger than the Virasoro algebra, they should be taken as the fields primary with respect to the larger algebra (later on we shall encounter examples of extended chiral algebras). The $N_{ij}{}^k$'s on the right hand side of (5.15) are integers that can be interpreted as the number of independent fusion paths from $\phi_i$ and $\phi_j$ to $\phi_k$ (the $k$ index is distinguished to allow for the possibility of non-self-conjugate fields). (5.4), on the other hand, symbolically indicates the conformal families that may occur in operator products of conformal families of operators with combined $z, \overline{z}$ dependence, but has no natural integral normalization. The algebra (5.15) together with its anti-holomorphic counterpart can always be used in any given theory to reconstruct less precise structures such as (5.4).

The $N_{ij}{}^k$'s are automatically symmetric in $i$ and $j$ and satisfy a quadratic condition due to associativity of (5.15). They can be analyzed extensively in a class of theories known as 'rational conformal field theories'. These are theories[29] that involve only a finite number of primary fields with respect to the (extended) chiral algebra. The $c < 1$ theories of section 4 are particular examples (in which there are a finite number of primaries with respect to the Virasoro algebra itself). The rationality condition means that the indices of the $N_{ij}{}^k$'s run only over a finite set of values, and summations over them are well-defined. If we use a matrix notation $(N_i)_j{}^k = N_{ij}{}^k$, then the $ij$ symmetry can be used to write the associativity condition either as

$$N_i N_\ell = N_\ell N_i , \quad \text{or as} \quad N_i N_j = \sum_k N_{ij}{}^k N_k .$$

The $N_i$'s themselves thus form a commutative associative matrix representation of the fusion rules (5.15). They can be simultaneously diagonalized and their eigenvalues $\lambda_i^{(n)}$ form one dimensional representations of the fusion rules. The algebra (5.15) is an algebra much like algebras that occur in finite group theory, such as for the multiplication of conjugacy classes or for the branching rules for representations. It is a generalization that turns out to embody these algebras in the orbifold models to be discussed in section 8. We shall see how the $N_{ij}{}^k$'s themselves may be determined[28][30] in section 9.

*5.4. More discrete series*

Since we have mentioned the idea of extended chiral algebras, we pause here to exhibit some specific examples of algebras larger than the Virasoro algebra. Supersymmetric extensions of the Virasoro algebra are obtained by generalizing conformal transformations to superconformal transformations of supercoordinates $\mathbf{z} = (z, \theta)$, where $\theta$ is an anticommuting coordinate ($\theta^2 = 0$). Superconformal transformations are generated by the moments of a super stress-energy tensor. If there is only a single anti-commuting coordinate ($N=1$ supersymmetry), then the super stress-energy tensor $\mathbf{T}(\mathbf{z}) = T_F(z) + \theta T(z)$ has components that satisfy the operator products[31][32]

$$T(z_1)T(z_2) \sim \frac{3\hat{c}/4}{(z_1-z_2)^4} + \frac{2}{(z_1-z_2)^2}T(z_2) + \frac{1}{z_1-z_2}\partial T(z_2) ,$$

$$T(z_1)T_F(z_2) \sim \frac{3/2}{(z_1-z_2)^2}T_F(z_2) + \frac{1}{z_1-z_2}\partial T_F(z_2) , \quad (5.16)$$

$$T_F(z_1)T_F(z_2) \sim \frac{\hat{c}/4}{(z_1-z_2)^3} + \frac{1/2}{z_1-z_2}T(z_2) ,$$

where $\hat{c} = \frac{2}{3}c$. The conventional normalization is such that a single free superfield $x(z) + \theta\psi(z)$ has central charge $\hat{c} = 1$ in (5.16), just as the stress-energy tensor for a single bosonic field $x(z)$ had central charge $c = 1$ in (3.1). The second equation in (5.16) is the statement that $T_F$ is a primary field of dimension 3/2.

In terms of the moments $L_n$ of $T$, and the moments

$$G_n = \oint \frac{dz}{2\pi i} z^{n+1/2} \, 2\, T_F(z) \qquad (5.17)$$

of $T_F$, the operator product expansions (5.16) are equivalent to the (anti-) commutation relations

$$[L_m, L_n] = (m-n)L_{m+n} + \frac{\hat{c}}{8}(m^3-m)\delta_{m+n,0}$$

$$[L_m, G_n] = \left(\frac{m}{2}-n\right)G_{m+n} \qquad (5.18)$$

$$\{G_m, G_n\} = 2L_{m+n} + \frac{\hat{c}}{2}\left(m^2-\frac{1}{4}\right)\delta_{m+n,0} .$$

The algebra (5.16) has a $\mathbf{Z}_2$ symmetry, $T_F \to -T_F$, so there are two possible modings for the $G_n$'s. For integer moding ($n \in \mathbf{Z}$) of $G_n$, the supersymmetric extension of the Virasoro algebra is termed the Ramond (R) algebra; for half-integer moding ($n \in \mathbf{Z} + \frac{1}{2}$), it is termed the Neveu-Schwarz (NS) algebra. Primary fields are again associated with highest weight states $|h\rangle$, satisfying $L_n|h\rangle = G_n|h\rangle = 0$, $n > 0$, and $L_0|h\rangle = h|h\rangle$. Note that (5.18) requires that a highest weight state in the Ramond sector have eigenvalue $h - \hat{c}/16$ under $G_0^2$. For $\hat{c} > 1$, the only restrictions imposed by unitarity are $h \geq 0$ (NS), and $h \geq \hat{c}/16$ (R), and the Verma modules again provide irreducible representations (no null states) except when the latter inequalities are saturated.

For $\hat{c} < 1$ ($c < \frac{3}{2}$), on the other hand, unitary representations of (5.16) can occur only at the discrete values

$$c = \frac{3}{2}\left(1 - \frac{8}{m(m+2)}\right) \qquad (5.19)$$

($m = 3, 4, \ldots$), and discrete values of $h$ from a formula analogous to (4.6b). Notice that the first value is $c = 7/10$, and coincides with the second member of the discrete series (4.6a), identified as the tricritical Ising model. Further discussion of the supersymmetry in this model may be found in [32][33].

There are also generalizations of (5.16) with more than one supersymmetry generator. In the case $N = 2$ [34], there is a discrete series [35]

$$c = 3\left(1 - \frac{2}{m}\right) \qquad (5.20)$$

($m = 3, 4, \ldots$) of allowed values for $c < 3$, and a continuum of allowed values for $c \geq 3$. The boundary value $c = 3$ can be realized in terms of a single free complex superfield. The first value, $c = 1$, coincides with the second non-trivial member of the series (5.19). The $N = 2$ superconformal algebra contains a $U(1)$ current algebra, under which the supersymmetry generators transform with non-zero charge. For $N = 3$ supersymmetry, unitary representations occur [36] only at the discrete set of values $c = \frac{3}{2}k$ ($k = 1, 2, \ldots$); and for $N = 4$ supersymmetry, only at the values $c = 6k$ ($k = 1, 2, \ldots$). In these last two cases unitarity allows no continuum of values for the central charge. This is related to the fact that the $N = 3, 4$ algebras contain an $SU(2)$ current algebra under which the supersymmetry generators transform non-trivially (we shall discuss affine $SU(2)$ in some detail in section 9).

## 6. Free bosons and fermions

Useful properties of conformal field theories can frequently be illustrated by means of free field realizations. In this section, we shall apply the general formalism of sections 1–3 to the cases of free bosons and free fermions, introduced in subsections 2.3 and 3.2. These will prove most useful in our applications of conformal field theory in succeeding sections.

*6.1. Mode expansions*

In section 3, we introduced mode expansions for general primary fields. In particular, for free bosons and fermions we have

$$i\partial_z x(z) = \sum_n \alpha_n z^{-n-1} \qquad i\psi(z) = \sum \psi_n z^{-n-1/2} \ . \qquad (6.1)$$

In what follows we shall take $n$ to run over either integers or half-integers, depending on the boundary conditions chosen for the fields. (The factors of $i$ have been inserted in (6.1) to give more familiar commutation relations for the modes. They compensate the choice of sign in (2.16).) The expansions (6.1) are easily inverted to give

$$\alpha_n = \oint \frac{dz}{2\pi i} z^n \, i\partial_z x(z) \qquad \psi_n = \oint \frac{dz}{2\pi i} z^{n-1/2} \, i\psi(z) \ . \qquad (6.2)$$

In section 3 we also saw how the operator product expansion (3.1) of the stress-energy tensor $T(z)$ implied commutation relations for the modes $L_n$ of the Virasoro algebra. In the case of the bosonic modes, we find that the short distance expansion (2.16) implies the commutation rules

$$\begin{aligned} [\alpha_n, \alpha_m] &= i^2 \left[ \oint \frac{dz}{2\pi i}, \oint \frac{dw}{2\pi i} \right] z^n \partial_z x(z) \, w^m \partial_w x(w) \\ &= i^2 \oint \frac{dw}{2\pi i} w^m \oint \frac{dz}{2\pi i} z^n \frac{-1}{(z-w)^2} = \oint \frac{dw}{2\pi i} n w^m w^{n-1} \\ &= n \delta_{n+m,0} \ , \end{aligned} \qquad (6.3)$$

where we have evaluated the commutator of integrals by first performing the $z$-integral with the contour drawn tightly around $w$, and then performing the $w$-integral.

Similarly, we find

$$\begin{aligned} \{\psi_n, \psi_m\} &= i^2 \left[ \oint \frac{dz}{2\pi i}, \oint \frac{dw}{2\pi i} \right] z^{n-1/2} w^{m-1/2} \psi(z)\psi(w) \\ &= i^2 \oint \frac{dw}{2\pi i} w^{m-1/2} \oint \frac{dz}{2\pi i} z^{n-1/2} \frac{-1}{z-w} \\ &= \oint \frac{dw}{2\pi i} w^{m-1/2} w^{n-1/2} = \delta_{n+m,0} \ , \end{aligned} \qquad (6.4)$$

although in this case we obtain an anti-commutator due to the fermionic nature of $\psi$ which gives an extra minus sign when we change the order of $\psi(z)$ and $\psi(w)$.

*6.2. Twist fields*

We shall choose to consider periodic (P) and anti-periodic (A) boundary conditions on the fermion $\psi(z)$ as $z$ rotates by $2\pi$ about the origin, $\psi(e^{2\pi i}z) = \pm\psi(z)$. Ultimately consideration of the two boundary conditions is dictated by the fact that spinors naturally live on a double cover of the punctured plane, and only bilinears in spinors, i.e. vectors, need transform as single-valued representations of the 2d Euclidean group. (On higher genus Riemann surfaces, spinors generally live in the spin bundle, i.e. the double cover of the principle frame bundle of the surface.) In the course of our discussion we shall also encounter other ways in which the twisted structure naturally emerges. From (6.1) we see that the two boundary conditions select respectively half-integer and integer modings

$$\psi(e^{2\pi i}z) = +\psi(z) \qquad n \in \mathbf{Z} + \tfrac{1}{2} \qquad \text{(P)}$$
$$\psi(e^{2\pi i}z) = -\psi(z) \qquad n \in \mathbf{Z} \qquad \text{(A)} \ . \qquad (6.5)$$

In preparation for the anti-periodic case, we first consider the calculation of the 2-point function in the periodic case $\psi(e^{2\pi i z}) = \psi(z)$. Then with $n \in \mathbf{Z}+\tfrac{1}{2}$, we find the expected result,

$$-\langle \psi(z)\psi(w)\rangle = \left\langle \sum_{n=1/2}^{\infty} \psi_n z^{-n-1/2} \sum_{m=-1/2}^{-\infty} \psi_m w^{-m-1/2} \right\rangle$$
$$= \sum_{n=1/2}^{\infty} z^{-n-1/2} w^{n-1/2} = \frac{1}{z}\sum_{n=0}^{\infty} \left(\frac{w}{z}\right)^n = \frac{1}{z-w} \ . \qquad (6.6)$$

For the anti-periodic case, it is useful to introduce the twist operator $\sigma(w)$ whose operator product with $\psi(z)$,

$$\psi(z)\sigma(w) \sim (z-w)^{-1/2}\mu(w) + \ldots \ , \qquad (6.7)$$

is defined to have a square-root branch cut. The field $\mu$ appearing in (6.7) is another twist field which by dimensional analysis has the same conformal weight as the field $\sigma$. Our immediate object is to infer the dimension of $\sigma$ by calculating the 2-point function of $\psi$. Due to the square-root in (6.7), when the field $\psi$ is transported around $\sigma$ it changes sign and the twist field $\sigma$ can be used to change the boundary conditions on $\psi$. We can thus view the combination $\sigma(0)$ and $\sigma(\infty)$ to create a cut (the precise location of which is unimportant) from the origin to infinity passing through which the fermion $\psi(z)$ flips sign. (The similarity with the Ising disorder operator described in Cardy's lectures, sec. 5.2, is not accidental.) Equivalently, we can view the state $\sigma(0)|0\rangle$ as a new incoming vacuum, and the operator product (6.7) allows only fermions with anti-periodic boundary conditions (half-integral modes) to be applied to this vacuum, resulting in overall single-valued states.

In either interpretation, the 2-point function of the fermion with anti-periodic boundary conditions is given by

$$\langle \psi(z)\,\psi(w)\rangle_A \equiv \langle 0|\sigma(\infty)\,\psi(z)\,\psi(w)\,\sigma(0)|0\rangle \qquad (6.8)$$

(see (3.10b) for what we mean by $\langle 0|\sigma(\infty)$ here). The evaluation of this quantity proceeds as in (6.6) except that now for anti-periodic fermions $\psi(e^{2\pi i z}) = -\psi(z)$, we take $n \in \mathbf{Z}$. That means the fermion mode algebra now has a zero mode $\psi_0$ that by (6.4) formally satisfies $\{\psi_0,\psi_0\} = 1$. We shall discuss the fermion zero mode algebra in some detail a bit later, but for the moment substituting $\psi_0^2 = \tfrac{1}{2}$ gives

$$-\langle \psi(z)\psi(w)\rangle_A = \left\langle \sum_{n=0}^{\infty} \psi_n z^{-n-1/2} \sum_{m=0}^{-\infty} \psi_m w^{-m-1/2} \right\rangle_A$$
$$= \sum_{n=1}^{\infty} z^{-n-1/2} w^{n-1/2} + \frac{1}{2}\frac{1}{\sqrt{zw}} \qquad (6.9)$$
$$= \frac{1}{\sqrt{zw}}\left(\frac{w}{z-w} + \frac{1}{2}\right) = \frac{\tfrac{1}{2}\left(\sqrt{\tfrac{z}{w}}+\sqrt{\tfrac{w}{z}}\right)}{z-w} \ .$$

This result has the property that it agrees with the result (6.6) in the $z \to w$ limit (the short distance behavior is independent of the global boundary conditions), and also changes sign as either $z$ or $w$ makes a loop around 0 or $\infty$. It could alternatively have been derived as the unique function with these properties.

We now wish to show how (6.9) may be used to infer the conformal weight $h_\sigma$ of the field $\sigma(w)$. This is extracted from the operator product with the stress-energy tensor

$$T(z)\sigma(0)|0\rangle \sim \frac{h_\sigma\,\sigma(0)}{z^2}|0\rangle + \ldots \ , \qquad (6.10)$$

where the stress-energy tensor is defined as the limit

$$T(z) = \frac{1}{2}\left(\psi(z)\partial_w\psi(w) + \frac{1}{(z-w)^2}\right)_{z\to w} \ .$$

The expectation value of the stress-energy tensor in the state $\sigma(0)|0\rangle$ may be evaluated from (6.9) by taking the derivative with respect to $w$ and then setting $z = w + \epsilon$ in the limit $\epsilon \to 0$,

$$\langle \psi(z)\partial_w \psi(w) \rangle_A = -\frac{\frac{1}{2}\left(\sqrt{\frac{z}{w}} + \sqrt{\frac{w}{z}}\right)}{(z-w)^2} + \frac{1}{4}\frac{1}{w^{3/2}z^{1/2}} = -\frac{1}{\epsilon^2} + \frac{1}{8}\frac{1}{w^2} ,$$

so that

$$\langle T(z) \rangle_A = \frac{1}{16}\frac{1}{z^2} .$$

If we now take the limit $z \to 0$ and compare with (6.10) we find that $h_\sigma = \frac{1}{16}$.

Before turning to the promised treatment of the fermion zero modes, we outline an analogous treatment for a bosonic twist field. As in (6.7), we write

$$\partial x(z)\sigma(w) \sim (z-w)^{-1/2}\tau(w) + \ldots , \qquad (6.11)$$

where now by dimensional analysis the "excited twist field" $\tau$ has $h_\tau = h_\sigma + \frac{1}{2}$. A twist field $\sigma(w,\overline{w})$ (with $h_\sigma = \overline{h}_\sigma$) that twists both $x(z)$ and $\overline{x}(\overline{z})$ can then be constructed as a product of separate holomorphic and anti-holomorphic pieces.

We define the 2-point function for the boson with anti-periodic boundary conditions as in (6.8),

$$\langle \partial x(z)\,\partial x(w) \rangle_A \equiv \langle 0|\sigma(\infty)\,\partial x(z)\,\partial x(w)\,\sigma(0)|0\rangle , \qquad (6.12)$$

and again evaluate using the mode expansion (6.1). Now the boson with anti-periodic boundary conditions requires $n \in \mathbf{Z} + \frac{1}{2}$, so that

$$-\langle \partial x(z)\partial x(w) \rangle_A = \left\langle \sum_{n=\frac{1}{2}}^{\infty} z^{-n-1}\alpha_n \sum_{m=-\frac{1}{2}}^{-\infty} w^{-m-1}\alpha_m \right\rangle_A$$

$$= \sum_{n=\frac{1}{2}}^{\infty} n\, z^{-n-1}w^{n-1} = \frac{1}{(zw)^{1/2}}\frac{1}{z}\sum_{n=0}^{\infty}\left(n+\tfrac{1}{2}\right)\left(\frac{w}{z}\right)^n \qquad (6.13)$$

$$= \frac{1}{(zw)^{1/2}}\left(\frac{w}{(z-w)^2} + \frac{1}{2}\frac{1}{z-w}\right) = \frac{\frac{1}{2}\left(\sqrt{\frac{z}{w}}+\sqrt{\frac{w}{z}}\right)}{(z-w)^2} .$$

This result could equally have been derived by requiring the correct short distance behavior (2.16) as $z \to w$, together with the correct sign change for $z$ or $w$ taken around $0$ or $\infty$.

We may now use (6.13) to evaluate the expectation value of the stress-energy tensor in the twisted sector

$$\langle T(z) \rangle_A = -\frac{1}{2}\lim_{z\to w}\left\langle \partial x(z)\partial x(w) + \frac{1}{(z-w)^2} \right\rangle_A = \frac{1}{16z^2} .$$

Taking $z \to 0$ we again infer from

$$T(z)\sigma(0)|0\rangle \sim \frac{h_\sigma\,\sigma(0)}{z^2}|0\rangle + \ldots$$

that the twist field for a single holomorphic boson has $h_\sigma = \frac{1}{16}$.

At first this result may seem strange, since a single $c=1$ boson is nominally composed of two $c=\frac{1}{2}$ fermions. The correspondence is given by

$$\psi_\pm(z) =: e^{\pm ix(z)}: , \qquad (6.14)$$

where by (2.19), $\psi_\pm(z)$ are seen to have conformal weight $h = \frac{1}{2}$ appropriate to fermions. Under the twist $x \to -x$ we see that $\psi_\pm \to \psi_\mp$. In terms of real fermions $\psi_{1,2}$ defined by $\psi_\pm = \frac{i}{\sqrt{2}}(\psi_1 \pm i\psi_2)$, we have $\psi_1 \to \psi_1$, $\psi_2 \to -\psi_2$. The bosonic twist $x \to -x$ thus corresponds to taking only one of the two fermions to minus itself, and it is natural that the twist operator for a boson have the same conformal weight as the twist operator for a single fermion. We can also understand this result by considering the current

$$\psi_1(z)\psi_2(z) = \lim_{z\to w}\frac{1}{i\sqrt{2}}\left((\psi_+(z)+\psi_-(z))\frac{-1}{\sqrt{2}}(\psi_+(w)-\psi_-(w))\right) = \partial x(z)$$

(here we have used

$$:e^{\pm ix(z)}::e^{\mp ix(w)}: \sim \frac{:e^{\pm ix(z)\mp ix(w)}:}{z-w} \sim \frac{:e^{\pm i(z-w)\partial x(w)}:}{z-w} \sim \pm i\partial x(w) ,$$

following from (2.19), and pulled out the leading term as $z \to w$). Again we see that twisting the $(1,0)$ current $\partial x \to -\partial x$ requires twisting only one of the two fermions $\psi_1$ or $\psi_2$.

There is a nice intuitive picture for calculating correlation functions involving twist fields (see e.g. [37]). A cut along which two fermions change sign is equivalent to an $SO(2)$ gauge field concentrated along the cut whose field strength, non-zero only at the endpoints of the cut, is adjusted to give a phase change of $\pi$ for parallel transport around them. In this language, the twist field looks like a point magnetic vortex, and changing the position of the cut just corresponds to a gauge transformation of its gauge potential. The physical spectrum of the model should consist only of operators that do not see the string of the vortex, so that the theory is local. If we bosonize the fermions, then correlations of twist fields can be calculated as ratios of partition functions of a free scalar field with and without these point sources of field strength. These ratios in turn are readily calculated correlation functions of exponentials of free scalars, and result in power law dependences for the correlators of twist fields. For their 2-point function, this reproduces in particular the conformal weight calculated earlier.

### 6.3. Fermionic zero modes

Now we return to a more careful treatment of the fermionic zero mode mentioned before (6.9). We begin by introducing an operator $(-1)^F$, defined to anticommute with the fermion field, $(-1)^F \psi(z) = -\psi(z)(-1)^F$, and to satisfy $\left((-1)^F\right)^2 = 1$. In terms of modes, this means that

$$\{(-1)^F, \psi_n\} = 0 \qquad \text{for all } n, \tag{6.15}$$

so $(-1)^F$ will have eigenvalue $\pm 1$ acting on states with even or odd numbers of fermion creation operators.

From (6.4) and (6.15) we thus have for $n \in \mathbf{Z}$ the anti-commutators

$$\{\psi_0, \psi_{n \neq 0}\} = 0, \qquad \{(-1)^F, \psi_0\} = 0, \quad \text{and} \quad \psi_0^2 = \frac{1}{2} \tag{6.16}$$

with the zero mode $\psi_0$. Since the mode $\psi_0$ acting on a state does not change the eigenvalue of $L_0$, in particular the ground state must provide a representation of the 2d clifford algebra consisting of $(-1)^F$ and $\psi_0$. The smallest irreducible representation of this algebra consists of two states that we label $\left|\frac{1}{16}\right\rangle_\pm$. The action of operators on these states can be represented in terms of Pauli matrices, defined to act as

$$\sigma_z \left|\tfrac{1}{16}\right\rangle_\pm = \pm \left|\tfrac{1}{16}\right\rangle_\pm \qquad \sigma_x \left|\tfrac{1}{16}\right\rangle_\pm = \left|\tfrac{1}{16}\right\rangle_\mp .$$

Then

$$(-1)^F = \sigma_z (-1)^{\sum \psi_{-n} \psi_n} \quad \text{and} \quad \psi_0 = \frac{1}{\sqrt{2}} \sigma_x (-1)^{\sum \psi_{-n} \psi_n} \tag{6.17}$$

provide a representation of (6.16) in a $(-1)^F$ diagonal basis. Since $\psi_0^2 \left|\tfrac{1}{16}\right\rangle_\pm = \tfrac{1}{2} \left|\tfrac{1}{16}\right\rangle_\pm$, if we identify the state $\sigma(0)|0\rangle$ in (6.9) with $\left|\tfrac{1}{16}\right\rangle_+$, the remaining steps in (6.9) are now justified. The state $\left|\tfrac{1}{16}\right\rangle_-$, on the other hand, can be identified with $\mu(0)|0\rangle$, where $\mu(z)$ is the conjugate twist field appearing in the right hand side of (6.7).

(If we are willing to give up having a well-defined $(-1)^F$, we could also use either of $\frac{1}{\sqrt{2}}\bigl(\left|\tfrac{1}{16}\right\rangle_+ \pm \left|\tfrac{1}{16}\right\rangle_-\bigr)$ as our ground state in (6.9). In terms of fields, this would mean trading the two fields $\sigma$ and $\mu$ for a single field $\widetilde\sigma$, taken as either of $\frac{1}{\sqrt{2}}(\sigma \pm \mu)$. Instead of the fusion rule $[\psi][\sigma] = [\mu]$ of (6.7), we would have $[\psi][\widetilde\sigma] = [\widetilde\sigma]$. The theories we consider later on here, however, will generally require a realization of $(-1)^F$ on the Hilbert space, so we have chosen to incorporate it into the formalism from the outset.)

An additional subtlety occurs when we consider both holomorphic fermions $\psi(z)$ and their anti-holomorphic partners $\overline\psi(\overline z)$. Then the $\overline\psi$'s satisfy the analog of (6.4), and as well

$$\{\psi_n, \overline\psi_m\} = 0 \qquad \forall\, n, m . \tag{6.18}$$

If we wish to realize separate operators $(-1)^{F_L}, (-1)^{F_R}$, satisfying $\{(-1)^{F_L}, \psi(z)\} = 0$, $\{(-1)^{F_R}, \overline\psi(\overline z)\} = 0$, then we simply duplicate the structure (6.17) for the $\psi$'s and $\overline\psi$'s to give four $\left|h = \tfrac{1}{16}, \overline h = \tfrac{1}{16}\right\rangle$ ground states of the form

$$\left|\tfrac{1}{16}\right\rangle_{L\pm} \otimes \left|\tfrac{1}{16}\right\rangle_{R\pm} . \tag{6.19}$$

But in general we need not require the existence of both chiral $(-1)^{F_L}$ and $(-1)^{F_R}$, but rather only the non-chiral combination $(-1)^F = (-1)^{F_L + F_R}$.

In fact (6.18) implies that $\psi_0$ and $\overline{\psi}_0$ already form a two dimensional Clifford algebra, so the combination $\psi_0 \overline{\psi}_0$ automatically serves to represent the non-chiral $(-1)^F$ restricted now to a two-dimensional ground state representation $\left|h = \frac{1}{16}, \overline{h} = \frac{1}{16}\right\rangle_\pm$. If we write the action of Pauli matrices on this basis as

$$\sigma_x \left|\tfrac{1}{16}, \tfrac{1}{16}\right\rangle_\pm = \left|\tfrac{1}{16}, \tfrac{1}{16}\right\rangle_\mp \qquad \sigma_y \left|\tfrac{1}{16}, \tfrac{1}{16}\right\rangle_\pm = \mp i \left|\tfrac{1}{16}, \tfrac{1}{16}\right\rangle_\mp$$
$$\sigma_z \left|\tfrac{1}{16}, \tfrac{1}{16}\right\rangle_\pm = \pm \left|\tfrac{1}{16}, \tfrac{1}{16}\right\rangle_\pm , \qquad (6.20a)$$

then it is easily verified that the zero mode representation

$$\psi_0 = \frac{\sigma_x + \sigma_y}{2} (-1)^{\sum_{n>0} \psi_{-n}\psi_n + \overline{\psi}_{-n}\overline{\psi}_n}$$
$$\overline{\psi}_0 = \frac{\sigma_x - \sigma_y}{2} (-1)^{\sum_{n>0} \psi_{-n}\psi_n + \overline{\psi}_{-n}\overline{\psi}_n} \qquad (6.20b)$$
$$(-1)^F = \sigma_z (-1)^{\sum_{n>0} \psi_{-n}\psi_n + \overline{\psi}_{-n}\overline{\psi}_n}$$

satisfies the algebra (6.16),(6.18). In (6.20$b$) we have chosen to represent the Clifford algebra in a rotated basis,

$$\frac{1}{\sqrt{2}}(\sigma_x \pm \sigma_y) = \begin{pmatrix} & e^{\mp i\pi/4} \\ e^{\pm i\pi/4} & \end{pmatrix} ,$$

since this is the representation we shall find induced by our choice of phase conventions (choice of gauge) for operator product expansions.

(The four dimensional representation (6.19), irreducible under the full chiral algebra including both $(-1)^{F_L}$ and $(-1)^{F_R}$, is reducible under the subalgebra that includes only the non-chiral $(-1)^F$. Explicitly the two two-dimensional irreducible representations of the non-chiral subalgebra are given by

$$\left|\tfrac{1}{16}, \tfrac{1}{16}\right\rangle_\pm = \left(\left|\tfrac{1}{16}\right\rangle_{L+} \otimes \left|\tfrac{1}{16}\right\rangle_{R\pm}\right) + \left(\left|\tfrac{1}{16}\right\rangle_{L-} \otimes \left|\tfrac{1}{16}\right\rangle_{R\mp}\right)$$
$$\left|\tfrac{1}{16}, \tfrac{1}{16}\right\rangle'_\pm = \left(\left|\tfrac{1}{16}\right\rangle_{L+} \otimes \left|\tfrac{1}{16}\right\rangle_{R\pm}\right) - \left(\left|\tfrac{1}{16}\right\rangle_{L-} \otimes \left|\tfrac{1}{16}\right\rangle_{R\mp}\right) .$$

We see that only the operators $(-1)^{F_L}$ and $(-1)^{F_R}$ act to connect the orthogonal Hilbert spaces built on $\left|\tfrac{1}{16}, \tfrac{1}{16}\right\rangle_\pm$ and $\left|\tfrac{1}{16}, \tfrac{1}{16}\right\rangle'_\pm$. Had we begun with the four dimensional representation (6.19), but required only the existence of the non-chiral $(-1)^{F_L+F_R}$, then we could consistently throw out all the states built say on $\left|\tfrac{1}{16}, \tfrac{1}{16}\right\rangle'_\pm$ and be left with the minimal two-dimensional representation (6.20) of the zero mode algebra. Similar considerations apply in the case of realizations of $N = 1$ superconformal algebras without chiral $(-1)^F$ [38].)

## 7. Free fermions on a torus

In this section we shall consider conformal theory not on the conformal plane, but rather on a torus, i.e., on a Riemann surface of genus one. Our motivation for doing this is both statistical mechanical and field theoretical. From the statistical mechanical point of view, it turns out that the fact that a given model admits a consistent formulation on the torus acts to constrain its operator content already on the plane. From the field theoretical point of view, conformal field theory achieves its full glamour when formulated on an arbitrary genus Riemann surface. Higher genus is also the natural arena for applications of conformal field theory to perturbative string theory. The torus is the first non-trivial step in this direction, and turns out to probe all of the essential consistency requirements for conformal field theory formulated on an arbitrary genus Riemann surface. We refer the reader to Friedan's lectures for more on the higher genus extension.

### 7.1. Back to the cylinder, on to the torus

Our strategy for constructing conformal field theory on the torus is to make use of the local properties of operators already constructed on the conformal plane, map them to the cylinder via the exponential map, and then arrive at a torus via discrete identification. While this procedure preserves all local properties of operators in a theory, it does not necessarily preserve all of their global properties. For example since the torus maps to an annulus on the plane, only the generators of dilatations and rotations, i.e. $L_0$ and $\overline{L}_0$, survive as global symmetry generators. On the torus, $L_{\pm 1}$ and $\overline{L}_{\pm 1}$ are reduced to playing the role of local symmetry generators, as played by the remaining $L_n$, $\overline{L}_n$ ($n \neq 0, \pm 1$) on the plane, and the global symmetry group is reduced to $U(1) \times U(1)$.

Another global property affected by the passage from the plane to the cylinder (or torus) is boundary conditions on fields. Let us consider the map $w \to z = e^w$, mapping the cylinder, coordinatized by $w$, to the plane, coordinatized by $z$. Since $\varphi(z,\overline{z})dz^h d\overline{z}^{\overline{h}}$ is invariant under this map, we find

$$\varphi_{\text{cyl}}(w,\overline{w}) = \left(\frac{dz}{dw}\right)^h \left(\frac{d\overline{z}}{d\overline{w}}\right)^{\overline{h}} \varphi(z,\overline{z}) = z^h \overline{z}^{\overline{h}} \varphi(z,\overline{z}) \ . \qquad (7.1)$$

This means that a field $\varphi(z,\overline{z})$ on the plane that is invariant under $z \to e^{2\pi i}z$, $\overline{z} \to e^{-2\pi i}\overline{z}$ corresponds to a field $\varphi_{\text{cyl}}(w,\overline{w})$ that picks up a phase $e^{2\pi i(h-\overline{h})}$ under $w \to w + 2\pi i$, $\overline{w} \to \overline{w} - 2\pi i$. Fields with integer spin $s = h - \overline{h}$ thus have the same boundary conditions on the plane and cylinder. Fields with half-integer spin having periodic boundary conditions become anti-periodic, and vice-versa, when passing from the plane to the cylinder.

We can see the same effect in terms of the mode expansion $\varphi(z) = \sum_n \varphi_n z^{-n-h}$ of a holomorphic field. The mode expansion induced on the cylinder,

$$\varphi_{\text{cyl}}(w) = \left(\frac{dz}{dw}\right)^h \varphi(z) = z^h \sum_n \varphi_n z^{-n-h} = \sum_n \varphi_n e^{-nw} \ , \qquad (7.2)$$

becomes an ordinary Fourier series. Again however a field moded as $n \in \mathbf{Z} - h$ so that it is non-singular at the origin of the conformal plane is no longer single-valued under $w \to w + 2\pi i$ on the cylinder.

For a fermion, with $h = \frac{1}{2}$, $\overline{h} = 0$, we have from (7.1) that $\psi_{\text{cyl}}(w) = z^{1/2}\psi(z)$ so A boundary conditions on the plane become P on the cylinder, and vice-versa. In terms of the mode expansion (7.2), we have

$$\psi_{\text{cyl}}(w) = \sum_n \psi_n e^{-nw}, \qquad n \in \begin{cases} \mathbf{Z} & (\text{P}) \\ \mathbf{Z} + \frac{1}{2} & (\text{A}) \end{cases} \ , \qquad (7.3)$$

opposite to the case (6.5) on the plane where the same modes $\psi_n$ give A for $n \in \mathbf{Z}$ and P for $n \in \mathbf{Z} + \frac{1}{2}$. On the cylinder it is thus the P sector that has ground state $L_0$ eigenvalue larger by $\frac{1}{16}$. We point out that even if we thought only one of the A or P boundary conditions the more natural, we would be forced to consider the other anyway in moving back and forth from plane to cylinder (giving a possible motivation for considering both on equal footing from the outset). (For superpartners $\psi^\mu$ of spacetime bosonic coordinates in string theory, the sectors corresponding to P and A on the cylinder, i.e. $n \in \mathbf{Z}$ and $n \in \mathbf{Z} + \frac{1}{2}$ respectively, are ordinarily termed the Ramond (R) and Neveu-Schwarz (NS) sectors.) Since the modes $\psi_n$ in our mode expansion (7.3) on the cylinder are identically those on the plane (6.1) (local operator products are not affected by conformal mapping), they satisfy the same anti-commutation rules (6.4),

$$\{\psi_n, \psi_m\} = \delta_{n+m,0} \ .$$

$\psi_{-n}$ and $\psi_n$ ($n > 0$) thus continue to be regarded as fermionic creation and annihilation operators acting on a vacuum state $|0\rangle$, defined to satisfy $\psi_n|0\rangle = 0$ ($n > 0$), and the Hilbert space of states $\psi_{-n_1} \ldots \psi_{-n_k}|0\rangle$ is built up by applying creation operators $\psi_{-n}$ to $|0\rangle$.

For a field such as the stress-energy tensor $T(z)$ that does not transform tensorially under conformal transformations, an additional subtlety arises in the transfer to the cylinder. Under conformal transformations $w \to z$, $T(z)$ in general picks up an anomalous piece proportional to the Schwartzian derivative $S(z,w) = (\partial_w z \, \partial_w^3 z - \frac{3}{2}(\partial_w^2 z)^2)/(\partial_w z)^2$ as in (3.3). For the exponential map $w \to z = e^w$, we have $S(e^w, w) = -1/2$, so

$$T_{\text{cyl}}(w) = \left(\frac{\partial z}{\partial w}\right)^2 T(z) + \frac{c}{12}S(z,w) = z^2 T(z) - \frac{c}{24} \ .$$

Substituting the mode expansion $T(z) = \sum L_n z^{-n-2}$, we find

$$T_{\text{cyl}}(w) = \sum_{n \in \mathbf{Z}} L_n z^{-n} - \frac{c}{24} = \sum_{n \in \mathbf{Z}} \left(L_n - \frac{c}{24}\delta_{n0}\right) e^{-nw} \ . \qquad (7.4)$$

The translation generator $(L_0)_{\text{cyl}}$ on the cylinder is thus given in terms of the dilatation generator $L_0$ on the plane as

$$(L_0)_{\text{cyl}} = L_0 - \frac{c}{24} \ .$$

Ordinarily one can always shift the energy of the vacuum by a constant (equivalently change the normalization of a functional integral), but in conformal field theory, scale and rotational invariance of the $SL(2,\mathbf{C})$ invariant vacuum on the plane naturally fixes $L_0$ and $\overline{L}_0$ to have eigenvalue zero on the vacuum, thereby fixing the zero of energy once and for all.

Conformal field field theory on a cylinder coordinatized by $w$ can now be transferred to a torus as follows. We let $H$ and $P$ denote the energy and momentum operators, i.e. the operators that effect translations in the "space" and "time" directions $\operatorname{Re} w$ and $\operatorname{Im} w$ respectively. On the plane we saw that $L_0 \pm \overline{L}_0$ respectively generated dilatations and rotations, so according to the discussion of radial quantization at the beginning of subsection 2.2, we have $H = (L_0)_{\text{cyl}} + (\overline{L}_0)_{\text{cyl}}$ and $P = (L_0)_{\text{cyl}} - (\overline{L}_0)_{\text{cyl}}$. To define a torus we need to identify two periods in $w$. It is convenient to redefine $w \to iw$, so that one of the periods is $w \equiv w + 2\pi$. The remaining period we take to be $w \equiv w + 2\pi\tau$, where $\tau = \tau_1 + i\tau_2$ and $\tau_1$ and $\tau_2$ are real parameters. This means that the surfaces $\operatorname{Im} w = 2\pi\tau_2$ and $\operatorname{Im} w = 0$ are identified after a shift by $\operatorname{Re} w \to \operatorname{Re} w + 2\pi\tau_1$ (see fig. 8). The complex parameter $\tau$ parametrizing this family of distinct tori is known as the modular parameter.

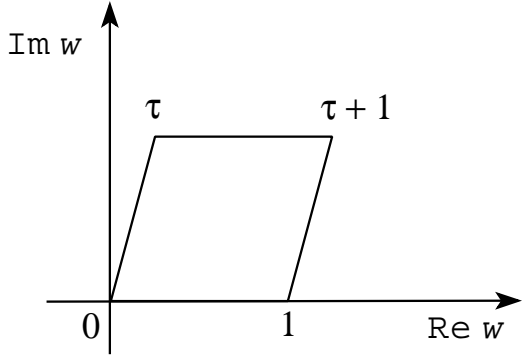

Fig. 8. Torus with modular parameter $\tau$.

Since we are defining (imaginary) time translation of $\operatorname{Im} w$ by its period $2\pi\tau_2$ to be accompanied by a spatial translation of $\operatorname{Re} w$ by $2\pi\tau_1$, the operator implementation for the partition function of a theory with action $S$ on a torus with modular parameter $\tau$ is

$$\begin{aligned}
\int e^{-S} &= \operatorname{tr} e^{2\pi i \tau_1 P} e^{-2\pi \tau_2 H}\\
&= \operatorname{tr} e^{2\pi i \tau_1 \left((L_0)_{\text{cyl}} - (\overline{L}_0)_{\text{cyl}}\right)} e^{-2\pi \tau_2 \left((L_0)_{\text{cyl}} + (\overline{L}_0)_{\text{cyl}}\right)}\\
&= \operatorname{tr} e^{2\pi i \tau (L_0)_{\text{cyl}}} e^{-2\pi i \overline{\tau} (\overline{L}_0)_{\text{cyl}}} = \operatorname{tr} q^{(L_0)_{\text{cyl}}} \overline{q}^{(\overline{L}_0)_{\text{cyl}}}\\
&= \operatorname{tr} q^{L_0 - \frac{c}{24}} \overline{q}^{\overline{L}_0 - \frac{\overline{c}}{24}} = q^{-\frac{c}{24}} \overline{q}^{-\frac{\overline{c}}{24}} \operatorname{tr} q^{L_0} \overline{q}^{\overline{L}_0} ,
\end{aligned} \tag{7.5}$$

where $q \equiv \exp(2\pi i \tau)$. For the $c = \overline{c} = \frac{1}{2}$ theory of a single holomorphic fermion $\psi(w)$ and a single anti-holomorphic fermion $\overline{\psi}(\overline{w})$ on the torus, we would thus find

$$\int e^{-S} = (q\overline{q})^{-\frac{c}{24}} \operatorname{tr} q^{L_0} \overline{q}^{\overline{L}_0} = (q\overline{q})^{-\frac{1}{48}} \operatorname{tr} q^{L_0} \overline{q}^{\overline{L}_0} . \tag{7.6}$$

Before turning to a treatment of free fermions in terms of the representation theory of the Virasoro algebra, we pause here to mention that the vacuum energies derived in section 6 can be alternatively interpreted to result from a vacuum normal ordering prescription on the cylinder. We find for example

$$\begin{aligned}
(L_0)_{\text{cyl}} &= \frac{1}{2} \sum_n n : \psi_{-n} \psi_n : = \sum_{n>0} n \psi_{-n} \psi_n - \frac{1}{2} \sum_{n>0} n \\
&= \sum_{n>0} n \psi_{-n} \psi_n + \begin{cases} -\frac{1}{2}\zeta(-1) = \frac{1}{24} & n \in \mathbf{Z} \\ -\frac{1}{2}(-\frac{1}{2}\zeta(-1)) = -\frac{1}{48} & n \in \mathbf{Z} + \frac{1}{2} \end{cases} ,
\end{aligned}$$

where we have used $\zeta$-function regularization to evaluate the infinite sums. We see that the result for $n \in \mathbf{Z} + \frac{1}{2}$ agrees with the result given earlier in this subsection for the A sector on the cylinder. For $n \in \mathbf{Z}$ we as well find correctly that the vacuum energy is shifted up by $\frac{1}{24} - (-\frac{1}{48}) = \frac{1}{16}$. The justification for this $\zeta$-function regularization procedure ultimately resides in its compatibility with conformal and modular invariance. For a boson on the cylinder we would instead find

$$\begin{aligned}
(L_0)_{\text{cyl}} &= \frac{1}{2} \sum_n : \alpha_{-n} \alpha_n : = \sum_{n>0} \alpha_{-n} \alpha_n + \frac{1}{2} \sum_{n>0} n \\
&= \sum_{n>0} \alpha_{-n} \alpha_n + \begin{cases} \frac{1}{2}\zeta(-1) = -\frac{1}{24} & n \in \mathbf{Z} \\ \frac{1}{2}(-\frac{1}{2}\zeta(-1)) = \frac{1}{48} & n \in \mathbf{Z} + \frac{1}{2} \end{cases} .
\end{aligned}$$

For $n \in \mathbf{Z}$ the result correctly gives $-\frac{c}{24}$, now with $c = 1$. For $n \in \mathbf{Z} + \frac{1}{2}$ we see that the vacuum energy is increased by $\frac{1}{16}$, again correctly giving the conformal weight of the bosonic twist field calculated in the previous section. (Note that the vacuum normal ordering constants for a single boson on the cylinder are simply opposite in sign from those for the fermion.) The anti-periodic boson parametrizes what is known as a $\mathbf{Z}_2$ orbifold, and will be treated in detail in the next section.

More generally this vacuum normal ordering prescription can be used to calculate the vacuum energy for a complex holomorphic fermion (i.e. two $c = \frac{1}{2}$ holomorphic fermions) with boundary condition twisted by a complex phase $\psi_{\text{cyl}}(w + 2\pi i) = \exp(2\pi i \eta) \psi_{\text{cyl}}(w)$, $0 \leq \eta \leq 1$. The resulting vacuum normal ordering constant calculated as above is $f(\eta) = \frac{1}{12} - \frac{1}{2}\eta(1-\eta)$. (As a consistency check, a single real fermion has one-half of $f$ as vacuum energy, and consequently we confirm that $\frac{1}{2}f(\frac{1}{2}) = -\frac{1}{48}$ and $\frac{1}{2}f(0) = \frac{1}{24}$ for vacuum energy in the A and P sectors respectively on the cylinder).

7.2. $c = \frac{1}{2}$ representations of the Virasoro algebra

Having introduced all of the necessary formalism for treating free fermions on the torus, we are now prepared to make contact with the general representation theory of the Virasoro algebra introduced in section 4. Since the stress-energy tensor for a single free fermion has $c = \frac{1}{2}$, we should expect to find free fermionic realizations of the three unitary irreducible representations allowed for this value of $c$, namely $h = \{h_{1,1}, h_{2,1}, h_{2,2}\} = \{0, \frac{1}{2}, \frac{1}{16}\}$.

We begin by considering the states built in the A sector of the fermion on the torus. In this case states take the form $\psi_{-n_1} \ldots \psi_{-n_k} |0\rangle$, with $n_i \in \mathbf{Z} + \frac{1}{2}$. The first few such states, ordered according to their eigenvalue under $L_0 = \sum_{n>0} n \psi_{-n} \psi_n$, are

| $L_0$ eigenvalue | state |
|---|---|
| 0 | $|0\rangle$ |
| 1/2 | $\psi_{-1/2}|0\rangle$ |
| 3/2 | $\psi_{-3/2}|0\rangle$ |
| 2 | $\psi_{-3/2}\psi_{-1/2}|0\rangle$ |
| 5/2 | $\psi_{-5/2}|0\rangle$ |
| 3 | $\psi_{-5/2}\psi_{-1/2}|0\rangle$ |
| 7/2 | $\psi_{-7/2}|0\rangle$ |
| 4 | $\psi_{-7/2}\psi_{-1/2}|0\rangle, \quad \psi_{-5/2}\psi_{-3/2}|0\rangle$ |
| $\ldots$ | . |

(7.7)

Denoting the trace in this sector by $\text{tr}_A$, we calculate

$$\text{tr}_A \, q^{L_0} = 1 + q^{1/2} + q^{3/2} + q^2 + q^{5/2} + q^3 + q^{7/2} + 2q^4 + \ldots \, .$$

In general traces of the form $\text{tr} \, q^{L_0} = \sum_n N_n q^n$ characterize the number of states $N_n$ that occur at a given level $n$ (eigenvalue of $L_0$). $q$ may thus be regarded as a formal parameter analogous to the Cartan angles that appear in character formulae for Lie groups. $q$ $(= e^{2\pi i \tau})$ obtains additional significance in terms of the modular parameter $\tau$ when these traces are regarded as the result of calculating functional integrals (7.5) for field theories on a torus.

The states (7.7) form a (not necessarily irreducible) representation of the Virasoro algebra with $c = \frac{1}{2}$. From the eigenvalues of $L_0$, we immediately identify the representation as the direct sum $[0] \oplus [\frac{1}{2}]$ of the highest weight representations with $h = 0$ and $h = \frac{1}{2}$. Since there is only a single state with $h = 0$ and only a single state with $h = \frac{1}{2}$ we see that each of these two representations appears with unit multiplicity. Moreover since states created by applying $L_{-n}$'s to a single highest weight state all have integrally spaced $L_0$ eigenvalues, we see that the states of the representation $[0]$ are identically those with even fermion number, and hence $L_0 \in \mathbf{Z}$; the states of $[\frac{1}{2}]$ are those with odd fermion number and hence $L_0 \in \mathbf{Z} + \frac{1}{2}$. These two sets of states are precisely distinguished by their opposite eigenvalues under the operator $(-1)^F$, i.e.

$$\mathrm{tr}_A \, (-1)^F q^{L_0} = 1 - q^{1/2} - q^{3/2} + q^2 - q^{5/2} + q^3 - q^{7/2} + 2q^4 + \ldots .$$

The projection operators $\frac{1}{2}(1 \pm (-1)^F)$ may therefore be used to disentangle the two representations, giving

$$\begin{aligned}
q^{-1/48} \, \mathrm{tr}_A \frac{1}{2} \big(1 + (-1)^F\big) q^{L_0} &= q^{-1/48}(1 + q^2 + q^3 + 2q^4 + \ldots) \\
&= q^{-1/48} \, \mathrm{tr}_{h=0} \, q^{L_0} \equiv \chi_0 \\
q^{-1/48} \, \mathrm{tr}_A \frac{1}{2} \big(1 - (-1)^F\big) q^{L_0} &= q^{-1/48}(q^{1/2} + q^{3/2} + q^{5/2} + q^{7/2} + \ldots) \\
&= q^{-1/48} \, \mathrm{tr}_{h=1/2} \, q^{L_0} \equiv \chi_{1/2} ,
\end{aligned} \quad (7.8)$$

where $\chi_{0,1/2}$ are the characters of the $h = 0, \frac{1}{2}$ representations of the $c = \frac{1}{2}$ Virasoro algebra (defined to include the offset of $L_0$ by $-c/24$).

In the periodic sector of the fermion on the torus, on the other hand, we have $L_0 = \sum_{n>0} \psi_{-n}\psi_n + \frac{1}{16}$ with $n \in \mathbf{Z}$. As seen in (6.17), the fermion zero mode algebra together with $(-1)^F$ requires two ground states $\big|\frac{1}{16}\big\rangle_\pm$, with eigenvalues $\pm 1$ under $(-1)^F$, that satisfy

$$\psi_0 \big|\tfrac{1}{16}\big\rangle_\pm = \frac{1}{\sqrt{2}} \big|\tfrac{1}{16}\big\rangle_\mp .$$

The states of the Hilbert space in this sector thus take the form

| $L_0$ eigenvalue | state |
|---|---|
| $\frac{1}{16} + 0$ | $\big|\frac{1}{16}\big\rangle_\pm$ |
| $\frac{1}{16} + 1$ | $\psi_{-1}\big|\frac{1}{16}\big\rangle_\pm$ |
| $\frac{1}{16} + 2$ | $\psi_{-2}\big|\frac{1}{16}\big\rangle_\pm$ |
| $\frac{1}{16} + 3$ | $\psi_{-3}\big|\frac{1}{16}\big\rangle_\pm , \quad \psi_{-2}\psi_{-1}\big|\frac{1}{16}\big\rangle_\pm$ |
| $\ldots$ | $.$ |

(7.9)

We find two irreducible representations of the $c = \frac{1}{2}$ Virasoro algebra with highest weight $h = \frac{1}{16}$. Again they can be disentangled by projecting onto $\pm 1$ eigenstates of $(-1)^F$,

$$\begin{aligned}
q^{-1/48} \, \mathrm{tr}_P \frac{1}{2} \big(1 \pm (-1)^F\big) q^{L_0} &= q^{1/24}(1 + q + q^2 + 2q^3 + \ldots) \\
&= q^{-1/48} \, \mathrm{tr}_{h=1/16} \, q^{L_0} \equiv \chi_{1/16} .
\end{aligned} \quad (7.10)$$

Although it happens that $\mathrm{tr}_P (-1)^F q^{L_0} = 0$ in this sector, due to a cancellation between equal numbers of states at each level with opposite $(-1)^F$, its insertion in (7.10) has the formal effect of assigning states with even numbers of fermions built on $\big|\frac{1}{16}\big\rangle_+$, or odd numbers on $\big|\frac{1}{16}\big\rangle_-$, to one representation $\big[\frac{1}{16}\big]_+$ with $(-1)^F = 1$, and vice-versa to the other representation $\big[\frac{1}{16}\big]_-$ with $(-1)^F = -1$.

### 7.3. The modular group and fermionic spin structures

We shall now introduce some essentials of the Lagrangian functional integral formalism for fermions $\psi(w)$ that live on a torus. (For the remainder of this section, $\psi$ will always mean $\psi_{\mathrm{cyl}}$.) This formalism will facilitate writing down and manipulating explicit forms for the characters of the $h = 0, \frac{1}{2}, \frac{1}{16}$ representations of the $c = \frac{1}{2}$ Virasoro algebra. In general a torus is specified by two periods which by rescaling coordinates we take as 1 and $\tau$, where $\tau$ is the modular parameter introduced in the previous subsection. Symbolically we write $w \equiv w + 1 \equiv w + \tau$, which means that fields that live on the torus must satisfy $\varphi(w+1) = \varphi(w+\tau) = \varphi(w)$. It is convenient to write the coordinate $w$ in terms of real coordinates $\sigma^{0,1} \in [0,1)$ as $w = \sigma^1 + \tau \sigma^0$.

To specify a fermionic theory, we now need to generalize the considerations of section 6 from a choice of P or A boundary conditions around the one non-trivial cycle on the cylinder, or punctured plane, to two such choices around the two non-trivial cycles of the torus. (This is known as choosing a spin structure for the fermion on a genus one Riemann surface.) In the coordinates $\sigma^0, \sigma^1$, this amounts to choosing signs $\psi(\sigma^0 + 1, \sigma^1) = \pm \psi(\sigma^0, \sigma^1)$, $\psi(\sigma^0, \sigma^1 + 1) = \pm \psi(\sigma^0, \sigma^1)$. As in section 6, we can view this sign ambiguity to result from spinors actually living on a double cover of the frame bundle, so that only bilinears, corresponding to two dimensional vector-like representations, need be invariant under parallel transport around a closed cycle.

We shall denote the result of performing the functional integral $\int \exp(-\int \psi\bar{\partial}\psi)$ over fermions with a given fixed spin structure by the symbol ${}_X\square_Y$. The result for the spin structure with periodic (P) boundary condition in the $\sigma^0$ (time) and anti-periodic boundary condition (A) in the $\sigma^1$ (space) direction, for example, we denote by ${}_P\square_A$. The result of the functional integral can also be regarded as calculating the square root of the determinant of the operator $\bar{\partial}$ for the various choices of boundary conditions. Due to the zero mode (i.e. the constant function) allowed by PP boundary conditions, we see for example that ${}_P\square_P = \left(\det_{PP}\bar{\partial}\right)^{1/2} = 0$.

In ordinary two-dimensional field theory on a torus, it would suffice to choose any particular spin structure and that would be the end of the story. But there is an additional invariance, modular invariance, that we shall impose on "good" conformal field theories on a torus that forces consideration of nontrivial combinations of spin structures. (In general a "really good" conformal theory is required to be sensible on an arbitrary Riemann surface, i.e. be modular invariant to all orders. This turns out to be guaranteed by duality of the 4-point functions on the sphere together with modular invariance of all 1-point functions on the torus[30][39]. Intuitively this results from the possibility of constructing all correlation functions on arbitrary genus surfaces by "sewing" together objects of the above form. Thus all the useful information about conformal field theories can be obtained by studying them on the plane and on the torus.)

The group of modular transformations is the group of disconnected diffeomorphisms of the torus, generated by cutting along either of the non-trivial cycles, then regluing after a twist by $2\pi$. Cutting along a line of constant $\sigma^0$, then regluing, gives the transformation $T: \tau \to \tau+1$, while cutting then regluing along a line of constant $\sigma^1$ gives the transformation $U: \tau \to \tau/(\tau+1)$. (This is the new ratio of periods (see fig. 9), and hence the new modular parameter after the coordinate rescaling $w \to w/(\tau+1)$.) These two transformations generate a group of transformations

$$\tau \to \frac{a\tau+b}{c\tau+d} \qquad \begin{pmatrix} a & b \\ c & d \end{pmatrix} \in SL(2,\mathbf{Z}) \qquad (7.11)$$

(i.e. $a,b,c,d \in \mathbf{Z}$, $ad-bc=1$), known as the modular group. Since reversing the sign of all of $a,b,c,d$ in (7.11) leaves the action on $\tau$ unchanged, the modular group is actually $PSL(2,\mathbf{Z}) = SL(2,\mathbf{Z})/\mathbf{Z}_2$. By a modular transformation one can always take $\tau$ to lie in the fundamental region $-\frac{1}{2} < \operatorname{Re}\tau \leq \frac{1}{2}$, $|\tau| \geq 1$ ($\operatorname{Re}\tau \geq 0$), $|\tau| > 1$ ($\operatorname{Re}\tau < 0$). Usually one uses $T: \tau \to \tau+1$ and $S = T^{-1}UT^{-1}: \tau \to -1/\tau$ to generate the modular group. They satisfy the relations $S^2 = (ST)^3 = 1$.

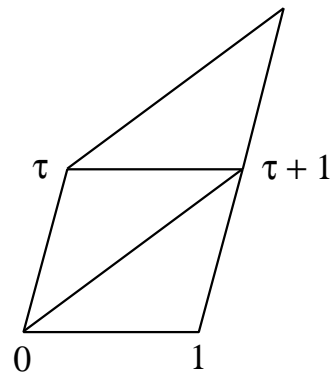

Fig. 9. The modular transformation $U: \tau \to \tau/(\tau+1)$.

Now we consider the transformation properties of fermionic spin structures under the modular group. Under $T$, we have for example

$$\tau \to \tau+1: \qquad {}_A\square_A \leftrightarrow {}_P\square_A. \qquad (7.12a)$$

We can see this starting from ${}_A\square_A$ since shifting the upper edge of the box one unit to the right means that the new "time" direction, from lower left to upper right, sees both the formerly anti-periodic boundary conditions, to give an overall periodic boundary condition. (see fig. 10) From $\mathrm{P}\square_\mathrm{A}$ the opposite occurs. The spin structures $\mathrm{A}\square_\mathrm{P}$ and $\mathrm{P}\square_\mathrm{P}$, on the other hand, transform into themselves under $T$.

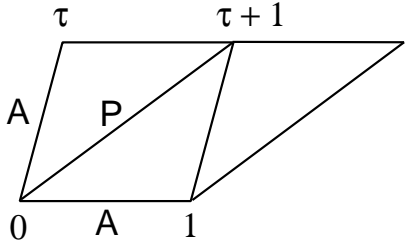

Fig. 10. The modular transformation $T : \tau \to \tau + 1$.

The action of $U : \tau \to \tau/(\tau + 1)$ on any spin structure can be determined similarly, and thence the action of $S = T^{-1}UT^{-1}$. We find that $S$ acts to interchange the boundary conditions in "time" and "space" directions, so that

$$\tau \to -1/\tau : \qquad \mathrm{P}\square_\mathrm{A} \leftrightarrow \mathrm{A}\square_\mathrm{P}, \qquad (7.12b)$$

while $\mathrm{A}\square_\mathrm{P}$ and $\mathrm{P}\square_\mathrm{P}$ transform into themselves. Since $S$ and $T$ generate the modular group, (7.12a,b) determine the transformation properties under arbitrary modular transformations (7.11). It is evident, for example, that the functional integral for the spin structure $\mathrm{P}\square_\mathrm{P}$ is invariant under all modular transformations (and in fact, as noted earlier, vanishes identically due to the zero mode). For the moment, (7.12a,b) are intended as symbolic representations of modular transformation properties of different fermionic spin structures. We shall shortly evaluate the functional integrals and find that (7.12a,b) become correct as equations, up to phases.

### 7.4. $c = \frac{1}{2}$ Virasoro characters

The $c = \frac{1}{2}$ Virasoro characters (7.8) and (7.10) introduced in the previous subsection may be written explicitly in terms of fermionic functional integrals over different spin structures. For example the result of the functional integral for a single holomorphic fermion with spin structure $\mathrm{A}\square_\mathrm{A}$, according to (7.5), is simply the trace in the anti-periodic sector $q^{-1/48} \mathrm{tr}_\mathrm{A}\, q^{L_0}$ (where the prefactor $q^{-1/48}$ results from the vacuum energy discussed earlier). The spin structure $\mathrm{P}\square_\mathrm{A}$ in Hamiltonian language corresponds to taking the trace of the insertion of an operator that anticommutes with the fermion (thereby flipping the boundary conditions in the time direction). Since $(-1)^F \psi = -\psi(-1)^F$, $(-1)^F$ is just such an operator and $\mathrm{P}\square_\mathrm{A} = q^{-1/48}\, \mathrm{tr}_\mathrm{A}\, (-1)^F q^{L_0}$. For the periodic sector, we define $\mathrm{A}\square_\mathrm{P} = \frac{1}{\sqrt{2}} q^{-1/48}\, \mathrm{tr}_\mathrm{P}\, q^{L_0}$ and $\mathrm{P}\square_\mathrm{P} = \frac{1}{\sqrt{2}} q^{-1/48}\, \mathrm{tr}_\mathrm{P}(-1)^F q^{L_0}\ (=0)$. (The factor $\frac{1}{\sqrt{2}}$ is included ultimately to simplify the behavior under modular transformations).

The calculation of these traces is elementary. In the $2 \times 2$ basis $(|0\rangle, \psi_{-n}|0\rangle)$ for the $n^\mathrm{th}$ fermionic mode, we have

$$q^{n\psi_{-n}\psi_n} = \begin{pmatrix} 1 & \\ & q^n \end{pmatrix},$$

and thus $\mathrm{tr}\, q^{n\psi_{-n}\psi_n} = 1 + q^n$, and similarly $\mathrm{tr}(-1)^F q^{n\psi_{-n}\psi_n} = 1 - q^n$. It follows that

$$q^{L_0} = q^{\sum_{n>0} n\psi_{-n}\psi_n} = \prod_{n>0} q^{n\psi_{-n}\psi_n} = \prod_{n>0} \begin{pmatrix} 1 & \\ & q^n \end{pmatrix}.$$

Since the trace of a direct product of matrices $\otimes_i M_i$ satisfies $\mathrm{tr} \otimes_i M_i = \prod_i \mathrm{tr}\, M_i$, we find $\mathrm{tr}_\mathrm{A}\, q^{L_0} = \prod_{n=0}^\infty (1 + q^{n+1/2})$, $\mathrm{tr}_\mathrm{A}(-1)^F q^{L_0} = \prod_{n=0}^\infty (1 - q^{n+1/2})$, and $\mathrm{tr}_\mathrm{P}\, q^{L_0} = q^{1/16} \prod_{n=0}^\infty (1 + q^n)$. Expanding out the first few terms, we can compare with (7.8) and (7.10) and see how these infinite products enumerate all possible occupations of modes satisfying Fermi-Dirac statistics. In the case with $(-1)^F$ inserted, each state is in addition signed according to whether it is created by an even or odd number of fermionic creation operators.

From (7.5), we may thus summarize the partition functions for a single $c = \frac{1}{2}$ holomorphic fermion as

$$\mathrm{A}\square_\mathrm{A} = q^{-1/48}\, \mathrm{tr}_\mathrm{A}\, q^{L_0} = q^{-1/48} \prod_{n=0}^\infty (1 + q^{n+1/2}) = \sqrt{\frac{\vartheta_3}{\eta}}, \qquad (7.13a)$$

$$\text{P}\boxed{\phantom{x}}_{\text{A}} = q^{-1/48}\,\text{tr}_A(-1)^F q^{L_0} = q^{-1/48}\prod_{n=0}^{\infty}(1-q^{n+1/2}) = \sqrt{\frac{\vartheta_4}{\eta}}\,, \quad (7.13b)$$

$$\text{A}\boxed{\phantom{x}}_{\text{P}} = \tfrac{1}{\sqrt{2}}\,q^{-1/48}\,\text{tr}_P\, q^{L_0} = \tfrac{1}{\sqrt{2}}\,q^{1/24}\prod_{n=0}^{\infty}(1+q^n) = \sqrt{\frac{\vartheta_2}{\eta}}\,, \quad (7.13c)$$

$$\text{P}\boxed{\phantom{x}}_{\text{P}} = \tfrac{1}{\sqrt{2}}\,q^{-1/48}\,\text{tr}_P(-1)^F q^{L_0} = \tfrac{1}{\sqrt{2}}\,q^{1/24}\prod_{n=0}^{\infty}(1-q^n) = 0$$

$$\text{``} = \text{''}\sqrt{\frac{\vartheta_1}{i\eta}} \quad (7.13d)$$

(where $\text{tr}_{A,P}$ continues to denote the trace in the anti-periodic and periodic sectors). In (7.13a–d) we have also indicated that these partition functions may be expressed directly in terms of standard Jacobi theta functions $\vartheta_i \equiv \vartheta_i(0,\tau)$ [40] and the Dedekind eta function $\eta(q) = q^{1/24}\prod_{n=1}^{\infty}(1-q^n)$.

It might seem strange that Jacobi and his friends managed to define functions including identically even the correct factor of $q^{-c/24}$ that we derived here physically as a vacuum energy on the torus. Their motivation, as we shall confirm a bit later, is that these functions have nice properties under modular transformations. (The connection between conformal invariance and modular transformations in this context is presumably due to the rescaling of coordinates involved in the transformation $\tau \to -1/\tau$.) With the explicit results (7.13) in hand, we can now reconsider the exact meaning of equations (7.12a,b). By inspection of (7.13) we find first of all under $\tau \to \tau+1$ that

$$\text{A}\boxed{\phantom{x}}_A \to e^{-\frac{i\pi}{24}}\,\text{P}\boxed{\phantom{x}}_A \qquad \text{P}\boxed{\phantom{x}}_A \to e^{-\frac{i\pi}{24}}\,\text{A}\boxed{\phantom{x}}_A$$
$$\text{A}\boxed{\phantom{x}}_P \to e^{\frac{i\pi}{12}}\,\text{A}\boxed{\phantom{x}}_P\,. \quad (7.14a)$$

The derivation of the transformation properties under $\tau \to -1/\tau$ uses the Poisson resummation formula, which we shall introduce at the end of this section. The even simpler (phase-free) result in this case is

$$\text{A}\boxed{\phantom{x}}_A \to \text{A}\boxed{\phantom{x}}_A \qquad \text{A}\boxed{\phantom{x}}_P \to \text{P}\boxed{\phantom{x}}_A \qquad \text{P}\boxed{\phantom{x}}_A \to \text{A}\boxed{\phantom{x}}_P\,. \quad (7.14b)$$

We also defer to the end of this section some other identities satisfied by these objects. For the time being, we point out that the definitions implicit in (7.13a–c) may be used to derive immediately one of the standard $\vartheta$-function identities,

$$\sqrt{\frac{\vartheta_2\vartheta_3\vartheta_4}{\eta^3}} = \sqrt{2}\prod_{n=1}^{\infty}(1-q^{2n-1})(1+q^n)$$
$$= \sqrt{2}\prod_{n=1}^{\infty}\left[\frac{1-q^n}{1-q^{2n}}\right](1+q^n) = \sqrt{2}\,,$$

usually written in the form

$$\vartheta_2\vartheta_3\vartheta_4 = 2\eta^3\,. \quad (7.15)$$

Equations (7.13a–d) can be regarded as basic building blocks for a variety of theories. They also provide a useful heuristic for thinking about Jacobi elliptic functions in terms of free fermions. This representation can be used to give a free fermionic realization of certain integrable models, where away from criticality $q$ acquires significance as a function of Boltzmann weights instead of as the modular parameter on a continuum torus.

Equations (7.13a–d) also have an interpretation as $(\det \overline{\partial})^{1/2}$ for the different fermionic spin structures, and indeed can be calculated from this point of view by employing a suitable regularization prescription such as $\zeta$-function regularization. In the next section we shall calculate the partition function for a single boson from this point of view. The generalization of the genus one results (7.13a–d) to partition functions (equivalently fermion determinants) on higher genus Riemann surfaces, as well as some of the later results to appear here, may be found in [41],[42].

Finally we can use (7.13a–d) to write the $c = \tfrac{1}{2}$ Virasoro characters defined in (7.8) and (7.10) as

$$\chi_0 = \frac{1}{2}\left(\text{A}\boxed{\phantom{x}}_A + \text{P}\boxed{\phantom{x}}_A\right) = \frac{1}{2}\left(\sqrt{\frac{\vartheta_3}{\eta}} + \sqrt{\frac{\vartheta_4}{\eta}}\right)$$

$$\chi_{1/2} = \frac{1}{2}\left(\text{A}\boxed{\phantom{x}}_A - \text{P}\boxed{\phantom{x}}_A\right) = \frac{1}{2}\left(\sqrt{\frac{\vartheta_3}{\eta}} - \sqrt{\frac{\vartheta_4}{\eta}}\right) \quad (7.16a)$$

$$\chi_{1/16} = \frac{1}{\sqrt{2}}\left(\text{A}\boxed{\phantom{x}}_P \pm \text{P}\boxed{\phantom{x}}_P\right) = \frac{1}{\sqrt{2}}\sqrt{\frac{\vartheta_2}{\eta}}\,,$$

or conversely we can write

$$\text{A}\boxed{\phantom{X}}_{\text{A}} = \chi_0 + \chi_{1/2} \qquad \text{A}\boxed{\phantom{X}}_{\text{P}} = \sqrt{2}\,\chi_{1/16}$$
$$\text{P}\boxed{\phantom{X}}_{\text{A}} = \chi_0 - \chi_{1/2} \qquad \text{P}\boxed{\phantom{X}}_{\text{P}} = 0 \, . \tag{7.16b}$$

7.5. *Critical Ising model on the torus*

We now proceed to employ the formalism developed thus far to describe the Ising model on the torus at its critical point. As explained in Cardy's lectures, this is a theory with $c = \overline{c} = \frac{1}{2}$ and a necessarily modular invariant partition function. (The role of modular invariance in statistical mechanical systems on a torus was first emphasized in [43].) Thus we should expect to be able to represent it in terms of a modular invariant combination of spin structures for two fermions $\psi(w), \overline{\psi}(\overline{w})$. It will turn out to be sufficient for (in fact required by) modular invariance to consider only those spin structures for which both fermions have the same boundary conditions on each of the two cycles. The calculation of the partition function for the various spin structures can then be read off directly from the purely holomorphic case. For anti-periodic boundary conditions for both fermions on the two cycles, for example, we use (7.13a) to write

$$\text{A}\overline{\text{A}}\boxed{\phantom{X}}_{\text{A}\overline{\text{A}}} \equiv \text{A}\boxed{\phantom{X}}_{\text{A}} \;\overline{\text{A}}\boxed{\phantom{X}}_{\overline{\text{A}}} = \sqrt{\frac{\vartheta_3}{\eta}}\sqrt{\frac{\overline{\vartheta_3}}{\overline{\eta}}} = \left|\frac{\vartheta_3}{\eta}\right| \, .$$

There is one minor subtlety in the $\text{P}\overline{\text{P}}$ Hamiltonian sector (i.e. with $\text{P}\overline{\text{P}}$ boundary conditions along the "spatial" ($\sigma^1$) direction), since we need to treat the zero mode algebra of $\psi_0$ and $\overline{\psi}_0$. Restricting to a non-chiral theory means that we allow no operator insertions of separate $(-1)^{F_L}$ or $(-1)^{F_R}$'s, i.e. we exclude boundary conditions for example of the form $\text{A}\overline{\text{P}}\boxed{\phantom{X}}$, and allow only $\text{A}\overline{\text{A}}\boxed{\phantom{X}}$ or $\text{P}\overline{\text{P}}\boxed{\phantom{X}}$. Then we need to represent only the non-chiral zero mode algebra $\{(-1)^F, \psi_0\} = \{(-1)^F, \overline{\psi}_0\} = \{\psi_0, \overline{\psi}_0\} = 0$.

According to (6.20), the representation of the non-chiral zero mode algebra requires only a two-dimensional ground state representation $\left|h = \frac{1}{16}, \overline{h} = \frac{1}{16}\right\rangle_{\pm}$, with eigenvalues $\pm 1$ under $(-1)^F$. These two states can be identified with two (non-chiral) primary twist fields $\sigma(w,\overline{w}), \mu(w,\overline{w})$ such that

$$\sigma(0)|0\rangle = \left|\tfrac{1}{16}, \tfrac{1}{16}\right\rangle_+ \qquad \text{and} \qquad \mu(0)|0\rangle = \left|\tfrac{1}{16}, \tfrac{1}{16}\right\rangle_- \, . \tag{7.17}$$

The exact form of the operator product expansions of $\psi$ and $\overline{\psi}$ with these two fields can be determined by considering 4-point correlation functions (as $C_{\sigma\sigma\varepsilon}$ was determined from (5.11)). The $x \to 0$ limit of (5.13) determines that the short distance operator product expansion of $\sigma$ and $\mu$ take the form

$$\sigma(z,\overline{z})\,\mu(w,\overline{w}) = \frac{1}{\sqrt{2}\,|z-w|^{1/4}}\left(e^{-i\pi/4}(z-w)^{1/2}\,\psi(w) \right. \\ \left. + e^{i\pi/4}(\overline{z}-\overline{w})^{1/2}\,\overline{\psi}(\overline{w})\right) . \tag{7.18}$$

Either by taking operator products on both sides with $\mu$ or by using permutation symmetry of operator product coefficients, we determine that the twist operators satisfy the operator product algebra*

$$\psi(z)\,\sigma(w,\overline{w}) = \frac{e^{i\pi/4}}{\sqrt{2}}\frac{\mu(w,\overline{w})}{(z-w)^{1/2}} \qquad \psi(z)\,\mu(w,\overline{w}) = \frac{e^{-i\pi/4}}{\sqrt{2}}\frac{\sigma(w,\overline{w})}{(z-w)^{1/2}}$$
$$\overline{\psi}(\overline{z})\,\sigma(w,\overline{w}) = \frac{e^{-i\pi/4}}{\sqrt{2}}\frac{\mu(w,\overline{w})}{(\overline{z}-\overline{w})^{1/2}} \qquad \overline{\psi}(\overline{z})\,\mu(w,\overline{w}) = \frac{e^{i\pi/4}}{\sqrt{2}}\frac{\sigma(w,\overline{w})}{(\overline{z}-\overline{w})^{1/2}} \, , \tag{7.19}$$

consistent with (6.20) under the identifications (7.17).

The remaining non-vanishing operator products of the Ising model can be used to complete the 'fusion rules' of (5.4) to

$$\begin{aligned}
[\varepsilon][\varepsilon] &= 1 & [\psi][\psi] &= 1 & [\overline{\psi}][\overline{\psi}] &= 1 \\
[\sigma][\sigma] &= 1+[\varepsilon] & [\mu][\mu] &= 1+[\varepsilon] & [\mu][\sigma] &= [\psi]+[\overline{\psi}] \\
[\sigma][\varepsilon] &= [\sigma] & [\mu][\varepsilon] &= [\mu] & [\psi][\overline{\psi}] &= [\varepsilon] \\
[\psi][\sigma] &= [\mu] & [\psi][\mu] &= [\sigma] & [\psi][\varepsilon] &= [\overline{\psi}] \\
[\overline{\psi}][\sigma] &= [\mu] & [\overline{\psi}][\mu] &= [\sigma] & [\overline{\psi}][\varepsilon] &= [\psi]
\end{aligned} \tag{7.20}$$

---

\* (7.18) was derived in [44] from the analog of (5.13) by correcting a sign (a misprint?) in the corresponding result in [1]. (7.19) here corrects the phases and normalizations (more misprints?) in eq. (1.13d) of [44]. I thank P. Di Francesco for guiding me through the typos.

for all the conformal families of the Ising model. We take this opportunity to point out that the analysis of such operator algebras has a long history in the statistical mechanical literature (see for example [15],[45]). As we noted near the end of section 3, the minimal models of [1] gave a class of examples that closed on only a finite number of fields. In [43], it was shown that modular invariance on the torus for models with $c \geq 1$ requires an infinite number of Virasoro primary fields. Thus the $c < 1$ discrete series described in section 4 exhausts all (unitary) cases for which the operator algebra can close with only a finite number of Virasoro primaries. Rational conformal field theories, whose operator algebras close on a finite number of fields primary under a larger algebra, however, can exist and be modular invariant at arbitrarily large values of $c$.

Given the two vacuum states (7.17), the analog of (7.13c) for the non-chiral case is thus

$$\text{A}\overline{\text{A}}\underset{\text{P}\overline{\text{P}}}{\square} = (q\overline{q})^{-1/48}\text{tr}\, q^{L_0}\overline{q}^{\overline{L}_0} = 2(q\overline{q})^{1/24} \prod_{n=0}^{\infty}(1+q^n)(1+\overline{q}^n)$$

$$= \sqrt{\frac{\vartheta_2}{\eta}}\sqrt{\frac{\overline{\vartheta_2}}{\overline{\eta}}} = \left|\frac{\vartheta_2}{\eta}\right|\,.$$

We see that the factor of $\frac{1}{\sqrt{2}}$ included in the definition (7.13c) together with the factor of $\frac{1}{2}$ reduction in ground state dimension for the non-chiral $(-1)^F$ zero mode algebra results in the simple relation $\text{A}\overline{\text{A}}\underset{\text{P}\overline{\text{P}}}{\square} = \text{A}\underset{\text{P}}{\square}\,\overline{\text{A}}\underset{\overline{\text{P}}}{\square}$.

From (7.14), we easily verify that the two combinations of spin structures,

$$\left(\text{A}\overline{\text{A}}\underset{\text{A}\overline{\text{A}}}{\square} + \text{P}\overline{\text{P}}\underset{\text{A}\overline{\text{A}}}{\square} + \text{A}\overline{\text{A}}\underset{\text{P}\overline{\text{P}}}{\square}\right) \quad \text{and} \quad \text{P}\overline{\text{P}}\underset{\text{P}\overline{\text{P}}}{\square}\,, \qquad (7.21)$$

for fermions $\psi(w)$, $\overline{\psi}(\overline{w})$ on the torus are modular invariant. Although it would seem perfectly consistent to retain only one of these two modular orbits to construct a theory, we shall see that both are actually required for a consistent operator interpretation. (In the path integral formulation of string theory this constraint, expressed from the point of view of factorization and modular invariance of amplitudes on a genus two Riemann surface, was emphasized in [46].)

As a contribution to the partition function, $\text{P}\overline{\text{P}}\underset{\text{P}\overline{\text{P}}}{\square}$ of course vanishes due to the fermion zero mode, but this spin structure does contribute to higher point functions. Hence we shall carry it along in what follows as a formal reminder of its non-trivial presence in the theory.

We thus take as our partition function

$$Z_{\text{Ising}} = \frac{1}{2}\left(\text{A}\overline{\text{A}}\underset{\text{A}\overline{\text{A}}}{\square} + \text{P}\overline{\text{P}}\underset{\text{A}\overline{\text{A}}}{\square} + \text{A}\overline{\text{A}}\underset{\text{P}\overline{\text{P}}}{\square} \pm \text{P}\overline{\text{P}}\underset{\text{P}\overline{\text{P}}}{\square}\right)$$

$$= (q\overline{q})^{-1/48}\text{tr}_{\text{A}\overline{\text{A}}}\frac{1}{2}\left(1+(-1)^F\right)q^{L_0}\overline{q}^{\overline{L}_0}$$

$$+ (q\overline{q})^{-1/48}\text{tr}_{\text{P}\overline{\text{P}}}\frac{1}{2}\left(1\pm(-1)^F\right)q^{L_0}\overline{q}^{\overline{L}_0} \qquad (7.22)$$

$$= \frac{1}{2}\left(\left|\frac{\vartheta_3}{\eta}\right| + \left|\frac{\vartheta_4}{\eta}\right| + \left|\frac{\vartheta_2}{\eta}\right| \pm \left|\frac{\vartheta_1}{\eta}\right|\right)$$

$$= \chi_0\overline{\chi}_0 + \chi_{1/2}\overline{\chi}_{1/2} + \chi_{1/16}\overline{\chi}_{1/16}\,.$$

The overall factor of $\frac{1}{2}$ is dictated by the operator interpretation of the sum over spin structures as a projection, as expressed in the second line of (7.22), and insures a unique ground state in each of the $\text{A}\overline{\text{A}}$ and $\text{P}\overline{\text{P}}$ sectors. We notice that the partition function (7.22) neatly arranges itself into a diagonal sum over three left-right symmetric Virasoro characters, corresponding to the three conformal families that comprise the theory.

The projection dictated by modular invariance of (7.21) is onto $(-1)^F = 1$ states in the $\text{A}\overline{\text{A}}$ sector, i.e. onto the states

$$\psi_{-n_1}\ldots\psi_{-n_\ell}\overline{\psi}_{-n_{\ell+1}}\ldots\overline{\psi}_{-n_{2k}}|0\rangle\,. \qquad (7.23)$$

In the $\text{P}\overline{\text{P}}$ sector the sign for the projection is not determined by modular invariance, and the two choices of signs, although giving the same partition function, lead to retention of two orthogonal sets of states, as discussed after (7.10). That these two choices lead to equivalent theories is simply the $\sigma \leftrightarrow \mu$ (order/disorder) duality of the critical Ising model.

In string theory projections onto states in each Hamiltonian sector with a given value of $(-1)^F$ go under the generic name of GSO projection[47]. Such a projection was imposed to insure spacetime supersymmetry, among other things, in superstring theory, and was later recognized as a general consequence of modular invariance of the theory on a genus one surface. In the spacetime context, the sign ambiguity in the P sector is simply related to the arbitrariness in conventions for positive and negative chirality spinors. A general discussion in the same notation employed here may be found in [48].

The partition function (7.22) corresponds to boundary conditions on the Ising spins $\sigma = \pm 1$ periodic along both cycles of the torus, i.e. to

$$Z_{PP} = P\boxed{\phantom{x}}_P$$

boundary conditions, where we use italic $A, P$ to denote boundary conditions for Ising spins (as opposed to the fermions $\psi, \overline{\psi}$). Depending on the choice of $(-1)^F$ projection, the operators that survive in the spectrum are either $\{1, \sigma, \varepsilon\}$ or $\{1, \mu, \varepsilon\}$, in each case providing a closed operator subalgebra of (7.20).

We can also consider the non-modular invariant case of Ising spins twisted along the "time" direction, which we denote

$$Z_{PA} = A\boxed{\phantom{x}}_P .$$

This case, as discussed in Cardy's lectures (section 5.2), corresponds to calculating the trace of an operator that takes the Ising spins $\sigma \to -\sigma$, but leaves the identity 1 and energy $\varepsilon$ invariant. In free fermion language, this is equivalent to an operation that leaves the $A\overline{A}$ sector invariant (the $(0,0)$ and $(\frac{1}{2},\frac{1}{2})$ families), and takes the $P\overline{P}$ sector (the $(\frac{1}{16},\frac{1}{16})$ family) to minus itself. The resulting partition function is thus

$$Z_{PA} = \frac{1}{2}\left(A\overline{A}\boxed{\phantom{x}}_{A\overline{A}} + P\overline{P}\boxed{\phantom{x}}_{A\overline{A}}\right) - \frac{1}{2}\left(A\overline{A}\boxed{\phantom{x}}_{P\overline{P}} \pm P\overline{P}\boxed{\phantom{x}}_{P\overline{P}}\right) \quad (7.24)$$
$$= |\chi_0|^2 + |\chi_{1/2}|^2 - |\chi_{1/16}|^2 .$$

The modular transformation $\tau \to -1/\tau$ then allows us to calculate the partition function for the boundary conditions

$$Z_{AP} = P\boxed{\phantom{x}}_A ,$$

with Ising spins now twisted in the "space" direction. Applying (7.14b) to (7.24), then using (7.16, ) we find

$$Z_{AP} = \frac{1}{2}\left(A\overline{A}\boxed{\phantom{x}}_{A\overline{A}} - P\overline{P}\boxed{\phantom{x}}_{A\overline{A}}\right) + \frac{1}{2}\left(A\overline{A}\boxed{\phantom{x}}_{P\overline{P}} \mp P\overline{P}\boxed{\phantom{x}}_{P\overline{P}}\right) \quad (7.25)$$
$$= \chi_0\overline{\chi}_{1/2} + \chi_{1/2}\overline{\chi}_0 + |\chi_{1/16}|^2 .$$

We see that the negative sign between the first two terms in (7.25) changes the choice of projection in the $A\overline{A}$ sector. Now we keep states with odd rather than even fermion number as in (7.23), i.e. states with $h - \overline{h} \in \mathbf{Z} + \frac{1}{2}$ rather than with $h - \overline{h} \in \mathbf{Z}$. This change is easily seen reflected in the off-diagonal combinations of 0 and $\frac{1}{2}$ characters in (7.25). Changing boundary conditions on the Ising spins thus allows us to focus on the operator content ($\psi$, $\overline{\psi}$, and $\mu$) of the theory that would not ordinarily survive the projection. Playing with boundary conditions is also a common practice in numerical simulations, so results such as these allow a more direct contact between theory and "experiment" in principle. Further analysis of partition functions with a variety of boundary conditions in $c < 1$ models, showing how the internal symmetries are tied in with their conformal properties, may be found in [49].

While neither $Z_{PA}$ nor $Z_{AP}$ is modular invariant, we note that the combination $Z_{PA} + Z_{AP} = A\overline{A}\boxed{\phantom{x}}_{A\overline{A}} = |\chi_0 + \chi_{1/2}|^2$ is invariant under a subgroup of the modular group, namely that generated by $\tau \to -1/\tau$ and $\tau \to \tau + 2$. The operator content surviving the projection for this combination is $\{1, \psi, \overline{\psi}, \epsilon\}$, again forming a closed operator subalgebra of (7.20).

Finally, from (7.25) the modular transformation $\tau \to \tau + 1$ can be used to determine the result for boundary conditions

$$Z_{AA} = A\boxed{\phantom{x}}_A ,$$

for anti-periodic Ising spins along both cycles of the torus. But from (7.14a) we see that this just exchanges the first two terms in (7.25),

$$Z_{AA} = -\frac{1}{2}\left(A\overline{A}\boxed{\phantom{x}}_{A\overline{A}} - P\overline{P}\boxed{\phantom{x}}_{A\overline{A}}\right) + \frac{1}{2}\left(A\overline{A}\boxed{\phantom{x}}_{P\overline{P}} \mp P\overline{P}\boxed{\phantom{x}}_{P\overline{P}}\right) \quad (7.26)$$
$$= -\chi_0\overline{\chi}_{1/2} - \chi_{1/2}\overline{\chi}_0 + |\chi_{1/16}|^2 ,$$

giving the $\mathbf{Z}_2$ transformation properties of the operators $\psi$, $\overline{\psi}$, and $\mu$ in the $A$ sector of the theory.

### 7.6. Recreational mathematics and $\vartheta$-function identities

In this subsection we detail some of the properties of Jacobi elliptic functions that will later prove useful. To illustrate the ideas involved, we begin with a proof of the Jacobi triple product identity,

$$\prod_{n=1}^{\infty}(1-q^n)(1+q^{n-1/2}w)(1+q^{n-1/2}w^{-1}) = \sum_{n=-\infty}^{\infty} q^{\frac{1}{2}n^2} w^n , \qquad (7.27)$$

for $|q| < 1$ and $w \neq 0$. (For $|q| < 1$ the expressions above are all absolutely convergent so naive manipulations of sums and products are all valid.)

Rather than the standard combinatorial derivation* of (7.27), we shall try to provide a more "physical" treatment. To this end, we consider the partition function for a free electron-positron system with linearly spaced energy levels $E = \varepsilon_0(n - \frac{1}{2})$, $n \in \mathbf{Z}$, and total fermion number $N = N_e - N_{\overline{e}}$. If we rewrite the energy $E$ and fugacity $\mu$ respectively in terms of $q = e^{-\varepsilon_0/T}$ and $w = e^{\mu/T}$, then the grand canonical partition function takes the form

$$Z(w,q) = \sum_{\substack{\text{fermion} \\ \text{occupations}}} e^{-E/T + \mu N/T} = \sum_{N=-\infty}^{\infty} w^N Z_N(q) \qquad (7.28)$$
$$= \prod_{n=1}^{\infty}(1+q^{n-1/2}w)(1+q^{n-1/2}w^{-1}) ,$$

where $Z_N(q)$ is the canonical partition function for fixed total fermion number $N$. The lowest energy state contributing to $Z_0$ has all negative energy levels filled (and by definition of the Fermi sea has energy $E = 0$). States excited to energy $E = M\varepsilon_0$ are described by a set of integers $k_1 \geq k_2 \geq \cdots \geq k_\ell > 0$, with $\sum_{i=1}^{\ell} k_i = M$ (these numbers specify the excitations of the uppermost $\ell$ particles in the Fermi sea, starting from the top). The total number of such states is just the number of partitions $P(M)$, so that

$$Z_0 = \sum_{M=0}^{\infty} P(M) q^M = \frac{1}{\prod_{n=1}^{\infty}(1-q^n)} .$$

---

\* following from the recursion relation $P(qw, q) = \frac{1+q^{-1/2}w^{-1}}{1+q^{1/2}w}P(w,q) = \frac{1}{q^{1/2}w}P(w,q)$, satisfied by the left hand side $P(w,q)$ of (7.27) (see e.g. [50]).

The lowest energy state in the sector with fermion number $N$, on the other hand, has the first $N$ positive levels occupied, contributing a factor

$$q^{1/2} \cdots q^{N-3/2} q^{N-1/2} = q^{\sum_{n=1}^{N}(j-\frac{1}{2})} = q^{N^2/2} .$$

Excitations from this state are described exactly as for $Z_0$, so that $Z_N = q^{N^2/2} Z_0$. Combining results gives

$$Z(w,q) = \sum_{N=-\infty}^{\infty} w^N Z_N(q) = \sum_{N=-\infty}^{\infty} w^N \frac{q^{N^2/2}}{\prod_{n=1}^{\infty}(1-q^n)} ,$$

thus establishing (7.27).

The basic result (7.27) can be used to derive a number of subsidiary identities. If we substitute $w = \pm 1, \pm q^{-1/2}$, (7.27) allows us to express the $\vartheta$-functions in (7.13a–d) as the infinite summations

$$\vartheta_3 = \sum_{n=-\infty}^{\infty} q^{n^2/2} \qquad \vartheta_2 = \sum_{n=-\infty}^{\infty} q^{\frac{1}{2}(n-\frac{1}{2})^2}$$
$$\vartheta_4 = \sum_{n=-\infty}^{\infty} (-1)^n q^{n^2/2} \qquad \vartheta_1 = i \sum_{n=-\infty}^{\infty} (-1)^n q^{\frac{1}{2}(n-\frac{1}{2})^2} (= 0) . \qquad (7.29)$$

We can also express the Dedekind $\eta$ function as an infinite sum. We substitute $q \to q^3$, $w \to -q^{-1/2}$ in (7.27) to find

$$\prod_{n=1}^{\infty}(1-q^{3n})(1-q^{3n-2})(1-q^{3n-1}) = \sum_{n=-\infty}^{\infty} q^{3n^2/2} (-1)^n q^{-n/2} ,$$

or equivalently

$$\prod_{n=1}^{\infty}(1-q^n) = \sum_{n=-\infty}^{\infty} (-1)^n q^{\frac{1}{2}(3n^2-n)} . \qquad (7.30)$$

Multiplying by $q^{1/24}$ then gives

$$\eta(q) = q^{1/24} \prod_{n=1}^{\infty}(1-q^n) = \sum_{n=-\infty}^{\infty}(-1)^n q^{\frac{3}{2}(n-1/6)^2} \ . \qquad (7.31)$$

The identity (7.30) is known as the Euler pentagonal number theorem. Someone invariably asks why. Those readers* with a serious interest in recreational mathematics will recall that there exists a series of $k$-gonal numbers given by
$$\frac{(k-2)n^2 - (k-4)n}{2} \ .$$
They describe the number of points it takes to build up successive embedded $k$-sided equilateral figures (see fig. 11 for the cases of triagonal ($k=3$) numbers, $(n^2+n)/2 = 1, 3, 6, \ldots$; square ($k=4$) numbers, $n^2 = 1, 4, 9, \ldots$; and pentagonal ($k=5$) numbers, $\frac{1}{2}(3n^2 - n) = 1, 5, 12, \ldots$). Generating functions for some of the other $k$-gonal numbers may be found in [50].

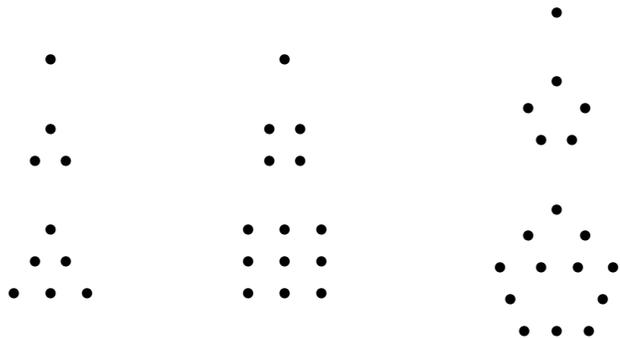

Fig. 11. First three triagonal, square, and pentagonal numbers.

(One of Euler's original interests in (7.30) was evidently its combinatorial interpretation. The left hand side is the generating function for $E(n) - U(n)$, where $E(n)$ is the number of partitions of $n$ into an even number of unequal parts, and $U(n)$ that into an odd number. Thus (7.30) states that $E(n) = U(n)$ except when $n = \frac{1}{2}(3k^2 \pm k)$, in which case $E(n) - U(n) = (-1)^k$.)

---
\* I am grateful to M. Peskin for initiation in these matters.

To treat modular transformation properties of the $\vartheta$'s and $\eta$ under $\tau \to -1/\tau$, we introduce the Poisson resummation formula in the form

$$\sum_{n=-\infty}^{\infty} f(nr) = \frac{1}{r}\sum_{m=-\infty}^{\infty} \widetilde{f}\!\left(\frac{m}{r}\right) \ , \qquad (7.32)$$

where the Fourier transform $\widetilde{f}$ is defined as

$$\widetilde{f}(y) = \int_{-\infty}^{\infty} dx \ e^{-2\pi i x y} f(x) \ .$$

(7.32) is easily established by substituting $\widetilde{f}$ on the right hand side. (The natural generalization of (7.32) to higher dimensions is

$$\sum_{v \in \Gamma} f(v) = \frac{1}{V}\sum_{w \in \Gamma^*} \widetilde{f}(w) \ ,$$

where $\Gamma$ is a lattice, $\Gamma^*$ its dual (reciprocal), and $V$ the volume of its unit cell.)

Using the sum form (7.31) of the $\eta$ function, we may apply (7.32) to find

$$\eta\bigl(q(-1/\tau)\bigr) = (-i\tau)^{1/2} \eta\bigl(q(\tau)\bigr) \ . \qquad (7.33)$$

Similarly, from (7.29) we find that under $\tau \to -1/\tau$,

$$\vartheta_2 \to (-i\tau)^{1/2}\vartheta_4 \qquad \vartheta_4 \to (-i\tau)^{1/2}\vartheta_2 \qquad \vartheta_3 \to (-i\tau)^{1/2}\vartheta_3 \ . \qquad (7.34)$$

We see that (7.12b) follows from (7.33) and (7.34). For completeness, we tabulate here as well the transformation properties under $\tau \to \tau + 1$,

$$\vartheta_3 \leftrightarrow \vartheta_4 \qquad \vartheta_2 \to \sqrt{i}\,\vartheta_2 \qquad \eta \to e^{\frac{i\pi}{12}}\eta \ , \qquad (7.35)$$

as already used in (7.14a).

We also note that the right hand side of (7.27) with $w = e^{2\pi i z}$ defines the function $\vartheta_3(z,\tau)$, in terms of which generalizations of all the $\vartheta_i \equiv \vartheta_i(0,\tau)$'s are

written

$$\vartheta_3(z,\tau) = \sum_{n=-\infty}^{\infty} q^{n^2/2} e^{2\pi i n z}$$

$$\vartheta_4(z,\tau) = \vartheta_3(z+\tfrac{1}{2},\tau) = \sum_{n=-\infty}^{\infty} (-1)^n q^{n^2/2} e^{2\pi i n z}$$

$$\vartheta_1(z,\tau) = -i e^{iz} q^{1/8} \vartheta_4(z+\tfrac{\tau}{2},\tau) = i \sum_{n=-\infty}^{\infty} (-1)^n q^{\frac{1}{2}(n-\frac{1}{2})^2} e^{i\pi(2n-1)z} \quad (7.36)$$

$$\vartheta_2(z,\tau) = \vartheta_1(z+\tfrac{1}{2},\tau) = \sum_{n=-\infty}^{\infty} q^{\frac{1}{2}(n-\frac{1}{2})^2} e^{i\pi(2n-1)z} \;.$$

The parameter $z$ is useful for expressing the functional integral for complex fermions with boundary conditions twisted by an arbitrary phase, as mentioned at the end of subsection 7.1. For representations of affine algebras in terms of free fermions, $z$ also plays the role of the Cartan angle in the affine characters. In string theory where spacetime gauge symmetries are realized as affine algebras on the worldsheet, the $z$ dependence would then provide the dependence of the partition function on background gauge fields. Properties of spacetime gauge and gravitational anomalies may then be probed via the modular transformation properties of the functions (7.36) (see [48] for more details). The $z$ dependence of the $\vartheta$-functions also provides the coordinate dependence of correlation functions on the torus (for the critical Ising model for example, see [44]).

Some other popular modular invariants are also readily constructed in terms of free fermions. For eight chiral fermions, $\psi^{\mu=1,8}(z)$, all with the same spin structure, we find

$$\frac{1}{2}\left(A^8 \underset{A^8}{\Box} - P^8 \underset{A^8}{\Box} - A^8 \underset{P^8}{\Box}\right) = \frac{1}{2}\frac{1}{\eta^4}\left(\vartheta_3^4 - \vartheta_4^4 - \vartheta_2^4\right) = 0 \;, \quad (7.37)$$

where the signs are determined by invariance under (7.34) and (7.35). A straightforward way to understand the vanishing of this quantity is to recognize that $\frac{1}{2}(\vartheta_3^4 - \vartheta_4^4) = \sum_{v\in\Gamma} q^{\frac{1}{2}v^2}$, where $\Gamma$ is the lattice composed of 4-vectors whose components $v^i \in \mathbf{Z}$ satisfy $\sum_{i=1}^4 v^i = 1 \bmod 2$. We also recognize

105

$\frac{1}{2}\vartheta_2^4 = \frac{1}{2}(\vartheta_2^4 + \vartheta_1^4) = \sum_{v\in\Gamma'} q^{\frac{1}{2}v^2}$ where $\Gamma'$ is composed of vectors with $v^i \in \mathbf{Z}+\frac{1}{2}$ and $\sum_{i=1}^4 v^i = 0 \bmod 2$. But these two lattices are related by $\Gamma' = M\Gamma$, where $M$ is the $SO(4)$ transformation

$$M = \begin{pmatrix} \frac{1}{2} & \frac{1}{2} & \frac{1}{2} & \frac{1}{2} \\ \frac{1}{2} & \frac{1}{2} & -\frac{1}{2} & -\frac{1}{2} \\ \frac{1}{2} & -\frac{1}{2} & -\frac{1}{2} & \frac{1}{2} \\ \frac{1}{2} & -\frac{1}{2} & \frac{1}{2} & -\frac{1}{2} \end{pmatrix},$$

so it follows that $\sum_{v\in\Gamma} q^{\frac{1}{2}v^2} = \sum_{v\in\Gamma'} q^{\frac{1}{2}v^2}$. (Acting on the weight lattice of $SO(8)$, the transformation $M$ above is the triality rotation that exchanges the vector with one of the two spinors.)

In superstring theory, the vanishing of (7.37) is the expression of spacetime supersymmetry at one-loop order. The first two terms represent the contribution to the spectrum of (GSO projected) spacetime bosons, and the last term the spacetime fermions. Another way to see that (7.37) has to vanish is to recall [51] that a basis for modular forms of weight $2k$ is given by $G_2^\alpha G_3^\beta$ ($\alpha, \beta \in \mathbf{Z}^+$, $2\alpha + 3\beta = k$), where the $G_k(\tau) = \sum_{\{m,n\}\neq\{0,0\}} (m\tau+n)^{-2k}$ are the Eisenstein series of weights 4 and 6 for $k = 2, 3$ respectively. (A modular form of weight $2k$ satisfies $f\left(\frac{a\tau+b}{c\tau+d}\right) = (c\tau+d)^{2k} f(\tau)$, so that $f(\tau)(d\tau)^k$ is invariant.) From the modular transformation properties (7.34) and (7.35), we see that $\vartheta_3^4 - \vartheta_4^4 - \vartheta_2^4$ is a modular form of weight 2, of which there are none non-trivial, and hence must vanish.

For 16 chiral fermions, $\psi^{\mu=1,16}(z)$, we find

$$\frac{1}{2}\left(A^{16}\underset{A^{16}}{\Box} + P^{16}\underset{A^{16}}{\Box} + A^{16}\underset{P^{16}}{\Box}\right) = \frac{1}{2}\frac{1}{\eta^8}\left(\vartheta_3^8 + \vartheta_4^8 + \vartheta_2^8\right) = \frac{\sum_{v\in\Gamma_8} q^{v^2/2}}{\eta^8} \;,$$

where the summation is over lattice vectors $v$ in $\Gamma_8$, the $E_8$ root lattice. This is a lattice composed of vectors whose components $v^i$ are either all integral, $v^i \in \mathbf{Z}$, or half-integral, $v^i \in \mathbf{Z} + \frac{1}{2}$, and in either case their sum is even, $\sum_{i=1}^8 v^i = 0 \bmod 2$ (the last a consequence of the GSO projection on even fermion number in the A and P sectors).

Actually, since 16 chiral fermions have $c = 8, \bar{c} = 0$, the above combination of spin structures has a leading $q$ behavior of $q^{-c/24} \sim q^{-1/3}$ so it is strictly

106

speaking only modular covariant. (In this case that means that it picks up a cube root of unity phase under $\tau \to \tau + 1$; since $S^2 = 1$, the only possible non-trivial phase for $S$ would be $-1$, but this is excluded here by the other relation $(ST)^3 = 1$.) To get a modular *invariant*, we cube the $E_8$ character to find

$$\frac{1}{2^3}\left(\mathrm{A}^{16}\underset{\mathrm{A}^{16}}{\Box} + \mathrm{P}^{16}\underset{\mathrm{A}^{16}}{\Box} + \mathrm{A}^{16}\underset{\mathrm{P}^{16}}{\Box}\right)^3 = \frac{1}{8}\frac{1}{\eta^{24}}(\vartheta_3^8 + \vartheta_4^8 + \vartheta_2^8)^3$$

$$= j(q) = \frac{1}{q} + 744 + 196884q + \ldots \; ,$$

where $j$ is the famous modular invariant function (the coefficients in whose $q$-expansion, excepting the constant term 744, are simply expressed in terms of the dimensions of the irreducible representations of the monster group (see [52] for a recent treatment with physicists in mind and for further references)).

We can also generalize this construction to $16k$ chiral fermions, $\psi^{\mu=1,16k}(z)$, to get

$$\frac{1}{2}\left(\mathrm{A}^{16k}\underset{\mathrm{A}^{16k}}{\Box} + \mathrm{P}^{16k}\underset{\mathrm{A}^{16k}}{\Box} + \mathrm{A}^{16k}\underset{\mathrm{P}^{16k}}{\Box}\right)$$

$$= \frac{1}{2}\frac{1}{\eta^{8k}}(\vartheta_3^{8k} + \vartheta_4^{8k} + \vartheta_2^{8k}) = \frac{\sum_{v \in \Gamma_{8k}} q^{v^2/2}}{\eta^{8k}} \; ,$$

where the lattice $\Gamma_{8k}$ is defined analogously to $\Gamma_8$, i.e. again a lattice composed of vectors whose components $v^i$ are either all integral, $v^i \in \mathbf{Z}$, or half-integral, $v^i \in \mathbf{Z} + \frac{1}{2}$, such that in either case their sum is even, $\sum_{i=1}^{8k} v^i = 0 \mod 2$. The $\Gamma_{8k}$ are examples of even self-dual integer lattices. (An integer lattice $\Gamma$ is such that vectors $v \in \Gamma$ have $v^2 \in \mathbf{Z}$. The dual lattice $\Gamma^*$ consists of all vectors $w$ such that $w \cdot v \in \mathbf{Z}$, and a self-dual lattice satisfies $\Gamma = \Gamma^*$. See [51] for more details.) Modular covariant fermionic partition functions of the form considered here generically bosonize to theories of chiral bosons compactified on such lattices.

## 8. Free bosons on a torus

We now continue our study of conformal field theory on the torus to the next simplest case, that of free bosons. This case affords a surprising richness of structure that begins to hint at the complexity of more general conformal field theories.

### 8.1. Partition function

In the previous section, we calculated the partition functions (7.13) for free fermions with assorted boundary conditions on a torus by means of the Hamiltonian interpretation in which the sum over Hilbert space states is implemented with appropriate operator insertions. A similar procedure could be employed to calculate free bosonic partition functions. To illustrate the alternative interpretation of partition functions as determinants of operators, however, we shall instead calculate the bosonic partition functions by means of a Lagrangian formulation in this section.

Since we are dealing with a free field theory with action

$$S = \frac{1}{2\pi}\int \partial X \overline{\partial} X \; , \tag{8.1}$$

we can calculate functional integral exactly simply by taking proper account of the boundary conditions. We assume a bosonic coordinate $X \equiv X + 2\pi r$ compactified on a circle of radius $r$. That means when we calculate the functional integral, we need to consider all "instanton" sectors $n'\underset{n}{\Box}$ with boundary conditions

$$X_0(z+\tau, \overline{z}+\overline{\tau}) = X_0(z,\overline{z}) + 2\pi r n' \qquad X_0(z+1, \overline{z}+1) = X_0(z,\overline{z}) + 2\pi r n \; .$$

The solutions to the classical equations of motion, $\partial\overline{\partial}X_0 = 0$, with the above boundary conditions, are

$$X_0^{(n',n)}(z,\overline{z}) = 2\pi r \frac{1}{2i\tau_2}\bigl(n'(z-\overline{z}) + n(\tau\overline{z} - \overline{\tau}z)\bigr) \; . \tag{8.2}$$

In each such sector, we also have a contribution from the fluctuations around the classical solution.

The functional integral is easily evaluated using the normalization conventions of [53].* (In general, functional integrals are defined only up to an infinite constant so only their ratios are well-defined, and any ambiguities are resolved via recourse to canonical quantization. The prescription here is chosen to give a $\tau_2$ dependence consistent with modular invariance, and an overall normalization consistent with the Hamiltonian interpretation. A related calculation may be found in [54].) To carry out the $\mathcal{D}X$ integration, we separate the constant piece by writing $X(z,\overline{z}) = \widetilde{X} + X'(z,\overline{z})$, where $X'(z,\overline{z})$ is orthogonal to the constant $\widetilde{X}$, and write $\mathcal{D}X = d\widetilde{X}\,\mathcal{D}X'$. We normalize the gaussian functional integral to $\int \mathcal{D}\delta X\, e^{-\frac{1}{2\pi}\int(\delta X)^2} = 1$, so that

$$\int \mathcal{D}\delta X'\, e^{-\frac{1}{2\pi}\int(\delta X)^2} = \left(\int dx\, e^{-\frac{1}{2\pi}\int x^2}\right)^{-1}$$

$$= \left(\frac{\pi}{\frac{1}{2\pi}\int 1}\right)^{-1/2} = \frac{\sqrt{2\tau_2}}{\pi} \;.$$

In (8.1), we have taken the measure to be $2i\,dz\wedge d\overline{z}$ ($=4\tau_2\,d\sigma^1\wedge d\sigma^0$ in coordinates $z = \sigma^1 + \tau\sigma^0$), so the integral on the torus is normalized to $\int 1 = 4\tau_2$. The integral over the constant piece $\widetilde{X}$, on the other hand, just gives $2\pi r$.

Now from (8.2), we have that $\partial X_0^{(n',n)} = \frac{2\pi r}{2i\tau_2}(n' - \overline{\tau}n)$. Substituting into the action (8.1), together with the above normalization conventions, allows us to express the functional integral in the form

$$\int e^{-S} = 2\pi r\frac{\sqrt{2\tau_2}}{\pi}\frac{1}{\det'^{1/2}\Box}\sum_{n,n'=-\infty}^{\infty} e^{-S[X_0^{(n',n)}]}$$

$$= 2r\sqrt{2\tau_2}\,\frac{1}{\det'^{1/2}\Box}\sum_{n,n'=-\infty}^{\infty} e^{4\tau_2\frac{1}{2\pi}\left(\frac{2\pi r}{2i\tau_2}\right)^2 (n'-\overline{\tau}n)(n'-\tau n)} \quad (8.3)$$

$$= 2r\sqrt{2\tau_2}\,\frac{1}{\det'^{1/2}\Box}\sum_{n,n'=-\infty}^{\infty} e^{-2\pi\left(\frac{1}{\tau_2}(n'r - \tau_1 nr)^2 + \tau_2 n^2 r^2\right)} \;,$$

where $\Box \equiv -\partial\overline{\partial}$, and $\det'$ is a regularized determinant.

---

* I am grateful to A. Cohen for his notes on the subject.

To evaluate $\det'\Box$ as a formal product of eigenvalues, we work with a basis of eigenfunctions

$$\psi_{nm} = e^{2\pi i\frac{1}{2i\tau_2}\left(n(z-\overline{z}) + m(\tau\overline{z} - \overline{\tau}z)\right)} \;,$$

single-valued under both $z \to z+1$, $z \to z+\tau$. The regularized determinant is defined by omitting the eigenfunction with $n = m = 0$,

$$\det'\Box \equiv \prod_{\{m,n\}\neq\{0,0\}} \frac{\pi^2}{\tau_2^2}(n - \tau m)(n - \overline{\tau}m) \;. \quad (8.4)$$

The infinite product may be evaluated using $\zeta$-function regularization (recall that $\zeta(s) = \sum_{n=1}^{\infty} n^{-s}$, $\zeta(-1) = -\frac{1}{12}$, $\zeta(0) = -\frac{1}{2}$, $\zeta'(0) = -\frac{1}{2}\ln 2\pi$). In this regularization scheme we have for example

$$\prod_{n=1}^{\infty} a = a^{\zeta(0)} = a^{-1/2} \quad \text{and} \quad \prod_{n=-\infty}^{\infty} a = a^{2\zeta(0)+1} = 1 \;,$$

so that in particular for the product in (8.4), with $m = n = 0$ excluded, we find $\prod'(\pi^2/\tau_2^2) = \tau_2^2/\pi^2$. Another identity in this scheme that we shall need is $\prod_{n=1}^{\infty} n^{\alpha} = e^{-\alpha\zeta'(0)} = (2\pi)^{\alpha/2}$.

The remainder of (8.4) is evaluated by means of the product formula $\prod_{n=-\infty}^{\infty}(n+a) = a\prod_{n=1}^{\infty}(-n^2)(1 - a^2/n^2) = 2i\sin\pi a$. The result is

$$\det'\Box = \prod_{\{m,n\}\neq\{0,0\}} \frac{\pi^2}{\tau_2^2}(n - m\tau)(n - m\overline{\tau})$$

$$= \frac{\tau_2^2}{\pi^2}\left(\prod_{n\neq 0} n^2\right)\prod_{m\neq 0,\,n\in\mathbf{Z}}(n - m\tau)(n - m\overline{\tau})$$

$$= \frac{\tau_2^2}{\pi^2}(2\pi)^2\prod_{m>0,\,n\in\mathbf{Z}}(n - m\tau)(n + m\tau)(n - m\overline{\tau})(n + m\overline{\tau})$$

$$= 4\tau_2^2\prod_{m>0}(e^{-\pi im\tau} - e^{\pi im\tau})^2(e^{-\pi im\overline{\tau}} - e^{\pi im\overline{\tau}})^2$$

$$= 4\tau_2^2\prod_{m>0}(q\overline{q})^{-m}(1 - q^m)^2(1 - \overline{q}^m)^2$$

$$= 4\tau_2^2(q\overline{q})^{1/12}\prod_{m>0}(1 - q^m)^2(1 - \overline{q}^m)^2 = 4\tau_2^2\,\eta^2\overline{\eta}^2 \;,$$

so the relevant contribution to (8.3) is

$$2r\sqrt{2\tau_2} \frac{1}{\det'^{1/2}\Box} = \sqrt{\frac{2}{\tau_2}} r \frac{1}{\eta\overline{\eta}} \ . \tag{8.5}$$

Since under the modular transformation $\tau \to -1/\tau$, we have $\tau_2 \to \tau_2/|\tau|^2$, we verify modular invariance of (8.5) from the modular transformation property (7.33) of $\eta$. Techniques identical to those used to derive (8.5) could also have been used to derive the fermion determinants (7.13). ((8.5) can also be compared with the result of section 4.2 of Cardy's lectures. For a general action $\frac{g}{4\pi}\int \partial\phi\overline{\partial}\phi$, with $\phi \equiv \phi + 2\pi R$, the "physical" quantity $r = \sqrt{\frac{g}{2}} R$ is independent of rescaling of $\phi$, and coincides with the usual radius for $g = 2$, as desired from the normalization of (2.14). We see that the right hand side of (8.5) takes the form $g^{1/2}R/(\tau_2^{1/2}\eta\overline{\eta})$, and for $R = 1$ agrees with Cardy's eq. (4.10)).

We have now to consider the effect of summing over the instanton sectors, or equivalently the interpretation of the momentum zero modes $p_L \equiv \alpha_0$, $p_R \equiv \overline{\alpha}_0$. As usual in making the comparison between Lagrangian and Hamiltonian formulations, the summation over the winding $n'$ in the "time" direction in (8.3) can be exchanged for a sum over a conjugate momentum by performing a Poisson resummation (7.32). Thus we first take the Fourier transform of $f(n'r) = e^{-(2\pi/\tau_2)(n'r - \tau_1 nr)^2}$,

$$\tilde{f}(p) = \int_{-\infty}^{\infty} dx\, e^{2\pi i x p} f(x) = \sqrt{\frac{\tau_2}{2}}\, e^{2\pi i \tau_1 nrp - \frac{1}{2}\pi\tau_2 p^2} \ .$$

Then we substitute (7.32) and (8.5) to express (8.3) as

$$\begin{aligned}\int e^{-S} &= \frac{1}{\eta\overline{\eta}} \sum_{n,m=-\infty}^{\infty} e^{-2\pi\tau_2 n^2 r^2 + 2\pi i \tau_1 nm - \frac{1}{2}\pi\tau_2(m/r)^2} \\ &= \frac{1}{\eta\overline{\eta}} \sum_{n,m=-\infty}^{\infty} q^{\frac{1}{2}(\frac{m}{2r}+nr)^2} \overline{q}^{\frac{1}{2}(\frac{m}{2r}-nr)^2} \\ &= \frac{1}{\eta\overline{\eta}} \sum_{n,m=-\infty}^{\infty} q^{\frac{1}{2}(\frac{p}{2}+w)^2} \overline{q}^{\frac{1}{2}(\frac{p}{2}-w)^2} \ .\end{aligned} \tag{8.6}$$

In the last line we have introduced the momentum $p = m/r$ and the winding $w = nr$. We see that this conjugate momentum is quantized in units of $1/r$. It is convenient to define as well $p_{L,R} = p/2 \pm w = m/2r \pm nr$, and express the result for the partition function in the form

$$Z_{\rm circ}(r) = \int e^{-S} = \frac{1}{\eta\overline{\eta}} \sum_{n,m=-\infty}^{\infty} q^{\frac{1}{2}p_L^2} \overline{q}^{\frac{1}{2}p_R^2} \ . \tag{8.7}$$

(Generalizations of (8.7) to higher dimensions and additional background fields are derived from the Hamiltonian and Lagrangian points of view in [55].)

To complete the identification with the Hamiltonian trace over Hilbert space states, we now recall the alternative interpretation of (8.7) as $(q\overline{q})^{-c/24}\, {\rm tr}\, q^{L_0}\overline{q}^{\overline{L}_0}$. We infer an infinite number of Hilbert space sectors $|m,n\rangle$, labeled by $m,n = -\infty, \infty$, for which

$$\begin{aligned}L_0 |m,n\rangle &= \frac{1}{2}\left(\frac{m}{2r} + nr\right)^2 |m,n\rangle \\ \text{and} \quad \overline{L}_0 |m,n\rangle &= \frac{1}{2}\left(\frac{m}{2r} - nr\right)^2 |m,n\rangle \ .\end{aligned} \tag{8.8}$$

We see that $L_0 = \sum \alpha_{-m}\alpha_m + \frac{1}{2}p_L^2$, with $\alpha_0 \equiv p_L = (\frac{p}{2} + w)$, and $\overline{L}_0 = \sum \overline{\alpha}_{-m}\overline{\alpha}_m + \frac{1}{2}p_R^2$, with $\overline{\alpha}_0 \equiv p_R = \frac{p}{2} - w$. We also see that the $|m,n\rangle$ state has energy and momentum eigenvalues

$$\begin{aligned}H &= L_0 + \overline{L}_0 = \frac{1}{2}(p_L^2 + p_R^2) = \frac{1}{4}p^2 + w^2 = \frac{m^2}{4r^2} + n^2 r^2 \\ P &= L_0 - \overline{L}_0 = \frac{1}{2}(p_L^2 - p_R^2) = pw = mn \in {\bf Z} \ .\end{aligned} \tag{8.9}$$

(We note briefly how the eigenvalues of $\alpha_0$ and $\overline{\alpha}_0$ can also be determined directly in the Hamiltonian point of view. Since $\alpha_0 + \overline{\alpha}_0$ is the zero mode of the momentum $\partial X$ conjugate to the coordinate $X$, with periodicity $2\pi r$, it has eigenvalues quantized as $p = m/r$ ($m \in {\bf Z}$). Mutual locality, i.e. integer eigenvalues of $L_0 - \overline{L}_0$, of the operators that create momentum/winding states then fixes the difference $\alpha_0 - \overline{\alpha}_0 = 2w = 2nr$.)

The factor of $(\eta\overline{\eta})^{-1}$ in (8.5) also has a straightforward Hamiltonian interpretation. The bosonic Fock space generated by $\alpha_{-n}$ consists of all states of the form $|m,n\rangle$, $\alpha_{-n}|m,n\rangle$, $\alpha_{-n}^2|m,n\rangle$, .... Calculating as for the fermionic case (before (7.13)) and ignoring for the moment the zero mode contribution, we find for the trace in the $|m,n\rangle$ Hilbert space sector

$$\text{tr} q^{L_0} = \text{tr} q^{\sum_{n=1}^{\infty} \alpha_{-n}\alpha_n} = \prod_{n=1}^{\infty}(1+q^n+q^{2n}+\ldots) = \prod_{n=1}^{\infty}\frac{1}{1-q^n},$$

as expected for Bose-Einstein statistics. Including the $\bar{\alpha}_{-n}$'s as well, we have

$$(q\bar{q})^{-c/24}\text{tr} q^{L_0}\bar{q}^{\bar{L}_0} = (q\bar{q})^{-1/24}\sum_{N,M=0}^{\infty} P(N)P(M) q^N \bar{q}^M = \frac{1}{\eta\bar{\eta}},$$

where the product $P(N)P(M)$ just counts the total number of states

$$\alpha_{-n_1}\ldots\alpha_{-n_m}\bar{\alpha}_{-m_1}\ldots\bar{\alpha}_{-m_k}|m,n\rangle$$

with $\sum_{i=1}^m n_i = N$, $\sum_{j=1}^k m_j = M$.

The result (8.6) is easily verified to be modular invariant. Under $\tau \to \tau+1$, each term in (8.6) acquires a phase $\exp 2\pi i \frac{1}{2}(p_L^2 - p_R^2)$, which is equal to unity by the second relation in (8.9). Under $\tau \to -1/\tau$, we note that the boundary conditions in the Lagrangian formulation transform as $n' \Box_n \to (-n) \Box_{n'}$, so we see how summation over $n'$ and $n$ may result in a modular invariant sum. We see moreover that the roles of "space" and "time" are interchanged by $\tau \to -1/\tau$, so it is clear that to verify modular invariance we should perform a Poisson resummation over both $m$ and $n$ in (8.6). Doing that and using the transformation property (7.33) of $\eta$ indeed establishes the modular invariance of (8.6).

(Modular invariance of (8.6) can be understood in a more general framework as follows[56]. Consider $(p_L, p_R)$ to be a vector in a two-dimensional space with Lorentzian signature, so that $(p_L, p_R) \cdot (p'_L, p'_R) \equiv p_L p'_L - p_R p'_R$. We may write arbitrary lattice vectors as

$$(p_L, p_R) = m\left(\frac{1}{2r}, \frac{1}{2r}\right) + n(r, -r) = mk + n\bar{k},$$

where the basis vectors $k, \bar{k}$ satisfy $k\bar{k} = 1$, $k^2 = \bar{k}^2 = 0$. $k$ and $\bar{k}$ generate what is known as an even self-dual Lorentzian integer lattice $\Gamma^{1,1}$. (Self-duality here is defined for Lorentzian signature just as was defined for Euclidean signature at the end of section 7.) The general statement is that partition functions of the form

$$Z_{\Gamma^{r,s}} = \frac{1}{\eta^r \bar{\eta}^s} \sum_{(p_L, p_R) \in \Gamma^{r,s}} q^{\frac{1}{2}p_L^2} \bar{q}^{\frac{1}{2}p_R^2}$$

are modular covariant provided that $\Gamma^{r,s}$ is an $r+s$ dimensional even self-dual Lorentzian lattice of signature $(r,s)$. The even property, $p_L^2 - p_R^2 \in 2\mathbf{Z}$, guarantees invariance under $\tau \to \tau+1$ (up to a possible phase from $\eta^{-r}\bar{\eta}^{-s}$ when $r - s \neq 0 \mod 24$), while the self-duality property guarantees invariance under $\tau \to -1/\tau$. Such lattices exist in every dimension $d = r - s = 0 \mod 8$, and for $r, s \neq 0$ are unique up to $SO(r,s)$ transformations. In the Euclidean case discussed at the end of section 7, on the other hand, there are a finite number of such lattices for every $d = r = 0 \mod 8$, unique up to $SO(d)$ transformations.)

We close here by pointing out that the partition function (8.7) can also be expressed in terms of $c = 1$ Virasoro characters. To see what these characters look like, we recall from the results of section 4 that there are no null states for $c > 1$ except at $h = 0$, and none at $c = 1$ except at $h = n^2/4$ ($n \in \mathbf{Z}$). For $c > 1$, this means that the Virasoro characters take the form

$$\chi_{h\neq 0}(q) = \frac{1}{\eta} q^{h-(c-1)/24} \qquad (8.10a)$$

$$\chi_0(q) = \frac{1}{\eta} q^{-(c-1)/24}(1-q) \qquad (8.10b)$$

(the extra factor of $(1-q)$ in the latter due to $L_{-1}|0\rangle = 0$). At $c = 1$ (8.10a) remains true for $h \neq n^2/4$ but for $h = n^2/4$, due to the null states the characters are instead

$$\chi_{n^2/4}(q) = \frac{1}{\eta}\left(q^{n^2/4} - q^{(n+2)^2/4}\right) = \frac{1}{\eta} q^{n^2/4}(1 - q^{n+1}). \qquad (8.11)$$

Unlike the Ising partition function (7.22), which was expressible in terms of a finite number of Virasoro characters, the expression for (8.7) would involve an infinite summation. This is consistent with result of [43] cited after (7.20), that for $c \geq 1$ modular invariance requires an infinite number of Virasoro primaries.

## 8.2. Fermionization

In earlier sections we have alluded to the fact that two chiral ($c = \frac{1}{2}$) fermions are equivalent to a chiral ($c = 1$) boson. In this subsection we shall illustrate this correspondence explicitly on the torus. Consider two Dirac fermions comprised of $\psi_1(z)$, $\psi_2(z)$ and $\overline{\psi}_1(\overline{z})$, $\overline{\psi}_2(\overline{z})$. By Dirac fermion on the torus [57], we mean to take all these fermions to have the same spin structure. The partition function for such fermions is consequently given by the modular invariant combination of spin structures

$$Z_{\text{Dirac}} = \frac{1}{2}\left( A^2\overline{A}^2 \boxed{\phantom{x}}_{A^2\overline{A}^2} + P^2\overline{P}^2 \boxed{\phantom{x}}_{A^2\overline{A}^2} + A^2\overline{A}^2 \boxed{\phantom{x}}_{P^2\overline{P}^2} + P^2\overline{P}^2 \boxed{\phantom{x}}_{P^2\overline{P}^2} \right)$$

$$= \frac{1}{2}\left( \left|\frac{\vartheta_3}{\eta}\right|^2 + \left|\frac{\vartheta_4}{\eta}\right|^2 + \left|\frac{\vartheta_2}{\eta}\right|^2 + \left|\frac{\vartheta_1}{\eta}\right|^2 \right), \quad (8.12)$$

where we have for convenience chosen the projection on $(-1)^F = +1$ states in the $P\overline{P}$ sector.

The partition functions (7.13) were all derived from the standpoint of the expressions of the $\vartheta$-functions as infinite products. In (7.29), however, we have seen that these functions also admit expressions as infinite sums via the Jacobi triple product identity. We shall now see that this equivalence is the expression of bosonization of fermions on the torus. Substituting the sum forms of the $\vartheta$-functions in (8.12), we find

$$Z_{\text{Dirac}} = \frac{1}{\eta\overline{\eta}} \sum_{n,m=-\infty}^{\infty} \left( q^{\frac{1}{2}n^2}\overline{q}^{\frac{1}{2}m^2} + q^{\frac{1}{2}(n+\frac{1}{2})^2}\overline{q}^{\frac{1}{2}(m+\frac{1}{2})^2} \right) \frac{1}{2}\left(1 + (-1)^{n+m}\right)$$

$$= \frac{1}{\eta\overline{\eta}} \sum_{n,m'=-\infty}^{\infty} \left( q^{\frac{1}{2}(n+m')^2}\overline{q}^{\frac{1}{2}(n-m')^2} + q^{\frac{1}{2}(n+\frac{1}{2}+m')^2}\overline{q}^{\frac{1}{2}(n+\frac{1}{2}-m')^2} \right) \quad (8.13)$$

$$= \frac{1}{\eta\overline{\eta}} \sum_{n,m=-\infty}^{\infty} q^{\frac{1}{2}(\frac{m}{2}+n)^2}\overline{q}^{\frac{1}{2}(\frac{m}{2}-n)^2} = Z_{\text{circ}}(r=1),$$

equal to the bosonic partition function (8.7) at radius $r = 1$. (In (8.13) we have used the property that $\frac{1}{2}(1 + (-1)^{n+m})$ acts as a projection operator, projecting onto terms in the summation with $n + m$ even, automatically implemented in the next line by the reparametrization of the summation in terms of $n$ and $m'$.) Recalling that the vertex operators $e^{\pm ix(z)}$ have conformal weight $h = \frac{1}{2}$, it is not surprising that (8.12) emerges as the bosonic partition function at radius $r = 1$. It is precisely at this radius that the vertex operators $e^{\pm ix(z)}$ are suitably single-valued under $x \to x + 2\pi/r = x + 2\pi$. The connection with the real fermions above is given, as in (6.14), by $e^{\pm ix(z)} = \frac{i}{\sqrt{2}}(\psi_1(z) \pm i\psi_2(z))$, $e^{\pm i\overline{x}(\overline{z})} = \frac{i}{\sqrt{2}}(\overline{\psi}_1(\overline{z}) \pm i\overline{\psi}_2(\overline{z}))$.

By comparing (8.12) and (8.13) we can identify the states in the bosonic form of the partition function that correspond to the states in the various sectors of the fermionic form. The partition function only includes states that survive the GSO projection onto $(-1)^F = +1$ (where $F = F_1 + F_2 + \overline{F}_1 + \overline{F}_2$ is the total fermion number). Thus we need to extend the range of $n$ in the last line of (8.13) to $n \in \mathbf{Z}/2$ to construct a non-local covering theory that includes as well the $(-1)^F = -1$ states prior to projection. Then the states of the $A^2\overline{A}^2$ fermionic sector with $(-1)^F = \pm 1$ are given respectively by $\{n \in \mathbf{Z}, m \in 2\mathbf{Z}\}$ and $\{n \in \mathbf{Z} + \frac{1}{2}, m \in 2\mathbf{Z} + 1\}$; while the states of the $P^2\overline{P}^2$ fermionic sector with $(-1)^F = \pm 1$ are given respectively by $\{n \in \mathbf{Z}, m \in 2\mathbf{Z} + 1\}$ and $\{n \in \mathbf{Z} + \frac{1}{2}, m \in 2\mathbf{Z}\}$. Thus we have seen how the classical identity (7.27) becomes the statement of bosonization of fermions on the torus. (The generalization of these results to arbitrary genus Riemann surfaces, including the interpretation of modular invariance at higher genus as enforcing certain projections, may be found in [41][58].)

If we relax the restriction in (8.12) that all fermions $\psi_{1,2}, \overline{\psi}_{1,2}$ have the same spin structure, then we can construct another obvious $c = \overline{c} = 1$ modular invariant combination,

$$Z^2_{\text{Ising}} = \frac{1}{2^2}\left( A\overline{A}\boxed{\phantom{x}}_{A\overline{A}} + P\overline{P}\boxed{\phantom{x}}_{A\overline{A}} + A\overline{A}\boxed{\phantom{x}}_{P\overline{P}} \right)\left( A\overline{A}\boxed{\phantom{x}}_{A\overline{A}} + P\overline{P}\boxed{\phantom{x}}_{A\overline{A}} + A\overline{A}\boxed{\phantom{x}}_{P\overline{P}} \right) \quad (8.14)$$

$$= \frac{1}{4}\left( \left|\frac{\vartheta_3}{\eta}\right| + \left|\frac{\vartheta_4}{\eta}\right| + \left|\frac{\vartheta_2}{\eta}\right| \right)^2.$$

Following [57], we refer to the choice of independent boundary conditions for $\psi_1, \overline{\psi}_1$ and $\psi_2, \overline{\psi}_2$ as specifying two Majorana fermions (as opposed to a single

Dirac fermion). The partition function (8.14) is of course the square of the Ising partition function (7.22).

It is natural to ask whether (8.14) as well has a representation in terms of a free boson. It is first of all straightforward to see that (8.14) does not correspond to (8.7) for any value of $r$. (For example, one may note that the spectrum of (8.14) has two $(\frac{1}{16}, \frac{1}{16})$ states. But (8.7) has two such states only for $r = \sqrt{2}$ and $r = 1/2\sqrt{2}$, at which points it is easy to see that there are no $(\frac{1}{2}, \frac{1}{2})$ states.) The distinction between (8.12) and (8.14) is the decoupling of the spin structures of the two Majorana fermions. Due to the correspondence $\psi_{1,2} \sim (e^{ix} \pm e^{-ix})$, we see that the bosonic operation $x \to -x$, taking $\psi_1 \to \psi_1$ and $\psi_2 \to -\psi_2$ (and similarly for $\overline{\psi}_{1,2}$), distinguishes between $\psi_1, \overline{\psi}_1$ and $\psi_2, \overline{\psi}_2$. The key to constructing a bosonic realization of (8.14), then, is to implement somehow the symmetry action $x \to -x$ on (8.7). This is provided by the notion of an orbifold, to which we now turn.

*8.3. Orbifolds in general*

Orbifolds arise in a purely geometric context by generalizing the notion of manifolds to allow a discrete set of singular points. Consider a manifold $\mathcal{M}$ with a discrete group action $G : \mathcal{M} \to \mathcal{M}$. This action is said to possess a fixed point $x \in \mathcal{M}$ if for some $g \in G$ ($g \neq$ identity), we have $gx = x$. The quotient space $\mathcal{M}/G$ constructed by identifying points under the equivalence relation $x \sim gx$ for all $g \in G$ defines in general an orbifold. If the group $G$ acts freely (no fixed points) then we have the special case of orbifold which is an ordinary manifold. Otherwise the points of the orbifold corresponding to the fixed point set have discrete identifications of their tangent spaces, and are not manifold points. (A slightly more general definition of orbifold is to require only that the above condition hold coordinate patch by coordinate patch.) A simple example is provided by the circle, $\mathcal{M} = S^1$, coordinatized by $x \equiv x + 2\pi r$, with group action $G = \mathbf{Z}_2 : S^1 \to S^1$ defined by the generator $g : x \to -x$. This group action has fixed points at $x = 0$ and $x = \pi r$, and we see in fig. 12 that the $S^1/\mathbf{Z}_2$ orbifold is topologically a line segment.

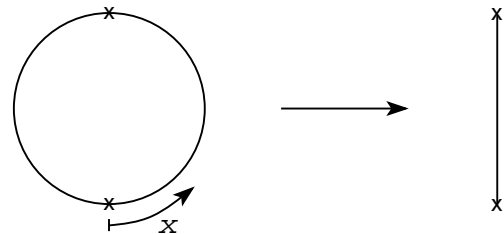

Fig. 12. The orbifold $S^1/\mathbf{Z}_2$.

In conformal field theory, the notion of orbifold acquires a more generalized meaning. It becomes a heuristic for taking a given modular invariant theory $\mathcal{T}$, whose Hilbert space admits a discrete symmetry $G$ consistent with the interactions or operator algebra of the theory, and constructing a "modded-out" theory $\mathcal{T}/G$ that is also modular invariant[59].

Orbifold conformal field theories occasionally have a geometric interpretation as $\sigma$-models whose target space is the geometrical orbifold discussed in the previous paragraph. This we shall confirm momentarily in the case of the $S^1/\mathbf{Z}_2$ example. We shall also see examples however where the geometrical interpretation is either ambiguous or non-existent. Consequently it is frequently preferable to regard orbifold conformal field theories from the more abstract standpoint of modding out a modular invariant theory by a Hilbert space symmetry. (Historically, orbifolds were introduced into conformal field theory [59] (see also [60]) via string theory as a way to approximate conformal field theory on "Calabi-Yau" manifolds. Even before the "phenomenological" interest in the matter subsided, orbifold conformal field theories were noted to possess many interesting features in their own right, and in particular enlarged the playground of tractable conformal field theories.)

The construction of an orbifold conformal field theory $\mathcal{T}/G$ begins with a Hilbert space projection onto $G$ invariant states. It is convenient to represent this projection in Lagrangian form as

$$\frac{1}{|G|} \sum_{g \in G} g \underset{1}{\square} , \qquad (8.15)$$

where $g\boxed{\phantom{x}}_1$ represents boundary conditions on any generic fields $x$ in the theory twisted by $g$ in the "time" direction of the torus, i.e. $x(z+\tau) = gx(z)$. In Hamiltonian language such twisted boundary conditions correspond to insertion of the operator realizations of group elements $g$ in the trace over states, and hence (8.15) corresponds to the insertion of the projection operator $P = \frac{1}{|G|}\sum_{g\in G} g$.

But (8.15) is evidently not modular invariant as it stands since under $S:$ $\tau \to -1/\tau$ for example we have $g\boxed{\phantom{x}}_1 \to 1\boxed{\phantom{x}}_g$ (this is easily verified by shifting appropriately along the two cycles of the torus using the representation $S = T^{-1}UT^{-1}$ given before (7.12)). Under $\tau \to \tau + n$ we have moreover that $1\boxed{\phantom{x}}_g \to g^n\boxed{\phantom{x}}_g$, so we easily infer the general result

$$g\boxed{\phantom{x}}_h \to g^a h^b\boxed{\phantom{x}}_{g^c h^d} \qquad \text{under} \qquad \tau \to \frac{a\tau + b}{c\tau + d}, \qquad (8.16)$$

for $g, h \in G$ such that $gh = hg$. (We note that there seems an ambiguity in (8.16) due to the possibility of taking $a, b, c, d$ to minus themselves. But for self-conjugate fields, for which charge conjugation C = 1 and the modular group is realized as $PSL(2,\mathbf{Z})$, $g\boxed{\phantom{x}}_h$ and $g^{-1}\boxed{\phantom{x}}_{h^{-1}}$ are equal. In a more general context one would have to implement $S^2 = (ST)^3 = $ C.)

To have a chance of recovering a modular invariant partition function, we thus need to consider as well twists by $h$ in the "space" direction of the torus, $x(z+1) = hx(z)$, and define

$$Z_{\mathcal{T}/G} \equiv \sum_{h \in G} \frac{1}{|G|} \sum_{g \in G} g\boxed{\phantom{x}}_h = \frac{1}{|G|} \sum_{g,h \in G} g\boxed{\phantom{x}}_h. \qquad (8.17)$$

The boundary conditions in individual terms of (8.17) are ambiguous for $x(z + \tau + 1)$ unless $gh = hg$. Thus in the case of non-abelian groups $G$, the summation in (8.17) should be restricted only to mutually commuting boundary conditions $gh = hg$. From (8.16) we see that modular transformations of such boundary conditions automatically preserve this property. Moreover we see that (8.17) contains closed sums over modular orbits so it is formally invariant under modular transformations. (In the following we shall consider for simplicity only symmetry actions that act symmetrically on holomorphic and anti-holomorphic fields, so modular invariance of (8.17) is more or less immediate. For more general asymmetric actions, additional conditions must be imposed on the eigenvalues of the realizations of the group elements to insure that no phase ambiguities occur under closed loops of modular transformations that restore the original boundary conditions [59][61][62].) We also note that the orbifold prescription, changing only boundary conditions of fields via a symmetry of the stress-energy tensor, always gives a theory with the same value of the central charge $c$.

For $G$ abelian, the operator interpretation of (8.17) is immediate. The Hilbert space decomposes into a set of twisted sectors labeled by $h$, and in each twisted sector there is a projection onto $G$ invariant states. A similar interpretation exists as well for the non-abelian case, although then it is necessary to recognize that twisted sectors should instead be labeled by conjugacy classes $C_i$ of $G$. This is because if we consider fields $hx(z)$ translated by some $h$, then the $g$ twisted sector, $hx(z+1) = ghx(z)$, is manifestly equivalent to the $h^{-1}gh$ twisted sector, $x(z+1) = h^{-1}ghx(z)$. Now the number of elements $g \in N_i \subset G$ that commute with a given element $h \in C_i \subset G$ depends only on the conjugacy class $C_i$ of $h$ (the group $N_i$ is known as the stabilizer group, or little group, of $C_i$ and is defined only up to conjugation). This number is given by $|N_i| = |G|/|C_i|$, where $|C_i|$ is the order of $C_i$. In the non-abelian case, we may thus rewrite the summation in (8.17) as

$$\frac{1}{|G|} \sum_{hg=gh} g\boxed{\phantom{x}}_h = \sum_i \frac{1}{|N_i|} \sum_{g \in N_i} g\boxed{\phantom{x}}_{C_i},$$

manifesting the interpretation of the summation over $g$ as a properly normalized projection onto states invariant under the stabilizer group $N_i$ in each twisted sector labeled by $C_i$.

While we have discussed here only the construction of the orbifold partition function (8.17), we point out that the orbifold prescription (at least in the abelian case) also allows one to construct all correlation functions in principle[63]. We also point out that we have been a bit cavalier in presenting the summation in (8.17). In general such a summation will decompose into distinct modular orbits, i.e. distinct groups of terms each of which is individually modular invariant. The full summation in (8.17) is nonetheless required for a consistent operator interpretation of the theory (or equivalently for modular invariance on higher genus Riemann surfaces). There may remain however distinct choices of relative phases between the different orbits in (8.17) (just as in the case of the Ising model (7.22)), corresponding in operator language to different choices of projections in twisted sectors. In [61], the different possible orbifold theories $\mathcal{T}/G$ that may result in this manner were shown to be classified by the second cohomology group $H^2(G, U(1))$, which equivalently classifies the projective representations of the group $G$. (Torsion-related theories can also be viewed to result from the existence of an automorphism of the fusion rules of the chiral algebra of a theory. Instead of a diagonal sesquilinear combination $\sum \chi_i \overline{\chi}_i$ of chiral characters as the partition function, we would have $\sum \chi_i P_{ij} \overline{\chi}_j$, where $P$ is a permutation of the chiral characters that preserves the fusion rules.)

8.4. $S^1/\mathbf{Z}_2$ orbifold

We now employ the general orbifold formalism introduced above to construct a $G = \mathbf{Z}_2$ orbifold conformal theory of the free bosonic field theory (8.1). We first note that the action (8.1) is invariant under $g : X \to -X$, under which $\alpha_n \to -\alpha_n$ and $\overline{\alpha}_n \to -\overline{\alpha}_n$. (Recall that $X(z,\overline{z}) = \frac{1}{2}(x(z) + \overline{x}(\overline{z}))$, and the $\alpha_n$'s and $\overline{\alpha}_n$'s are respectively the modes of $i\partial x(z)$ and $i\overline{\partial}\overline{x}(\overline{z})$.) These include the momentum zero modes $p_L = \alpha_0$ and $p_R = \overline{\alpha}_0$ so the action of $g$ on the Hilbert space sectors $|m,n\rangle$ of (8.8) is given by $|m,n\rangle \to |-m,-n\rangle$.

The general prescription (8.17) for the $\mathcal{T}/G$ orbifold partition function reduces for $G = \mathbf{Z}_2$ to

$$\begin{aligned} Z_{\text{orb}}(r) &= \frac{1}{2}\left( +\underset{+}{\boxed{\phantom{x}}} + -\underset{+}{\boxed{\phantom{x}}} + +\underset{-}{\boxed{\phantom{x}}} + -\underset{-}{\boxed{\phantom{x}}} \right) \\ &= (q\overline{q})^{-1/24}\text{tr}_{(+)}\frac{1}{2}(1+g)q^{L_0}\overline{q}^{\overline{L}_0} \\ &\quad + (q\overline{q})^{-1/24}\text{tr}_{(-)}\frac{1}{2}(1+g)q^{L_0}\overline{q}^{\overline{L}_0} . \end{aligned} \tag{8.18}$$

In the first line of (8.18), we use $\pm$ to represent periodic and anti-periodic boundary conditions on the free boson $X$ along the two cycles of the torus. In the second line $\text{tr}_{(+)}$ denotes the trace in the untwisted Hilbert space sector $H_{(+)}$ (corresponding to $X(z+1,\overline{z}+1) = X(z,\overline{z})$), and $\text{tr}_{(-)}$ denotes the trace in the twisted sector $H_{(-)}$ (corresponding to $X(z+1,\overline{z}+1) = -X(z,\overline{z})$).

The above symmetry actions induced by $g : X \to -X$ imply that the untwisted Hilbert space $H_{(+)}$ decomposes into $g = \pm 1$ eigenspaces $H_{(+)}^{\pm}$ as

$$\begin{aligned} H_{(+)}^+ &= \left\{ \alpha_{-n_1} \cdots \alpha_{-n_\ell} \overline{\alpha}_{-n_{\ell+1}} \cdots \overline{\alpha}_{-n_{2k}} \big(|m,n\rangle + |-m,-n\rangle\big) \right\} \\ &\quad + \left\{ \alpha_{-n_1} \cdots \alpha_{-n_\ell} \overline{\alpha}_{-n_{\ell+1}} \cdots \overline{\alpha}_{-n_{2k+1}} \big(|m,n\rangle - |-m,-n\rangle\big) \right\} , \\ H_{(+)}^- &= \left\{ \alpha_{-n_1} \cdots \alpha_{-n_\ell} \overline{\alpha}_{-n_{\ell+1}} \cdots \overline{\alpha}_{-n_{2k+1}} \big(|m,n\rangle + |-m,-n\rangle\big) \right\} \\ &\quad + \left\{ \alpha_{-n_1} \cdots \alpha_{-n_\ell} \overline{\alpha}_{-n_{\ell+1}} \cdots \overline{\alpha}_{-n_{2k}} \big(|m,n\rangle - |-m,-n\rangle\big) \right\} , \end{aligned} \tag{8.19}$$

where $n_i \in \mathbf{Z}^+$. We see that in each sector with $\{m,n\} \neq \{0,0\}$, exactly half the states at each level of $L_0$ and $\overline{L}_0$ have eigenvalue $g = +1$. To calculate $\text{tr}_{(+)}\frac{1}{2}(1+g)q^{L_0}\overline{q}^{\overline{L}_0}$, we note that $g|m,n\rangle = |-m,-n\rangle$, so that the trace with $g$ inserted receives only contributions from the states built with $\alpha$'s and $\overline{\alpha}$'s on $|0,0\rangle$. The overall trace over states with eigenvalue $g = +1$ in the untwisted

sector is thus given by

$$\begin{aligned}(q\bar{q})^{-1/24} \operatorname{tr}_{H_{(+)}^+} q^{L_0}\bar{q}^{\bar{L}_0} &= (q\bar{q})^{-1/24} \operatorname{tr}_{(+)} \frac{1}{2}(1+g) q^{L_0}\bar{q}^{\bar{L}_0} \\ &= \frac{1}{2}\frac{1}{\eta\bar{\eta}} \sum_{m,n=-\infty}^{\infty} q^{\frac{1}{2}(\frac{m}{2r}+nr)^2}\bar{q}^{\frac{1}{2}(\frac{m}{2r}-nr)^2} \\ &\quad + \frac{1}{2}\frac{(q\bar{q})^{-1/24}}{\prod_{n=1}^{\infty}(1+q^n)(1+\bar{q}^n)} \\ &= \frac{1}{2}Z_{\text{circ}}(r) + \left|\frac{\eta}{\vartheta_2}\right| \,.\end{aligned} \tag{8.20}$$

Next we need to construct the twisted Hilbert space $H_{(-)}$. The first subtlety is that there are actually two dimension $(\frac{1}{16}, \frac{1}{16})$ twist operators $\sigma_{1,2}$, satisfying

$$\begin{aligned}\partial x(z)\, \sigma_{1,2}(w,\overline{w}) &\sim (z-w)^{-1/2} \tau_{1,2}(w,\overline{w}) \\ \overline{\partial}\overline{x}(\overline{z})\, \sigma_{1,2}(w,\overline{w}) &\sim (\overline{z}-\overline{w})^{-1/2} \widetilde{\tau}_{1,2}(w,\overline{w})\end{aligned} \tag{8.21}$$

as in (6.11). (Here the dimensions of the excited twist operators $\tau_{1,2}$ and $\widetilde{\tau}_{1,2}$ are given respectively by $(\frac{9}{16}, \frac{1}{16})$ and $(\frac{1}{16}, \frac{9}{16})$. The states identified with $\tau_{1,2}(0)|0\rangle$ and $\widetilde{\tau}_{1,2}(0)|0\rangle$ can also be written $\alpha_{-1/2}\left|\frac{1}{16},\frac{1}{16}\right\rangle_{1,2}$ and $\overline{\alpha}_{-1/2}\left|\frac{1}{16},\frac{1}{16}\right\rangle_{1,2}$.) Geometrically the existence of two twist operators results from the two fixed points of the symmetry action $g: X \to -X$, as depicted in fig. 12, and two distinct Hilbert spaces are built on top of each of these two fixed point sectors. Equivalently, we note two ways of realizing $g$, either as $x \to -x$ or as $x \to 2\pi - x$, and each realization is implemented by a different twist operator. The multiplicity is also easily understood in terms of the fermionic form of the current, $\partial x \sim \psi_1\psi_2$. Then the two twist operators may be constructed explicitly in terms of the individual twist operators for each of the two fermions. Finally the multiplicity of vacuum states can also be verified by performing the modular transformation

$$\tau \to -1/\tau: \quad -\boxed{\phantom{x}}_{+} \;\to\; +\boxed{\phantom{x}}_{-}$$

to construct the trace $+\boxed{\phantom{x}}_{-}$ over the spectrum of the unprojected twisted sector from the trace $-\boxed{\phantom{x}}_{+}$ over the untwisted sector with the operator insertion of $g$.

123

Denoting the two $(\frac{1}{16}, \frac{1}{16})$ twisted sector ground states by $\left|\frac{1}{16},\frac{1}{16}\right\rangle_{1,2}$, we find that the twisted Hilbert space $H_{(-)}$ decomposes into $g = \pm 1$ eigenspaces $H_{(-)}^{\pm}$ as

$$\begin{aligned}H_{(-)}^+ &= \left\{\alpha_{-n_1}\cdots\alpha_{-n_\ell}\overline{\alpha}_{-n_{\ell+1}}\cdots\overline{\alpha}_{-n_{2k}}\left|\tfrac{1}{16},\tfrac{1}{16}\right\rangle_{1,2}\right\} \\ H_{(-)}^- &= \left\{\alpha_{-n_1}\cdots\alpha_{-n_\ell}\overline{\alpha}_{-n_{\ell+1}}\cdots\overline{\alpha}_{-n_{2k+1}}\left|\tfrac{1}{16},\tfrac{1}{16}\right\rangle_{1,2}\right\},\end{aligned} \tag{8.22}$$

where the moding is now $n_i \in (\mathbf{Z} + \tfrac{1}{2})^+$. The overall trace over states with eigenvalue $g = +1$ in the twisted $\mathbf{Z}_2$ sector is thus given by

$$\begin{aligned}(q\bar{q})^{-1/24} \operatorname{tr}_{H_{(-)}^+} q^{L_0}\bar{q}^{\bar{L}_0} &= (q\bar{q})^{-1/24} \operatorname{tr}_{(-)} \frac{1}{2}(1+g) q^{L_0}\bar{q}^{\bar{L}_0} \\ &= 2\frac{1}{2}\Bigg(\frac{(q\bar{q})^{1/48}}{\prod_{n=1}^{\infty}(1-q^{n-1/2})(1-\bar{q}^{n-1/2})} \\ &\quad + \frac{(q\bar{q})^{1/48}}{\prod_{n=1}^{\infty}(1+q^{n-1/2})(1+\bar{q}^{n-1/2})}\Bigg) \\ &= \left|\frac{\eta}{\vartheta_4}\right| + \left|\frac{\eta}{\vartheta_3}\right| \,.\end{aligned} \tag{8.23}$$

Now if we substitute (8.20) and (8.23) into (8.18), and use the identity $\vartheta_2\vartheta_3\vartheta_4 = 2\eta^3$, we find that the orbifold partition function satisfies

$$\begin{aligned}Z_{\text{orb}}(r) &= \frac{1}{2}\left(+\boxed{\phantom{x}}_+ + -\boxed{\phantom{x}}_+ + +\boxed{\phantom{x}}_- + -\boxed{\phantom{x}}_-\right) \\ &= \frac{1}{2}\left(Z_{\text{circ}}(r) + \frac{|\vartheta_3\vartheta_4|}{\eta\bar{\eta}} + \frac{|\vartheta_2\vartheta_3|}{\eta\bar{\eta}} + \frac{|\vartheta_2\vartheta_4|}{\eta\bar{\eta}}\right) \,.\end{aligned} \tag{8.24}$$

We note that modular invariance of (8.24) can be easily verified from the transformation properties (7.14).

We may now at last return to the point left open earlier, namely the bosonic realization of the Ising$^2$ partition function (8.14). From (8.12) and (8.24) we evaluate $Z_{\text{orb}}(r=1)$,

$$\begin{aligned}Z_{\text{orb}}(1) &= \frac{1}{2}\left(\frac{|\vartheta_3|^2 + |\vartheta_4|^2 + |\vartheta_2|^2}{2|\eta|^2}\right) + \frac{1}{2}\left(\frac{|\vartheta_3\vartheta_4|}{|\eta|^2} + \frac{|\vartheta_2\vartheta_3|}{|\eta|^2} + \frac{|\vartheta_2\vartheta_4|}{|\eta|^2}\right) \\ &= \frac{1}{4}\left(\left|\frac{\vartheta_3}{\eta}\right| + \left|\frac{\vartheta_4}{\eta}\right| + \left|\frac{\vartheta_2}{\eta}\right|\right)^2 = Z_{\text{Ising}}^2 \,.\end{aligned}$$

124

We thus see that two Majorana fermions bosonize onto an $S^1/\mathbf{Z}_2$ orbifold at radius $r = 1$. The $Z^2_{\text{Ising}}$ theory can also be constructed directly as an orbifold from the $Z_{\text{Dirac}}$ theory by modding out by the $\mathbf{Z}_2$ symmetry $\psi_2 \to -\psi_2$, $\overline{\psi}_2 \to -\overline{\psi}_2$.

It is useful to consider the generic symmetry possessed by the family of theories (8.24). The two twist operators $\sigma_{1,2}$ of (8.21) and their operator algebras are unaffected by changes in the radius $r$. The theory consequently admits a generic symmetry generated by separately taking either $\sigma_1 \to -\sigma_1$ or $\sigma_2 \to -\sigma_2$, or interchanging the two, $\sigma_1 \leftrightarrow \sigma_2$. The group so generated is isomorphic to $\mathbf{D}_4$, the eight element symmetry group of the square. (This group may also be represented in terms of Pauli matrices as $\{\pm 1, \pm \sigma_x, \pm i\sigma_y, \pm \sigma_z\}$, with the order four element $i\sigma_y$, say, corresponding to $\sigma_1 \to -\sigma_2$, $\sigma_2 \to \sigma_1$).

$\mathbf{D}_4$ is also the generic symmetry group of a lattice model constructed by coupling together two Ising models, known as the Ashkin-Teller model. If we denote the two Ising spins by $\sigma$ and $\sigma'$, then the Ashkin-Teller action is given by

$$S_{\text{AT}} = -K_2 \sum_{\langle ij \rangle} \left( \sigma_i \sigma_j + \sigma'_i \sigma'_j \right) - K_4 \sum_{\langle ij \rangle} \sigma_i \sigma_j \sigma'_i \sigma'_j \,, \qquad (8.25)$$

where the summation is over nearest neighbor sites $\langle ij \rangle$ on a square lattice. The $\mathbf{D}_4$ symmetry group in this case is generated by separately taking either $\sigma \to -\sigma$ or $\sigma' \to -\sigma'$, or interchanging the two, $\sigma \leftrightarrow \sigma'$, on all sites. Since there are now two parameters, (8.25) has a line of critical points, given by the self-duality condition $\exp(-2K_4) = \sinh 2K_2$. As shown in [64], the critical partition function for the Ashkin-Teller model on a torus takes identically the form (8.24), with $\sin(\pi r^2/4) = \frac{1}{2} \coth 2K_2$. For $K_4 = 0$, (8.25) simply reduces to two uncoupled copies of the Ising model, with critical point partition function (8.14). That is the point $r = 1$ on the orbifold line. Calculations of the critical correlation functions in the Ashkin-Teller model from the bosonic point of view may be found in [65].

In general the Ashkin-Teller model can be regarded as two Ising models coupled via their energy densities $\varepsilon_1$ and $\varepsilon_2$. On the critical line this interaction takes the form of a four-fermion interaction $\varepsilon_1 \varepsilon_2 = \psi_1 \overline{\psi}_1 \psi_2 \overline{\psi}_2$. This four-fermion interaction defines what is known as the massless Thirring model. Although seemingly an interacting model of continuum fermions, properly described it is really just a free theory since in bosonic form we see that the interaction simply changes the radius of a free boson. (A recent pedagogical treatment with some generalizations and references to the earlier literature may be found in [66].) At radius $r = \sqrt{2}$ the partition function $Z_{\text{orb}}(\sqrt{2})$ turns out to have a full $S_4$ permutation symmetry and coincides with the critical partition function of the 4-state Potts model on the torus [67][68].

*8.5. Orbifold comments*

It may seem that an orbifold theory is somehow less fundamental than the original theory. In the case of abelian orbifolds we shall now see that a theory and its orbifold stand on equal footing. Let us first consider the case of a $G = \mathbf{Z}_2$ orbifold. Then the orbifold theory always possesses as well a $\mathbf{Z}_2$ symmetry, generated by taking all states in the $\mathbf{Z}_2$ twisted sectors (or equivalently the operators that create them) to minus themselves, i.e.

$$\widetilde{g}: \quad \pm\boxed{\phantom{x}}_{-} \quad \to \quad -\pm\boxed{\phantom{x}}_{-}\,.$$

From the geometrical point of view, for example, it is clear that acting twice with the twist $X \to -X$ takes us back to the untwisted sector. This is reflected in the interactions (operator products) of twist operators.

If we denote the partition function for the orbifold theory by $+\boxed{\phantom{x}}_{+}{}'$, then we can mod out the orbifold theory by its $\mathbf{Z}_2$ symmetry by constructing in turn,

$$+\boxed{\phantom{x}}_{+}{}' = \frac{1}{2}\left( +\boxed{\phantom{x}}_{+} + -\boxed{\phantom{x}}_{+} + +\boxed{\phantom{x}}_{-} + -\boxed{\phantom{x}}_{-} \right),$$

$$-\boxed{\phantom{x}}_{+}{}' = \frac{1}{2}\left( +\boxed{\phantom{x}}_{+} + -\boxed{\phantom{x}}_{+} - +\boxed{\phantom{x}}_{-} - -\boxed{\phantom{x}}_{-} \right),$$

$$\tau \to -\frac{1}{\tau} \quad \Rightarrow \quad +\boxed{\phantom{x}}_{-}{}' = \frac{1}{2}\left( +\boxed{\phantom{x}}_{+} + +\boxed{\phantom{x}}_{-} - -\boxed{\phantom{x}}_{+} - -\boxed{\phantom{x}}_{-} \right),$$

$$\tau \to \tau + 1 \quad \Rightarrow \quad -\boxed{\phantom{x}}_{-}{}' = \frac{1}{2}\left( +\boxed{\phantom{x}}_{+} + -\boxed{\phantom{x}}_{-} - -\boxed{\phantom{x}}_{+} + +\boxed{\phantom{x}}_{-} \right).$$

The second line follows from the definition of the operator insertion of the symmetry generator $\widetilde{g}$, and the third and fourth lines follow by performing the indicated modular transformations. The result of orbifolding the orbifold is thus

$$\frac{1}{2}\left( +\boxed{\phantom{x}}'_+ + -\boxed{\phantom{x}}'_+ + +\boxed{\phantom{x}}'_- + -\boxed{\phantom{x}}'_- \right) = +\boxed{\phantom{x}}_+ ,$$

and we see that the original theory $+\boxed{\phantom{x}}_+$ and the orbifold theory $+\boxed{\phantom{x}}'_+$ stand on symmetrical footing, each a $\mathbf{Z}_2$ orbifold of the other.

It is easy to generalize this to a $\mathbf{Z}_n$ orbifold, and consequently to an arbitrary abelian orbifold. If we let the $\mathbf{Z}_n$ be generated by an element $g \in \mathbf{Z}_n$, with $g^n$=identity, then the spectrum of the orbifold theory is constructed by projecting onto $\mathbf{Z}_n$ invariant states in each of the $n$ twisted sectors labeled by $g^j$ ($j = 0, \ldots, n-1$). The orbifold theory in this case has an obvious $\mathbf{Z}_n$ symmetry, given by assigning the phase $\omega^j$ to the $g^j$ twisted sector, where $\omega^n = 1$. The statement that this is a symmetry of the operator algebra of the orbifold theory is just the fact that the selection rules allow three point functions for a $g^{j_1}$ twist operator and a $g^{j_2}$ twist operator only with a $g^{-j_1-j_2}$ twist operator. Straightforward generalization of the argument given above for the $G = \mathbf{Z}_2$ case shows that modding out a $\mathbf{Z}_n$ orbifold by this $\mathbf{Z}_n$ symmetry gives back the original theory. For a non-abelian orbifold, on the other hand, the symmetry group is only $G/[G,G]$, where $[G,G]$ is the commutator subgroup (generated by all elements of the form $ghg^{-1}h^{-1} \in G$), so in general this procedure cannot be used to undo a non-abelian orbifold (except if the group is solvable).

As another class of examples of $\mathbf{Z}_2$ orbifolds, this time without an obvious geometrical interpretation, we consider conformal field theories built from any member of the $c < 1$ discrete series. To identify the $\mathbf{Z}_2$ symmetry of their operator algebras, it is convenient to retain the operators of the (double-counted) conformal grid with $p + q =$ even, as indicated by $\pm$ in the checkerboard pattern of fig. 13. We indicate the operators $\varphi_{(+)}$ with both $p$ and $q$ even by $+$, and operators $\varphi_{(-)}$ with both $p$ and $q$ odd by $-$. The operators left blank are redundant in the conformal grid. The only non-vanishing operator product coefficients allowed by the selection rules described in subsection 5.3 are of the form $C_{+++}$ and $C_{+--}$ (i.e. with an even number of $(-)$-type operators, in accord with their "spinorial" nature). The conformal field theories built from these models therefore possess an automatic $\mathbf{Z}_2$ symmetry $\varphi_{(\pm)} \to \pm\varphi_{(\pm)}$.

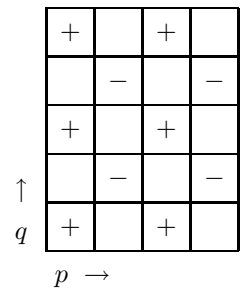

Fig. 13. $\mathbf{Z}_2$ symmetry of $c < 1$ fusion rules.

We can thus take for example any of the $c < 1$ theories with partition function given by the diagonal modular invariant combination of characters, i.e. any member of what is known as the $A$ series, and mod out by this $\mathbf{Z}_2$ symmetry acting say only on the holomorphic part. That means we throw out the odd $p, q$ operators, non-invariant under the symmetry, and then use a $\tau \to -1/\tau$ transformation to construct the twisted sector. The resulting orbifold theory turns out to have a non-diagonal partition function, representing the corresponding member of the $D$ series. The $D$ series models equally have $Z_2$ symmetries, modding out by which takes us back to the corresponding $A$ series models. Further discussion of the $A$ and $D$ series may be found in Zuber's lectures and in section 9.

*8.6. Marginal operators*

A feature that distinguishes the $c = 1$ models $Z_{\text{circ}}(r)$ and $Z_{\text{orb}}(r)$ considered here from the $c < 1$ models is the existence of a parameter $r$ that labels a continuous family of theories. This is related to the possession by the former models of dimension $(1,1)$ operators, known as marginal operators. (More generally, operators of conformal weight $(h, \overline{h})$ are said to be relevant if $h + \overline{h} < 2$ and irrelevant if $h + \overline{h} > 2$.) Deformations of a conformal field theory, preserving the infinite conformal symmetry and central charge $c$, are generated by fields $V_i$ of conformal dimension (1,1) [69]. To first order, the perturbations they generate can be represented in the path integral as an addition to the action, $\delta S = \delta g_i \int dz d\bar{z} V_i(z,\bar{z})$, or equivalently in the correlation function of products of operators $\mathcal{O}$ as $\delta\langle\mathcal{O}\rangle = \delta g_i \int dz d\bar{z}\langle V_i(z,\bar{z})\mathcal{O}\rangle$. It is clear that a conformal weight (1,1) operator is required to preserve conformal invariance of the action at least at the classical level.

In the case of the circle theory (8.1), we have the obvious (1,1) operator $V = \partial X \bar{\partial} X$. We see that perturbing by this operator, since it is proportional to the Lagrangian, just changes the overall normalization of the action, which by a rescaling of $X$ can be absorbed into a change in the radius $r$. The operator $V$, invariant under $X \to -X$, evidently survives the $\mathbf{Z}_2$ orbifold projection in the untwisted sector, and remains to generate changes in the radius of the orbifold theory (8.24). (See [70] for further details concerning the marginal operators in $c = 1$ theories.) (In the Ashkin-Teller language of (8.25), the marginal operator at the two Ising decoupling point is given by $V = \varepsilon_1 \varepsilon_2$. This is the Ashkin-Teller interaction coupling the two Ising energy operators.)

In general whenever there exists a generic symmetry of a continuous family of modular invariant conformal field theories, modding out by the symmetry gives another continuous family of (orbifold) theories. From the operator point of view, this may be expressed as the fact the marginal operators generating the original family of theories are invariant under the symmetry. Hence they survive the projection in the untwisted sector of the orbifold theory and continue to generate a family of conformal theories.

The mere existence of (1,1) operators is not sufficient, however, to result in families of conformal theories. An additional "integrability condition" must be satisfied [69] to guarantee that the perturbation generated by the marginal operator does not act to change its own conformal weight from (1,1). In the case of a single marginal operator $V$ as above, this reduces in leading order to the requirement that there be no term of the form $C_{VVV}(z-w)^{-1}(\bar{z}-\bar{w})^{-1}V$ in the operator product of $V$ with itself. Otherwise the two-point function $\langle V(z,\bar{z})V(w,\bar{w})\rangle = (z-w)^{-2}(\bar{z}-\bar{w})^{-2}$ varies according to

$$\delta\langle V(z,\bar{z})V(w,\bar{w})\rangle = \delta g \int d^2z' \langle V(z,\bar{z})V(w,\bar{w})V(z',\bar{z}')\rangle$$
$$= \delta g\, 2\pi C_{VVV}(z-w)^{-2}(\bar{z}-\bar{w})^{-2}\log|z-w|^2,$$

showing that the conformal weight of $V$ is shifted to $(h,\bar{h}) = (1-\delta g\,\pi C_{VVV}, 1-\delta g\,\pi C_{VVV})$ under the perturbation generated by $V$. $V$ would therefore not remain marginal away from the point of departure, and could not be used to generate a one-parameter family of conformal theories.

To higher orders, we need to require as well the vanishing of integrals of $(n+2)$-point functions $(\delta g)^n \langle V(z,\bar{z})V(w,\bar{w}) \prod_i \int d^2z'_i V(z'_i,\bar{z}'_i)\rangle$ to insure that the 2-point function remains unperturbed. If this is the case, so the operator $V$ generates a one-parameter family of conformal theories, then it is called either exactly marginal, truly marginal, critical, persistent, or integrable, etc. In general, it is difficult to verify by examination of $(n+2)$-point functions that an operator remains marginal to all orders. In some cases, however, it is possible[71] to show integrability to all orders just by verifying that the 4-point function takes the form of that of the marginal operator $\partial X \bar{\partial} X$ for a free boson.

8.7. *The space of $c = 1$ theories*

It can be verified from (8.7) and (8.24) that the circle and orbifold partition functions coincide at

$$Z_{\text{orb}}\left(r = \frac{1}{\sqrt{2}}\right) = Z_{\text{circ}}\left(r = \sqrt{2}\right). \tag{8.26}$$

Although such an analysis of the partition functions shows the two theories at the above radii have identical spectra, it is not necessarily the case that they are identical theories, i.e. that their operator algebras are as well identical (although two conformal field theories whose partition functions coincide on arbitrary genus Riemann surfaces can probably be shown to be equivalent in this sense). We shall now proceed to show that the equivalence (8.26) does indeed hold at the level of the operator algebras of the theories by making use of a higher symmetry, in this case an affine $SU(2) \times SU(2)$ symmetry, possessed by the circle theory at $r = 1/\sqrt{2}$. Equivalences such as (8.26) show that geometrical interpretations of the target spaces of these models, as alluded to earlier, can be ambiguous at times. The geometrical data of a target space probed by a conformal field theory (or a string theory) can be very different from the more familiar point geometry probed by maps of a point (as opposed to loops) into the space.

We first note from (8.6) that $Z_{\text{circ}}(r)$ possesses a duality symmetry $Z_{\text{circ}}(r) = Z_{\text{circ}}(1/2r)$, in which the roles of winding and momentum are simply interchanged. (From (8.24), we recognize this as a symmetry also of the orbifold theory $Z_{\text{orb}}(r)$.) At the self-dual point $r = 1/\sqrt{2}$, we read off from (8.8) the eigenvalues of $L_0$ and $\overline{L}_0$ for the $|m, n\rangle$ states as $\frac{1}{4}(m \pm n)^2$. For $m = n = \pm 1$ we thus find two (1,0) states, and for $m = -n = \pm 1$ two (0,1) states. In operator language these states are created by the operators

$$J^{\pm}(z) = e^{\pm i\sqrt{2}\, x(z)} \qquad \text{and} \qquad \overline{J}^{\pm}(\overline{z}) = e^{\pm i\sqrt{2}\, \overline{x}(\overline{z})} \, , \qquad (8.27a)$$

with conformal weights (1,0) and (0,1). They become suitably single-valued under $x \to x + 2\pi r$ only at the radius $r = 1/\sqrt{2}$. At arbitrary radius, on the other hand, we always have the (1,0) and (0,1) oscillator states $\alpha_{-1}|0\rangle$ and $\overline{\alpha}_{-1}|0\rangle$, created by the operators

$$J^3(z) = i\partial x(z) \qquad \text{and} \qquad \overline{J}^3(\overline{z}) = i\overline{\partial}\overline{x}(\overline{z}) \, . \qquad (8.27b)$$

The operators $J^{\pm}, J^3$ in (8.27a,b) are easily verified to satisfy the operator product algebra

$$J^+(z)\, J^-(w) \sim \frac{e^{i\sqrt{2}(x(z)-x(w))}}{(z-w)^2} \sim \frac{1}{(z-w)^2} + \frac{i\sqrt{2}}{z-w}\, \partial x(w) \, ,$$

$$J^3(z)\, J^{\pm}(w) \sim \frac{\sqrt{2}}{z-w}\, J^{\pm}(w) \, ,$$

and similarly for $\overline{J}^{\pm}, \overline{J}^3$. If we define $J^{\pm} = \frac{1}{\sqrt{2}}(J^1 \pm iJ^2)$, then this algebra can be written equivalently as

$$J^i(z)\, J^j(w) = \frac{\delta^{ij}}{(z-w)^2} + \frac{i\sqrt{2}\,\epsilon^{ijk}}{z-w}\, J^k(w) \, . \qquad (8.28)$$

(8.28) defines what is known as the algebra of affine Kac-Moody $SU(2)$ at level $k = 1$ (level $k$ would be given by substituting $\delta^{ij} \to k\delta^{ij}$ in the first term on the right hand side of (8.28)).

For the terms in the mode expansions

$$J^i(z) = \sum_{n \in \mathbf{Z}} J^i_n\, z^{-n-1} \, , \qquad \text{where} \quad J^i_n = \oint \frac{dz}{2\pi i}\, z^n\, J^i(z) \, ,$$

we find by the standard method (as employed to determine (3.8)) the commutation relations

$$[J^i_n, J^j_m] = i\sqrt{2}\,\epsilon^{ijk}\, J^k_{n+m} + n\,\delta^{ij}\,\delta_{n+m,0} \, .$$

We see that the zero modes $J^i_0$ satisfy an ordinary $su(2)$ algebra (in a slightly irregular normalization of the structure constants corresponding to length-squared of highest root equal to 2), and the remaining modes $J^i_n$ provide an infinite dimensional generalization (known as an affinization) of the algebra. The generalization of this construction to arbitrary Lie algebras will be discussed in detail in the next section.

So we see that the circle theory $Z_{\text{circ}}(r)$ at radius $r = 1/\sqrt{2}$ has an affine $SU(2) \times SU(2)$ symmetry. It possesses at this point nine marginal operators, corresponding to combinations of the $SU(2) \times SU(2)$ currents $J^i\overline{J}^j$ ($i, j = 1, 2, 3$). But these are all related by $SU(2) \times SU(2)$ symmetry to the single marginal operator $J^3\overline{J}^3 = \partial X \overline{\partial} X$, which simply changes the compactification radius $r$. In fact, it is no coincidence that the enhanced symmetry occurs at the self-dual point since either of the chiral $SU(2)$ symmetries also relates the marginal operator $\partial X \overline{\partial} X$ to minus itself, rendering equivalent the directions of increasing and decreasing radius at $r = 1/\sqrt{2}$. (So one might say that there is only "half" a marginal operator at this point.)

To return to establishing the equivalence (8.26), we consider two possible ways of constructing a $\mathbf{Z}_2$ orbifold of the theory $Z_{\text{circ}}(1/\sqrt{2})$. Under the symmetry $X \to -X$ (so that $x \to -x$, $\overline{x} \to -\overline{x}$) discussed in detail earlier, we see that the affine $SU(2)$ generators (8.27) transform as $J^{\pm} \to J^{\mp}$, $J^3 \to -J^3$. The shift $X \to X + 2\pi/(2\sqrt{2})$ (shifting $x$ and $\overline{x}$ by the same amount) is also a symmetry of the action (8.1), and instead has the effect $J^\pm \to -J^\pm$, $J^3 \to J^3$. The effect of these two $\mathbf{Z}_2$ symmetry actions thus can be expressed as

$$\begin{array}{cccc} J^1 \to J^1 & \overline{J}^1 \to \overline{J}^1 & J^1 \to -J^1 & \overline{J}^1 \to -\overline{J}^1 \\ J^2 \to -J^2 & \overline{J}^2 \to -\overline{J}^2 \quad \text{and} \quad & J^2 \to -J^2 & \overline{J}^2 \to -\overline{J}^2 \\ J^3 \to -J^3 & \overline{J}^3 \to -\overline{J}^3 & J^3 \to J^3 & \overline{J}^3 \to \overline{J}^3 \ . \end{array}$$

But by affine $SU(2)$ symmetry, we see that these two symmetry actions are equivalent, one corresponding to rotation by $\pi$ about the 1-axis, the other to rotation by $\pi$ about the 3-axis.

The final step in demonstrating (8.26) is to note that modding out the circle theory at radius $r$ by a $\mathbf{Z}_n$ shift $X \to X + 2\pi r/n$ in general reproduces the circle theory, but at a radius decreased to $r/n$. Geometrically, the $\mathbf{Z}_N$ group generated by a rotation of the circle by $2\pi/n$ is an example of a group action with no fixed points, hence the resulting orbifold $S^1/\mathbf{Z}_n$ is a manifold — in this case topologically still $S^1$, but at the smaller radius. From the Hilbert space point of view, the projection in the untwisted sector removes the momentum states allowed at the larger radius, and the twisted sectors provide the windings appropriate to the smaller radius.

Modding out $Z_{\text{circ}}(1/\sqrt{2})$ by the $\mathbf{Z}_2$ shift $X \to X + 2\pi/(2\sqrt{2})$ thus decreases the radius by a factor of 2, giving $Z_{\text{circ}}(1/2\sqrt{2})$, which by $r \leftrightarrow 1/2r$ symmetry is equivalent to $Z_{\text{circ}}(\sqrt{2})$. Modding out $Z_{\text{circ}}(1/\sqrt{2})$ by the reflection $X \to -X$, on the other hand, by definition gives $Z_{\text{orb}}(1/\sqrt{2})$. Affine $SU(2) \times SU(2)$ symmetry thus establishes the equivalence (8.26) as a full equivalence between the two theories at the level of their operator algebras.

The picture[70][72][73] of the moduli space of $c = 1$ conformal theories that emerges is depicted in fig. 14. The horizontal axis represents compactification on a circle $S^1$ with radius $r_{\text{circle}}$, and the vertical axis represents compactification on the $S^1/\mathbf{Z}_2$ orbifold with radius $r_{\text{orbifold}}$. As previously mentioned, the former is also known as the gaussian model, and the latter is equivalent to the critical Ashkin-Teller model (which also encompasses two other of the models described in Cardy's lectures, namely the 6-vertex model and the 8-vertex model on its critical line). The regions represented by dotted lines are determined by the duality $r \leftrightarrow \frac{1}{2r}$.

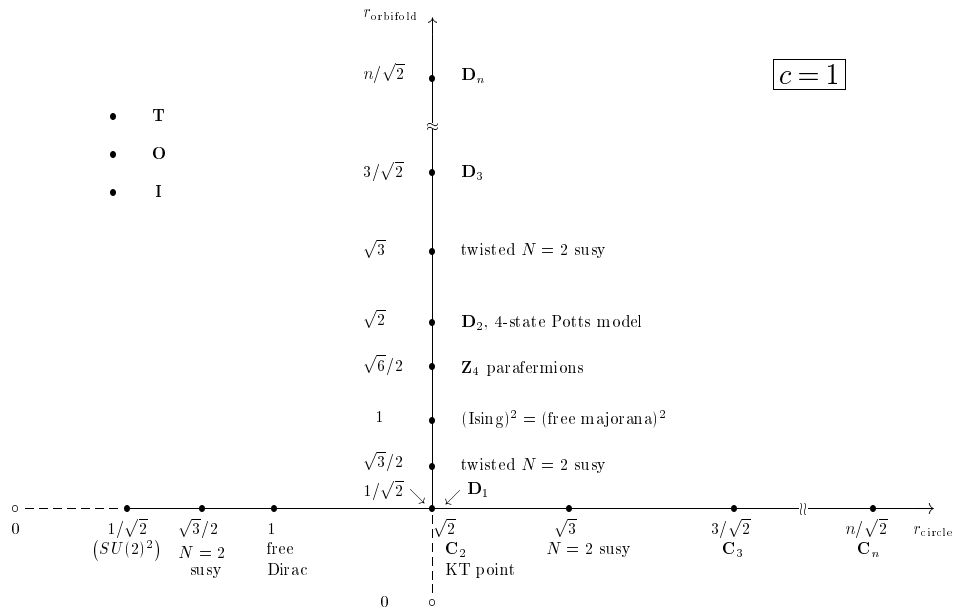

Fig. 14. Survey of conformal field theory at $c = 1$.

In fig. 14, we have indicated some of the special radii $r = 1/\sqrt{2}, 1, \sqrt{2}$ that we have discussed. The partition function at the common point (8.26) of the two lines turns out to correspond to the continuum limit Kosterlitz-Thouless point of the X-Y model on the torus[69]. At this point there are five marginal operators, $J_1 \overline{J}_1$ and $J_i \overline{J}_j$ ($i, j = 2, 3$), that survive the projection under the group action $x \to -x$. In this language, $J_3 \overline{J}_3$ again generates changes in the circle radius $r$, and the remaining 4 operators, all equivalent to one another due to the $U(1) \times U(1)$ symmetry generated by $J_3$ and $\overline{J}_3$, instead deform the theory in the orbifold direction of fig. 14. This is the only such multicritical point in the figure where there exist inequivalent directions of deformation[70].

Two other special radii for circle compactifications are $r = \sqrt{3}/2$ and $\sqrt{3}$, where four operators of dimension $(\frac{3}{2}, \frac{3}{2})$ appear, corresponding to a GSO projected system with $N = 2$ supersymmetry[74][75]. (The chiral spin-3/2 vertex operators take the form $\exp(\pm i\sqrt{3}x(z))$, $\exp(\pm i\sqrt{3}\overline{x}(\overline{z}))$.) The corresponding points $r = \sqrt{3}/2, \sqrt{3}$ on the orbifold line realize a twisted $N = 2$ supersymmetry algebra[35][36] that contains an $N = 1$ supersymmetry surviving the $\mathbf{Z}_2$ projection[75][76]. (Actually the partition functions at the points $r = \sqrt{3}/2$ and $r = \sqrt{3}$ on the circle line differ by a constant, equal to 2 (and for the same points on the orbifold line the difference of the partition functions is 1). This is because these theories are actually $\mathbf{Z}_2$ orbifolds of one another[77], and the difference of their partition functions is $\text{tr}(-1)^F$ in the Ramond sector, which is a constant due to superconformal invariance. By examination of the partition functions (8.7), this relationship can be used to provide a simple superconformal proof of the Euler pentagonal number theorem (7.30).) $r = \sqrt{6}/2$ on the orbifold line realizes a modular invariant combination of $\mathbf{Z}_4$ parafermions[78]. (Other properties of $c = 1$ models have also been considered in [79].)

The $\mathbf{Z}_2$ orbifolding that took us from the affine $SU(2) \times SU(2)$ point to the multicritical point at $r = \sqrt{2}$ on the circle line can be generalized. Indeed we can mod out by any of the discrete subgroups $\Gamma$ of the diagonal $SU(2)$. It is easiest think of this in terms of subgroups of $SO(3)$ acting simultaneously on the vectors $J^i(z), \overline{J}^i(\overline{z})$. Then the generator of the symmetry group $\mathbf{C}_n$, the cyclic group of rotations of order $n$ about the 3-axis, corresponds to the action $X \to X + 2\pi/(n\sqrt{2})$ (i.e. $J^\pm \to e^{\pm 2\pi i/n} J^\pm$, $J^3 \to J^3$, and similarly for $\overline{J}$'s). The additional generator adjoined to give the dihedral group $\mathbf{D}_n$ corresponds to $X \to -X$ ($J_3 \to -J_3$, $J^\pm \to J^\mp$). Modding out by the $\mathbf{C}_n$'s thus gives points on the circle line at radius $r = n/\sqrt{2}$, and modding out by the $\mathbf{D}_n$'s gives the corresponding points on the orbifold line, as indicated in fig. 14.

Something special happens, however, for the tetrahedral, octahedral, and icosahedral groups, $\mathbf{T}$, $\mathbf{O}$, and $\mathbf{I}$. For these it is easy to see that the only (1,1) operator that is invariant under the full discrete group is $V = \sum_{i=1}^{3} J_i \overline{J}_i$, which is hence the only marginal operator that survives the projection. But recalling that our affine $SU(2)$ currents satisfy (8.28), we easily verify that $C_{VVV} = -2$ for $V = \sum_{i=1}^{3} J_i \overline{J}_i$. This means[72] that the $SU(2)/\Gamma$ orbifold models for $\Gamma = \mathbf{T}, \mathbf{O}, \mathbf{I}$ are isolated points in the moduli space for $c = 1$ conformally invariant theories, as depicted in fig. 14. This absence of truly marginal operators is intuitively satisfactory for these cases since we are modding out by symmetries that exist only at a given fixed radius, the $SU(2) \times SU(2)$ radius $r = 1/\sqrt{2}$, and hence modding out by the symmetries effectively freezes the radius. Further properties of the $SU(2)$ orbifold models are discussed in [80], and an identification of critical RSOS-type models that have the same partition functions is included in [72].

Part of the motivation for studying $c = 1$ systems is that they represent the first case beyond the classification methods discussed in section 4. For systems with $N = 1$ superconformal symmetry (5.16), the corresponding boundary case between the (classified) discrete series and (unclassified) continuum lies at $\hat{c} = 1$. The analog of fig. 14 for this case may be found in [77].

## 9. Affine Kac-Moody algebras and coset constructions

### 9.1. Affine algebras

In the previous section, we saw the important role played by affine $SU(2)$ at level $k = 1$ in characterizing the enhanced symmetry at the point $r = 1/\sqrt{2}$ on the circle line. We now wish to consider the generalization of this construction to arbitrary groups and arbitrary level. We begin by considering a set of $(1,0)$ conformal fields $J^a(z)$, called currents (where $a$ labels the different currents). Dimensional analysis constrains their operator products to take the form

$$J^a(z) J^b(w) = \frac{\widetilde{k}^{ab}}{(z-w)^2} + \frac{i f^{abc}}{z-w} J^c(w) + \dots ,  \qquad (9.1)$$

where the $f^{abc}$'s are necessarily antisymmetric in $a$ and $b$. Furthermore, associativity of the operator products can be used to show that the $f^{abc}$'s satisfy as well a Jacobi identity. That means that they constitute the structure constants of some Lie algebra $\mathcal{G}$, which we shall assume in what follows to be that associated to a compact Lie group $G$ (i.e. to have a positive definite Cartan metric). For each simple component of the algebra we can choose a basis in which the central extension $\widetilde{k}^{ab} = \widetilde{k}^a \delta^{ab}$. The operator product (9.1) is the operator product for what is known as an affine, or affine Kac-Moody, algebra (for a recent review, see [3]), or a 2d current algebra. Affine algebras play an important role in closed string theory, where they provide the worldsheet realization of spacetime gauge symmetries. They also provide many new non-trivial examples of exactly solvable quantum field theories in two dimensions, and may ultimately play a role in the classification program of two dimensional conformal field theories at arbitrary $c$.

In terms of the mode expansion $J^a(z) = \sum_{n\in \mathbf{Z}} J^a_n z^{-n-1}$, we find from (9.1) the commutators

$$[J^a_m, J^b_n] = if^{abc} J^c_{m+n} + \widetilde{k}\, m\, \delta^{ab}\, \delta_{m+n,0}\,, \tag{9.2}$$

where we have restricted for simplicity to the case that the $f^{abc}$ are the structure constants associated to a simple Lie group $G$. (9.2) by definition defines the untwisted affine algebra $\widehat{\mathcal{G}}$ associated with a compact finite-dimensional lie algebra $\mathcal{G}$, where $m, n \in Z$; and $a, b, c$ run over the values 1 to $|G| \equiv \dim G$. We see that the subalgebra of zero modes $J^a_0$ constitutes an ordinary Lie algebra, known as the horizontal Lie subalgebra, in which the $c$-number central extension $\widetilde{k}$ does not appear. The full infinite set of $J^a_n$'s provides what is known as an "affinization" of the finite dimensional subalgebra of $J^a_0$'s. As in (7.1), we can pull back $J(z)$ to the cylinder, so that we have the Fourier series $J^a_{\mathrm{cyl}}(w) = \sum_n J^a_n\, e^{-nw}$. With $w$ real, we recognize the modes $J^a_n$ as the infinitesimal generators of the group of gauge transformations $g(\sigma): S^1 \to G$ on the circle.

The representation theory of affine algebras shares many features with that of the Virasoro algebra. For example, regularity of $J(z)|0\rangle$ at $z = 0$ requires that

$$J^a_n|0\rangle = 0 \quad \text{for} \quad n \geq 0\,.$$

There also exists a notion of primary field $\varphi^\ell_{(r)}$ (actually a multiplet of fields) with respect to the affine algebra, for which the operator product has the leading singularity

$$J^a(z)\,\varphi_{(r)}(w) \sim \frac{t^a_{(r)}}{z - w}\,\varphi_{(r)}(w) + \ldots\,. \tag{9.3}$$

This should be recognized as the statement that $\varphi_{(r)}$ transforms as some representation $(r)$ of $G$, where the right hand side is shorthand for $(t^a_{(r)})^{\ell k}\varphi^k_{(r)}$, and

$t^a_{(r)}$ are representation matrices for $G$ in the representation $(r)$. These primary fields create states, called highest weight states,

$$|(r)\rangle \equiv \varphi_{(r)}(0)|0\rangle \tag{9.4}$$

(again a multiplet of states), that provide a representation of the zero mode algebra

$$J^a_0|(r)\rangle = t^a_{(r)}|(r)\rangle\,, \quad \text{with} \quad J^a_n|(r)\rangle = 0 \quad (n > 0)\,. \tag{9.5}$$

The Ward identities for affine symmetry take the form

$$\begin{aligned}\langle J^a(z)\,\varphi_{(r_1)}(w_1,\overline{w}_1)\ldots\varphi_{(r_n)}(w_n,\overline{w}_n)\rangle \\ = \sum_{j=1}^n \frac{t^a_{(r_j)}}{z - w_j}\langle \varphi_{(r_1)}(w_1,\overline{w}_1)\ldots\varphi_{(r_n)}(w_n,\overline{w}_n)\rangle\,.\end{aligned} \tag{9.6}$$

These are derived as was (2.22) by computing the contour integral $\int \frac{dz}{2\pi i}\alpha^a(z)J^a(z)$ inserted in a correlation function of $\varphi_{(r_j)}$'s, where the contour encloses all of the points $w_j$ (as in fig. 3) and the $\alpha^a(z)$'s parametrize an infinitesimal local $G$-transformation. Then by deforming the contour to a sum of small contours around each of the $w_j$'s we find from (9.3)

$$\begin{aligned}\int \frac{dz}{2\pi i}\,\alpha^a(z)\,\langle J^a(z)\,\varphi_{(r_1)}(w_1,\overline{w}_1)\cdots\varphi_{(r_n)}(w_n,\overline{w}_n)\rangle \\ = \sum_{j=1}^n \langle\varphi_{(r_1)}(w_1,\overline{w}_1)\cdots\delta_\alpha\varphi_{(r_j)}(w_j,\overline{w}_j)\cdots\varphi_{(r_n)}(w_n,\overline{w}_n)\rangle\,,\end{aligned}$$

where $\delta_\alpha\varphi_{(r_j)} = \alpha^a t^a_{(r_j)}\varphi_{(r_j)}$ is by definition the change in $\varphi_{(r_j)}$ under the infinitesimal $G$ transformation parametrized by $\alpha$. We shall see a bit later how (9.6) may be used to derive first-order differential equations for Green functions involving primary fields $\varphi_{(r_j)}$.

9.2. Enveloping Virasoro algebra

The algebraic structure (9.1), characterizing an affine or current algebra, turns out to incorporate as well a natural definition of a stress-energy tensor $T(z)$. Equivalently, we may construct generators $L_n$ of a Virasoro algebra in terms of the modes $J_n^a$, thereby making contact with the Virasoro representation theory detailed earlier.

Recall that for a single boson, the natural $(2,0)$ object was $T(z) = -\frac{1}{2}{:}\partial x(z)\partial x(z){:} = \frac{1}{2}{:}J^3(z)J^3(z){:}$, where $J^3 = i\partial x$. (In the language of affine algebras, this is the case $G = U(1)$, with central charge $c = 1$.) The natural group invariant generalization is

$$T(z) = \frac{1}{\beta}\sum_{a=1}^{|G|} {:}J^a(z)J^a(z){:} = \lim_{z\to w}\sum_{a=1}^{|G|} J^a(z)J^a(w) - \frac{\widetilde{k}|G|}{(z-w)^2} . \qquad (9.7)$$

The constant $\beta$ above is fixed either by requiring that $T(z)$ satisfy the canonical operator product (3.1), or by requiring that the $J^a(z)$'s indeed transform as dimension $(1,0)$ primary fields.

Implementing the latter approach, we write the singular terms in the operator product expansion

$$T(z)J^a(w) = \frac{J^a(w)}{(z-w)^2} + \frac{\partial J^a(w)}{z-w} , \qquad (9.8a)$$

implying the commutations relations

$$[L_m, J_n^a] = -nJ_{m+n}^a \qquad (9.8b)$$

for the modes of $T$ and $J$. From (9.7), we have

$$L_n = \frac{1}{\beta}\sum_{m=-\infty}^{\infty} {:}J_{m+n}^a J_{-m}^a{:} , \qquad (9.9)$$

so that applying $L_{-1}$ to a highest weight state and using (9.5) gives

$$L_{-1}|(r)\rangle = \frac{2}{\beta} J_{-1}^a t_{(r)}^a |(r)\rangle .$$

We next apply $J_1^b$ to both sides and use (9.2) and (9.8b) to get

$$t_{(r)}^b|(r)\rangle = \frac{2}{\beta}(if^{bac} J_0^c + \widetilde{k}\delta^{ab}) t_{(r)}^a |(r)\rangle$$
$$= \frac{2}{\beta}(if^{bac}\tfrac{1}{2}if^{dca} t_{(r)}^d + \widetilde{k}t_{(r)}^b)|(r)\rangle$$
$$= \frac{2}{\beta}\left(\tfrac{1}{2}C_A + \widetilde{k}\right) t_{(r)}^b |(r)\rangle ,$$

139

where the quadratic casimir $C_A$ of the adjoint representation is defined by $f^{acd}f^{bcd} = C_A \delta^{ab}$. We conclude that consistency of (9.7) with (9.8) requires that

$$\beta = 2\widetilde{k} + C_A . \qquad (9.10)$$

At this point it is now straightforward to check that the stress-energy tensor

$$T(z) = \frac{1/2}{\widetilde{k} + C_A/2}\sum_{a=1}^{|G|} {:}J^a(z)J^a(z){:} \qquad (9.11)$$

satisfies as well the canonical operator product expansion (3.1), with leading singularity

$$T(z)T(w) \sim \frac{c_G/2}{(z-w)^4} + \ldots$$

given by the central charge

$$c_G = \frac{\widetilde{k}\,|G|}{\widetilde{k} + C_A/2} . \qquad (9.12)$$

The stress-energy tensor (9.11), quadratic in the currents, is known as the Sugawara form of the stress-energy tensor. Historically, the normalization (9.10) was the culmination of effort by numerous parties (see [3] for extensive references).

The number $C_A/2$ depends in general on the normalization chosen for the structure constants $f^{abc}$. Since its value plays an important role in what follows, we digress briefly to introduce some of the necessary group theoretic notation. If we write

$$\operatorname{tr} t_{(r)}^a t_{(r)}^b = \ell_r \delta^{ab} \qquad (9.13)$$

for an arbitrary representation $(r)$ of $G$ of dimension $d_r$, then summing over $a, b = 1, \ldots, |G|$ gives

$$C_r d_r = \ell_r |G| , \qquad (9.14)$$

where $C_r$ is the quadratic Casimir of the representation. Summing only over the Cartan subalgebra of $G$ ($a, b = 1, \ldots, r_G$), on the other hand, gives

$$\sum_{j=1}^{d_r} \mu_{(j)}^2 = \ell_r r_G , \qquad (9.15)$$

140

where $r_G$ is the rank of the group $G$ and the $\mu$ are the weights of the representation $(r)$.

For the adjoint representation, we have $d_A = |G|$ and $C_A = \ell_{(A)} = r_G^{-1} \sum_{a=1}^{|G|} \alpha_{(a)}^2$, where $\alpha$ are the roots. If we let $\psi$ denote the highest root, then the normalization independent quantity $\widetilde{h}_G \equiv C_A/\psi^2$, known as the dual Coxeter number, satisfies

$$\widetilde{h}_G \equiv \frac{C_A}{\psi^2} = \frac{1}{r_G} \left( n_L + \left(\frac{S}{L}\right)^2 n_S \right) . \tag{9.16}$$

In (9.16), $n_{S,L}$ are the number of short and long roots of the algebra (the highest root $\psi$ is always a long root), and $(S/L)^2$ is the ratio of their squared lengths (roots of simple Lie algebras come at most in two lengths). Those algebras associated to Dynkin diagrams with only single lines, i.e. $SU(n), SO(2n), E_{6,7,8}$, are called "simply-laced", and have roots all of the same length. (In more mathematical circles these are known as the $(A, D, E)$ series of algebras. In general, the Coxeter number itself is the order of the Coxeter element of the Weyl group, by definition the product of the simple Weyl reflections. The Coxeter number is also equal to the number of (non-zero) roots divided by the rank of the algebra, and coincides with the dual Coxeter number only for the simply-laced algebras.) The remaining algebras have roots of two lengths, their ratio $(L/S)$ either $\sqrt{2}$ (for $SO(2n+1)$, $Sp(2n)$, $F_4$) or $\sqrt{3}$ (for $G_2$).

Equation (9.16) allows us to tabulate the dual Coxeter numbers for all the compact simple Lie algebras:

$$\begin{aligned}
&SU(n) \ (n \geq 2): \ \widetilde{h}_{SU(n)} = n, \quad \ell_{(n)} = \tfrac{1}{2}\psi^2 \\
&SO(n) \ (n \geq 4): \ \widetilde{h}_{SO(n)} = n-2, \quad \ell_{(n)} = \psi^2 \\
&E_6: \ \widetilde{h}_{E_6} = 12, \quad \ell_{(27)} = 3\psi^2 \qquad E_7: \ \widetilde{h}_{E_7} = 18, \quad \ell_{(56)} = 6\psi^2 \\
&\qquad E_8: \ \widetilde{h}_{E_8} = 30, \quad \ell_{(248)} = 30\psi^2 \\
&Sp(2n) \ (n \geq 1): \ \widetilde{h}_{Sp(2n)} = n+1, \quad \ell_{(2n)} = \tfrac{1}{2}\psi^2 \\
&G_2: \ \widetilde{h}_{G_2} = 4, \quad \ell_{(7)} = \psi^2 \qquad F_4: \ \widetilde{h}_{F_4} = 9, \quad \ell_{(26)} = 3\psi^2 \ .
\end{aligned} \tag{9.17}$$

We see that the dual Coxeter number is always an integer. In (9.17) we have also tabulated the index $\ell_r$, as defined in (9.13), for the lowest dimensional representations as a function of $\psi^2$.

### 9.3. Highest weight representations

In what follows, we shall be interested in so-called irreducible unitary highest weight representations of the algebra (9.2). This means that the highest weight states transform as an irreducible representation of the ordinary Lie algebra of zero modes $J_0^a$ (the horizontal subalgebra), as in (9.5). Since these are also the states in a given irreducible representation of the affine algebra with the smallest eigenvalue of $L_0$, we shall frequently refer to the multiplet of states (9.4) as the vacuum states, and $(r)$ as the vacuum representation. The states at any higher level, i.e. higher $L_0$ eigenvalue, will also transform as some representation of the horizontal subalgebra, although only the lowest level necessarily transforms irreducibly.

Unitarity is implemented as the condition of hermiticity on the generators, $J^{a\dagger}(z) = J^a(z)$. By the same argument leading to (3.12) in the case of the Virasoro algebra, we see that this implies $J_n^{a\dagger} = J_{-n}^a$. In a Cartan basis the $J^a(z)$'s are written $H^i(z)$ and $E^{\pm\alpha}(z)$, where $i = 1, \ldots, r_G$ labels the mutually commuting generators, and the positive roots $\alpha$ label the raising and lowering operators. In this basis the truly highest weight state $|\lambda\rangle \equiv |(r), \lambda\rangle$ of the vacuum representation satisfies

$$H_n^i|\lambda\rangle = E_n^{\pm\alpha}|\lambda\rangle = 0, \quad n > 0,$$

$$H_0^i|\lambda\rangle = \lambda^i|\lambda\rangle, \quad \text{and} \quad E_0^\alpha|\lambda\rangle = 0, \ \alpha > 0 \ .$$

New states are created by acting on the state $|\lambda\rangle$ with the $E_0^{-\alpha}$'s or any of the $J_{-n}^a$'s for $n > 0$.

Now we wish to consider the quantization condition on the central extension $\widetilde{k}$ in (9.2). It is evident that $\widetilde{k}$ depends on the normalization of the structure constants. We shall show that the normalization independent quantity $k \equiv 2\widetilde{k}/\psi^2$, known as the level of the affine algebra, is quantized as an integer in a highest weight representation. (Equivalently, in a normalization in which the highest root $\psi$ satisfies $\psi^2 = 2$, we have $\widetilde{k} = k \in \mathbf{Z}$. The normalization condition $\psi^2 = 2$ on the structure constants is easily translated into a condition on the index $\ell_r$ for the lowest dimensional representations listed in (9.17).) In terms of the integer quantities $k$ and $\widetilde{h}_G$, we may rewrite the formula (9.12) for the central charge as

$$c_G = \frac{k\,|G|}{k + \widetilde{h}_G}\ . \tag{9.18}$$

As an example, we see from (9.17) that $\widetilde{h}_{SU(2)} = 2$, so for the lowest level $k = 1$ we find from (9.18) that $c_{SU(2)} = 3/(1+2) = 1$. Thus we infer that the realization of affine $SU(2)$ provided at radius $r = 1/\sqrt{2}$ on the ($c = 1$) circle line is at level $k = 1$.

To establish the quantization condition on $k$, we first consider the case $G = SU(2)$. Note that the normalization of structure constants, $f^{ijk} = \sqrt{2}\epsilon^{ijk}$, in (8.28) corresponds to the aforementioned $\psi^2 = 2$. Because of the $\sqrt{2}$ in the commutation rules, we need to take

$$I^{\pm} = \frac{1}{\sqrt{2}}(J_0^1 \pm iJ_0^2) \quad \text{and} \quad I^3 = \frac{1}{\sqrt{2}}J_0^3 \tag{9.19a}$$

to give a conventionally normalized $su(2)$ algebra $[I^+, I^-] = 2I^3$, $[I^3, I^{\pm}] = \pm I^{\pm}$, in which $2I^3$ has integer eigenvalues in any finite dimensional representation. But from (9.2) we find that

$$\widetilde{I}^+ = \frac{1}{\sqrt{2}}(J_{+1}^1 - iJ_{+1}^2)\,,\ \ \widetilde{I}^- = \frac{1}{\sqrt{2}}(J_{-1}^1 + iJ_{-1}^2)\,,\ \ \text{and}\ \ \widetilde{I}^3 = \tfrac{1}{2}k - \frac{1}{\sqrt{2}}J_0^3 \tag{9.19b}$$

as well satisfy $[\widetilde{I}^+, \widetilde{I}^-] = 2\widetilde{I}^3$, $[\widetilde{I}^3, \widetilde{I}^{\pm}] = \pm\widetilde{I}^{\pm}$, so $2\widetilde{I}^3 = k - 2I^3$ also has integer eigenvalues. It follows that $k \in \mathbf{Z}$ for unitary highest weight representations.

This argument is straightforwardly generalized by using the canonical $su(2)$ subalgebra

$$I^{\pm} = E_0^{\pm\psi}\,, \quad I^3 = \psi \cdot H_0/\psi^2 \tag{9.20a}$$

generated by the highest root $\psi$ of any Lie algebra. From (9.2),

$$\widetilde{I}^{\pm} = E_{\pm 1}^{\mp\psi}\,, \quad \widetilde{I}^3 = (\widetilde{k} - \psi \cdot H_0)/\psi^2 \tag{9.20b}$$

also form an $su(2)$ subalgebra, implying that the level $k = 2\widetilde{k}/\psi^2 = 2\widetilde{I}^3 + 2I^3$ is quantized for unitary highest weight representations of affine algebras based on arbitrary simple Lie algebras.

We pause here to remark that the quantization condition on $k$ also follows [81] from the quantization of the coefficient of the topological term $\Gamma = \frac{1}{24\pi}\int \mathrm{tr}(g^{-1}dg)^3$ in the Wess-Zumino-Witten lagrangian,

$$S = \frac{1}{4\lambda^2}\int d^2\xi\,\mathrm{tr}(\partial_\mu g)(\partial^\mu g^{-1}) + k\Gamma = k\left(\frac{1}{16\pi}\int \mathrm{tr}(\partial_\mu g)(\partial^\mu g^{-1}) + \Gamma\right)\,, \tag{9.21}$$

for a two dimensional $\sigma$-model with target space the group manifold of $G$. In (9.21) we have substituted the value of the coupling $\lambda$ for which the model becomes conformally invariant. The currents $J = J^a t^a \sim \partial g g^{-1}$, $\overline{J} = \overline{J}^a t^a \sim g^{-1}\overline{\partial}g$, derived from the above action, satisfy the equations of motion $\overline{\partial}J = \partial\overline{J} = 0$. This factorization of the theory was shown in [81] to imply an affine $G\times\overline{G}$ symmetry, and theories of the form (9.21) were analyzed extensively from this point of view in [82][83]. More details and applications of these theories may be found in Affleck's lectures.

Before turning to other features of the representation theory of (9.2), we continue briefly the discussion of the conformal Ward identities (9.6). First we recall from (9.11) that

$$L_{-1} = \frac{1}{\widetilde{k} + C_A/2}(J_{-1}^a J_0^a + J_{-2}^a J_1^a + \ldots)$$

(where the factor of $1/2$ in the numerator of (9.11) is compensated by the appearance of each term exactly twice in the normal ordered sum (9.9)). Acting on a primary field, we thus find the null field

$$\left(L_{-1} - \frac{\sum_a J_{-1}^a t_{(r)}^a}{\widetilde{k} + C_A/2}\right)\varphi_{(r)} = 0\ . \tag{9.22}$$

(9.22) implies that correlation functions involving $n$ primary fields satisfy $n$ first-order differential equations. To derive them, we multiply (9.6) by $t_{(r_k)}^a$, take $z \to w_k$ and use the operator product expansion (9.1), giving finally[82]

$$\left((\widetilde{k} + C_A/2)\frac{\partial}{\partial w_k} + \sum_{\substack{j\neq k \\ a}}\frac{t_{(r_j)}^a t_{(r_k)}^a}{w_j - w_k}\right)\langle\varphi_{r_1}(w_1)\ldots\varphi_{r_n}(w_n)\rangle = 0\ . \tag{9.23}$$

The first-order equations (9.23) for each of the $w_k$, together with their anti-holomorphic analogs, can be solved subject to the constraints of crossing symmetry, monodromy conditions, and proper asymptotic behavior. The simplest solution involves a symmetric holomorphic/anti-holomorphic pairing, and corresponds to the correlation functions of the $\sigma$-model (9.21).

Returning now to (9.11), we observe that the vacuum state (9.4) in general has $L_0$ eigenvalue

$$L_0|(r)\rangle = \frac{1/2}{\widetilde{k} + C_A/2} \sum_{a,m} :J^a_m J^a_{-m}: |(r)\rangle$$
$$= \frac{1/2}{\widetilde{k} + C_A/2} \sum_a t^a_{(r)} t^a_{(r)} |(r)\rangle = \frac{C_r/2}{\widetilde{k} + C_A/2} |(r)\rangle , \quad (9.24a)$$

where $C_r$ is the quadratic Casimir of the representation $(r)$. The conformal weight of the primary multiplet $\varphi_{(r)}(z)$ is thus

$$h_r = \frac{C_r/2}{\widetilde{k} + C_A/2} = \frac{C_r/\psi^2}{k + \widetilde{h}_G} . \quad (9.24b)$$

For the case $G = SU(2)$ with ground state transforming as the spin-$j$ representation of the horizontal $su(2)$, (9.24) gives

$$L_0|(j)\rangle = \frac{j(j+1)}{k+2}|(j)\rangle \quad (9.25)$$

(where the quadratic Casimir satisfies $C_{(j)} = 2j(j+1)$ in a normalization of $su(2)$ with $\psi^2 = 2$). For affine $SU(2)$ at level $k = 1$ we find the values $h = 0, \frac{1}{4}$ for $j = 0, \frac{1}{2}$.

We can easily see how these conformal weights enter into the partition function at the $SU(2) \times SU(2)$ point $r = 1/\sqrt{2}$ of the circle theory considered in the previous section. By steps similar to those in (8.13), we can write the partition function (8.7) in the form

$$Z_{\text{circ}}\left(\frac{1}{\sqrt{2}}\right) = \chi_{(0),1}\overline{\chi}_{(0),1} + \chi_{(1/2),1}\overline{\chi}_{(1/2),1} , \quad (9.26)$$

where

$$\chi_{(0),1}(q) = \frac{1}{\eta}\sum_{n=-\infty}^{\infty} q^{n^2} , \qquad \chi_{(1/2),1}(q) = \frac{1}{\eta}\sum_{n=-\infty}^{\infty} q^{(n+\frac{1}{2})^2} . \quad (9.27)$$

We see that the values $h = 0, \frac{1}{4}$ emerge as the conformal weights of the leading terms of the quantities $\chi_{(0),1}$ and $\chi_{(1/2),1}$. (9.26) corresponds to a decomposition of the partition function in terms of characters of an extended chiral algebra, here affine $SU(2) \times SU(2)$. A bit later we will discuss affine characters at arbitrary level.

There exists a simple constraint on the possible vacuum representations $(r)$ allowed in a unitary highest weight realization of (9.2) at a given level $k$. To see this most easily, we return again to $G = SU(2)$. We take our "vacuum" $|(r)\rangle$ in the spin-$j$ representation of $SU(2)$. The $2j+1$ states of this representation are labeled as usual by their $I^3$ eigenvalue, $I^3|(j), m\rangle = m|(j), m\rangle$, where $I^3$ is as defined in (9.19a). Using the other $su(2)$ generators (9.19b), we derive the most stringent condition by considering the state $|j\rangle \equiv |(j), j\rangle$ with highest isospin $m = j$,

$$0 \leq \langle j|\widetilde{I}^+\widetilde{I}^-|j\rangle = \langle j|[\widetilde{I}^+, \widetilde{I}^-]|j\rangle = \langle j|k - 2I^3|j\rangle = k - 2j . \quad (9.28)$$

It follows that only ground state representations with

$$2j \leq k \quad (9.29)$$

are allowed. For a given $k$, these are the $k+1$ values $j = 0, \frac{1}{2}, 1, \ldots, \frac{k}{2}$. Thus it is no coincidence that the $SU(2)$ level $k = 1$ partition function (9.26) is composed of only $j = 0, \frac{1}{2}$ characters.

The generalization of (9.29) to arbitrary groups is more or less immediate. Instead of $|j\rangle$ we consider $|\lambda\rangle$, where $\lambda$ is highest weight of the vacuum representation. Then from (9.28) using instead the $\widetilde{I}^i$'s of (9.20b), we find

$$2\psi \cdot \lambda/\psi^2 \leq k . \quad (9.30)$$

(For $SU(n)$ this condition on allowed vacuum representations turns out in general to coincide with the condition that the width of their Young tableau be less than the level $k$. For $SU(2)$, for which the spin-$j$ representation is the symmetric combination of $2j$ spin-$\frac{1}{2}$ representations, this is already manifest in (9.29).)

The assemblage of states created by acting on the highest weight states $|(r)\rangle$ with the $J^a_{-n}$'s again constitutes a Verma module. As was the case for the $c<1$ representations of the Virasoro algebra, this module will in general contain null states which must be removed to provide an irreducible representation of the affine algebra. In the case at hand, it can be shown that all the null states are descendants of a single primitive null state. This state is easily constructed for a general affine algebra by using the generators (9.20$b$) of the (non-horizontal) $su(2)$ subalgebra. Note that the eigenvalue of $2\widetilde{I}^3$ acting on the highest weight state $|(r),\lambda\rangle$ of the vacuum representation is given by $M = k - 2\psi\cdot\lambda/\psi^2$. For the affine representations of interest, the set of states generated by acting with successive powers of $\widetilde{I}^-$ on $|(r),\lambda\rangle$ forms a finite dimensional irreducible representation of the $su(2)$ subalgebra (9.20$b$). Thus $M$ is an integer and

$$\left(\widetilde{I}^-\right)^{M+1}|(r),\lambda\rangle = 0 \ .$$

This is the primitive null state mentioned above, whose associated null field $\left(\widetilde{I}^-\right)^{M+1}\phi_{(r),\lambda}$ can be used to generate all non-trivial selection rules[82][83] in the theory. In the case of a level $k$ representation of affine $SU(2)$, the above null state becomes $\left(J^+_{-1}\right)^{k+1}|0\rangle = 0$ for the basic representation, or more generally $\left(J^+_{-1}\right)^{k-2j+1}|(j),j\rangle = 0$ for the spin-$j$ representation.

### 9.4. Some free field representations

In the case of the Virasoro algebra, we found a variety of useful representations afforded by free bosons and fermions. Free systems can also be used to realize particular representations of affine algebras. For example, we take $N$ free fermions $\psi^i$ with operator product algebra

$$\psi^i(z)\psi^j(w) = -\frac{\delta^{ij}}{z-w} \ .$$

We consider these fermions to transform in the vector representation of $SO(N)$, with representation matrices $t^a$. Then for $N \geq 4$, the currents

$$J^a(z) = \psi(z)t^a\psi(z) \qquad (9.31)$$

are easily verified to satisfy (9.1) for $SO(N)$ at level $k=1$. We also verify from (9.17) and (9.18) that

$$c_{SO(N),k=1} = \frac{1\frac{1}{2}N(N-1)}{1+(N-2)} = \tfrac{1}{2}N \ , \qquad (9.32)$$

consistent with the central charge for $N$ free fermions. (For $N=3$, we would find instead a level $k=2$ representation of $SU(2)$ with $c=\frac{3}{2}$). The free fermion representation (9.31) provides the original context in which affine algebras arose as two dimensional current algebras.

We could equivalently use $N$ complex fermions taken to transform in the vector representation of $SU(N)$, and construct currents $J^a(z) = \psi^*(z)t^a\psi(z)$ analogous to (9.31). These realize affine $SU(N)\times U(1)$, with the $SU(N)$ at level $k=1$. (The notion of level for an abelian $U(1)$ current algebra is more subtle than we need to discuss here — for our purposes it will suffice to recall that it always has $c=1$, and the current has the free bosonic realization $J=i\partial x$.) The central charge comes out as

$$c_{U(1)} + c_{SU(N),k=1} = 1 + \frac{1(N^2-1)}{1+N} = N \ ,$$

consistent with the result for $N$ free complex fermions.

Another example is to take $r_G$ free bosons, where $r_G$ is the rank of some simply-laced Lie algebra (i.e., as mentioned earlier, $SU(n)$, $SO(2n)$, or $E_{6,7,8}$). Generalizing the affine $SU(2)$ construction (8.27), we let $H^i(z) = i\partial x^i(z)$ represent the Cartan subalgebra and $J^{\pm\alpha}(z) = c_\alpha : e^{\pm i\alpha\cdot x(z)}:$ represent the remaining currents, where $\alpha$ are the positive roots all normalized to $\alpha^2 = \psi^2 = 2$. $c_\alpha$ is a cocycle (Klein factor), in general necessary to give correct signs in the commutation relations (for more details see [3]). This realization of simply-laced affine algebras is known as the 'vertex operator' construction[84] (and was anticipated for the case $SU(n)$ in [85]). From (9.16) we infer the general relation

$$\widetilde{h}_G = \frac{|G|}{r_G} - 1 \qquad (9.33)$$

for simply-laced groups, and from (9.18) the central charge $c_G = r_G$ thus comes out appropriate to $r_G$ free bosons.

There is a generalization of this construction that works for any algebra at any level, but no longer involves only free fields. We begin again with $r_G$ free bosons, but now take $H^i(z) = i\sqrt{k}\,\partial x^i(z)$ to represent the Cartan currents (with the factor of $\sqrt{k}$ inserted to get the level correct). Now the exponential $:e^{\pm i\alpha \cdot x(z)/\sqrt{k}}:$ has the correct operator product with the Cartan currents, but no longer has dimension $h = 1$ in general. For the full current we write instead

$$J^{\pm\alpha}(z) = :e^{\pm i\alpha \cdot x(z)/\sqrt{k}}: \chi_\alpha(z) , \qquad (9.34)$$

where $\chi_\alpha$ is an operator of dimension $h = 1 - \alpha^2/2k$ whose operator products[86] mirror those of the exponentials so as to give overall the correct operator products (9.1). The $\chi_\alpha$'s are known as 'parafermions' and depend on $G$ and its level $k$. Since the affine algebra is constructed from $r_G$ free bosons and the parafermions, the central charge of the parafermion system is given by $c_G(k) - r_G$.

A final free example is take $|G|$ free fermions to transform in the adjoint representation of some group $G$. Then the currents (in a normalization of structure constants with highest root $\psi^2 = 2$)

$$J^a(z) = \frac{i}{2} f^{abc} \psi^b(z) \psi^c(z) \qquad (9.35)$$

give a realization of affine $G$ at level $k = \widetilde{h}_G$. The central charge comes out to be $c_G = \widetilde{h}_G |G|/(\widetilde{h}_G + \widetilde{h}_G) = \frac{1}{2}|G|$. This case of $\dim G$ free Majorana fermions in fact realizes[87][88][19] what is known as a super-affine $G$ algebra with an enveloping super Virasoro algebra. In general, a super-affine algebra has, in addition to the structure (9.1) and (9.8), a spin-3/2 super stress tensor $T_F$ satisfying (5.16) and superfield affine generators $\mathbf{J}^a = \mathcal{J}^a + \theta J^a$, whose components satisfy

$$T_F(z) J^a(w) = \frac{1/2}{(z-w)^2} \mathcal{J}^a(w) + \frac{1/2}{z-w} \partial \mathcal{J}^a(w)$$

$$T_F(z) \mathcal{J}^a(w) = \frac{1/2}{z-w} J^a(w)$$

$$J^a(z) \mathcal{J}^b(w) = \frac{i f^{abc}}{z-w} \mathcal{J}^c(w)$$

$$\mathcal{J}^a(z) \mathcal{J}^b(w) = \frac{k \delta^{ab}}{z-w} .$$

In the free fermionic representation, these operator products are satisfied (at affine level $k = \widetilde{h}_G$) by the super stress tensor $T_F = -\frac{1}{12\sqrt{C_A/2}} f^{abc} \psi^a \psi^b \psi^c$, and superpartners $\mathcal{J}^a = i\sqrt{k}\psi^a$ of the affine currents (9.35).

A modular invariant super-affine theory on the torus can be constructed by taking left and right fermions $\psi^a$ and $\overline{\psi}^a$ and summing over the same spin structure for all the fermions (GSO projecting on $(-1)^{F_L + F_R} = +1$ states). At $c = 3/2$, for example, three free fermions $\psi^i$ taken to transform as the adjoint of $SU(2)$ (vector of $SO(3)$) can be used to represent an $N = 1$ superconformal algebra with a super-affine $SU(2)$ symmetry at level $k = 2$. The supersymmetry generator is given by $T_F = -\frac{1}{12}\epsilon_{ijk}\psi^i\psi^j\psi^k = -\frac{1}{2}\psi^1\psi^2\psi^3$, and similarly for $\overline{T}_F$. (For an early discussion of supersymmetric systems realized by three fermions, see [89].) The sum over fully coupled spin structures gives a theory that manifests the full super-affine $SU(2)^2$ symmetry. It has partition function

$$\frac{1}{2}\left( {}_{A^3 \overline{A}^3}^{A^3 \overline{A}^3}\square + {}_{A^3 \overline{A}^3}^{P^3 \overline{P}^3}\square + {}_{P^3 \overline{P}^3}^{A^3 \overline{A}^3}\square + {}_{P^3 \overline{P}^3}^{P^3 \overline{P}^3}\square \right)$$

$$= \frac{1}{2}\left( \left|\frac{\vartheta_3}{\eta}\right|^3 + \left|\frac{\vartheta_4}{\eta}\right|^3 + \left|\frac{\vartheta_2}{\eta}\right|^3 \right) \qquad (9.36)$$

$$= \chi_{(0),2} \overline{\chi}_{(0),2} + \chi_{(1/2),2} \overline{\chi}_{(1/2),2} + \chi_{(1),2} \overline{\chi}_{(1),2} ,$$

which we have also expressed in terms of the level 2 affine $SU(2)$ characters $\chi_{(j=0,1/2,1),k=2}$. From (9.25), we see that the associated primary fields have conformal weights $h = j(j+1)/(2+2) = 0, \frac{3}{16}, \frac{1}{2}$. The characters themselves may be calculated just as the $c = \frac{1}{2}$ characters of (7.16a), with the result

$$\chi_{(0),2} = \frac{1}{2}\left(A^3\underset{A^3}{\Box} + P^3\underset{A^3}{\Box}\right) = \frac{1}{2}\left(\left(\frac{\vartheta_3}{\eta}\right)^{3/2} + \left(\frac{\vartheta_4}{\eta}\right)^{3/2}\right)$$

$$\chi_{(1),2} = \frac{1}{2}\left(A^3\underset{A^3}{\Box} - P^3\underset{A^3}{\Box}\right) = \frac{1}{2}\left(\left(\frac{\vartheta_3}{\eta}\right)^{3/2} - \left(\frac{\vartheta_4}{\eta}\right)^{3/2}\right) \quad (9.37)$$

$$\chi_{(1/2),2} = \frac{1}{\sqrt{2}}\left(A^3\underset{P^3}{\Box} \pm P^3\underset{P^3}{\Box}\right) = \frac{1}{\sqrt{2}}\left(\frac{\vartheta_2}{\eta}\right)^{3/2} ,$$

We also point out that we can bosonize two of the fermions of this construction, say $\psi^1$ and $\psi^2$, so that $J^3 = i\partial x$. Then the remaining fermion can be regarded as an $SU(2)$ level 2 parafermion, providing the simplest non-trivial example of the general parafermionic construction (9.34).

For the free fermion constructions (9.31) and (9.35) of affine currents, we noted that the central charge came out equal to a contribution of $c = \frac{1}{2}$ from each real fermion. This was not necessarily guaranteed, since we were considering theories defined not by a free stress-energy tensor, $T = \frac{1}{2}\sum_i \psi^i \partial \psi^i$, but rather by the stress-energy tensor $T$ of (9.7), which is quadratic in the $J$'s and thus looks quadrilinear in the fermions. The conditions under which the seemingly interacting stress tensor of (9.7) turns out to be equivalent to a free fermion stress tensor were determined in [87]. If we take fermions in (9.31) to transform as some representation (not necessarily irreducible) of $G$, then the result is that the Sugawara stress tensor is equivalent to that for free fermions if and only if there exists a group $G' \supset G$ such that $G'/G$ is a symmetric space whose tangent space generators transform under $G$ in the same way as the fermions. (This was shown in [87] by a careful evaluation of the normal ordering prescription in the definition (9.7), finding that it reduces to a free fermion form if and only if a quadratic condition on the representation matrices $t^a$ of (9.31) is satisfied. The condition turns out to be equivalent to the Bianchi identity for the Riemann tensor of $G'/G$ when the $t^a$'s are in the representation of the tangent space generators.) The three free fermion examples considered earlier here correspond to the symmetric spaces $S^N = SO(N+1)/SO(N)$, where the tangent space transforms as the $N$ of $SO(N)$; $CP^N = SU(N+1)/U(N)$, where the tangent space transforms as the $N$ of $U(N)$; and $G \times G/G$, where the tangent space transforms as the adjoint of $G$. Later we will encounter some other interesting examples of symmetric spaces.

### 9.5. Coset construction

The question that naturally suggests itself at this point is whether the enveloping Virasoro algebras associated to affine algebras are also related to any of the other representations of the Virasoro algebra discussed here. In particular we wish to focus on the $c < 1$ discrete series of unitary Virasoro representations. First of all for $SU(2)$ we see from (9.18) that

$$c_{SU(2)} = \frac{3k}{k+2} \quad (9.38)$$

satisfies $1 \leq c_{SU(2)} \leq 3$ as $k$ ranges from 1 to $\infty$, so there is no possibility to get $c < 1$. From the expression (9.16), we can easily show furthermore for any group that

$$\text{rank}\,G \leq c_G \leq \dim G ,$$

so $c < 1$ is never obtainable directly via the Sugawara stress-tensor (9.11) of an affine algebra. (The lower bound in the above, $c_G = \text{rank}\,G$, is saturated identically by simply-laced groups $G$ at level $k = 1$, i.e. identically the case allowing the vertex operator construction of an affine algebra in terms of $r_G$ free bosons.)

To increase in an interesting way the range of central charge accessible by affine algebra constructions, we need somehow to break up the stress-tensor (9.11) into pieces each with smaller central charge. This is easily implemented by means of a subgroup $H \subset G$. We denote the $G$ currents by $J_G^a$, and the $H$ currents by $J_H^i$, where $i$ runs only over the adjoint representation of $H$, i.e. from 1 to $|H| \equiv \dim H$. We can now construct two stress-energy tensors (for the remainder we shall take all structure constants to be normalized to $\psi^2 = 2$)

$$T_G(z) = \frac{1/2}{k_G + \widetilde{h}_G}\sum_{a=1}^{|G|} :J_G^a(z)J_G^a(z): , \quad (9.39a)$$

and also

$$T_H(z) = \frac{1/2}{k_H + \widetilde{h}_H} \sum_{i=1}^{|H|} :J_H^i(z)J_H^i(z): \ . \qquad (9.39b)$$

Now from (9.8) we have that

$$T_G(z)\, J_H^i(w) \sim \frac{J_H^i(w)}{(z-w)^2} + \frac{\partial J_H^i(w)}{z-w} \ ,$$

but as well that

$$T_H(z)\, J_H^i(w) \sim \frac{J_H^i(w)}{(z-w)^2} + \frac{\partial J_H^i(w)}{z-w} \ .$$

We see that the operator product of $(T_G - T_H)$ with $J_H^i$ is non-singular. Since $T_H$ above is constructed entirely from $H$-currents $J_H^i$, it also follows that $T_{G/H} \equiv T_G - T_H$ has a non-singular operator product with all of $T_H$. This means that

$$T_G = (T_G - T_H) + T_H \equiv T_{G/H} + T_H \qquad (9.40)$$

gives an orthogonal decomposition of the Virasoro algebra generated by $T_G$ into two mutually commuting Virasoro subalgebras, $[T_{G/H}, T_H] = 0$.

To compute the central charge of the Virasoro subalgebra generated by $T_{G/H}$, we note that the most singular part of the operator expansion of two $T_G$'s decomposes as

$$T_G T_G \sim \frac{\tfrac{1}{2}c_G}{(z-w)^4} \sim T_{G/H}T_{G/H} + T_H T_H \sim \frac{\tfrac{1}{2}c_{G/H} + \tfrac{1}{2}c_H}{(z-w)^4} \ .$$

The result is[19][90]

$$c_{G/H} = c_G - c_H = \frac{k_G |G|}{k_G + \widetilde{h}_G} - \frac{k_H |H|}{k_H + \widetilde{h}_H} \ ,$$

and we see that a central charge less than the rank of $G$ may be obtained. (Early examples of related algebraic structures may be found in [91].) Further insight into $G/H$ models is provided by their realization as Wess-Zumino-Witten models (9.21) with the $H$ currents coupled to a gauge field[73][92]; their correlation functions are moreover computable in terms of those of WZW models.

If it turns out that $c_{G/H} = 0$, then the argument of subsection 3.5 shows that $T_{G/H}$ must act trivially on any highest weight representation. From (9.40) there follows[3] the quantum equivalence $T_G = T_H$ between two superficially very different stress-energy tensors. Classifications of embeddings which generate $c_{G/H} = 0$, known as 'conformal embeddings', are considered in [93]. A particularly simple example is provided by a group divided by its Cartan subgroup, $G/U(1)^{r_G}$. If $G$ is simply-laced, then we saw from (9.33) that its affine algebra realized at level 1 has $c_G = r_G$. This means that $T_G$ in this case is equivalent to $T_{U(1)^{r_G}}$, i.e. to the stress-energy tensor for $r_G$ free bosons, motivating the vertex operator construction. For $G$ not simply-laced or at level $k \geq 1$, $T_{G/U(1)^{r_G}}$ is the (non-trivial) stress-energy tensor of level-$k$ $G$ parafermions.

Now we turn to the specific case of coset spaces of the form $G \times G/G$, where the group $G$ in the denominator is the diagonal subgroup. If we call the generators of the two groups in the numerator $J_{(1)}^a$ and $J_{(2)}^a$, the generators of the denominator are $J^a = J_{(1)}^a + J_{(2)}^a$. The most singular part of their operator product expansion is

$$J^a(z)J^b(w) \sim J_{(1)}^a(z)J_{(1)}^b(w) + J_{(2)}^a(z)J_{(2)}^b(w) \sim \frac{(k_1+k_2)\delta^{ab}}{(z-w)^2} + \cdots \ ,$$

so that the level of the $G$ in the denominator is determined by the diagonal embedding to be $k = k_1 + k_2$.

A simple example of this type is provided by

$$G/H = SU(2)_k \times SU(2)_1 \big/ SU(2)_{k+1} \ ,$$

in which case

$$c_{G/H} = \frac{3k}{k+2} + 1 - \frac{3(k+1)}{(k+1)+2} = 1 - \frac{6}{(k+2)(k+3)} \ . \qquad (9.41)$$

We recognize these as precisely the values of the $c < 1$ discrete series (4.6a) where $m = k+2 = 3, 4, 5, \ldots$. Using the known unitarity [94] of the representations of affine $SU(2)$, this construction allowed the authors of [19] to deduce the existence of unitary representations for all the discrete values of $c$ and $h$ allowed for $c < 1$ by the analysis of the Kac determinant formula (4.5). (Unitary coset constructions for which $c_{G/H} < 1$ must of course always coincide with some member of the unitary discrete series (4.6a).)

Another example is to take $G/H = SU(2)_k \times SU(2)_2/SU(2)_{k+2}$, giving instead

$$c_{G/H} = \frac{3k}{k+2} + \frac{3}{2} - \frac{3(k+2)}{(k+2)+2} = \frac{3}{2}\left(1 - \frac{8}{(k+2)(k+4)}\right) . \quad (9.42)$$

These values of the central charge coincide with those of the $N=1$ superconformal discrete series (5.19), with $m = k + 2 = 3, 4, 5, \ldots$. Again this shows[19] that unitary representations of the superconformal algebra (5.16) indeed exist at all these values of $c$. More generally, the coset construction $G/H = SU(2)_k \times SU(2)_\ell/SU(2)_{k+\ell}$ gives other discrete series associated to more extended chiral algebras[95]. Algebras of this form have been considered for a bewildering variety of groups and levels. Their unitary representation theory is discussed in [96].

To understand better the states that arise in the $G/H$ theory, we need to consider how the representations of $G$ decompose under (9.40). We denote the representation space of affine $G$ at level $k_G$ by $|c_G, \lambda_G\rangle$, where $c_G$ is the central charge appropriate to $k_G$, and $\lambda_G$ is the highest weight of the vacuum representation. (For a coset space of the form $G \times G/G$, for example, we would write $k_G \to (k_{(1)}, k_{(2)})$, and $\lambda_G \to (\lambda_{(1)}, \lambda_{(2)})$, where 1,2 denote the two groups in the numerator.) Under the orthogonal decomposition of the Virasoro algebra $T_G = T_{G/H} + T_H$, this space must decompose as some direct sum of irreducible representations,

$$|c_G, \lambda_G\rangle = \oplus_j |c_{G/H}, h^j_{G/H}\rangle \otimes |c_H, \lambda^j_H\rangle , \quad (9.43)$$

where $|c_{G/H}, h^i_{G/H}\rangle$ denotes an irreducible representation of $T_{G/H}$ with lowest $L_0$ eigenvalue $h^i_{G/H}$.

For the case $G/H = SU(2)_k \times SU(2)_1/SU(2)_{k+1}$ mentioned above, (9.43) takes the explicit form[19]

$$(j)_k \times (\epsilon)_1 = \oplus_q \left(h^{(c)}_{p,q}\right) \otimes \left(\frac{1}{2}[q-1]\right)_{k+1} ,$$

where $c$ is given by (9.41), $p = 2j + 1$ ($1 \leq p \leq k+1$), and the sum is over $1 \leq q \leq k+2$ with $p - q$ even (odd) for $\epsilon = 0$ ($\frac{1}{2}$). We are thus able to obtain via the coset construction all representations (4.6b) of the Virasoro algebra at the values of $c$ in (4.6a) (with $m = k + 2$). For the first non-trivial case $k = 1$, for example, the coset construction $SU(2)_1 \times SU(2)_1/SU(2)_2$ has $c = \frac{1}{2}$. The products of $SU(2)_1$ representations decompose as

$$(0)_1 \times (0)_1 = (h^{(1/2)}_{1,1})(0)_2 \oplus (h^{(1/2)}_{2,1})(1)_2$$
$$(0)_1 \times (\tfrac{1}{2})_1 = (h^{(1/2)}_{1,2})(\tfrac{1}{2})_2$$
$$(\tfrac{1}{2})_1 \times (\tfrac{1}{2})_1 = (h^{(1/2)}_{2,1})(0)_2 \oplus (h^{(1/2)}_{1,1})(1)_2 .$$

The three allowed Virasoro representations, with conformal weights $h^{(1/2)}_{p,q} = 0, \frac{1}{16}, \frac{1}{2}$, all appear in the decompositions consistent with the affine $SU(2)$ conformal weights $h_{(0),k} = 0$, $h_{(1/2),1} = \frac{1}{4}$, $h_{(1),2} = \frac{1}{2}$, $h_{(1/2),2} = \frac{3}{16}$, and the integer spacing of the levels.

As a final example, we consider $G/H = SO(N)_1 \times SO(N)_1/SO(N)_2$, with central charge

$$c_{G/H} = \frac{N}{2} + \frac{N}{2} - \frac{2\frac{1}{2}N(N-1)}{2+(N-2)} = N - (N-1) = 1 .$$

This case turns out to be related to specific points $r = \sqrt{N}/2$ on the $c = 1$ circle and orbifold lines discussed in section 8. The holomorphic weights that enter into the circle line partition function (8.6) at this radius are

$$h(m, n) = \frac{1}{2}\left(\frac{m}{2(\sqrt{N}/2)} + n\frac{\sqrt{N}}{2}\right)^2 = \frac{1}{8N}(2m + nN)^2 . \quad (9.44)$$

To give a flavor for how to analyze these constructions more generally, we compare some of the weights inferred from (9.43) with these $h$ values.* For $SO(N)$, the representations allowed at level 1 are the adjoint, vector, and spinor(s). The representations allowed at level 2 include all of these together with other representations present in the decompositions of their direct products. We will concentrate here only on the rank $r$ antisymmetric tensor representations, denoted $[r]$, which appear in the product of two spinors. From (9.14) and (9.15), we find $C_v/\psi^2 = \frac{1}{2}(N-1)$, $C_s/\psi^2 = N(N-1)/16$, and $C_{[r]}/\psi^2 = \frac{1}{2}r(N-r)$. (9.24b) gives

$$h_{v,1} = \frac{1}{2} \qquad h_{s,1} = \frac{N}{16}$$
$$h_{v,2} = \frac{N-1}{2N} \qquad h_{s,2} = \frac{N-1}{16} \qquad h_{[r],2} = \frac{r(N-r)}{2N} ,$$

and of course $h_{(0),k} = 0$. The values of $h_{G/H}^j$ obtainable from (9.43) may be determined by picking specific representations $\lambda_G$ and $\lambda_H$ at the appropriate levels and taking the difference of their conformal weights. In the case under consideration, $\lambda_G$ is specified by two $SO(N)_1$ representations, and $\lambda_H$ by any $SO(N)_2$ representation allowed in their product. Using $v \times 1 = v$, for example, gives the coset conformal weight $h_{v,1} - h_{v,2} = 1/(2N) = h(\pm 1, 0)$. From $s \times s \supset [r] + \ldots$, we calculate $2h_{s,1} - h_{[r],2} = (2r-N)^2/(8N) = h(r,-1)$, giving a variety of the weights of (9.44). $s \times 1 = s$, on the other hand, gives $2h_{s,1} - h_{s,2} = \frac{1}{16}$, the dimension of the twist field in the $S^1/\mathbf{Z}_2$ orbifold model. In fact, taking appropriate modular invariant combinations of $SO(N)_1 \times SO(N)_1/SO(N)_2$ characters, we can realize either the circle or orbifold partition functions at $r = \sqrt{N}/2$. These partition functions are thereby organized into characters of the extended algebras that exist at these points.

---

* I thank L. Dixon for his notes on the subject.

*9.6. Modular invariant combinations of characters*

We now turn to discuss the decomposition of affine algebra representations with respect to the coset space decomposition (9.40) of the stress-energy tensor. To this end, we begin by introducing more formally the notion of a character of a representation of an affine algebra, analogous to that considered earlier for the Virasoro algebra. In the case of affine $SU(2)$ for example, if we consider the level $k$ representation built on the spin-$j$ vacuum state $|(j)\rangle$, then the trace

$$\chi^k_{(j)}(\theta, \tau) \equiv q^{-c_{SU(2)}/24} \, \mathrm{tr}_{(j),k} \, q^{L_0} \, e^{i\theta J_0^3} \qquad (9.45)$$

characterizes the number of states at any given level (as explained before (7.8)). The group structure also allows us to probe additional information, namely the $J_0^3$ eigenvalues, by means of the parameter $\theta$. In (9.27), we have given the explicit forms for the $k = 1$ characters $\chi^{k=1}_{(j=0,1/2)}(0,\tau)$ and in (9.37) for the $k = 2$ characters $\chi^{k=2}_{(j=0,1/2,1)}(0,\tau)$.

The generalization to arbitrary group $G$, at level $k$ and vacuum representation with highest weight $\lambda$, is given by

$$\chi^k_{(\lambda)}(\theta^i, \tau) = q^{-c_G/24} \, \mathrm{tr}_{(\lambda),k} \, q^{L_0} \, e^{i\theta^i H_0^i} . \qquad (9.46)$$

(9.46) should be recognized as the natural generalization of ordinary character formulae except with the Cartan subalgebra, i.e. the maximal set of commuting generators $H_0^i$, extended to include $L_0$ as well. For cases realizable in terms of free bosons or fermions, the characters take simple forms as in (9.27) and (9.37). In other cases, they can be built up from bosonic and parafermionic characters (see e.g. [86]). In general there exists a closed expression for these characters (see e.g. [97][83]), known as the Weyl-Kac formula, which generalizes the Weyl formula for the characters of ordinary Lie groups.

It follows immediately from the decomposition (9.43) that the character of an affine $G$ representation with highest weight $\lambda^a$ satisfies

$$\chi^{k_G}_{\lambda^a_G}(\theta^i, \tau) = \sum_j \chi^{c_{G/H}}_{h_{G/H}(\lambda^a_G, \lambda^j_H)}(\tau) \, \chi^{k_H}_{\lambda^j_H}(\theta^i, \tau) \equiv \chi_{G/H} \cdot \chi^H_{\lambda_H} \qquad (9.47)$$

(where the $\theta^i$'s are understood restricted to the Cartan subalgebra of $H$). In (9.47) the $L_0$ eigenvalues $h_{G/H}$ characterizing the $T_{G/H}$ Virasoro representations depend implicitly on the highest weights $\lambda^a_G$ and $\lambda^j_H$ characterizing the associated $G$ and $H$ affine representations. On the right hand side of (9.47) we have introduced a matrix notation (see for example [98]) in which the $G$ and $H$ characters, $\chi^{k_G}_{\lambda^a_G}$ and $\chi^{k_H}_{\lambda^j_H}$, are considered vectors labelled by $a$ and $j$ respectively, and $\chi_{G/H}$ is considered a matrix in $a, j$ space.

Under modular transformations

$$\gamma: \tau \to \frac{a\tau + b}{c\tau + d} ,$$

the characters allowed at any given fixed level $k_G$ of an affine algebra transform as a unitary representation

$$\chi^{k_G}(\tau') = M^{k_G}(\gamma)\,\chi^{k_G}(\tau), \qquad (9.48)$$

with $(M^{k_G})_a{}^b$ a unitary matrix (see e.g. [97][83]). But from (9.47) we also have

$$\chi^{k_G}(\tau') = \chi_{G/H}(\tau')\,M^{k_H}(\gamma)\,\chi^{k_H}(\tau)\ .$$

Linear independence of the $G$ and $H$ characters then allows us to solve for the modular transformation properties of the $T_{G/H}$ characters, as

$$\chi_{G/H}(\tau') = M^{k_G}(\gamma)\,\chi_{G/H}(\tau)\,M^{k_H}(\gamma)^{-1}\ . \qquad (9.49)$$

For example for $SU(2)$ level $k$ characters, the modular transformation matrices for $\gamma = S : \tau \to -1/\tau$ are

$$S^{(k)}_{jj'} = \left(\frac{2}{k+2}\right)^{1/2} \sin\frac{\pi(2j+1)(2j'+1)}{k+2}\ , \qquad (9.50a)$$

with $j, j' = 0, \ldots, \frac{k}{2}$ (and we use the notation $S \equiv M(\gamma : \tau \to -1/\tau)$). In particular for $k=1$, this gives

$$S^{(1)} = \frac{1}{\sqrt{2}}\begin{pmatrix} 1 & 1 \\ 1 & -1 \end{pmatrix}\ . \qquad (9.50b)$$

Using these results, we can derive the modular transformation properties of the characters $\chi_{p,r}(q)$ for the $c<1$ discrete series. These characters were derived in [99] by careful analysis of null states, but we will never need their explicit form here. (The characters can also be derived as solutions of differential equations induced by inserting null vectors, a method that generalizes as well to higher genus[100].) The matrix $S$ for the transformation $\chi_{p,r}\bigl(q(-1/\tau)\bigr) = \sum_{p',r'} S^{p'r'}_{pr}\chi_{p',r'}(q)$ is determined by substituting (9.50a,b) in (9.49). The result is

$$S^{p'r'}_{pr} = \left(\frac{8}{m(m+1)}\right)^{1/2} (-1)^{(p+r)(p'+r')} \sin\frac{\pi pp'}{m} \sin\frac{\pi rr'}{m+1}\ , \qquad (9.51)$$

where $m = k+2$ (see eq. (4.27) of Cardy's lectures, also [43][54][101]).

(9.49) allows us to use known modular invariant combinations of $G$ and $H$ characters to construct modular invariant combinations of $T_{G/H}$ characters. For example the fact that $M^{k_G}$ is unitary (i.e. that $\chi^{G\dagger}\chi^G$ is modular invariant), and similarly for $M^{k_H}$, implies that $\operatorname{tr}\chi^\dagger_{G/H}\chi_{G/H}$ is modular invariant. More generally given any two modular invariants for $G$ and $H$ characters at levels $k_G$ and $k_H$,

$$\chi^{k_G\dagger} I^{k_G}_G \chi^{k_G} = \chi^{k_G\dagger}_\lambda I^{k_G}_{\lambda\lambda'} \chi^{k_G}_{\lambda'} \quad\text{and}\quad \chi^{k_H\dagger} I^{k_H}_H \chi^{k_H} = \chi^{k_H\dagger}_\lambda I^{k_H}_{\lambda\lambda'} \chi^{k_H}_{\lambda'}\ ,$$

we see that the combination

$$\operatorname{tr} I^{k_H\dagger}_H \chi^\dagger_{G/H}(\tau) I^{k_G}_G \chi_{G/H}(\tau) \qquad (9.52)$$

is a modular invariant combination of $G/H$ characters.

### 9.7. The A-D-E classification of $SU(2)$ invariants

It follows from (9.41) and (9.52) that modular invariants for $SU(2)$ at levels $1$, $k$, and $k+1$ can be used to construct modular invariants for the $(m=k+2)^{\text{th}}$ member of the $c<1$ discrete series. Arguments of [102] also combine to show that all such modular invariants can be so constructed. Thus the challenge of constructing all possible modular invariant combinations of the characters of a particular member of the $c<1$ discrete series, originally posed in [43], is reduced to the classification of modular invariant combinations of $SU(2)$ characters for arbitrary level $k$. For physical applications, we are specifically interested in modular invariant combinations that take the form of partition functions all of whose states have positive integer multiplicities.

The problem of finding all such affine $SU(2)$ invariants was solved in [103] and is discussed further in Zuber's lectures. The result is that the $SU(2)$ modular invariants are classified by the same $ADE$ series that classifies the simply-laced Lie algebras. The invariant associated to a given $G = A, D, E$ occurs for affine $SU(2)$ at level $k = \widetilde{h}_G - 2$. The invariant associated to $A_{\ell-1} = SU(\ell)$, for example, is just the diagonal $SU(2)$ invariant at level $k = \ell - 2$. The modular invariant combinations of $c < 1$ characters for the $(m = k+2)^{\text{th}}$ member of the unitary discrete series are given by pairs

$$(G, G') \qquad (9.53)$$

with Coxeter numbers $m$ and $m + 1$. Using the coset construction (9.42), modular invariant combinations of the characters of the $N = 1$ superconformal discrete series (5.19) have been similarly classified[104].

Although it is not immediately obvious why there should be a relation between affine $SU(2)$ invariants and the $ADE$ classification of simply-laced Lie algebras, some insight is given by an argument of [105]. First we recall that an embedding $H \subset G$ induces a realization of affine $H$ at some integer multiple of the level of affine $G$. One way of seeing this is to recall that the level satisfies $k = 2\widetilde{k}/\psi^2$, so the level of $H$ will be related to the level of $G$ by the ratio of highest roots $\psi_G^2/\psi_H^2$ induced by the embedding. This integer is known as the index of embedding. It can also be calculated by working in a fixed normalization, and comparing the $\ell$ of (9.13) for a given representation of $G$ with that for its decomposition into $H$ representations. For example consider the embedding $G \subset SO(d_G)$, $d_G = \dim G$, defined such that the vector of $SO(d_G)$ decomposes to the adjoint representation of $G$. From (9.17), $\ell_{(d_G)}/\psi^2 = 1$ for the vector representation of $SO(d_G)$, whereas $\ell_A/\psi^2 = \widetilde{h}_G$ for the adjoint representation of $G$. The index of the embedding is the ratio $\ell_A/\ell_{d_G} = \widetilde{h}_G$, and the embedding $G \subset SO(d_G)$ thus induces a level $k\widetilde{h}_G$ representation of affine $G$ from a level $k$ representation of affine $SO(d_G)$.

For any subgroup $H \subset G$ of index 1, $H \subset SO(d_G)$ is also index $\widetilde{h}_G$. This means that

$$\sum_{r_i} \ell_{r_i}/\psi^2 = \frac{1}{\psi^2} \sum_{r_i,j} \frac{\mu_{(j),r_i}^2}{r_H} = \widetilde{h}_G \ ,$$

where the sum is over the weights of all representations $r_i$ of $H$ in the decomposition of the vector of $SO(d_G)$. Now consider the coset space $G/H$, of dimension $d_{G/H} = \dim G - \dim H$. With the canonical $H$-invariant metric and torsion-free connection, this space has holonomy group $H$ so there is a natural embedding $H \subset SO(d_{G/H})$ in the tangent space group. The $H$ representations in the decomposition of the vector of $SO(d_{G/H})$ are the same as for the vector of $SO(d_G)$, except for the removal of one occurrence of the adjoint representation of $H$. It is easy to calculate the index of the embedding $H \subset SO(d_{G/H})$ in the case that $H$ is simply-laced, for which from (9.33) we have $\sum_{(\text{adj } H)} \mu^2/r_H = \widetilde{h}_H \psi^2$. Removing a single adjoint representation of $H$ from the equation above, we find

$$\sideset{}{'}\sum_{r_i} \ell_{r_i}/\psi^2 = \frac{1}{\psi^2}\left(\sum_{r_i,j} \frac{\mu_{(j),r_i}^2}{r_H} - \sum_j \frac{\mu_{(j),\text{adj}}^2}{r_H}\right) = \widetilde{h}_G - \widetilde{h}_H \ , \qquad (9.54)$$

and the index of $H \subset SO(d_{G/H})$ is $\widetilde{h}_G - \widetilde{h}_H$.

Now recall that every simply-laced algebra $G$ has a distinguished $SU(2)$ subalgebra (9.20a), generated by its highest root $\psi$. (We sloppily use $G$ to refer both to the Lie group and to its algebra.) If we take $H = SU(2) \times K$, where $K$ is the maximal commuting subalgebra, then $G/H$ is a symmetric space. Consider a level 1 representation of affine $SO(d_{G/H})$ given by free fermions in the vector representation as in (9.31). This vector representation transforms under $H \subset SO(d_{G/H})$ exactly as do the tangent space generators of $G/H$ under $H \subset G$. This is the symmetric space condition[87] cited at the end of subsection 9.4, for which $c_H = c_{SO(d_{G/H})}$, and for which the Virasoro algebras based on the two affine algebras coincide. (There are actually two steps here: first $T_H$ is equivalent to the stress-energy tensor for $d_{G/H}$ free fermions, second that the latter is equivalent to $T_{SO(d_{G/H})_1}$.)

As an example, we consider the case $G = E_8$, for which $H = SU(2) \times E_7$ and $d_{G/H} = 248 - 3 - 133 = 112$. From (9.32), the level 1 representation of $SO(112)$ has $c_{SO(112),1} = 56$. From (9.54), we find that the indices of the embeddings of $SU(2)$ and $E_7$ in $SO(112)$ are $\widetilde{h}_{E_8} - \widetilde{h}_{SU(2)} = 30 - 2 = 28$ and $\widetilde{h}_{E_8} - \widetilde{h}_{E_7} = 30 - 18 = 12$. It follows from (9.18) that

$$c_{SU(2),28} + c_{E_7,12} = \frac{28 \cdot 3}{28 + 2} + \frac{12 \cdot 133}{18 + 12} = 56 \ .$$

The diagonal modular invariant for $SO(d_{G/H})_1$ characters thus decomposes into a modular invariant combination of $SU(2)_{\widetilde{h}_G - \widetilde{h}_{SU(2)}} \times K_{\widetilde{h}_G - \widetilde{h}_K}$ characters.

This combination always contains a piece proportional to the diagonal invariant for the $K_{\widetilde{h}_G - \widetilde{h}_K}$ characters, whose coefficient is necessarily an $SU(2)$ invariant at level $\widetilde{h}_G - \widetilde{h}_{SU(2)} = \widetilde{h}_G - 2$. It turns out[105] that this induced invariant is identically the one labeled by the simply-laced algebra $G = A, D, E$ in the classification of [103]. It thus becomes natural that there should be an $SU(2)$ invariant at level $k = \widetilde{h}_G - 2$ associated to each of the $G = A, D, E$ algebras: each has a canonical $SU(2)$ generated by its highest root and the above construction associates to it a particular affine invariant at the required level. It is not yet obvious from this point of view, however, why *all* the invariants should be generated this way (unless the construction could somehow always be run backwards to start from an invariant to reconstruct an appropriate symmetric space). A similar construction has been investigated further in [73][106] to give realizations of the $c < 1$ unitary series directly in terms of free fermions.

We mentioned before (9.53) that the $A$ series corresponds to the diagonal invariants. The first non-diagonal case is the $D_4 = SO(8)$ invariant that occurs at $SU(2)$ level $\widetilde{h}_{SO(8)} - 2 = 4$. It is given by

$$|\chi_{(0),4} + \chi_{(2),4}|^2 + 2|\chi_{(1),4}|^2 \; , \qquad (9.55)$$

and involves only integer spin ($SO(3)$) representations. According to the discussion surrounding (9.53), there are thus two possible modular invariants for the $(m = 5)^{\text{th}}$ member of the $c < 1$ discrete series: $(A_5, A_4)$ and $(D_4, A_4)$. From (4.6a,b), $m = 5$ gives $c = 4/5$ and characters that we label $\chi_a$, $a = 0, 2/5, 1/40, 7/5, 21/40, 1/15, 3, 13/8, 2/3, 1/8$. The $(A_5, A_4)$ invariant is just the diagonal sum $\sum_a \chi_a \overline{\chi}_a$, and gives the critical partition function on the torus for the fifth member of the RSOS series of [20] (described in subsection 4.4). From (9.52) and (9.55), we calculate the $(D_4, A_4)$ invariant

$$|\chi_0 + \chi_3|^2 + |\chi_{2/5} + \chi_{7/5}|^2 + 2|\chi_{1/15}|^2 + 2|\chi_{2/3}|^2 \; , \qquad (9.56)$$

identified in [43] as the critical partition function for the 3-state Potts model on the torus.

In general the RSOS models of [20] at criticality on the torus are described by the diagonal invariants $(A_m, A_{m-1})$. The restriction on the heights in these models can be regarded as coded in the Dynkin diagram of $A_m$, with the nodes specifying the height values and linked nodes representing pairs of heights allowed at nearest neighbor lattice points. Generalized versions[22] of these models, defined in terms of height variables that live on the Dynkin diagrams of any of the $ADE$ algebras, turn out to have critical points whose partition functions realize the remaining invariants.

In the extended chiral algebra game, we encounter a variety of coincidences. For example, one can easily check from (9.18) that the central charge $c = 2(k-1)/(k+2)$ for $SU(2)_k/U(1)$ coincides with that for $SU(k)_1 \times SU(k)_1/SU(k)_2$. One can also check that $(E_8)_1 \times (E_8)_1/(E_8)_2$ and $SO(n)_1/SO(n-1)_1$ each have $c_{G/H} = 1/2$, giving alternative realizations of the critical Ising model. Another coincidence that we omitted to mention is that the $N = 2$ superconformal discrete series (9.38) and the $SU(2)$ level $k$ series (5.20) coincide (with $m = k + 2$). This is more or less explained by the construction of [107], in which the $N = 2$ superconformal algebra is realized in terms of $SU(2)$ level $k$ parafermions and a single free boson (at a radius different from what would be used to construct level $k$ $SU(2)$ currents).

In the present context, we note that the partition function (9.56), which looks off-diagonal in terms of Virasoro characters, is actually diagonal in terms of a larger algebra, the spin-3 $W$ algebra of [108]. This algebra can also be realized as the coset algebra $SU(3)_1 \times SU(3)_1/SU(3)_2$ (from (9.18), we find central charge $c = 2 + 2 - 16/5 = 4/5$), the diagonal combination of whose characters turns out to coincide with (9.56). (By the comments of the preceding paragraph, there is also a relation to $SU(2)_3/U(1)$, i.e. to $SU(2)$ level 3 parafermions.) The spin-3 $W$ algebra is generated by the stress-energy tensor $T$ together with the operator $\phi_{4,1}$, with $h_{4,1} = 3$ (see fig. 7). These two operators transform in a single representation of the chiral algebra, so that the identity character with respect to this larger algebra is $\chi'_0 = \chi_0 + \chi_3$. The fields with $h_{3,1} = 7/5$ and $h_{3,5} = 2/5$ also transform as a single representation. This is a special case of a general phenomenon[12][109] (see also [110]): modular invariant partition functions of rational conformal field theories (mentioned briefly in subsection (5.3)), when expressed in terms of characters $\chi_i$ of the largest chiral algebra present, are either diagonal, $\sum \chi_i \overline{\chi}_i$, or of the form $\sum \chi_i P_{ij} \overline{\chi}_j$, where $P$ is a permutation of the chiral characters that preserves the fusion rules.

*9.8. Modular transformations and fusion rules*

We close our treatment of coset theories with a discussion of some other information that can be extracted from the modular transformation properties of the characters. To place the discussion in a more general context, we first point out that the modular transformation matrix $M(\gamma)$ of (9.48) generalizes to other rational conformal field theories. Recall that for these theories there are by definition a finite number of fields primary with respect to a possibly extended chiral algebra. All coset models are examples of rational conformal field theories (and, in fact, all rational conformal field theories known at this writing are expressible either as coset models or orbifolds thereof). The characters $\chi_i(q)$ are given by tracing over the Hilbert space states in the (extended) family of primary field $i$, and are acted on unitarily by the matrix $M(\gamma)$. For convenience we continue to denote the matrix $M(S)$, representing the action of $S: \tau \to -1/\tau$ on the characters, by $S_i{}^j$.

There is an extremely useful relation (conjectured in [28], proven in [30] (see also [109]), and discussed further in Dijkgraaf's seminar) between this matrix and the fusion algebra (5.15). The statement is that $S$ diagonalizes the fusion rules, i.e. $N_{ij}{}^k = \sum_n S_j{}^n \lambda_i^{(n)} S_n^{\dagger\,k}$ (where the $\lambda_i^{(n)}$'s are the eigenvalues of the matrix $N_i$). This relation can be used to solve for the (integer) $N_{ij}{}^k$'s in terms of the matrix $S$. If we use $i=0$ to specify the character for the identity family, then we have $N_{0j}{}^k = \delta_j^k$. It follows that the eigenvalues satisfy $\lambda_i^{(n)} = S_i{}^n / S_0{}^n$, so that

$$N_{ij}{}^k = \sum_n \frac{S_j{}^n S_i{}^n S_n^{\dagger\,k}}{S_0{}^n} \ . \tag{9.57}$$

We stress that it is not at all obvious a priori that there should be a relation such as (9.57) between the fusion rules and the modular transformation properties of the characters of the algebra. Applied to (9.50a), for example, we derive the fusion rules for affine $SU(2)$,

$$\phi_{j_1} \times \phi_{j_2} = \sum_{j_3 = |j_1 - j_2|}^{\min(j_1+j_2,\ k-(j_1+j_2))} \phi_{j_3} \ ,$$

in agreement with the result derived alternatively by considering the differential equations induced by null states as in [83].

We sketched a similar differential equation method before stating the fusion rules (5.14) for the $c<1$ theories. We are now in a position to see how the fusion rules for these theories can instead be inferred directly from the coset construction: the result (5.14) is easily derived directly from (9.51) by using (9.57). Since the matrix $S$ is effectively factorized into the product of $S$ matrices for $SU(2)$ at levels $k=m-2$ and $k+1=m-1$, we see that the fusion rules similarly factorize. This derivation thus explains our earlier observations concerning the resemblance of (5.14) to two sets of $SU(2)$ branching rules.

# 10. Advanced applications

Lecture 10, in which further extensions and likely directions for future progress would have been discussed, was cancelled due to weather.

[103] A. Cappelli, C. Itzykson, and J.-B. Zuber, "The A-D-E classification of minimal and $A_1^{(1)}$ conformal invariant theories," Comm. Math. Phys. 113 (1987) 1, and further references therein.

[104] A. Cappelli, "Modular invariant partition functions of superconformal theories," Phys. Lett. 185B (1987) 82;
D. Kastor, "Modular invariance in superconformal models," Nucl. Phys. B280[FS18] (1987) 304.

[105] W. Nahm, "Quantum field theories in one-dimension and two-dimensions," Duke Math. J. 54 (1987) 579;
W. Nahm, "Lie group exponents and $SU(2)$ current algebras," Comm. Math. Phys. 118 (1988) 171.

[106] D. Altschuler, K. Bardakci, and E. Rabinovici, "A construction of the $c < 1$ modular invariant partition functions," Comm. Math. Phys. 118 (1988) 241.

[107] Z. Qiu, "Nonlocal current algebra and $N = 2$ superconformal field theory in two-dimensions," Phys. Lett. 188B (1987) 207.

[108] A.B. Zamolodchikov, "Infinite additional symmetries in two-dimensional conformal quantum field theory," Theo. Math. Phys. 65 (1986) 1205.

[109] R. Dijkgraaf and E. Verlinde, "Modular invariance and the fusion algebra," Utrecht preprint ThU-88/25, to appear in the proceedings of the Annecy conference on conformal field theory (March 1988).

[110] P. Bouwknegt and W. Nahm, "Realizations of the exceptional modular invariant $A(1)$ partition functions," Phys. Lett. 184B (1987) 359.
177    178